\documentclass[sort, compress,11pt]{elsarticle}

\usepackage{amssymb}
\usepackage{amsthm}
\usepackage{amsmath}
\usepackage{mathtools}
\usepackage{mathrsfs}
\usepackage{algorithm}
\usepackage[algo2e]{algorithm2e}
\usepackage{algpseudocode}
\usepackage{array}
\usepackage{multirow}
\usepackage{listings}
\usepackage{tabu}
\usepackage{tabularx}
\usepackage{enumerate}
\usepackage{lineno}
\usepackage{fullpage}
\usepackage{xcolor}
\usepackage[colorinlistoftodos]{todonotes}
\usepackage{bm}
\usepackage{bbm}
\usepackage[colorlinks=true]{hyperref}
\usepackage{url}
\usepackage{textcomp}
\usepackage{gensymb}
\usepackage{soul}
\usepackage{graphicx}
\usepackage{subfig}
\usepackage{float}
\usepackage{setspace}
\usepackage{booktabs}
\usepackage{tikz}
\usetikzlibrary{shapes}
\usetikzlibrary{shapes.geometric}

\definecolor{myred}{RGB}{214,39,40}
\definecolor{mygray}{RGB}{176,176,176}
\definecolor{myorange}{RGB}{255,127,14}
\definecolor{myorangeT}{RGB}{221,171,100}
\definecolor{mygreen}{RGB}{44,160,44}
\definecolor{mylightgray}{RGB}{204,204,204}
\definecolor{mypurple}{RGB}{148,103,189}
\definecolor{mybrown}{RGB}{140,86,75}
\definecolor{steelblue}{RGB}{31,119,180}

\def\newname{CoNFiLD-inlet}
\def\oldname{CoNFiLD}

\DeclareRobustCommand\dashedblackline{\raisebox{2pt}{\tikz{\draw[black,line width = 0.8pt, dashed] (0,0)--(0.5,0);}}}
\DeclareRobustCommand\blackcircle{\raisebox{0.8pt}{\tikz{\draw[draw=black,fill=none] (0,0) circle (.4ex);}}}

\DeclareRobustCommand\bluedotline{\raisebox{0.8pt}{\tikz{\draw[draw=none,fill=steelblue] (0,0) circle (.4ex);\draw[steelblue,line width = 0.8pt] (-0.15,0)--(0.15,0);}}}
\DeclareRobustCommand\reddotline{\raisebox{0.8pt}{\tikz{\draw[draw=none,fill=myred] (0,0) circle (.4ex);\draw[myred,line width = 0.8pt] (-0.15,0)--(0.15,0);}}}
\DeclareRobustCommand\orangedotline{\raisebox{0.8pt}{\tikz{\draw[draw=none,fill=myorange] (0,0) circle (.4ex);\draw[myorange,line width = 0.8pt] (-0.15,0)--(0.15,0);}}}
\DeclareRobustCommand\greendotline{\raisebox{0.8pt}{\tikz{\draw[draw=none,fill=mygreen] (0,0) circle (.4ex);\draw[mygreen,line width = 0.8pt] (-0.15,0)--(0.15,0);}}}

\DeclareRobustCommand\blueline{\raisebox{2pt}{\tikz{\draw[steelblue,line width = 0.8pt] (-0.15,0)--(0.15,0);}}}
\DeclareRobustCommand\redline{\raisebox{2pt}{\tikz{\draw[myred,line width = 0.8pt] (-0.15,0)--(0.15,0);}}}
\DeclareRobustCommand\orangeline{\raisebox{2pt}{\tikz{\draw[myorange,line width = 0.8pt] (-0.15,0)--(0.15,0);}}}
\DeclareRobustCommand\greenline{\raisebox{2pt}{\tikz{\draw[mygreen,line width = 0.8pt] (-0.15,0)--(0.15,0);}}}

\DeclareRobustCommand\greendot{\raisebox{0.8pt}{\tikz{\draw[draw=none,fill=mygreen] (0,0) circle (.4ex);}}}
\tikzset{cross/.style={cross out, draw=mypurple, minimum size=2*(#1-\pgflinewidth), line width=2pt, inner sep=1pt, outer sep=1pt}, cross/.default={1pt}
}
\DeclareRobustCommand\purplecross{\raisebox{-0.8pt}{\tikz{\draw[] (0,0) node[cross=.6ex] {}}}}
\DeclareRobustCommand\blackstar{\raisebox{0.5pt}{\tikz{ \node [star, star point ratio=2., minimum size=1em, scale=0.2, draw, fill=black] {} ;}}}
\DeclareRobustCommand\bluetri{\raisebox{0.5pt}{\tikz{\node[regular polygon, regular polygon sides=3, minimum size=1em, scale=0.3, fill=steelblue] {};}}}

\theoremstyle{definition}

\theoremstyle{remark}

\biboptions{numbers,comma,round,square}
\graphicspath{ {./figs/} }

\journal{Elsevier}

\begin{document}
\makeatletter
\def\ps@pprintTitle{%
  \let\@oddhead\@empty
  \let\@evenhead\@empty
  \let\@oddfoot\@empty
  \let\@evenfoot\@oddfoot
}

\begin{frontmatter}

\title{CoNFiLD-inlet: Synthetic Turbulence Inflow Using Generative Latent Diffusion Models with Neural Fields}



\author[ndAME]{Xin-Yang Liu\corref{contrib}}
\author[ndAME]{Meet Hemant Parikh\corref{contrib}}
\author[ndAME]{Xiantao Fan}
\author[ndAME]{Pan Du}
\author[google]{Qing Wang}
\author[google]{Yi-Fan Chen}
\author[ndAME,cornellMAE]{Jian-Xun Wang\corref{corr}}

\address[ndAME]{Department of Aerospace and Mechanical Engineering, University of Notre Dame, Notre Dame, IN, USA}
\address[google]{Google Research, Mountain View, CA, USA} 
\address[cornellMAE]{Sibley School of Mechanical and Aerospace Engineering, Cornell University, Ithaca, NY, USA}
\cortext[contrib]{XYL and MHP contributed equally}

\cortext[corr]{Corresponding author. Tel: +1 540 3156512}
\ead{jw2837@cornell.edu}

\begin{abstract}
Eddy-resolving turbulence simulations require stochastic inflow conditions that accurately replicate the complex, multi-scale structures of turbulence. Traditional recycling-based methods rely on computationally expensive precursor simulations, while existing synthetic inflow generators often fail to reproduce realistic coherent structures of turbulence. Recent advances in deep learning (DL) have opened new possibilities for inflow turbulence generation, yet many DL-based methods rely on deterministic, autoregressive frameworks prone to error accumulation, resulting in poor robustness for long-term predictions.
In this work, we present {\newname}, a novel DL-based inflow turbulence generator that integrates diffusion models with a conditional neural field (CNF)-encoded latent space to produce realistic, stochastic inflow turbulence. By parameterizing inflow conditions using Reynolds numbers, {\newname} generalizes effectively across a wide range of Reynolds numbers ($Re_\tau$ between $10^3$ and $10^4$) without requiring retraining or parameter tuning. Comprehensive validation through \emph{a priori} and \emph{a posteriori} tests in Direct Numerical Simulation (DNS) and Wall-Modeled Large Eddy Simulation (WMLES) demonstrates its high fidelity, robustness, and scalability, positioning it as an efficient and versatile solution for inflow turbulence synthesis.

\end{abstract}

\end{frontmatter}


\section{Introduction}
\label{sec:intro}

Transient turbulence simulations are indispensable for understanding and predicting the complex, unsteady behaviors of turbulent flows, which are critical in various engineering and scientific applications. These simulations provide detailed insights into turbulent dynamics under different conditions, supporting the design, optimization, and analysis of systems in aerospace, naval, and environmental engineering. The rapid advancement in computational power has significantly enhanced the feasibility and popularity of eddy-resolving computational fluid dynamics (CFD) techniques, such as Direct Numerical Simulation (DNS), Large Eddy Simulation (LES), and hybrid LES and Reynolds Averaged Navier-Stokes (RANS) models.

For these eddy-resolving simulations, providing proper inflow boundary conditions is crucial yet highly challenging~\cite{wu2017inflow, tabor2010inlet}. A time-varying flow field that accurately reflects the complex, stochastic nature of turbulence is needed to capture both random and coherent turbulent structures across relevant scales, ensuring realistic flow development downstream. Insufficiently realistic inflows often dissipate before reaching the region of interest, and it often require extensive spatial development to transition into realistic turbulence, significantly increasing computational costs.

Recycling methods can address this challenge for simple flow configurations with periodic conditions by reusing velocity data from a downstream location at the upstream boundary to avoid direct turbulence synthesis~\cite{chung1997comparative}. For non-periodic flows, rescaling is often applied to ensure the turbulence intensity and other statistical properties align with the desired inflow conditions~\cite{wu1995large,na1998direct,pierce1998method}. For more complex flows, precursor/concurrent simulations using recycling methods are often employed~\cite{lee1993direct,chung1997comparative,lund1998generation}. These simulations run either before or simultaneously with the main simulation at the same space-time resolution, generating developed turbulence serving as inflow conditions. Although these approaches are very effective and accurate due to the direct use of simulated turbulence from Navier-Stokes equations, they are computationally expensive and labor intensive. Managing and coupling multiple high-resolution simulations adds complexity to the workflow, significantly increasing the total computational cost.

In response to these challenges, an alternative strategy is to directly synthesize the stochastic turbulent inflows as the inlet boundary condition, referred to as synthetic turbulence generation methods. These methods aim to efficiently produce realistic inflow turbulence without requiring extensive precursor or concurrent simulations. The simplest approach in this category is to superimpose a desired mean velocity profile with random white noise to introduce unsteadiness and randomness~\cite{aider2006large, aider2007large, mathey2006assessment}. While straightforward, this naive method fails to capture the correct statistics and coherence structures of turbulent flows. As a result, the unphysical fluctuations prescribed at the inlet are quickly damped and fail to develop into realistic turbulence downstream. Over the past several decades, substantial efforts have been made to develop more advanced methods for generating realistic turbulent inflows that satisfy one or more desired properties of the flow being studied, such as Reynolds stress tensors, energy spectra, and phase coherence~\cite{lee1992simulation,  guo2023efficient, di2006synthetic, touber2009large, kempf2005efficient, druault2004generation}. 
One widely used technique is the digital filter method~\cite{klein2003digital, veloudis2007novel, kempf2012efficient}, which imposes spatiotemporal coherence on a random field by applying spatial and temporal filtering to reproduces some desired turbulence statistics. 
Synthetic random Fourier method~\cite{karweit1991simulation, bechara1994stochastic}, on the other hand, generates turbulence by prescribing random phase shifts to each mode of the spectral representation of the velocity field. By carefully adjusting the amplitudes and phases of these Fourier modes, this method can replicate certain statistical properties of turbulence, particularly preserving the energy spectrum of the flow. 
The synthetic eddy method~\cite{jarrin2006synthetic,  sandham2003large} takes a different approach by utilizing synthetic eddies with varying positions, sizes, and velocities to match the statistical properties of turbulence. This method involves randomly placing synthetic eddies within the inflow plane and assigning velocity perturbations based on the desired turbulence intensity and length scales. 
From another perspective, volumetric forcing method~\cite{spille2001generation, keating2006interface} introduces a forcing term into the governing equations to inject energy and generate turbulence. This approach is often coupled with a feedback control mechanism to ensure the generated turbulence develops effectively downstream. However, all these traditional synthetic methods still fall short of generating realistic turbulence that replicates the complex, multi-scale interactions and coherent structures of real flows. Therefore, significant downstream distances from the inlet are often required for unphysical fluctuations to transition into a fully developed, realistic turbulent state. This recovery process increases the required computational domain length and, consequently, the overall cost of simulations. 

With the ever-increasing availability of data, deep learning (DL) has shown great promise in modeling and emulating fluid dynamics~\cite{brunton2020machine,wang2017physics,kochkov2021machine,liu2024multi,fan2025neural}, particularly in generating inflow turbulence. Advanced DL techniques such as convolutional neural networks (CNN)~\cite{mohan2020embedding,gao2021super}, conditional neural fields (NF)~\cite{pan2023neural}, recurrent neural networks (RNN)~\cite{maulik2021reduced}, and attention models~\cite{han2022predicting} have demonstrated the potential to capture complex flow behavior from high-fidelity simulation/experiment data. 
In the context of inflow turbulence generation, Fukami et al.~\cite{fukami2019synthetic} combined a CNN autoencoder with a multilayer perceptron (MLP) to build a one-step model for learning velocity fluctuations in a single cross-section of a turbulent channel flow at a specific Reynolds number. This approach effectively captured turbulent structures to a certain extend and initially demonstrated the feasibility of using DL for inflow turbulence generation. Similarly, Yousif et al.~\cite{yousif2022physics} employed a long short-term memory (LSTM) network to learn the temporal evolution in the CNN-encoded latent space, introducing statistics-based loss terms to regularize the training process. They trained separate models to learn spatiotemporal DNS inlet data at various Reynolds numbers. In a subsequent study, Yousif et al.~\cite{yousif2023transformer} extended their model by introducing adversarial training and temporal attention mechanisms, which were tested on generating inflow for a flat-plate turbulent boundary layer. 
Despite these advances, existing models remain deterministic, while turbulence is inherently stochastic. Most DL-based turbulence generators rely on autoregressive learning architectures, which predict the temporal evolution of instantaneous turbulent flow deterministically from labeled training data. Consequently, these models cannot capture the underlying probabilistic distribution of turbulence data, rendering them incapable of randomly generating instantaneous flow realizations as a stochastic process. Moreover, the deterministic autoregressive setting is prone to cumulative error propagation, compromising their long-term robustness and leading to potential failure, especially when matching the instantaneous turbulence data that are chaotic by nature.

Some recent works have explored the potential of generative artificial intelligence (GenAI) to better capture the stochastic nature of turbulence without requiring labeled training data. 
For example, Kim and Lee~\cite{kim2020deep} utilized a Wasserstein generative adversarial network (WGAN) to generate instantaneous flow fields, and an RNN was then employed to ensure temporal coherence of the generated snapshots in the latent space encoded by the WGAN, effectively decoupling the spatial and temporal generation processes. 
In addition to GANs, Geneva and Zabaras~\cite{geneva2020multi} applied normalizing flows (NFlows) for super-resolving (SR) very large eddy simulation data, enabling SR turbulence generation. Sun et al.~\cite{sun2024unifying} coupled NFlows with transformers to synthesize 2D turbulence in irregular domains using graph neural network (GNN)-based autoencoder. These approach have demonstrated the potential of generative models for turbulence, but challenges remain. GANs are notorious for their training instability, while the scalability of NFlows poses limitations for complex turbulent flow generation.
A significant advancement in generative AI has been the introduction of probabilistic diffusion models, which have garnered attention for their stable training and ability to generate high-quality samples. Diffusion models operate by learning to reverse a noise-injection process, progressively denoising data through a series of steps. This iterative process enables diffusion models to capture the intricate, multi-scale details of turbulence, making them particularly well-suited for spatiotemporal turbulence generation~\cite{shu2023physics,gao2024bayesian,gao2024generative1,gao2024generative2,fan2025neural,du2024confild}. For example, Gao et al.~\cite{gao2024bayesian} built a video diffusion model based on 3D convolutions to directly synthesize spatiotemporal turbulence fields in the physical space. However, due to the high-dimensional nature of turbulence, GPU memory constraints, and significant inference costs associated with diffusion models, directly generating long-term and coherent 4D (space-time) turbulence fields remains computationally prohibitive, particularly for inflow turbulence generation for large-scale DNS/LES simulations, where such temporal coherence is often required. To address this limitation, Du et al.~\cite{du2024confild} proposed the Conditional Neural Field Latent Diffusion Model (\oldname), which efficiently generates turbulence by sampling in the latent space encoded by conditional neural fields (CNFs). This approach significantly reduces the computational burden while preserving the essential fidelity, making it a promising framework for scalable inflow turbulence generation.

In this work, we build upon the concept of integrating latent diffusion models with conditional neural fields (CNFs) to develop a parametric turbulent inflow generator, {\newname}, extending the {\oldname} framework to incorporate parameterized inflow conditions across different Reynolds numbers ($Re$). A novel auto-decoding hyper-network conditioning method is introduced for neural fields, enabling effective dimensionality reduction, while a $Re$-conditioned latent diffusion model is designed to generate highly accurate and realistic inflow turbulence over a significantly broader range of Reynolds numbers than existing DL-based turbulence generators. The proposed {\newname} is rigorously validated through both \emph{a priori} and \emph{a posteriori} tests using eddy-resolving simulations such as DNS and Wall-Modeled LES (WMLES). Results demonstrate that {\newname} consistently reproduces inflow conditions that align closely with the ``ground truth'' extracted from high-fidelity precursor simulations, achieving excellent fidelity and robustness. Moreover, {\newname} generalizes effectively to unseen Reynolds numbers without the need for extensive parameter tuning, relying only on minimal and easily accessible input data. This study represents the first comprehensive application of a conditional latent diffusion model for inflow turbulence generation, marking a transformative step forward in the field. Unlike traditional synthetic turbulence generators, which demand intricate manual tuning and exhibit limited generalizability, {\newname} offers an automated, scalable, and efficient solution. By addressing key limitations of both conventional and DL-based methods, this work establishes a robust, flexible, and computationally efficient framework for inflow turbulence generation, with wide-ranging implications for advanced eddy-resolving simulations in engineering and scientific domains.

The rest of the paper is organized as follows. Section~\ref{sec:meth} introduces the methodology of the proposed {\newname} framework . Section~\ref{sec:result} 
presents the \emph{a pirori} and \emph{a posteriori} testing results of {\newname}, compared with both traditional and existing DL-based synthetic inflow turbulence generators. The generalizability of {\newname} in both time-forecasting and unseen flow conditions are discussed in Section~\ref{sec:discussion}. Finally, Section~\ref{sec:conclusion} 
concludes the paper.

\section{Methodology}
\label{sec:meth}
\newcommand*{\vertbar}{\rule[-1ex]{0.5pt}{2.5ex}}
\newcommand*{\horzbar}{\rule[.5ex]{2.5ex}{0.5pt}}

\subsection{Overview of {\newname} framework}
\begin{figure}
    \centering
    \includegraphics[width=\linewidth]{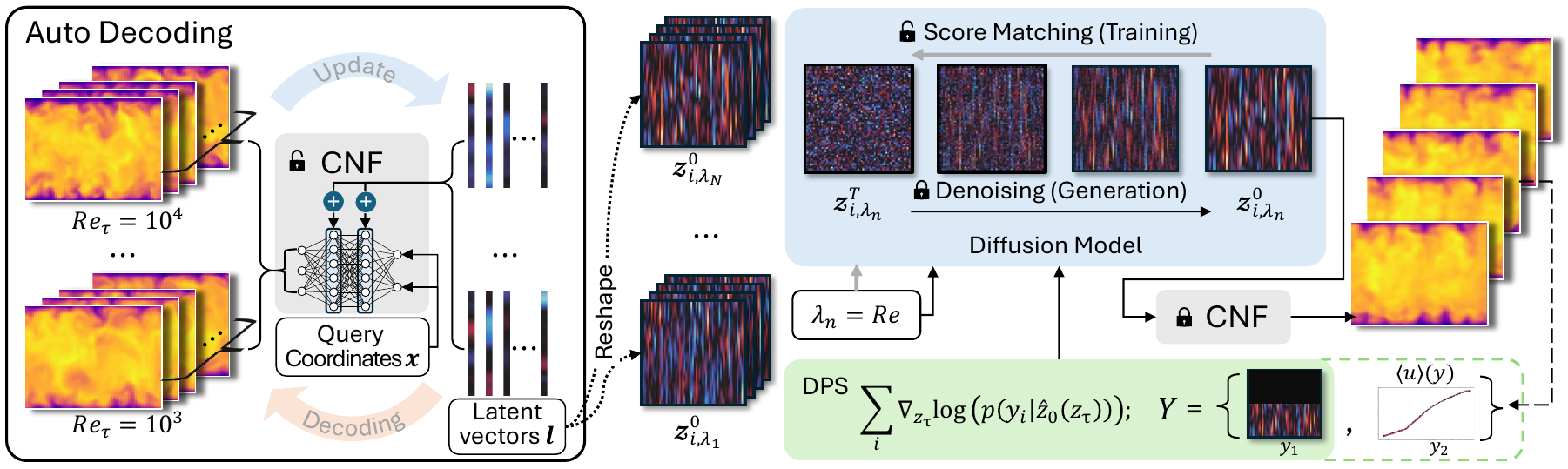}
    \caption{Overview of the proposed \newname\ model. The dashed lines denote the extra conditioning information (the mean velocity $\langle u\rangle(y)$) that can be provided to guide the conditional generation process.}
    \label{fig:schem}
\end{figure}

In this work, we present {\newname}, a parametric inflow turbulence generator accommodating varying Reynolds numbers, {as illustrated in Fig.~\ref{fig:schem}}. Building upon the framework of {\oldname}, {\newname} leverages an auto-decoding based CNF to encode high-dimensional turbulence fields into compact latent vectors at each time step. These latent vectors serve as a reduced-dimensional representation of the turbulence dynamics, capturing multiscale features and enabling efficient turbulence generation. Within the latent space, a conditional latent diffusion model is employed to model the stochastic evolution of turbulence. To facilitate its training, long sequences of latent vectors are rearranged into shorter, overlapping time segments, augmenting the training data. This augmentation ensures that the model effectively learns the temporal evolution of multiscale turbulence structures while maintaining coherence across scales. Once trained, {\newname} can efficiently generate realistic, stochastic inflow boundary conditions across a wide range of Reynolds numbers, providing a scalable and effective inflow solution for transient turbulence simulations. 
\subsubsection{Full-Projected Auto-Decoding CNF for Latent Compression}
The scalability of {\newname} is rooted in the powerful compression capabilities of Conditional Neural Fields (CNF), which effectively encode high-dimensional turbulence fields from the physical space into a compact latent space. This process enables efficient storage and manipulation of turbulence data while preserving critical multiscale features. To enhance this capability, {\newname} introduces a novel auto-decoder framework with a full-projected hypernetwork, replacing the FiLM-based modulation approach~\cite{perez2018film} used in {\oldname}. The full-projected hypernetwork improves the conditioning mechanism by directly projecting the auto-decoded latent vectors into all layers of the primary neural field container, implemented as a SIREN network~\cite{sitzmann2020implicit} in this work. Unlike FiLM, which modulates intermediate features via scaling and shifting parameters at selected layers, the full-projected approach applies globally consistent parameter conditioning across all layers. This ensures a more expressive and unified representation of auxiliary parameters, such as time steps and Reynolds numbers, within the CNF. This improvement is particularly effective for capturing the complex dynamics for inflow turbulence, enabling the model to handle diverse temporal snapshots and a broad range of Reynolds numbers with greater accuracy and robustness.

The data compression process of the fully-projected CNF model is described as follows: Given high-dimensional flow data $\bm{\Phi}$, consisting of spatiotemporal field solutions at multiple Reynolds numbers, the CNF encoder $\mathcal{E}$ compresses each spatial flow field $\bm{\Phi}(\mathbf{x}, t_j, Re_k)$ into a latent vector $\bm{l}_{j, Re_k}$, where $j$ represents the snapshot index and $Re_k$ denotes the corresponding Reynolds number. The encoding process is conducted in an auto-decoding fashion: latent vectors $\bm{l}$ (initialized with zero) and point coordinates are fed into the CNF encoder to produce the model output. Subsequently, the CNF parameters and latent vectors $\bm{l}$ are optimized iteratively to minimize the discrepancy (measured as the $L_2$ norm) between the model output and the flow field data. This process is mathematically expressed as:
\begin{equation}
\begin{aligned}
\mathbf{L}^*, \zeta^*, \gamma^* = \arg \min_{L,\zeta,\gamma} \sum_k^{N_{Re}}\sum_j^{N_t}\sum_i^{N_m} \Big\|\bm{\Phi}(\mathbf{X}_i, t_j, Re_k) - \mathcal{E}\Big(\bm{l}_{j,Re_k}, \mathbf{X}_i; \zeta,\gamma \Big) \Big\|_{L_2},
\end{aligned} 
\end{equation}
where $\mathbf{L}^* = \{\bm{l}^{j, Re_k} \;|\; j, k \in [1, N_t] \times [1, N_{Re}]\}$ is the set of optimized latent vectors, $\zeta^*$ and $\gamma^*$ denote the trainable parameters of the SIREN network and the modulation network within the CNF model. The forward process of the CNF module is given as:
\begin{equation}
\begin{gathered}
\mathcal{E}(\mathbf{x}, \bm{l}) = \mathbf{W}_p \left(\sigma_{p-1}\circ\sigma_{p-2}\circ\dots\circ\sigma_1\right)\left(s_0\mathbf{W}_0\mathbf{x} + s_0\mathbf{\Delta W}_0\mathbf{x} + \mathbf{B}_0 + \mathbf{\Delta B}_0\right) + \mathbf{B}_p,  \\
\sigma_i(\mathbf{h}_{i-1}, \mathbf{\Delta W}_{i}, \mathbf{\Delta B}_i) = \sin(\mathbf{W}_i\mathbf{x} + \mathbf{\Delta W}_i\mathbf{x} + \mathbf{B}_i + \mathbf{\Delta B}_i), \;\;\; i \in [1,2,\dots,p], \\ 
\mathbf{\Delta W}_i(\bm{l})= \mathbf{W}^{w}_i\bm{l}+\mathbf{B}^{w}_i, \\
\mathbf{\Delta B}_i(\bm{l})= \mathbf{W}^{b}_i\bm{l}+\mathbf{B}^{b}_i,
\end{gathered}
\end{equation}
where $\zeta = \{\mathbf{W}_i, \mathbf{B}_i\}_{i=0}^p$ and $\gamma = \{\mathbf{W}^w_i, \mathbf{B}^w_i, \mathbf{W}^b_i, \mathbf{B}^b_i\}_{i=0}^p$ represent all the trainable parameters of the CNF model. The activation function $\sigma$ denotes the sinusoidal function, and $\mathbf{h}_{i-1}$ refers to the output of the $(i-1)^\text{th}$ layer. A scaling factor $s_0$ controls the frequency of the initial input signal to ensure a standard normal distribution across all layer outputs. The SIREN network employs a specialized weight initialization scheme: each weight $\mathbf{W}_i$ in the matrices is initialized as $\mathbf{W}_i \sim \mathcal{U}(-r / \sqrt{n}, r / \sqrt{n})$, where $\mathcal{U}(\cdot)$ represents a uniform distribution with specified bounds. Typically, $s_0 = 30$ and $r = \sqrt{6}$ are chosen to ensure stable and robust performance.

The obtained latents $\mathbf{L}^*$ compactly encode the turbulence field, drastically reducing the dimensionality while retaining the multiscale dynamics of the flow. To model the temporal evolution of turbulence, these latent vectors are rearranged into shorter overlapping time segments. Specifically, a sequence of latent vectors is segmented into blocks $\bm{z}^0_{i, Re} = [\bm{l}_{i+1,Re}, \bm{l}_{i+2, Re}, ..., \bm{l}_{i+N_t, Re}]$, where $N_t$ denotes the number of snapshots in each segment, and $i$ represents the starting index of the segment. These segments preserve temporal coherence and enable the diffusion model to effectively learn the temporal evolution of the flow. For simplicity, we use $\bm{z}_0$ to denote an arbitrary latent spatiotemporal segment $\bm{z}^0_{i, Re}$ in subsequent sections.


\subsubsection{Conditional Latent Diffusion Model}
After compressing the high-dimensional turbulence data into spatiotemporal latent segments $\bm{z}_0$, the next step is to model their stochastic evolution, capturing the underlying statistical distribution $p(\bm{z}_0)$. To achieve this, we employ a diffusion-based generative model, denoising diffusion probabilistic model (DDPM)~\cite{ho2020denoising}, that learns to stochastically synthesize spatiotemporal latent states by leveraging a forward and reverse diffusion process, which are explained as follows. 

The forward diffusion process incrementally adds Gaussian noise to the latent vectors $\bm{z}_0$ over a series of discrete diffusion steps $\tau \in [1, N_\tau]$, where $N_\tau$ denotes the total number of diffusion steps. Each step perturbs the spatiotemporal latent segments using a transition kernel,
\begin{equation}
    p(\bm{z}_\tau|\bm{z}_{\tau-1}) = \mathcal{N}\Big(\bm{z}_\tau; \sqrt{1-\beta_\tau}\bm{z}_{\tau-1}, \beta_\tau\bm{I}\Big),
\end{equation}
where $\beta_\tau$ represents the variance schedule for noise addition, and $\bm{I}$ is the identity matrix. Through this process, the latent segment gradually transitions toward an isotropic Gaussian distribution as $\tau \to N_\tau$. Since the transition kernels are Gaussian, we can use the re-parameterization trick to sample perturbed latent segments $\bm{z}_\tau$ from the clean ones $\bm{z}_0$, which can be expressed as,
\begin{equation}\label{eq:para}
    p(\bm{z}_\tau|\bm{z}_0) = \mathcal{N}\Big(\bm{z}_\tau; \sqrt{\bar{\alpha}_\tau}\bm{z}_0, (1-\bar{\alpha}_\tau)
    \bm{I}\Big),
\end{equation}
where $\alpha_\tau = 1 - \beta_\tau$ and $\bar{\alpha}_\tau = \prod_{s=1}^\tau \alpha_s$. These equations show that the model's generative capability depends critically on two hyperparameters: the number of diffusion steps $N_\tau$ and the variance schedule $\beta_\tau|^{N_{\tau}}_{\tau=1}$. A larger $N_\tau$ allows finer perturbations, ensuring smoother transitions, while the choice of ${\beta_\tau}$ affects the balance between noise injection and reconstruction fidelity.

The reverse diffusion process reconstructs the clean latent segments $\bm{z}_0$ from the noisy segments $\bm{z}\tau$ by approximating the reverse transition kernel $p(\bm{z}_{\tau-1}|\bm{z}_\tau)$. If the forward diffusion process employs a sufficiently large number of diffusion steps ($N_\tau \gg 1$), the perturbations introduced at each step are minimal. This ensures that the reverse step transition kernel $p(\bm{z}_{\tau-1}|\bm{z}_\tau)$ can be well-approximated as a Gaussian distribution. Although computing $p(\bm{z}_{\tau-1}|\bm{z}_\tau)$ directly is intractable, the reverse kernel can be approximated in its conditional form, $p(\bm{z}_{\tau-1}|\bm{z}_\tau, \bm{z}_0)$, using neural network-based parameterization with trainable parameters $\bm{\theta}$,
\begin{equation}
    p_{\bm{\theta}}(\bm{z}_{\tau-1}|\bm{z}_\tau, \bm{z}_0) = \mathcal{N}\Big(\bm{z}_{\tau-1}; \bm{\mu}_{\bm{\theta}}, \bm{\Sigma}_{\bm{\theta}}\Big),
\end{equation}
with the mean $\bm{\mu}_{\bm{\theta}}$ and variance $\bm{\Sigma}_{\bm{\theta}}$ parameterized as, 
\begin{align*}
    \bm{\mu}_{\bm{\theta}} &= \frac{1}{\sqrt{\alpha_\tau}}\left(\bm{z}_\tau - \frac{1-\alpha_\tau}{\sqrt{1-\bar{\alpha}_\tau}}\bm{\epsilon}_{\bm{\theta}}(\bm{z}_\tau, \tau, \bm{\lambda}; \bm{\theta})\right), \\
    \bm{\Sigma_\theta} &= (1-\bar{\alpha}_{\tau-1})\beta_\tau/(1-\bar{\alpha}_\tau) \bm{I}.
\end{align*}
Practically, we use a neural network to directly parameterize the noise $\bm{\epsilon}_{\bm{\theta}}$ added at each step during the forward process. This formulation simplifies the reverse diffusion process by allowing $\bm{\mu}_{\bm{\theta}}$ to be computed using the predicted noise $\bm{\epsilon}_{\bm{\theta}}$. The variance $\bm{\Sigma}_\theta$ is pre-defined and non-trainable, further stabilizing the reverse process. Specifically, a modernized U-Net $\bm{\epsilon}_{\bm{\theta}}(\bm{z}_\tau, \tau, \bm{\lambda}; \bm{\theta})$ conditioned on $\bm{\lambda}$ is built for the noise prediction, equipped with convolutional residual blocks, spatial self-attention mechanisms, and group normalization layers to enhance feature representation. The input layer includes the perturbed latent spatiotemporal segments at $\tau$, diffusion step $\tau$, as well as flow condition parameters $\bm{\lambda}$ (e.g., Reynolds number $Re$ in this work). In order to allow the model to generate inflow given different Reynolds numbers, the $Re$ scalar is transformed into an embedding vector via a trainable sub-network and integrated into the noise prediction network. This sub-network has the same architecture as the one used to condition the U-Net on the diffusion time index $\tau$, ensuring consistent conditioning across both input parameters. 

To optimize neural network parameters $\bm{\theta}$, we minimize a simplified variational lower bound on the negative log-likelihood of the clean data $\bm{z}_0$, 
\begin{equation} 
    L^\mathrm{cond}_\mathrm{simple} = \mathbb{E}_{\tau \sim U[1,N_\tau], (\bm{z}_0, \bm{\lambda}) \sim p(\bm{z}_0, \bm{\lambda}), \bm{\epsilon}\sim\mathcal{N}(0,\bm{I})}\bigg[\Big\| \bm{\epsilon} - \bm{\epsilon}_{\bm{\theta}}\big(\sqrt{\bar{\alpha}_\tau} \bm{z}_0 + \sqrt{1-\bar{\alpha}_\tau}\bm{\epsilon}, \tau, \bm{\lambda} \big) \Big\|^2
    \bigg].
\end{equation}
By reparameterizing $\bm{z}_\tau$ in terms of $\bm{z}_0$ and $\bm{\epsilon}$, the model efficiently learns to denoise the latent segments through iterative reverse transition.

\subsubsection{Autoregressive Posterior Sampling for Long-Horizon Flow Generation}

To provide accurate inflow conditions for turbulence simulations, it is essential to generate long, statistically accurate spatiotemporal flow sequences. This is achieved using the pre-trained diffusion model in a Bayesian conditional auto-regressive sampling framework based on diffusion posterior sampling (DPS)~\cite{chung2023diffusion}, enabling zero-shot generation that satisfies the prescribed conditions. The key is to modify the score function (i.e., the gradient of the log-probability of the data distribution $\bm{s} = \nabla_{\bm{\Phi}}\log p(\bm{\Phi})$ with conditional information $\bm{Y}$ in a Bayesian manner. Specifically, the unconditional score function is obtained from the trained unconditional noise network based on the following relationship,
\begin{equation}
    \bm{s_\theta}(\bm{z}_\tau, \tau, \bm{\lambda}; \bm{\theta}^*) = -\frac{\bm{\epsilon_\theta}(\bm{z}_\tau, \tau, \bm{\lambda}; \bm{\theta}^*)}{\sqrt{(1-\bar{\alpha}_\tau)}}, 
\end{equation}
where $\bm{\epsilon_\theta}(\bm{z}_\tau, \tau, \bm{\lambda}; \bm{\theta}^*)$ is the pre-trained diffusion model. The conditional information $\bm{Y}$ can be incorporated by modifying the score function using Bayes' rule,
\begin{equation}
    \bm{s_\theta}(\bm{z}_\tau, \tau, \bm{\lambda}, \bm{Y}; \bm{\theta}^*) = \bm{s_\theta}(\bm{z}_\tau, \tau, \bm{\lambda}; \bm{\theta}^*) + \sum_i \nabla_{\bm{z}_\tau}\log\Big(p( \bm{y}_i|\bm{z}_\tau)\Big),
\end{equation}
where $\bm{Y} = \{\bm{y}_1,  \bm{y}_2, ...,  \bm{y}_i, ...\}$ is the set of different conditions. The likelihood term involves a marginalization integral,
\begin{equation} \label{eq:margin}
    p(\bm{y}_i|\bm{z}_\tau) = \int p(\bm{y}_i|\bm{z}_0,\bm{z}_\tau)p(\bm{z}_0|\bm{z}_\tau)d\bm{z}_0,
\end{equation}
which is computationally infeasible. Instead, a point estimate is used~\cite{chung2023diffusion}, leading to the modified formulation:
\begin{equation}
        \bm{s_\theta}(\bm{z}_\tau, \tau, \bm{\lambda}, \bm{Y}; \bm{\theta}^*) \approx \bm{s_\theta}(\bm{z}_\tau, \tau, \bm{\lambda}; \bm{\theta}^*) + \sum_i \nabla_{\bm{z}_\tau}\log \Big(p(\bm{y}_i|\bm{\hat{z}}_0(\bm{z}_\tau))\Big)
\end{equation}
where $\bm{\hat{z}}_0(\bm{z}_\tau)$ is the clean latent approximation of $\bm{z}_\tau$, which can be efficiently calculated using Tweedie's formula \cite{kim2021noise2score, efron2011tweedie}
\begin{equation}
    \bm{\hat{z}}_0 \simeq \frac{1}{\sqrt{\bar{\alpha}_\tau}}\Big(\bm{z}_\tau + (1 - \bar{\alpha}_\tau)\bm{s_\theta}(\bm{z}_\tau, \tau, \bm{\lambda}; \bm{\theta}^*)\Big)
\end{equation}

To generate a continuous spatiotemporal trajectory for a given $Re$, the generation process starts by sampling a flow conditioned only on $Re$. The trajectory is extended auto-regressively by overlapping the latter half of the previously generated sequence with the first half of the new one. Formally, for a generated $(i-1)^{th}$ latent segment $\bm{z}_{0}^{(i-1)}$, the latter half of the sequence serves as conditioning information $\bm{y}_1$, defined as $\bm{y}_1 = \{\bm{l}_{i+(N_t-1)/2}, \cdots, \bm{l}_{i+N_t}\}$. The newly sampled latent $(i)^{th}$ $\bm{z}_{0}^{(i)}$ is generated conditioned on $\bm{y}_1$ (i.e., latter half of $\bm{z}_{0}^{(i-1)}$), where the log-likelihood term can be expressed as a quadratic term by assuming a Gaussian distributed process error. The modified score function for auto-regressive generation is,
\begin{equation}
        \bm{s_\theta}(\bm{z}_\tau, \tau, Re, \bm{y}_1; \bm{\theta}^*) \approx \bm{s_\theta}(\bm{z}_\tau, \tau, Re; \bm{\theta}^*) + \nabla_{\bm{z}_\tau}\log \Big(p(\bm{y}_1|\bm{\hat{z}}_0(\bm{z}_\tau))\Big).
\end{equation}

In a similar vein, additional conditioning information such as a mean velocity profile $\langle u \rangle(y)$ can be introduced, denoted as $\bm{y}_2$. To enforce this constraint, the clean latent segment $\bm{\hat{z}}_0$ is decoded via the CNF to reconstruct the flow field $u(t, y, z)$, which is then averaged over time and spanwise dimensions. The log-likelihood term for $\bm{y}_2$ is expressed as, 
\begin{equation}
    \log\Big(p(\bm{y}_2|\bm{\hat{z}}_0)\Big) \propto \left\|\frac{1}{N_T N_{span}}\sum\sum \mathrm{CNF}(\bm{\hat{z}}_0) - \langle u\rangle(y)\right\|^2,
\end{equation}
where $N_T$ is number of snapshots in one sampled segment and $N_{span}$ is the number of query points in the spanwise direction. The revised score function for conditional sampling in this scenario is given as,
\begin{equation}
        \bm{s_\theta}(\bm{z}_\tau, \tau, Re, \bm{y}_1, \bm{y}_2; \bm{\theta}^*) \approx \bm{s_\theta}(\bm{z}_\tau, \tau, Re; \bm{\theta}^*) + \nabla_{\bm{z}_\tau}\log \Big(p(\bm{y}_1|\bm{\hat{z}}_0(\bm{z}_\tau))\Big) + \nabla_{\bm{z}_\tau} \log\Big(p(\bm{y}_2|\bm{\hat{z}}_0(\bm{z}_\tau))\Big).
\end{equation}

\section{Numerical Results}
\label{sec:result}

\subsection{Data-efficient turbulent inlet generator for DNS}
\label{sec:DNS}

To evaluate the performance of {\newname} as a data-efficient synthetic inlet generator for DNS, we test it on a turbulent channel flow at a given Reynolds number. Three baseline methods are selected for comparison, including two state-of-the-art DL-based synthetic turbulence inflow generators: CNN-LSTM, proposed by Yousif et al.~\cite{yousif2022physics} and ConvLSTM (details provided in~\ref{appen:baselines}), as well as a traditional synthetic inflow generator, digital filtering method (DFM). The precursor simulation results are used as the reference data for training and validation.

\subsubsection{Data generation}
\label{sec:dns_datagen}
The training dataset is generated by solving the incompressible Naiver-Stokes equations using our in-house GPU-accelerated solver~\cite{xiantao2024diffsi},
\begin{equation}
\begin{split}
    \frac{\partial \bm{u}}{\partial t} + (\bm{u}\cdot\nabla)\,\bm{u} &=  - \frac{1}{\rho}\nabla p + \nu \nabla^2 \bm{u}  \\ 
    \nabla \cdot \bm{u} &= 0  
\end{split} 
\label{eq:ns-dns}
\end{equation}
where $\bm{u} = [u,v,w]$ represents the velocity vector, with $u, v, w$ denote the streamwise, wall-normal, and spanwise directional velocity components, respectively. The pressure field is denoted by $p$, and the kinematic viscosity $\nu$ is set to $\nu = 2.5\times10^{-4}$. This configuration generates a turbulent channel flow with a friction Reynolds number $Re_\tau = u_\tau\delta/\nu = 180$, where $u_\tau$ is the friction velocity. The simulation is conducted on a periodic channel domain with dimensions $l_x\times l_y\times l_z = 2\pi\delta\times 2\delta\times \pi\delta$, where $\delta = 1$ is the half-width of the channel, and $x,y,z$ denote the streamwise, wall-normal, and spanwise directions, respectively. The computational domain is discretized into $N_x\times N_y\times N_z = 160\times 400 \times 100$ uniform staggered mesh grid points. 

From the DNS simulation, $1200$ fully developed (i.e. statistical stationary) velocity snapshots $\bm{u}\in\mathbb{R}^{400\times100}$ are collected at a single $y-z$ cross-section. These snapshots are sampled at a fixed time interval $\Delta t^+ = 0.405$ to ensure temporal resolution that aligns with the statistical stationarity of the flow. Each snapshot retains the original DNS resolution ($N_y \times N_z = 400\times100$), capturing the multiscale turbulent structures. The dataset is well validated to ensure statistical consistency with previous DNS studies. The validation includes comparisons of key turbulence statistics between the generated dataset and the reference DNS data. Further details on the case configuration and dataset validation procedures are provided in \ref{appen:dnsData}.

\subsubsection{{A priori} test}
To evaluate the ability of {\newname} to capture the statistical distribution of turbulence represented in the training dataset, we first use the well-trained {\newname} to generate $1200$ temporally coherent velocity field $\bm{u}$ at a single cross-section. All the DL-based baselines are trained on the same dataset and tested on unseen inputs, while the hyperparameters required by DFM (e.g. mean velocity profile and Reynolds stress) are calculated based on the same training set. 

\begin{figure}[!ht]
\captionsetup[subfloat]{farskip=-2pt,captionskip=-6pt}
    \centering
    \subfloat[]{\includegraphics[width=0.33\textwidth]{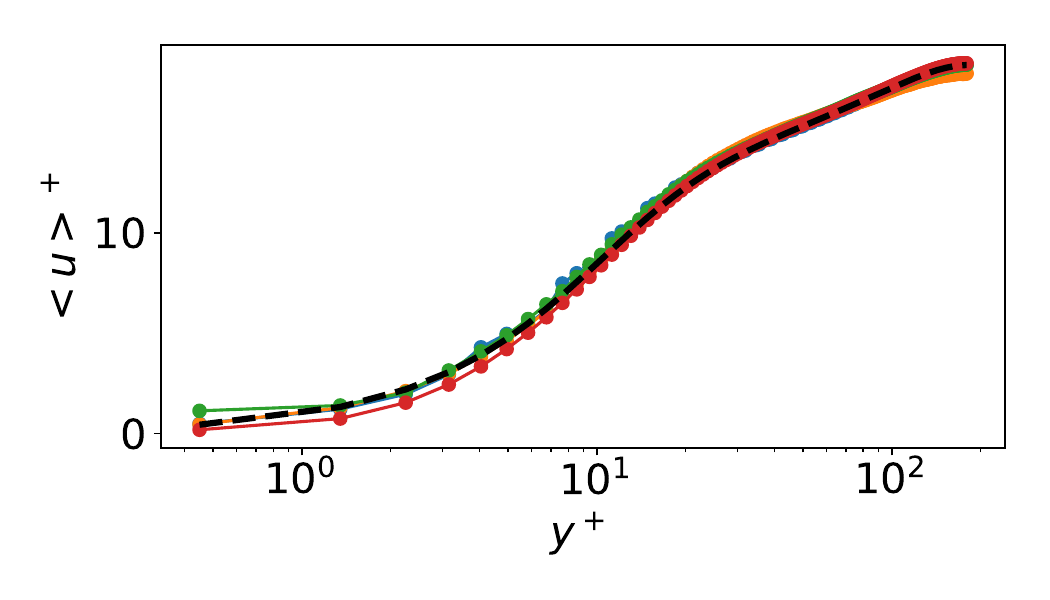}}\hfill
    \subfloat[]{\includegraphics[width=0.33\textwidth]{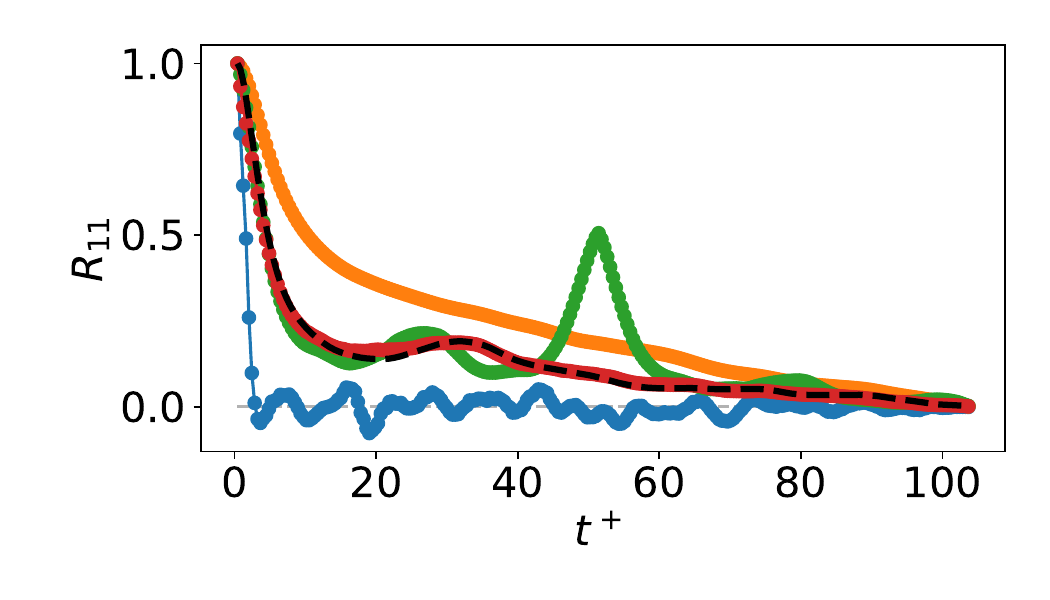}}\hfill
    \subfloat[]{\includegraphics[width=0.33\textwidth]{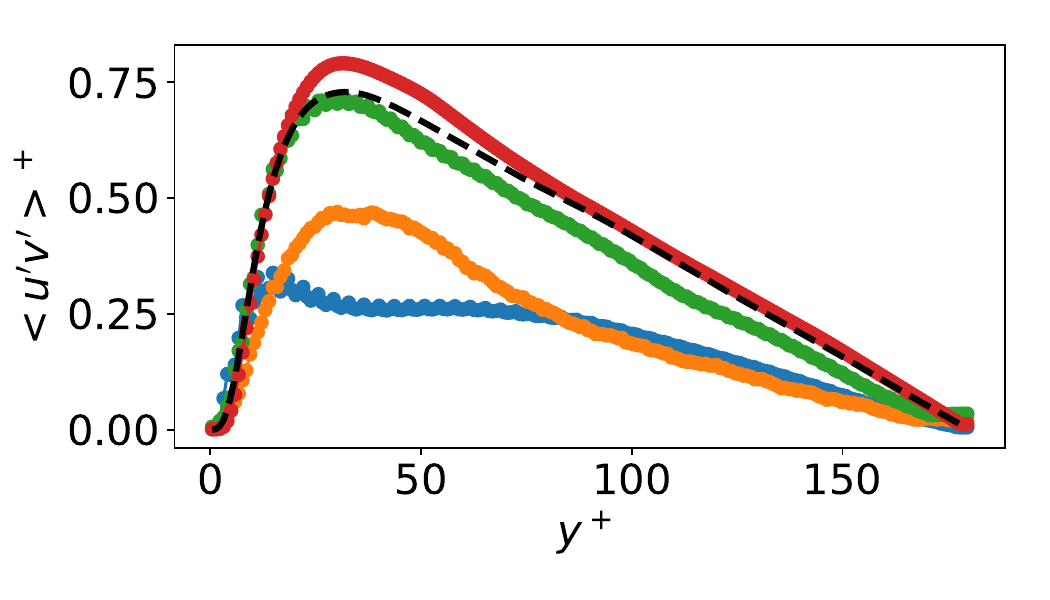}}\\
    \subfloat[]{\includegraphics[width=0.33\textwidth]{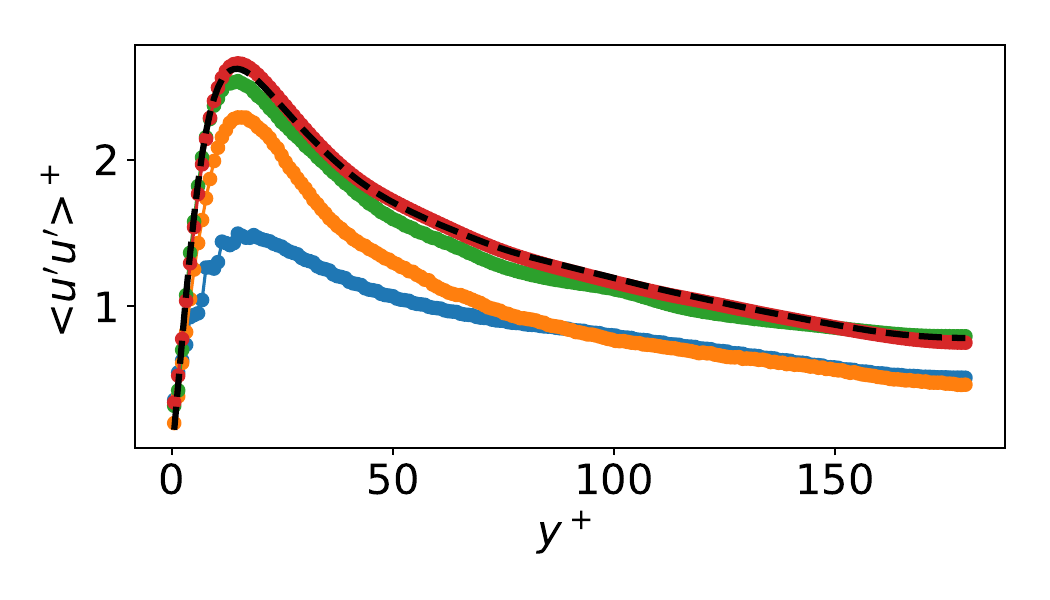}}\hfill
    \subfloat[]{\includegraphics[width=0.33\textwidth]{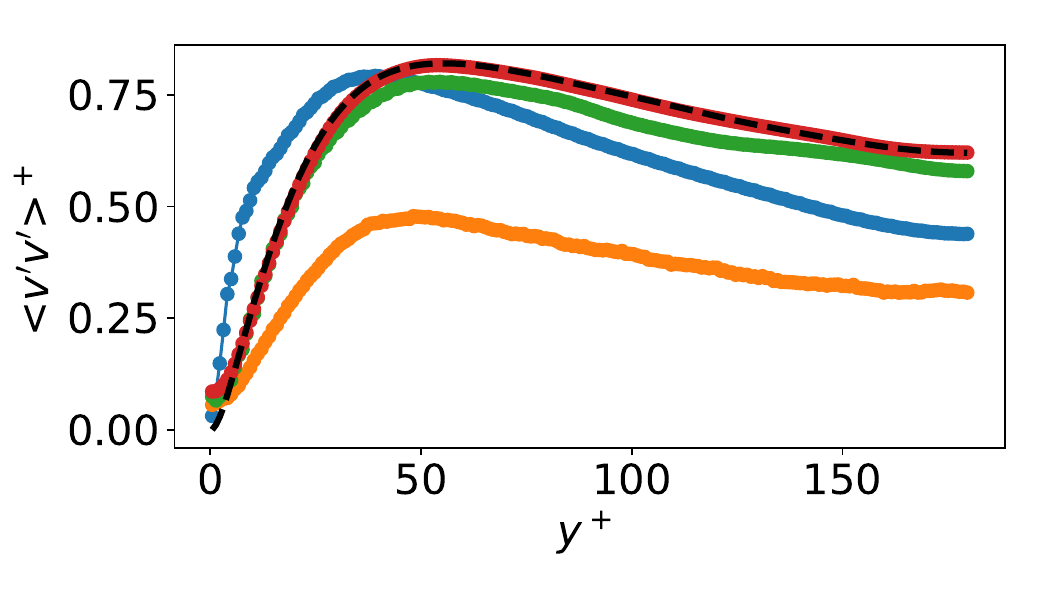}}\hfill
    \subfloat[]{\includegraphics[width=0.33\textwidth]{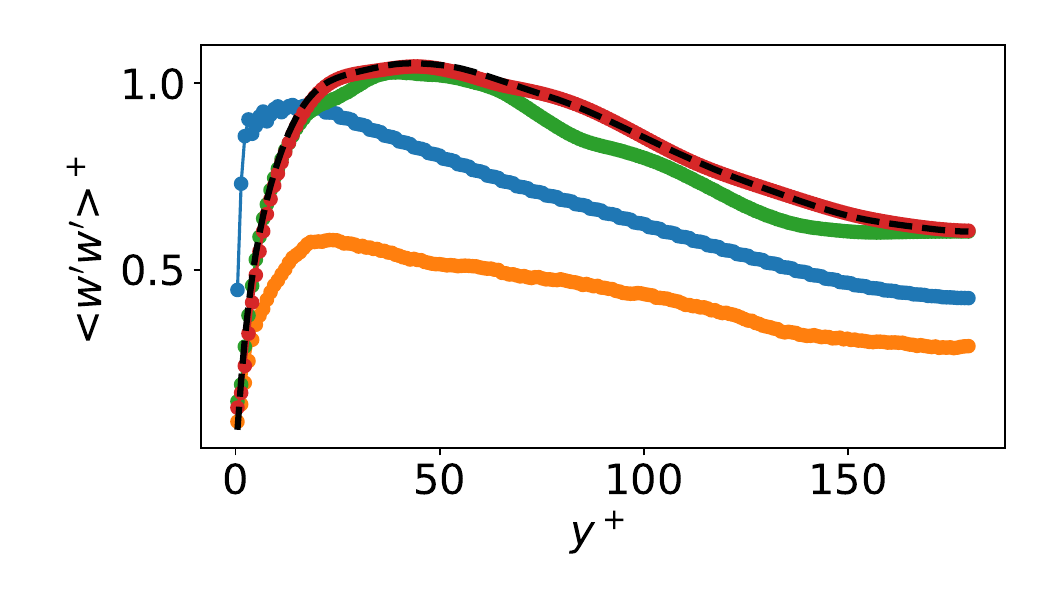}}\\
    \subfloat[]{\includegraphics[width=0.33\textwidth]{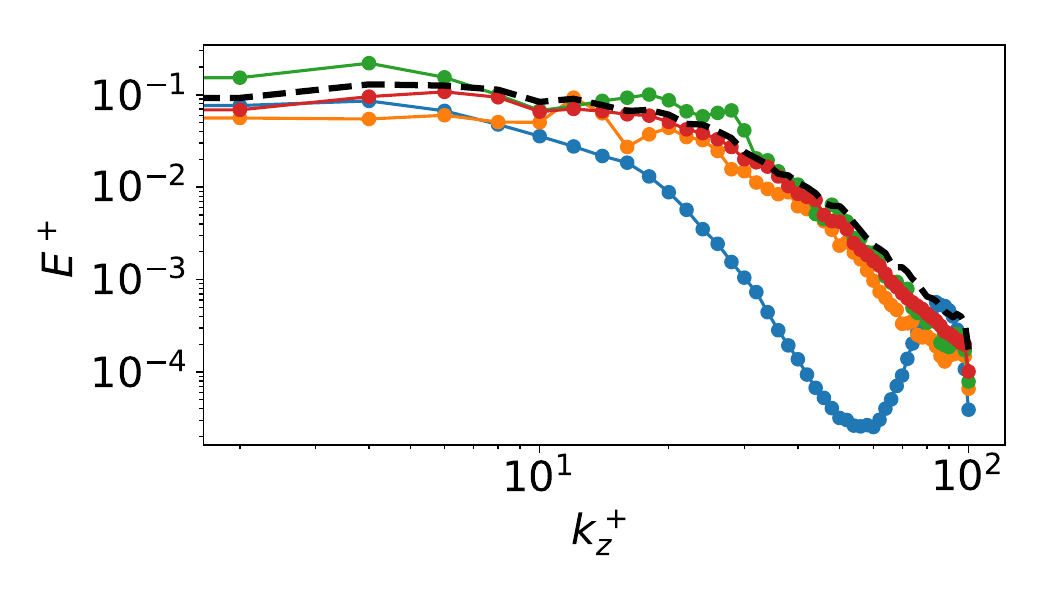}}\hfill
    \subfloat[]{\includegraphics[width=0.33\textwidth]{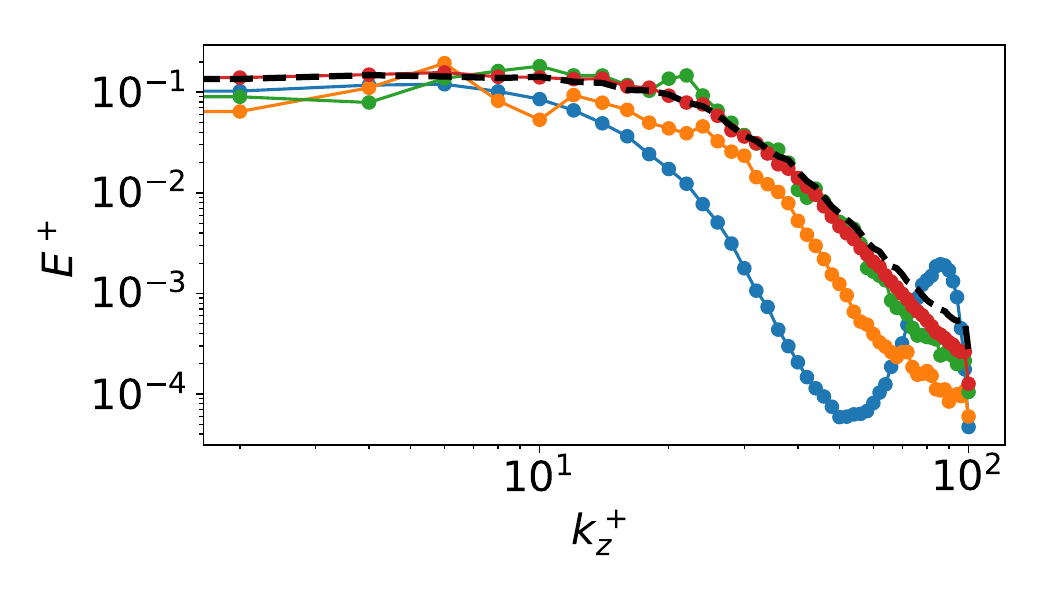}}\hfill
    \subfloat[]{\includegraphics[width=0.33\textwidth]{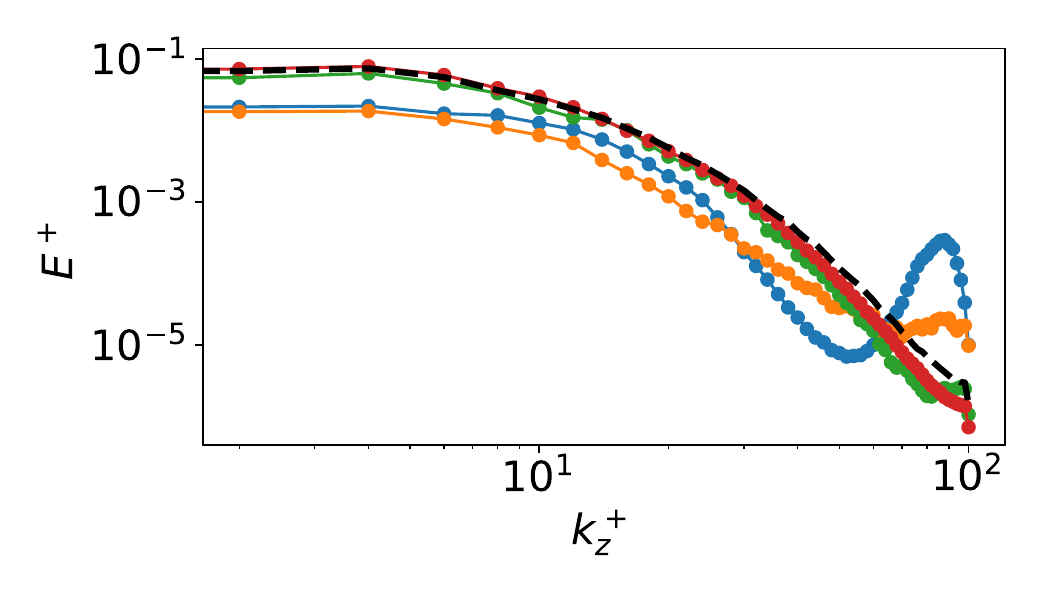}}\\
    \caption{\emph{A priori} test comparision of various turbulence statistics between {\newname} (\reddotline), CNN-LSTM (\greendotline), Digital Filtering (\bluedotline), ConvLSTM (\orangedotline), and the training data set (\dashedblackline, i.e. recycling simulation). (a) Mean velocity profile along wall-normal ($y$) direction. (b) Auto correlation of $u'$ ($R_{11}$) at $y^+\approx5$, (c) Reynolds stress ($\langle u'v'\rangle$) (d-f) Turbulence intensity of $u'$ (d), $v'$ (e), and $w'$ (f), respectively. (g-i) Turbulence kinetic energy distribution along spanwise ($z$) direction at different wall normal locations $y^+\approx$ 5 (g), 30 (h), and 180 (i), respectively.}
    \label{fig:dns_prior}
\end{figure}

Figure~\ref{fig:dns_prior} compares the turbulence statistics of the spatiotemporal inflow generated by the various methods against the training dataset (i.e., the precursor simulation with the periodic boundary).
Across all evaluated metrics, {\newname} exhibits excellent agreement with the reference data and outperforms the baseline methods, particularly for higher-order statistics. For the first-order statistic, the mean velocity profile $\langle u\rangle^+$ along the wall-normal direction, all methods achieve a reasonable match with the reference, as shown in Fig.~\ref{fig:dns_prior}(a). 
However, for higher order statistics (\ref{fig:dns_prior}(b-i)), {\newname} demonstrates significantly better accuracy compared to all baselines. As shown in Fig.~\ref{fig:dns_prior}(b), the auto-correlation of velocity fluctuations of the velocity fluctuation $u'$ ($R_{11}$) in near-wall region ($y^+\approx5$) closely matches the reference data at all time delays $\Delta t^+$. In contrast, CNN-LSTM introduces an artificial peak at $\Delta t^+\approx50$, indicating periodic repetition in the generated flow by CNN-LSTM, which has a fixed period of $\Delta t^+\approx 50$. This behavior implies that CNN-LSTM memorizes training data rather than generating novel flow samples. Additional evidence and analysis of this artificial peak are provided in Sec.~\ref{sec:analyse}. The other two baselines even fail to provide reasonable statistics for both long-term and short-term auto-correlation. Specifically, ConvLSTM overpredicts short-term correlations, while DFM fails to generate sufficiently temporally correlated fields. Figures~\ref{fig:dns_prior}(c-f) show Reynolds stress $\langle u'v'\rangle$ and turbulence intensity ($u', v', w'$) at various wall-normal locations. {\newname} maintains strong agreement with the reference dataset across all locations, while DFM and ConvLSTM exhibit substantial discrepancies. Although CNN-LSTM captures these metrics reasonably well in most regions, {\newname} achieves greater statistical fidelity in reflecting the target distribution.
The turbulent kinetic energy (TKE) along spanwise wavenumber ($k_z^+$) at different wall-locations is shown in last three sub-figures (g-i) of Fig.\ref{fig:dns_prior}. Consistently, {\newname} accurately reproduces the TKE distribution across different scales, though it slightly underpredicts energy at high wavenumbers. This underprediction can be attributed to the smoothing effects of the conditional neural field~\cite{du2024confild}. However, for turbulent inlet generation, small-scale structures are less critical as they can be restored in downstream simulations. Larger-scale structures (low wavenumbers), which are essential for accurate inflow generation, are well captured by {\newname}. The CNN-LSTM model, while performing better than DFM and ConvLSTM, introduces extra fluctuations at low wavenumbers for the first two wall-normal locations and exhibits larger deviations at high wavenumbers compared to {\newname}. 

Overall, the results confirm that our {\newname} achieves high fidelity in reproducing both first-order and higher-order turbulence statistics, demonstrating clear advantages over traditional and DL-based inflow generator baselines in the \emph{a priori} tests.

\subsubsection{{A posteriori} test}
\emph{A posteriori} tests are more critical for evaluating the performance of inlet turbulence generators, as they directly assess the capability of generated inflow conditions to drive physically consistent turbulence downstream in eddy-resolving simulations. While \emph{a priori} tests may show promising statistical accuracy, nonphysical or incorrect coherent structures in the synthetic inflow can lead to rapid dissipation of turbulence or other inaccuracies in downstream flow development. In this section, the synthetic inflows generated by {\newname}, the three baseline models (CNN-LSTM, ConvLSTM, and DFM), and the reference recycling DNS data are used as inlet boundary conditions (BC) for a channel flow DNS simulation with inflow-outflow BCs as \emph{a posteriori} test. The computational domain of the \emph{a posteriori} test has dimensions $l_x \times l_y \times l_z = 8\pi\delta\times2\delta\times\pi\delta$, where $\delta = 1$, discretized into a uniform grid of size $N_x \times N_y \times N_z = 640\times400\times100$. The domain shares the same cross-section geometry and mesh resolution as the training dataset. For the tests, $1800$ snapshots of the generated velocity field ($\bm{u} \in \mathbb{R}^{N_t\times N_y\times N_z\times3}, N_t = 1800$) are applied at the inlet, with the same temporal interval ($\Delta t^+=0.405$) as the training dataset. OpenFOAM is used to perform \emph{a posteriori} DNS simulations. To eliminate the influence of initial conditions, only the last $600$ time steps are analyzed for statistics computations. Additional details about the \emph{a posteriori} test configuration are provided in \ref{appen:posterior}.

\begin{figure}[!ht]
\captionsetup[subfloat]{farskip=-2pt,captionskip=-6pt}
    \centering
    \subfloat[]{\includegraphics[width=\textwidth]{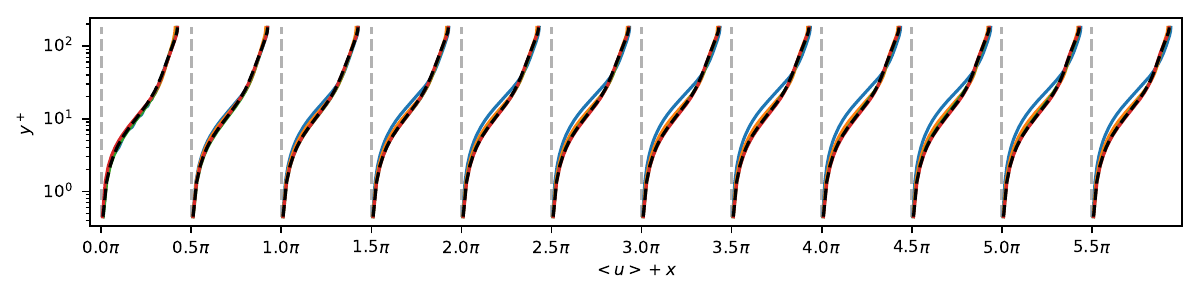}}\\
    \subfloat[]{\includegraphics[width=\textwidth]{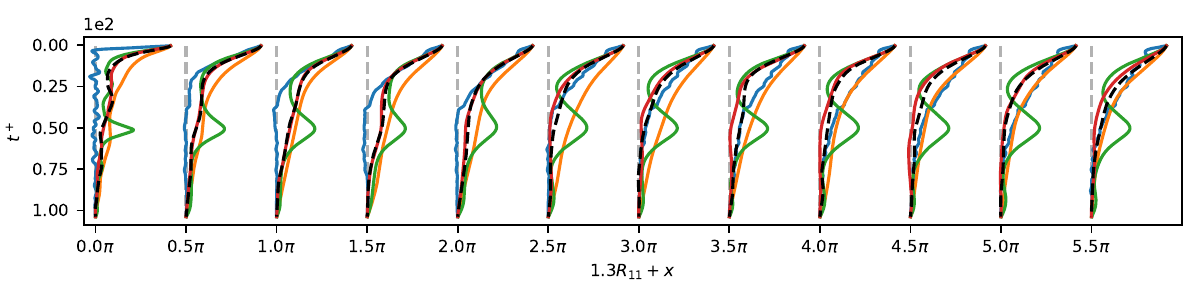}}\\
    \subfloat[]{\includegraphics[width=\textwidth]{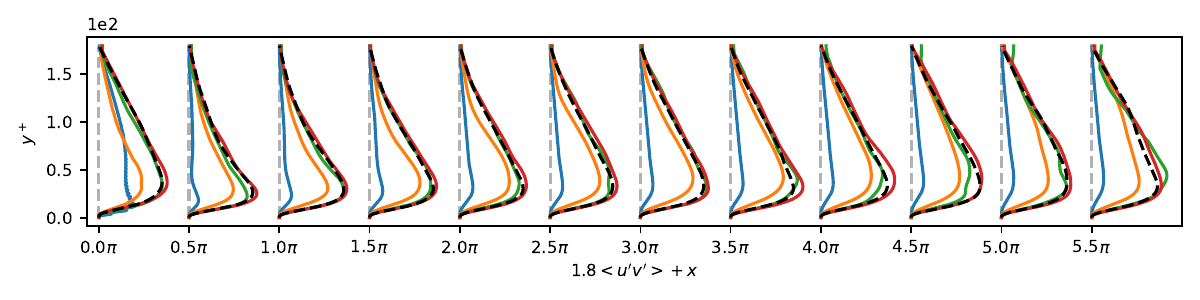}}\\
    \subfloat[$x=0.04$]{\includegraphics[width=0.33\textwidth]{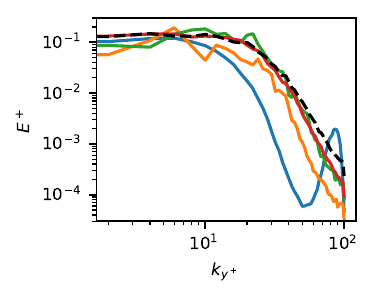}}\hfill
    \subfloat[$x=0.5\pi$]{\includegraphics[width=0.33\textwidth]{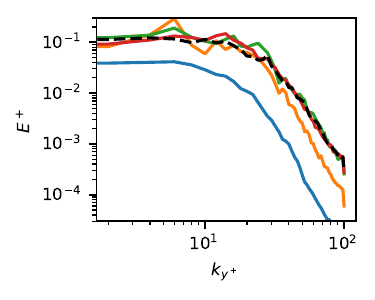}}\hfill
    \subfloat[$x=5.5\pi$]{\includegraphics[width=0.33\textwidth]{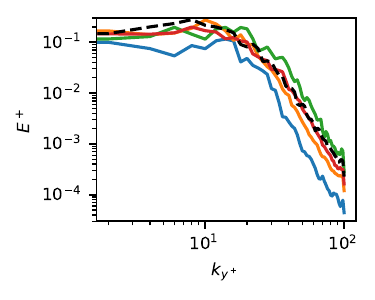}}
    \caption{Statistics comparison of \emph{A posteriori} tests between {\newname} (\redline), CNN-LSTM (\greenline), Digital Filtering (\blueline), ConvLSTM (\orangeline), and the reference data (\dashedblackline, i.e. precursor simulation) at different stream wise location. (a) Mean stream wise velocity ($<u>$) profile along wall normal direction. (b) Auto correlation of velocity component $u$ at wall normal location $y^+ \approx 5$. (c) Reynolds share stress $<u'v'>$. (d-f) Turbulence kinetic energy (TKE) distribution along spanwise wavenumber $k_z^+$ at $y^+\approx30$ at different stream wise locations $x = 0.04$ (d), $0.5\pi$ (e), and $5.5\pi$ (f), respectively}
    \label{fig:dns_post}
\end{figure}

Figure~\ref{fig:dns_post} presents the results of the \emph{a posteriori} tests, showing various turbulence statistics at different streamwise locations. The statistics are evaluated within the domain segment $x\in[0, \,5.5\pi]$ to avoid possible artifacts caused by the outlet boundary condition.
The performance of each turbulence generator is analyzed based on its ability to sustain turbulence as it develops downstream, where poor-quality synthetic inflows exhibit pronounced decay and dissipation. Specifically, the mean velocity profiles along the wall-normal direction, shown in Fig.~\ref{fig:dns_post}(a), reveal significant differences in performance across these models. While all DL-based inflow generators, {\newname} (\redline), ConvLSTM (\orangeline), and CNN-LSTM (\greenline), maintain good agreement with the reference data (\dashedblackline) throughout the computational domain, the DFM baseline (\blueline) exhibits increasing deviations along the streamwise direction, particularly around $y^+\approx20$. ConvLSTM (\orangeline) also deviates slightly from the reference data at the very end of downstream locations.  

The auto-correlation of velocity fluctuations $R_{11}$, shown in Fig.~\ref{fig:dns_post}(b), further differentiates the performance of these models. Similar to the \emph{a priori} tests, the velocity fields obtained with CNN-LSTM inlet exhibit an artificial periodic peak around $t^+=50$, indicative of data repetition. Although this artifact slightly decays as the turbulence develops downstream, it remains evident even at the last streamwise location $x=5.5\pi$, leading to significant deviations from the reference data. The other two baselines perform even worse, failing to reproduce accurate short- or long-term correlations. In contrast, {\newname} shows a much closer match to the reference data, with only very minor discrepancies. Notably, even the recycling DNS inflows experience some decay and redevelopment of turbulence, attributable to the unphysical recycling behavior imposed by the periodic BC in the relatively short channel.

The Reynolds shear stress profiles in Fig.~\ref{fig:dns_post}(c) further illustrate the distinctions between the models. At the first downstream grid point ($x=0.04$) {\newname} slightly deviates from the reference at $y^+\approx 50$, while CNN-LSTM slightly underpredicts the turbulent shear stresses at $y^+\approx100$. Both methods show minor overpredictions near the channel center. However, as the flow develops downstream, the turbulence generated by {\newname} quickly converges to the reference statistics. Although minor discrepancies appear at intermediate locations, {\newname} stays closest to the reference data at stream-wise locations compared to all the synthetic inlets. On contrary, CNN-LSTM fails to converge to the reference at downstream locations even though its Reynolds shear stress matches with the reference very well at the inlet. Particularly, CNN-LSTM increasingly diverges from the reference data at downstream locations, particularly near the channel center beyond $x=3.5\pi$. This divergence indicates that the CNN-LSTM-generated inflows are not fully compatible with the underlying flow physics. Both DFM and ConvLSTM exhibit substantial turbulence decay and redevelopment, showing significantly poorer agreement with the reference data compared to {\newname}.

Figures~\ref{fig:dns_post}(d-f) compare the turbulent kinetic energy (TKE) spectra along the spanwise wavenumber $k_z^+$ at a wall-normal location $y^+=30$ for three streamwise locations. At the inlet, similar to what was observed in the \emph{a priori} test, our {\newname} outperforms other baselines, and it matches well with the reference data at lower wavenumbers, with a slight under-prediction of TKE in the higher wavenumber range ($k_z^+ > 60$). This indicates that {\newname} effectively captures large-scale features while exhibiting minor smoothing effects on small-scale structures. Among all the baseline methods, CNN-LSTM performs better than the other two, achieving a rough match at intermediate wavenumbers ($k_z^+ \in (8, 60)$) but showing similar under-predicted TKE at higher wavenumbers. For the remaining two baselines, both significantly under-predict TKE across nearly all wavenumbers. As the flow develops along the streamwise direction (e.g., at $x = 0.5\pi$), {\newname} and CNN-LSTM rapidly recover small-scale features, while the TKE values of the other two baselines marginally move towards the reference data. At the furthest streamwise location shown ($x = 5.5\pi$), the reference data itself exhibits noticeable changes around $k_z^+ = 10$. Nonetheless, {\newname} successfully captures this trend and maintains much closer agreement with the reference data compared to all the baseline methods. More results of the \emph{a posteriori} tests can be found in Section~\ref{appen:more_stat_dns}.

\subsection{Robust inlet generator for WMLES with a wide range of Reynolds number}
\label{sec:wmles}
In the previous section, we demonstrate the capability of {\newname} as a synthetic inlet generator for DNS simulations at single Reynolds number (extrapolation in time). However, for practical applications, it is more desirable to generalize to a wide range of Reynolds numbers / flow conditoins as well. In this section, we present the use of {\newname} to generate inflow data for a variety of Reynolds numbers. Given the prohibitive computational cost of collecting a comprehensive DNS dataset spanning multiple high Reynolds numbers and training {\newname} on it, we demonstrate its generalizability to varying $Re$ using a dataset obtained by wall-modeled large eddy simulation (WMLES). The training data was generated on the same computational domain described in the Sec.\ref{sec:dns_datagen}, but with a coarser mesh of size $N_x \times N_y \times N_z = 64 \times 64 \times 64$, encompassing 10 different Reynolds numbers ranging from 1000 to 10,000 ($Re_{\tau} \in {1, 2, \cdots, 10} \times 10^3$). Additional details of the simulation setup are provided in \ref{appen:wmles_data}.
%
\subsubsection{Performance on training flow conditions}

\paragraph{\textbf{A priori test}}

\begin{figure}[!ht]
{\scriptsize
    \begin{tabularx}{\textwidth}{XXX}
    \centering
       $Re_\tau = 1000$  &\centering $Re_\tau = 5000$ &\centering $Re_\tau = 10000$
    \end{tabularx}}
    \captionsetup[subfloat]{farskip=-2pt,captionskip=-6pt}
    \centering
    \subfloat[]{\includegraphics[width=0.33\textwidth]{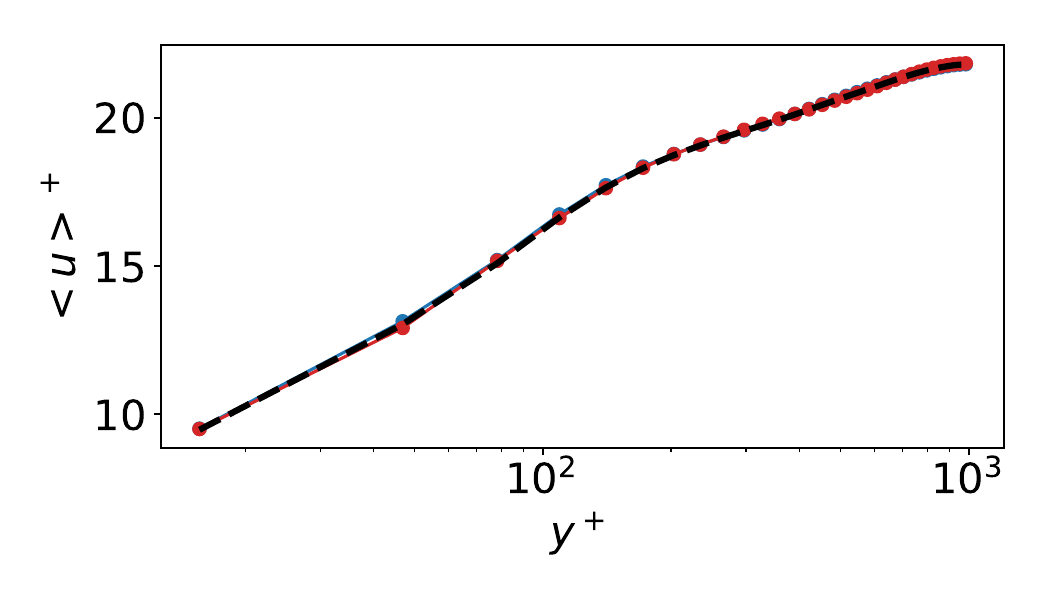}}\hfill
    \subfloat[]{\includegraphics[width=0.33\textwidth]{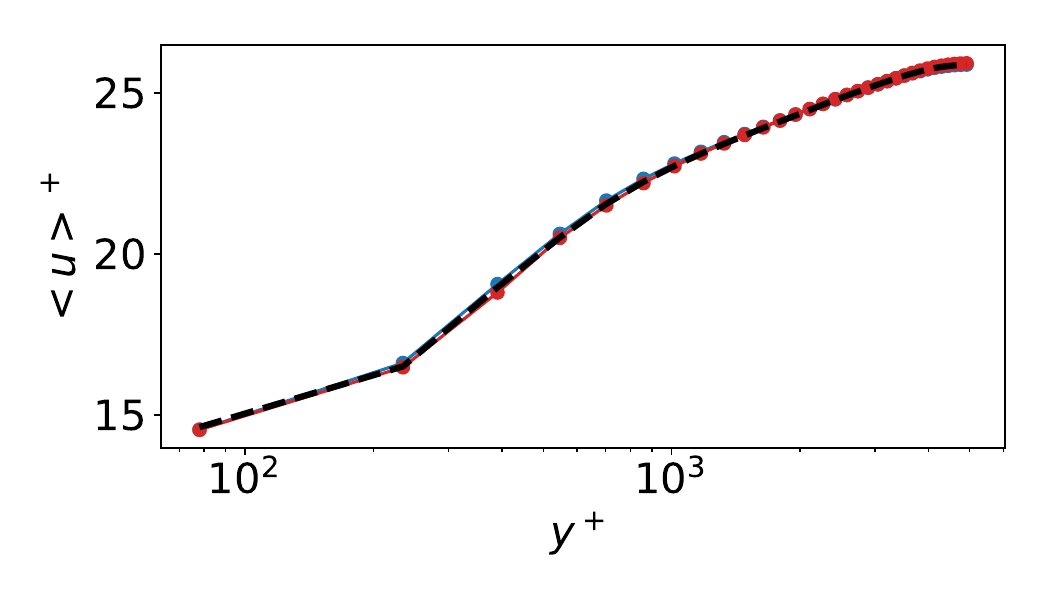}}\hfill
    \subfloat[]{\includegraphics[width=0.33\textwidth]{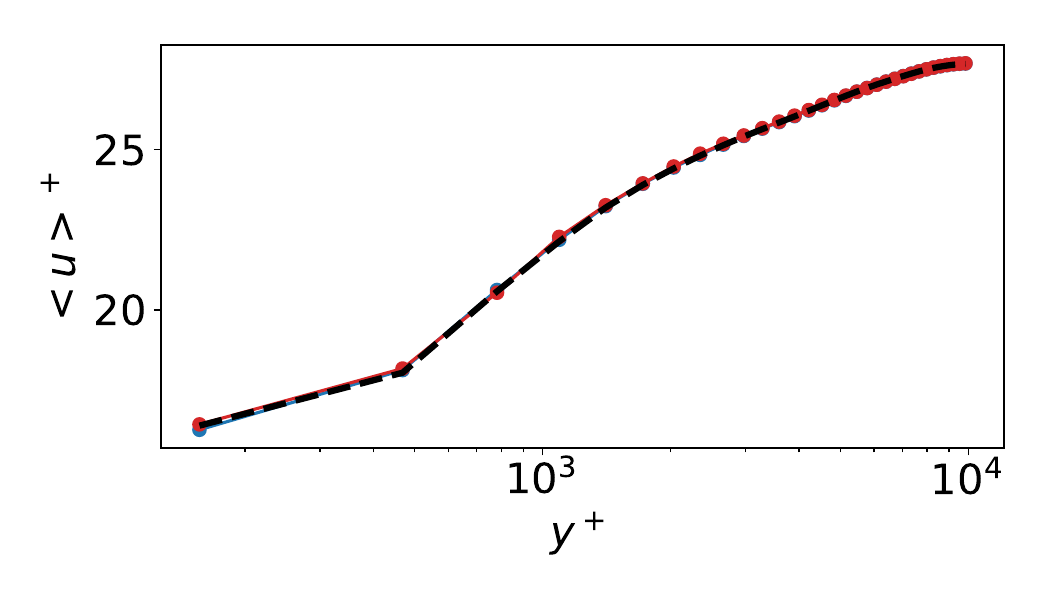}}\\
    \subfloat[]{\includegraphics[width=0.33\textwidth]{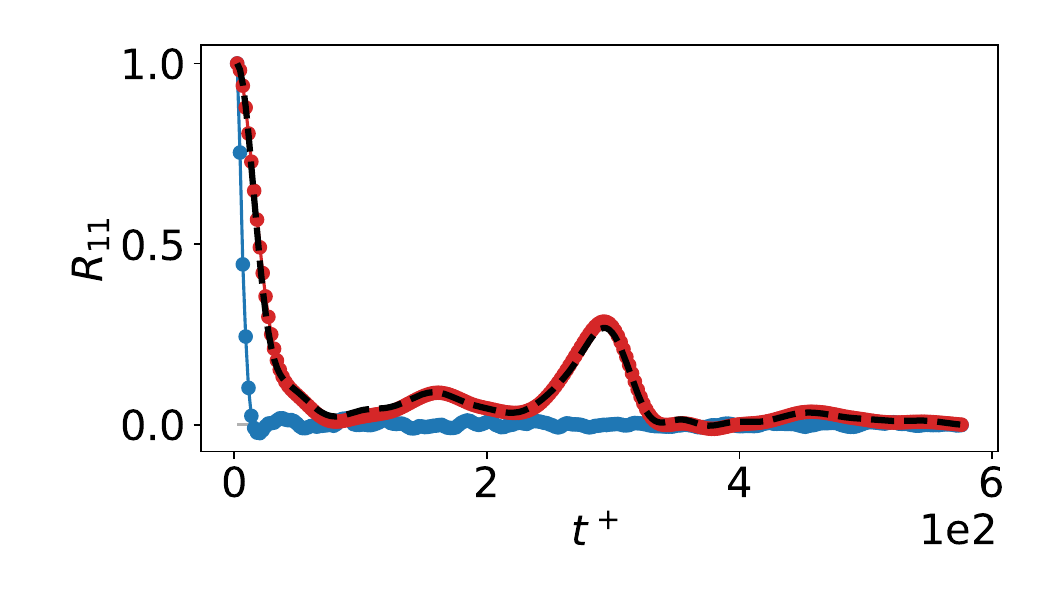}}\hfill
    \subfloat[]{\includegraphics[width=0.33\textwidth]{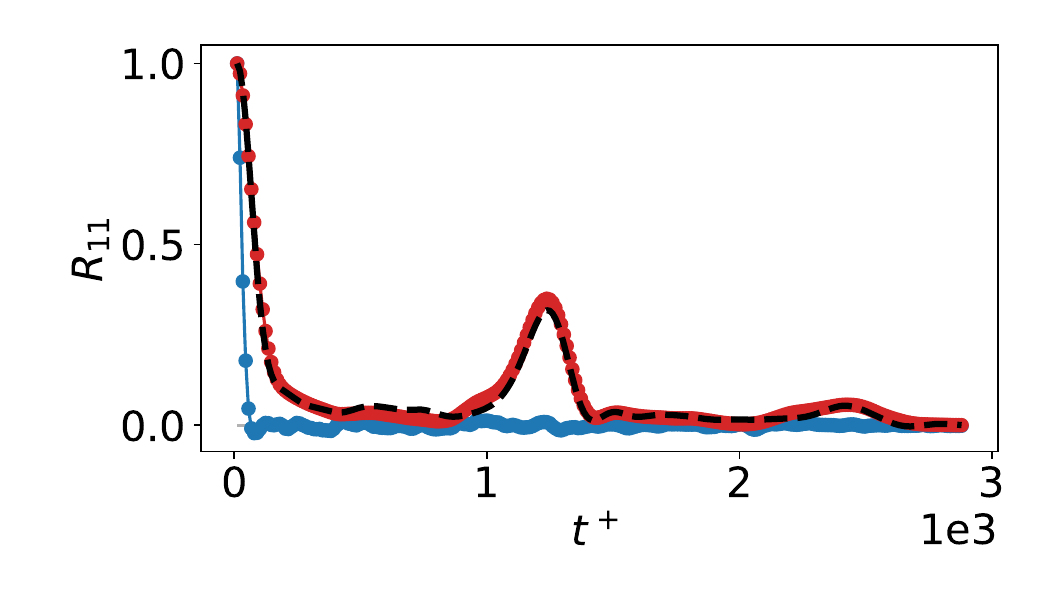}}\hfill
    \subfloat[]{\includegraphics[width=0.33\textwidth]{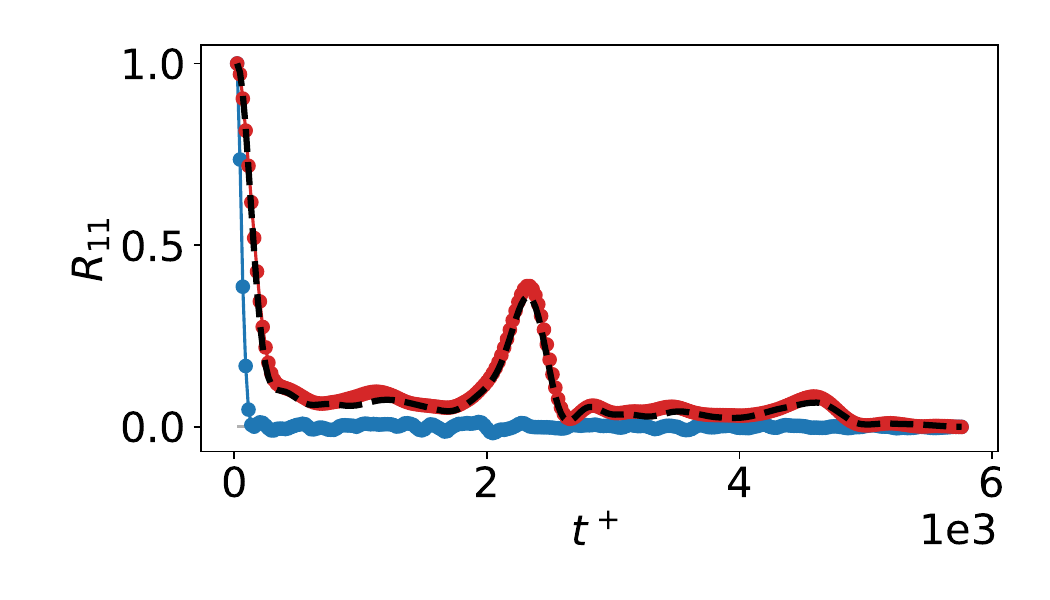}}\\
    \subfloat[]{\includegraphics[width=0.33\textwidth]{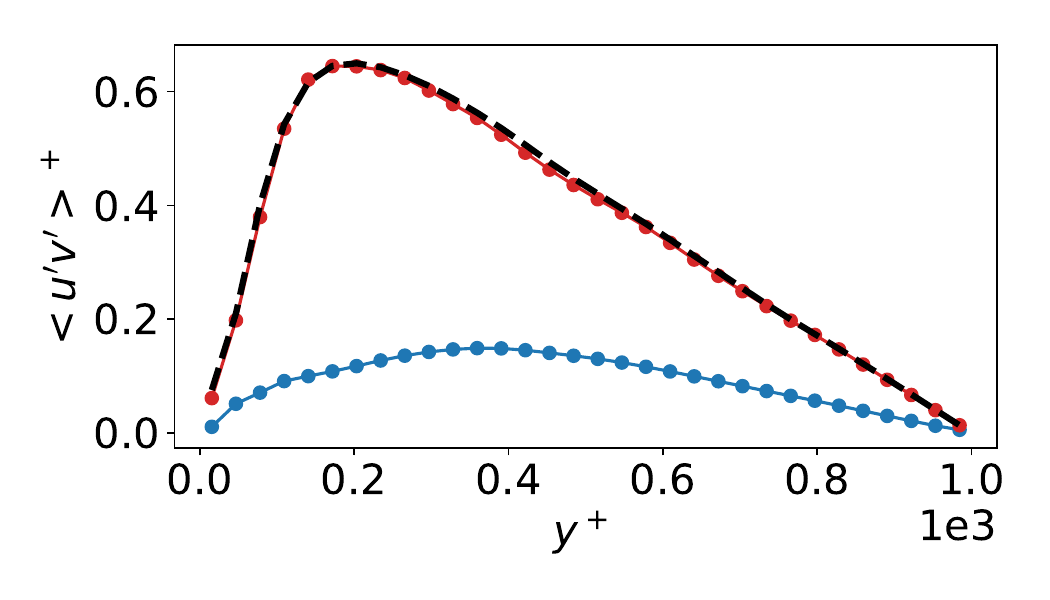}}\hfill
    \subfloat[]{\includegraphics[width=0.33\textwidth]{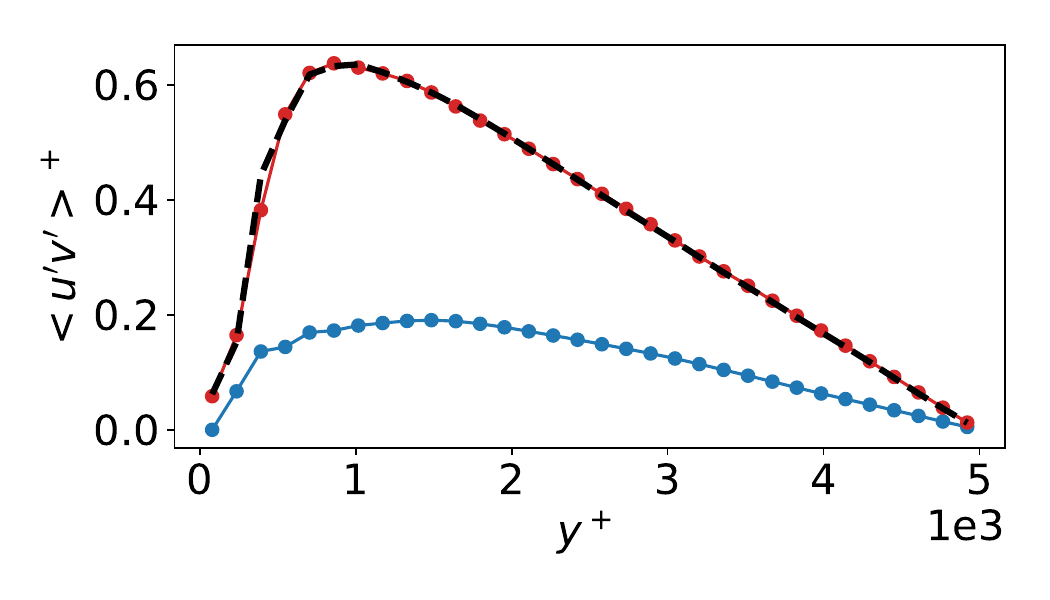}}\hfill
    \subfloat[]{\includegraphics[width=0.33\textwidth]{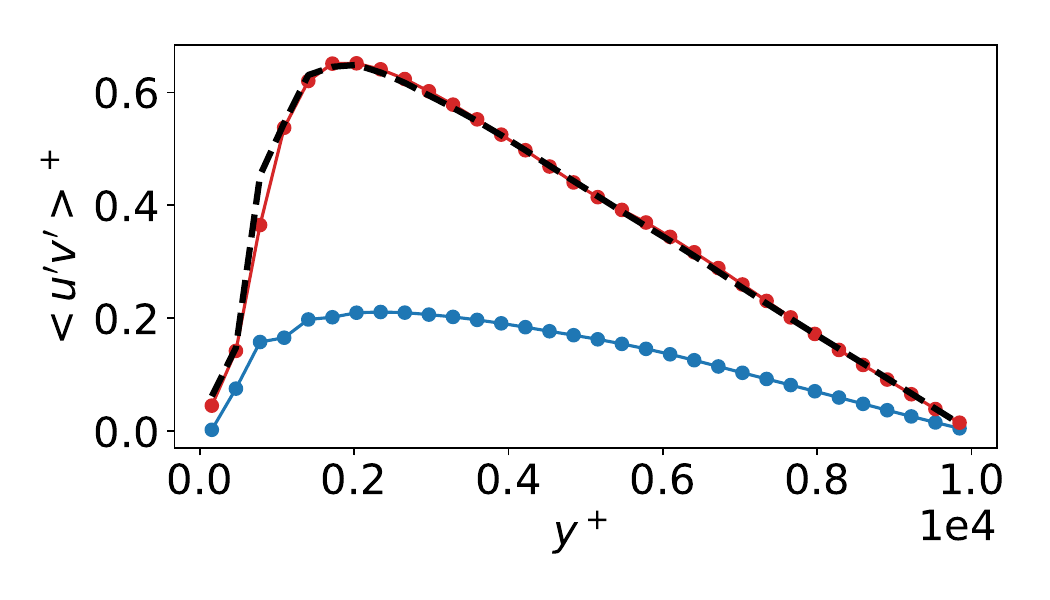}}\\
    \subfloat[]{\includegraphics[width=0.33\textwidth]{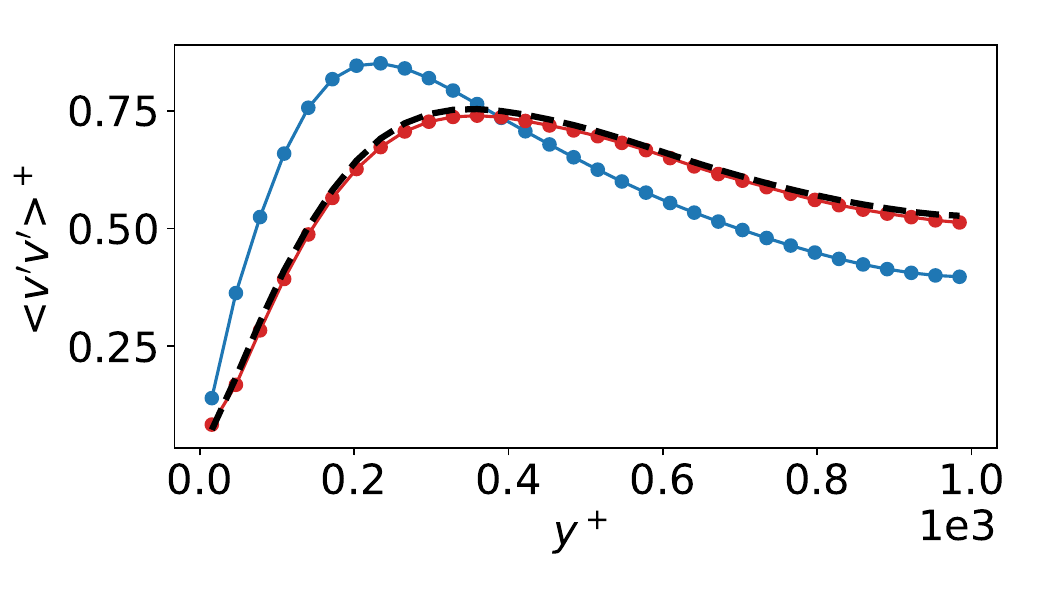}}\hfill
    \subfloat[]{\includegraphics[width=0.33\textwidth]{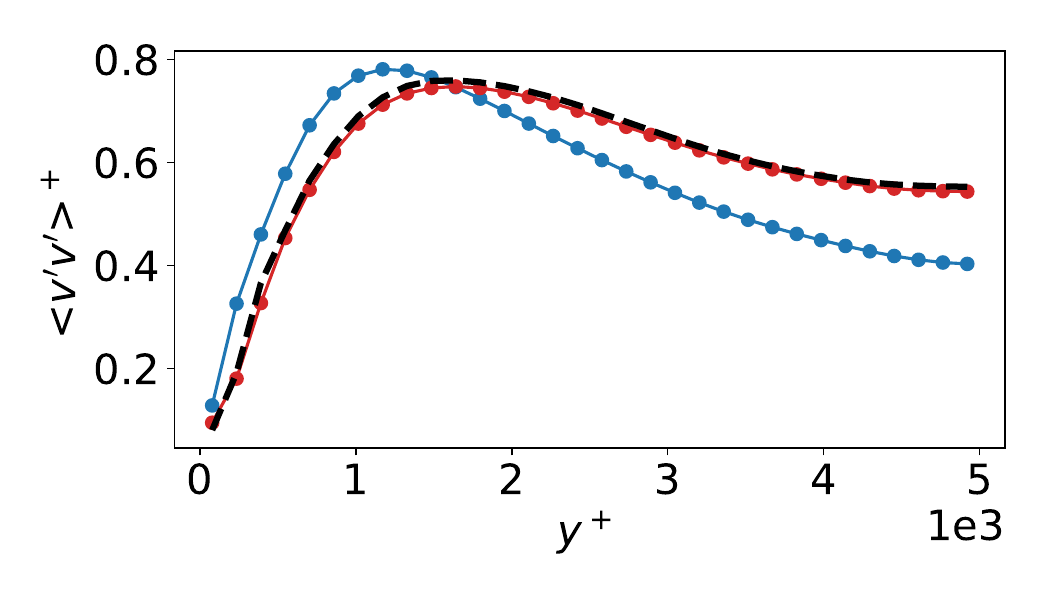}}\hfill
    \subfloat[]{\includegraphics[width=0.33\textwidth]{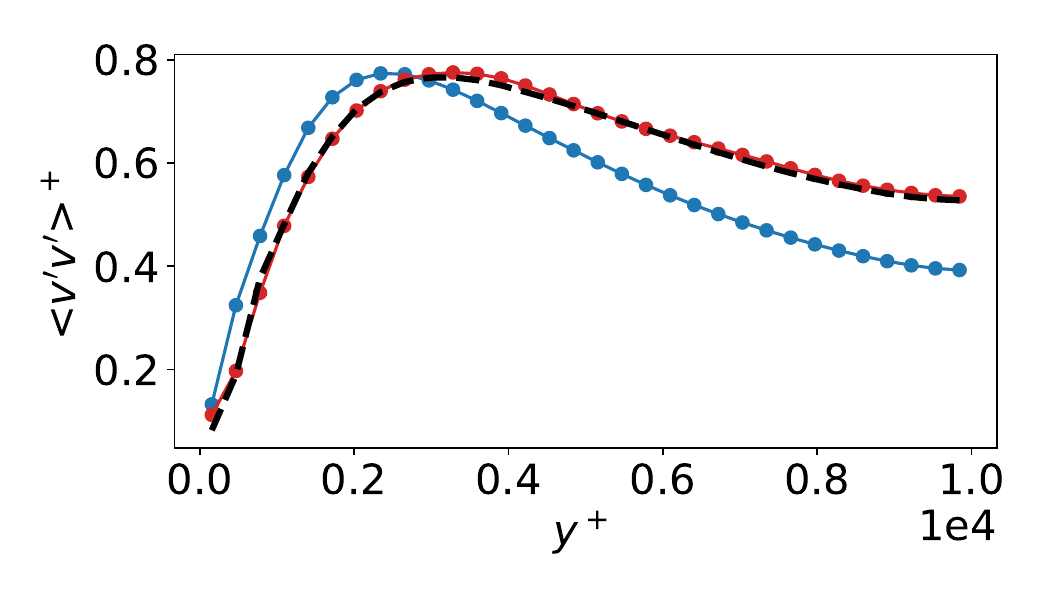}}\\
    \subfloat[]{\includegraphics[width=0.33\textwidth]{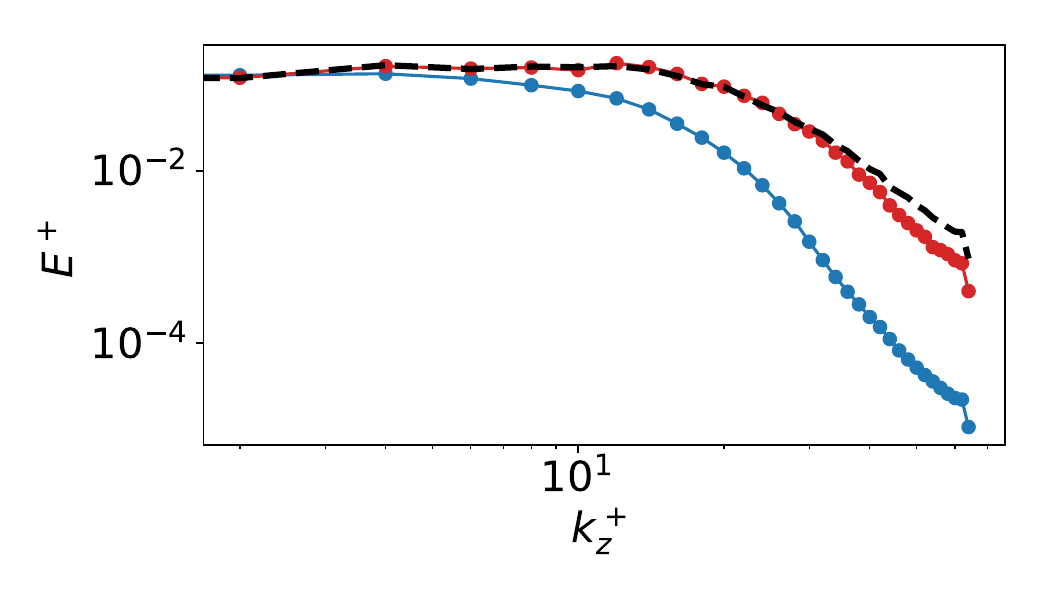}}\hfill
    \subfloat[]{\includegraphics[width=0.33\textwidth]{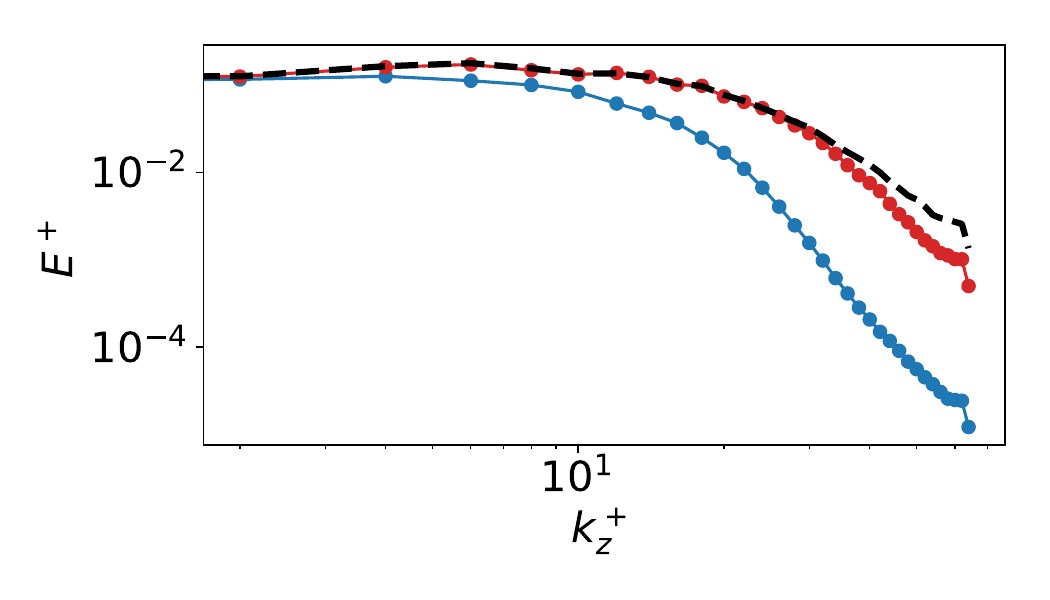}}\hfill
    \subfloat[]{\includegraphics[width=0.33\textwidth]{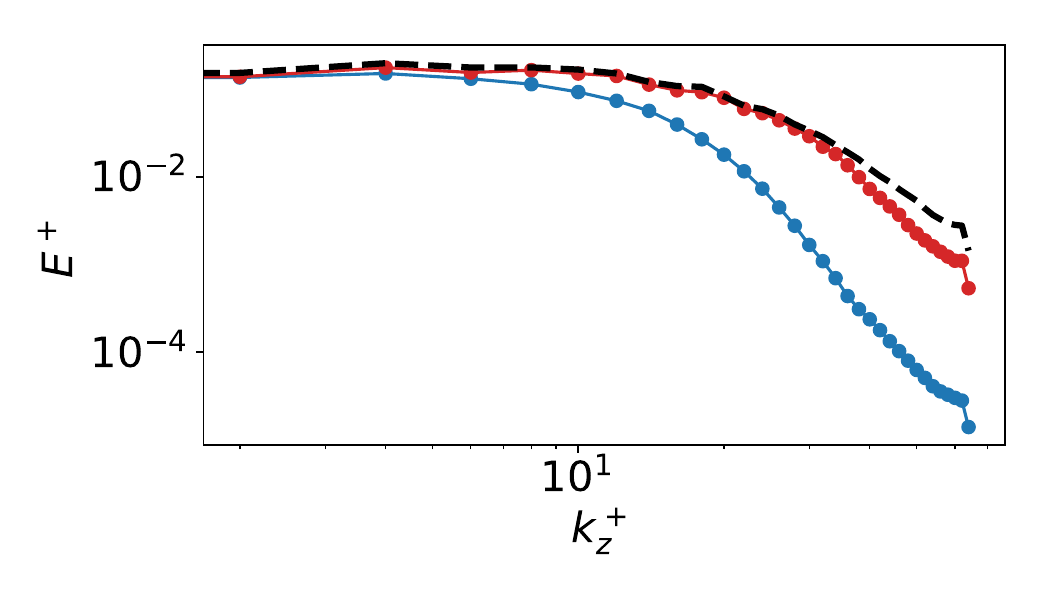}}\\
    \caption{Comparison of multiple statistics between {\newname} (\reddotline), DFM (\bluedotline) and recycling simulation (reference) data (\dashedblackline) at selected Reynolds numbers $Re_\tau = 1000$ (left column), 5000 (middle column), and 10000 (right column), respectively. (a-c) Mean stream wise velocity profile along wall normal direction. (d-f) Auto correlation of stream wise velocity ($R_{11}$) at the fifth off-wall grid point ($y^+\approx172, 859, 1718$ for (d), (e), and (f) respectively). (g-i) Reynolds shear stress along wall normal direction. (j-l) Turbulence intensity of the spanwise velocity component ($<v'v'>^+$). (m-o) Turbulence kinetic energy (TKE) at the fifth off-wall grid point ($y^+\approx172$, 859, and 1718 for (m), (n), and (o), respectively} 
    \label{fig:les_prior}
\end{figure}
We evaluate the performance of the well-trained {\newname} model by generating $1800$ continuous-in-time velocity snapshots $\bm{u}\in\mathbb{R}^{64\times64}$ at the inlet cross-section. The \emph{a priori} test results of {\newname} are compared against the DFM and reference data from the recycling simulation. For conciseness, we present results for three representative Reynolds numbers, $Re_\tau \in \{1, 5, 10\} \times 10^3$. Additional statistical comparisons for other Reynolds numbers can be found in \ref{appen:more_stat_les}. In Fig.~\ref{fig:les_prior}, each column corresponds to statistics from a specific Reynolds number: $Re_\tau = 1000$, $Re_\tau = 5000$, and $Re_\tau = 10,000$, arranged from left to right. The first row of Fig.~\ref{fig:les_prior} shows the mean streamwise velocity profiles along the wall-normal direction. Both {\newname} and DFM align well with the reference across all Reynolds numbers, demonstrating their ability to reproduce first-order statistics. However, consistent with the results in previous DNS cases, DFM struggles to provide acceptable higher-order statistics. As shown in Fig.~\ref{fig:les_prior}(d-f), DFM significantly underpredicts the auto-correlation ($R_{11}$) of the streamwise velocity at the fifth off-wall grid across all Reynolds numbers, whereas {\newname} closely follows the reference data with high accuracy. Similarly for the Reynolds shear stress ($\langle u'v'\rangle$, Fig.\ref{fig:les_prior}(g-i)) and the turbulence intensity of the spanwise velocity component ($\langle v'v'\rangle$, Fig.\ref{fig:les_prior}(j-l)), {\newname} consistently agree very well with the reference data, exhibiting negligible discrepancies. In contrast, despite utilizing the Reynolds stress tensor profiles extracted directly from the reference data, DFM generated flow fields still deviate significantly from the expected values in the \emph{a priori} tests. 
Figures~\ref{fig:les_prior}(m-o) show the TKE distribution at the fifth off-wall grid along the spanwise wavenumber ($k_z^+$). At lower wavenumbers, {\newname} matches the reference data very well. However, in the higher wavenumber regions ($k_z^+ > 40$), {\newname} exhibits a slightly faster decay in TKE, primarily due to the smoothing effect inherent to the neural field. Despite this, {\newname} shows significantly better agreement with the reference data compared to the DFM baseline across all Reynolds numbers and wall-normal locations.

\paragraph{\textbf{A posteriori test}}

\begin{figure}[!ht]
{
    \scriptsize
    \begin{tabularx}{\textwidth}{XXX}
    \centering
       $Re_\tau = 1000$  &\centering $Re_\tau = 5000$ &\centering $Re_\tau = 10000$
    \end{tabularx}
}
\captionsetup[subfloat]{farskip=-2pt,captionskip=-6pt}
    \centering
    \subfloat[$x\approx0.1$, $y^+\approx984$]{\includegraphics[width=0.33\textwidth]{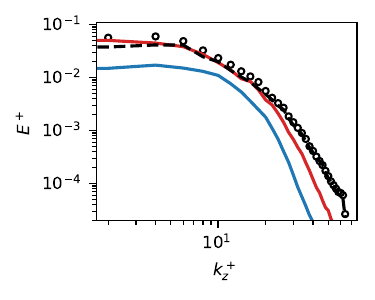}}
    \subfloat[$x\approx0.1$, $y^+\approx4920$]{\includegraphics[width=0.33\textwidth]{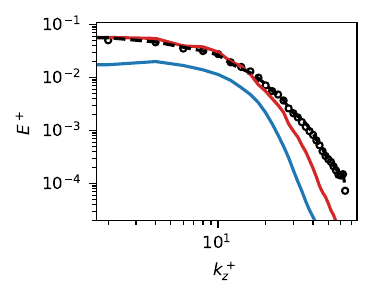}}
    \subfloat[$x\approx0.1$, $y^+\approx9844$]{\includegraphics[width=0.33\textwidth]{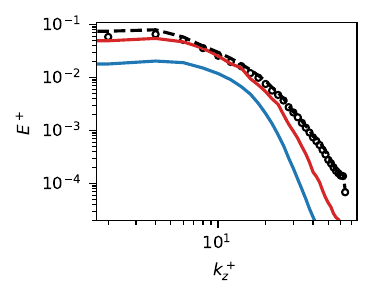}}\\
    \subfloat[$x\approx5\pi$, $y^+\approx984$]{\includegraphics[width=0.33\textwidth]{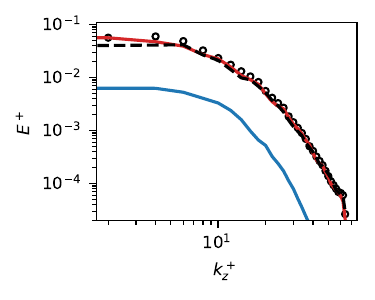}}
    \subfloat[$x\approx5\pi$, $y^+\approx4920$]{\includegraphics[width=0.33\textwidth]{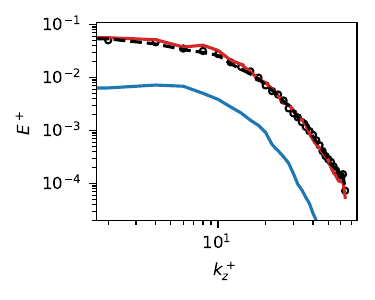}}
    \subfloat[$x\approx5\pi$, $y^+\approx9844$]{\includegraphics[width=0.33\textwidth]{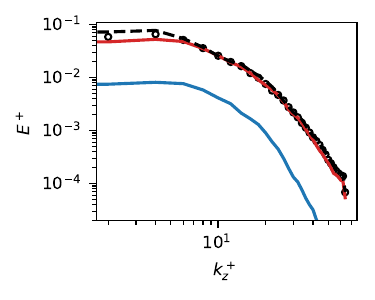}}\\
    \caption{Comparison of turbulence kinetic energy (TKE) spectra in \emph{a posteriori} tests between {\newname} (\redline), DFM (\blueline), the reference data (using recycling simulation as inlet boundary condition, \dashedblackline), and the training data (\blackcircle) near channel center ($y^+=984$, 4920, and 9844 for $Re_\tau = 1000$, 5000, and 10000, respectively.) Under selected Reynolds numbers $Re_\tau = 1000$ (left column), $5000$ (middle column), and $10000$ (right column), respectively. Subfigures in the same column shows the TKE evolution along stream wise. (a-c) TKE spectra near the inlet ($x=0.1$). (d-f) TKE spectra at $x=5\pi$}
    \label{fig:les_post_tke}
\end{figure}

We then use the generated inflows from these model as the inlet boundary condition for WMLES simulations to run \emph{a posteriori} test.  The computational domain is identical in size to that used in the DNS \emph{a posteriori} test ($l_x \times l_y \times l_z = 8\pi\times2\times\pi$), which is discretized into a grid of $N_x\times N_y\times N_z = 256\times64\times64$ mesh points.
\begin{figure}[!ht]
    {\scriptsize
    \begin{tabularx}{\textwidth}{XXX}
    \centering
       $Re_\tau = 1000$  &\centering $Re_\tau = 5000$ &\centering $Re_\tau = 10000$
    \end{tabularx}}
    \captionsetup[subfloat]{farskip=-2pt,captionskip=-6pt}
    \centering
    \subfloat[]{\includegraphics[width=0.33\textwidth]{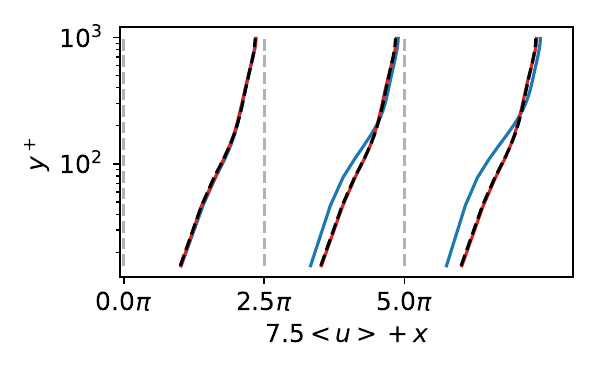}}
    \subfloat[]{\includegraphics[width=0.33\textwidth]{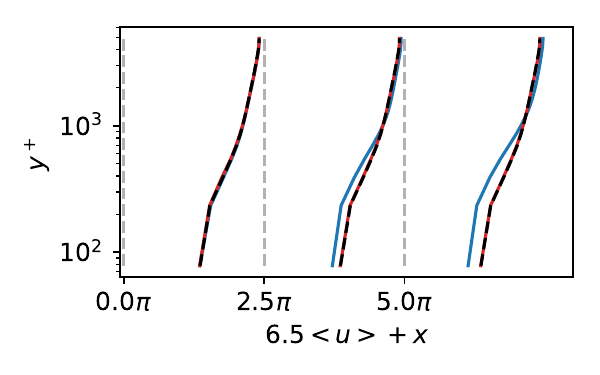}}
    \subfloat[]{\includegraphics[width=0.33\textwidth]{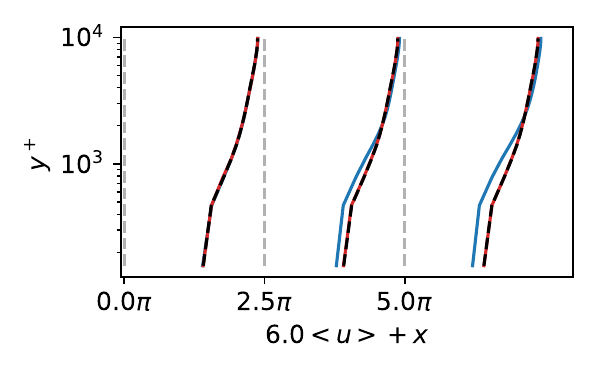}}\\
    \subfloat[]{\includegraphics[width=0.33\textwidth]{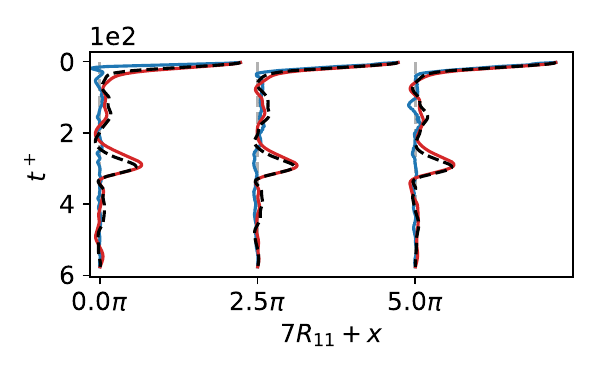}}
    \subfloat[]{\includegraphics[width=0.33\textwidth]{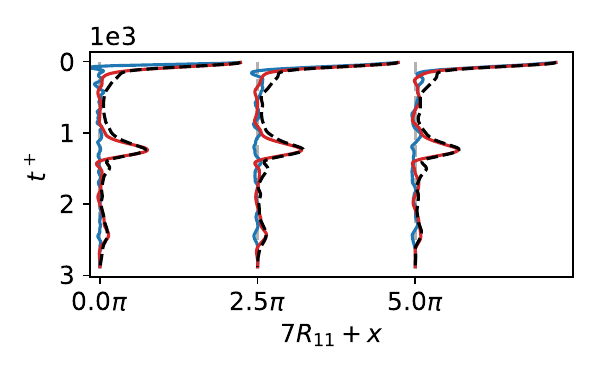}}
    \subfloat[]{\includegraphics[width=0.33\textwidth]{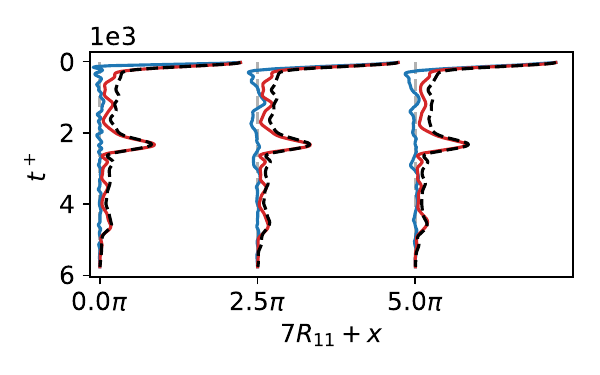}}\\
    \subfloat[]{\includegraphics[width=0.33\textwidth]{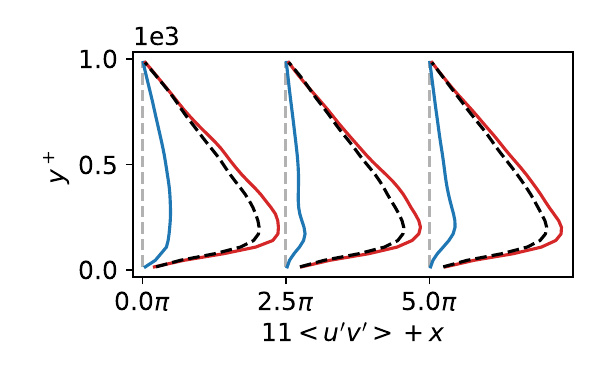}}
    \subfloat[]{\includegraphics[width=0.33\textwidth]{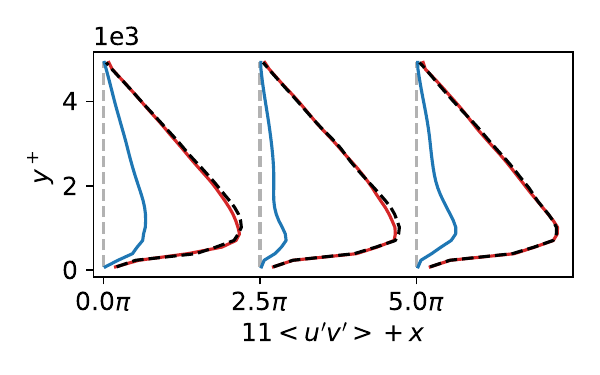}}
    \subfloat[]{\includegraphics[width=0.33\textwidth]{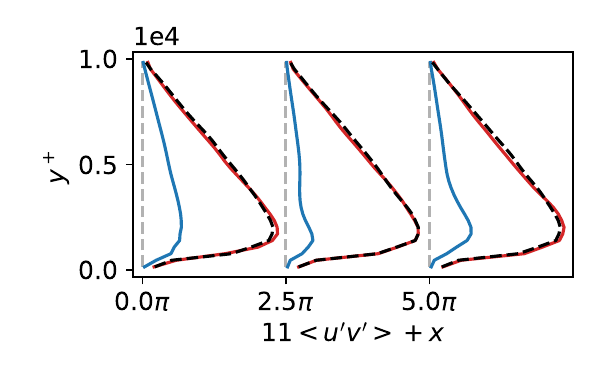}}\\
    \subfloat[]{\includegraphics[width=0.33\textwidth]{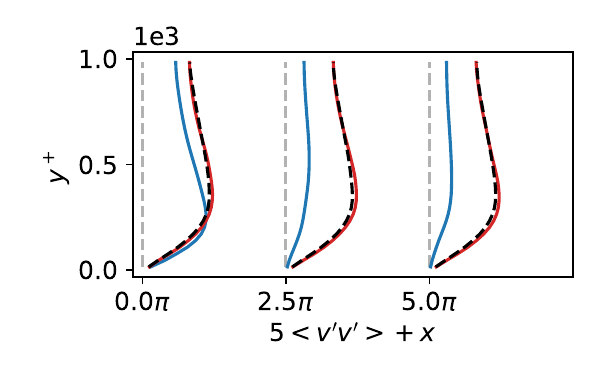}}
    \subfloat[]{\includegraphics[width=0.33\textwidth]{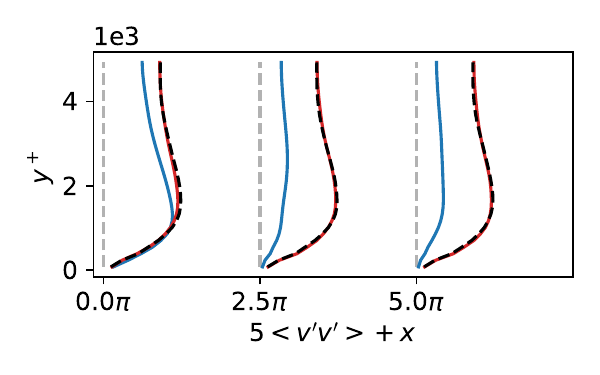}}
    \subfloat[]{\includegraphics[width=0.33\textwidth]{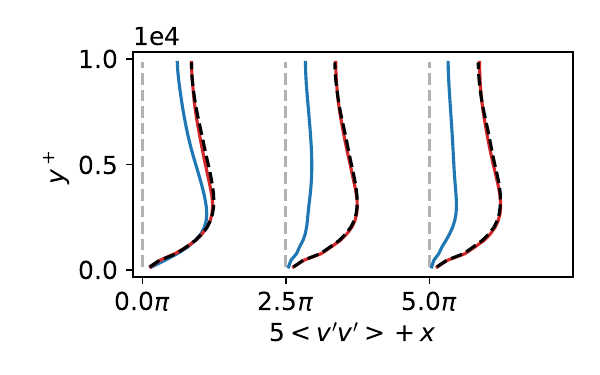}}\\
    \caption{Comparison of multiple statistics between {\newname} (\redline), DFM (\blueline) and the reference data (using recycling simulation as inlet boundary condition, \dashedblackline) at selected Reynolds numbers $Re_\tau = 1000$ (left column), $5000$ (middle column), and $10000$ (right column), respectively, at $3$ different stream wise locations (namely $x=0.098$, $2.5\pi$, $5\pi$). (a-c) Mean stream wise velocity profile. (d-f) Auto correlation of stream wise velocity ($R_{11}$) at the fifth off-wall grid point ($y^+\approx172, 859, 1718$ for (d), (e), and (f) respectively). (g-i) Reynolds shear stress along wall normal direction. (j-l) Turbulence intensity of the spanwise velocity component ($\langle v'v'\rangle^+$).}
    \label{fig:les_post}
\end{figure}
 Both the \emph{a posteriori} and \emph{a priori} tests share the same grid resolution and time step size. The simulation starts from a uniform zero-velocity initial condition and is run for $1800$ time steps. To eliminate the potential influence of initial conditions and the outlet boundary, statistics are calculated excluding the first flow-through time and the downstream region near the outlet ($x\in(5.5\pi, 8\pi)$). The results are summarized in Figs.~\ref{fig:les_post} and \ref{fig:les_post_tke}.
The evolution of the mean velocity profile along the streamwise direction is shown in Figs.~\ref{fig:les_post}(a-c). The results for {\newname} (\reddotline) align closely with the reference data (\dashedblackline) across all streamwise locations and Reynolds numbers. In contrast, DFM (\blueline) exhibits a growing deviation from the reference as the flow develops for all Reynolds numbers, indicating the significant decay of prescribed perturbations generated by DFM. The auto-correlations of the streamwise velocity components ($R_{11}$) at at the fifth off-wall grid ($y^+ \approx 172$, $859$, and $1718$ for $Re_\tau = 1000, 5000, 10000$, respectively) are presented in Figs.~\ref{fig:les_post}(d-f). At lower Reynolds number, e.g., $Re_\tau = 1000$, both {\newname} and DFM show a progressively improved alignment with the reference data along the downstream directly, but {\newname} exhibits significantly better agreement, particularly at $t^+ \approx 280$. At higher Reynolds numbers, {\newname} maintains superior consistency with the reference data across all streamwise locations, whereas DFM demonstrates much larger discrepancies, especially at longer time delays. The Reynolds shear stress ($\langle u'v' \rangle$) along the wall-normal direction is shown in Figs.~\ref{fig:les_post}(g-i). {\newname} effectively captures the correct magnitude and distribution of shear stress, maintaining close agreement with the reference data across all Reynolds numbers. In contrast, DFM significantly underpredicts the shear stress and exhibits distorted profiles, indicating its inability to model turbulent momentum transfer accurately. Finally, Figs.~\ref{fig:les_post}(j-l) illustrate the turbulence intensity of the spanwise velocity component ($\langle v'v' \rangle^+$). {\newname} accurately replicates the spanwise turbulent fluctuations at all streamwise locations and Reynolds numbers, while DFM exhibits reasonable agreement near the inlet but fails to sustain the turbulence intensity as the flow develops downstream. This decay underscores the importance of capturing all aspects of turbulence statistics to ensure sustained perturbations downstream.

The comparison of turbulence kinetic energy (TKE) spectra along the spanwise wavenumber ($k_z^+$) is shown in Fig.~\ref{fig:les_post_tke} for {\newname} (\redline), DFM (\blueline), the reference data (using training data as the inlet boundary condition, \dashedblackline), and the training data (\blackcircle) at varying Reynolds numbers ($Re_\tau = 1000$, 5000, and 10000). These comparisons focus on the channel center ($y^+ = 984$, 4920, and 9844 for $Re_\tau = 1000$, 5000, and 10000, respectively) at two streamwise locations: near the inlet ($x \approx 0.1$, top row) and near the outlet ($x \approx 5\pi$, bottom row).
In the near-inlet region ($x \approx 0.1$), shown in Figs.~\ref{fig:les_post_tke}(a-c), {\newname} (\redline) achieves significantly better alignment with the reference data (\dashedblackline) compared to the DFM baseline, across all wavenumbers and Reynolds numbers. Although minor discrepancies remain in the high wavenumber regimes, {\newname} demonstrates excellent accuracy in capturing low-wavenumber components, indicative of its ability to effectively model large-scale turbulence structures. The DFM, by contrast, struggles to replicate the TKE spectra at both low and high wavenumbers.
The downstream TKE spectra at $x \approx 5\pi$ are shown in Figs.~\ref{fig:les_post_tke}(d-f). Here, {\newname} significantly outperforms DFM with a much better recovery, closely matching the reference data across all Reynolds numbers, particularly at high wavenumbers where small-scale turbulent structures are prevalent. This highlights the capability of {\newname} to preserve and accurately reconstruct small-scale features throughout the development of the flow. While minor discrepancies between {\newname} and the reference data persist at low wavenumbers, it is notable that the reference data also deviates from the recycling simulation (\blackcircle, i.e., the training dataset). This suggests that these discrepancies are likely due to differences in numerical settings between the \emph{a priori} and \emph{a posteriori} tests, rather than shortcomings of {\newname}. In contrast, DFM continues to fail in recovering the TKE spectra and it shows continuous delay after considerable downstream development, consistently underpredicting TKE across all scales. This underscores the failure of DFM in generating high-quality turbulence at the inlet compared to {\newname}. Additional statistical comparisons at other Reynolds numbers can be found in \ref{appen:more_stat_les}.

\subsubsection{Performance on unseen Reynolds numbers}
To further evaluate the robustness and generalizability of the proposed methods, we perform \emph{a posteriori} tests on nine interpolated Reynolds numbers $Re_\tau\in\{1500, 2500,\cdots,  9500\}$ using inflow generated by {\newname}. For comparison, we also use recycling simulation data at these Reynolds numbers as well as DFM generated synthetic inlet to conduct the same \emph{a posteriori} test. In practice, incorporating the mean velocity profile as additional conditioning through DPS proves crucial for improving performance. It is worth noting that the mean velocity profile required by {\newname} is simply an algebraic average of the profiles from two adjacent Reynolds numbers within the training set. In contrast, the DFM requires much more input flow information (e.g., mean velocity, Reynolds stress) directly extracted from the recycling simulation at the target Reynolds numbers.
\begin{figure}[!ht]
    {\scriptsize
    \begin{tabularx}{\textwidth}{XXX}
    \centering
       $Re_\tau = 1500$  &\centering $Re_\tau = 5500$ &\centering $Re_\tau = 9500$
    \end{tabularx}}
    \captionsetup[subfloat]{farskip=-2pt,captionskip=-6pt}
    \centering
    \subfloat[]{\includegraphics[width=0.33\textwidth]{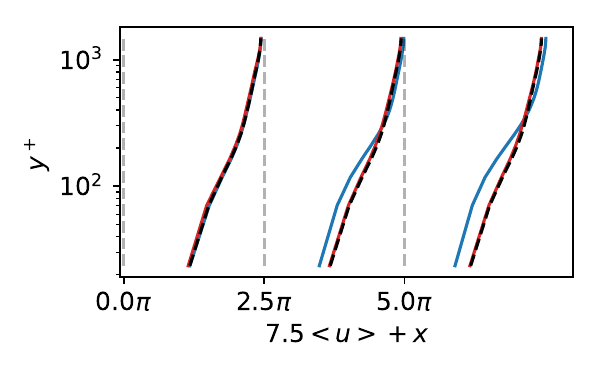}}
    \subfloat[]{\includegraphics[width=0.33\textwidth]{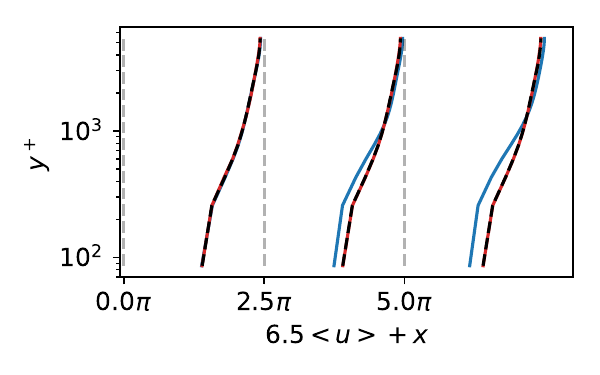}}
    \subfloat[]{\includegraphics[width=0.33\textwidth]{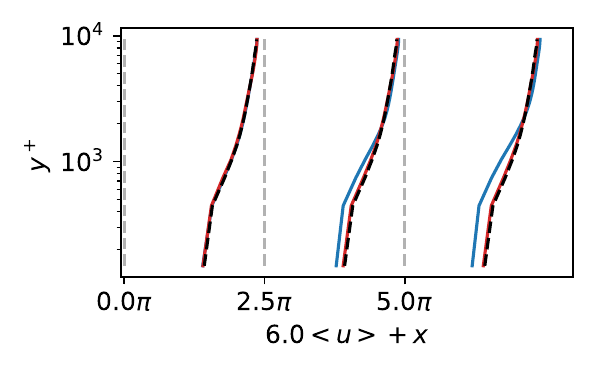}}\\
    \subfloat[]{\includegraphics[width=0.33\textwidth]{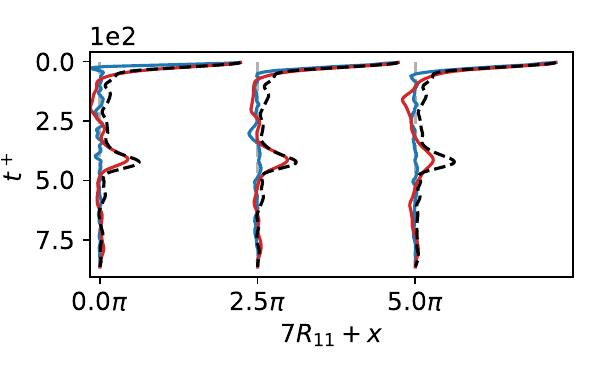}}
    \subfloat[]{\includegraphics[width=0.33\textwidth]{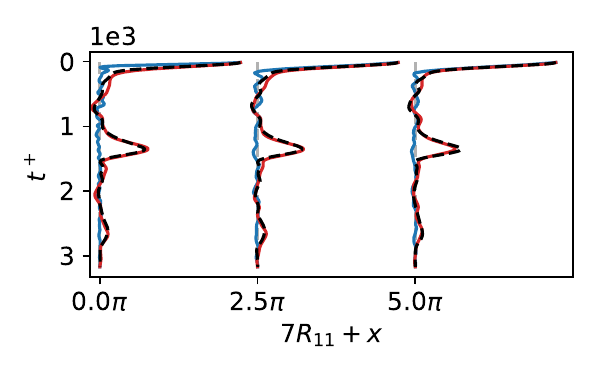}}
    \subfloat[]{\includegraphics[width=0.33\textwidth]{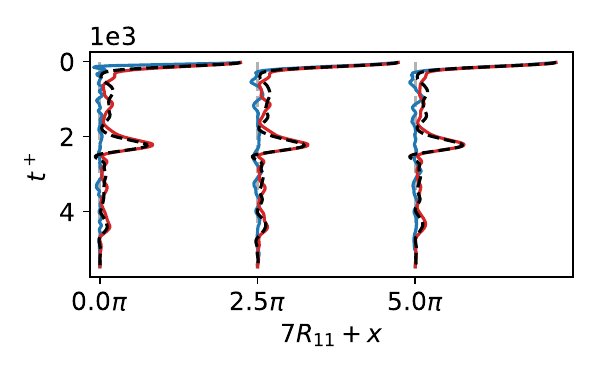}}\\
    \subfloat[]{\includegraphics[width=0.33\textwidth]{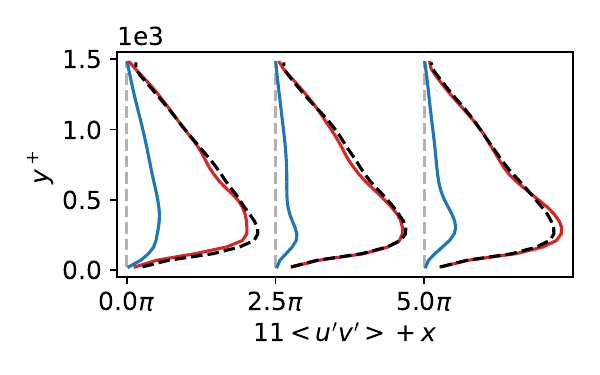}}
    \subfloat[]{\includegraphics[width=0.33\textwidth]{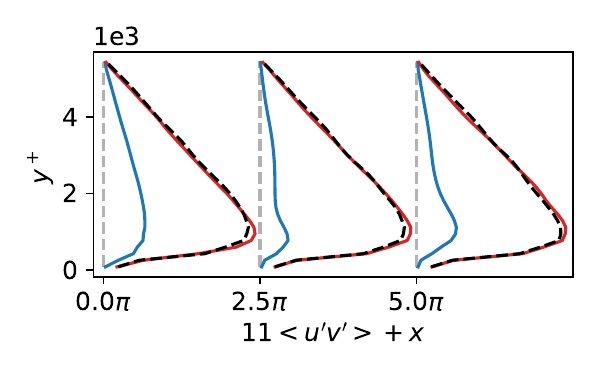}}
    \subfloat[]{\includegraphics[width=0.33\textwidth]{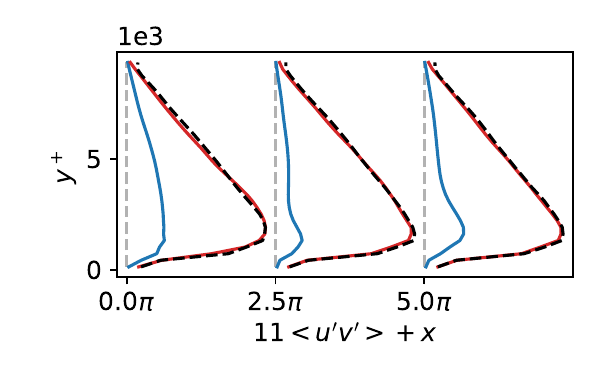}}\\
    \subfloat[]{\includegraphics[width=0.33\textwidth]{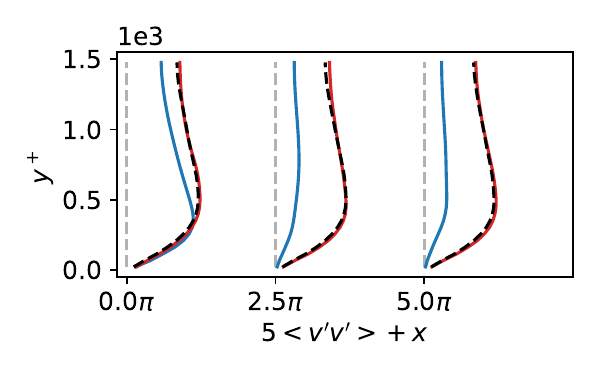}}
    \subfloat[]{\includegraphics[width=0.33\textwidth]{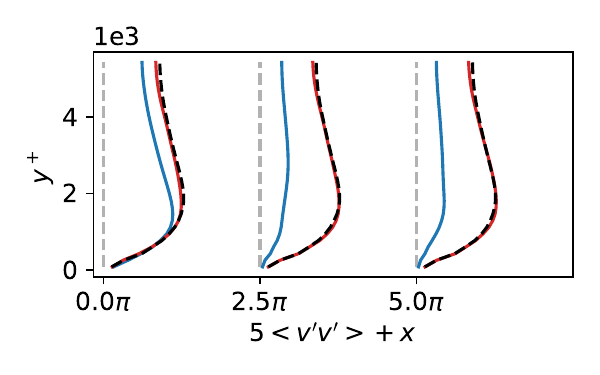}}
    \subfloat[]{\includegraphics[width=0.33\textwidth]{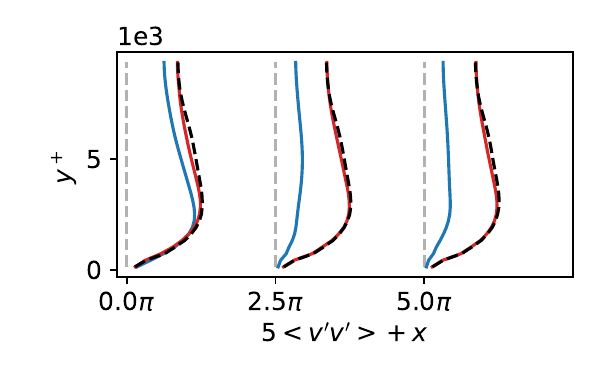}}\\
    \caption{Comparison of multiple statistics between {\newname} (\redline), DFM (\blueline) and the reference data (using recycling simulation as inlet boundary condition, \dashedblackline) at unseen Reynolds numbers $Re_\tau = 1500$ (left column), $5500$ (middle column), and $9500$ (right column), respectively, at $3$ different stream wise locations (namely $x=0.098$, $2.5\pi$, $5\pi$). (a-c) Mean stream wise velocity profile. (d-f) Auto correlation of stream wise velocity ($R_{11}$) at the fifth off-wall grid point ($y^+\approx258, 945, 1633$ for (d), (e), and (f) respectively). (g-i) Reynolds shear stress along wall normal direction. (j-l) Turbulence intensity of the spanwise velocity component ($\langle v'v'\rangle^+$).}
    \label{fig:int_re_post}
\end{figure}
Figure~\ref{fig:int_re_post} presents comparisons of multiple turbulence statistics between {\newname} (\redline), DFM (\blueline), and the reference (using recycling simulation data as the inlet boundary condition, \dashedblackline) at selected unseen Reynolds numbers ($Re_\tau = 1500$, 5500, and 9500). Similar to previous section, these comparisons are performed at three different streamwise locations: $x = 0.098$, $2.5\pi$, and $5\pi$.
The first row of Fig.~\ref{fig:int_re_post} (a-c), showing the mean streamwise velocity profiles, {\newname} (\redline) closely matches the reference (\dashedblackline) across all examined Reynolds numbers and locations, even with slightly biased mean velocity conditioning. In contrast, although the DFM (\blueline) method utilized the some ground truth statistics of the reference, the \emph{a posteriori} simulation results still progressively diverges from the reference, particularly in the near-wall regions, as the flow develops along the streamwise direction.
The second row of Fig.~\ref{fig:int_re_post} (d-f) shows the auto-correlation of the streamwise velocity component ($R_{11}$) at the fifth off-wall grid point ($y^+ \approx 258$, 945, and 1633 for (d), (e), and (f), respectively). {\newname} effectively captures the temporal coherence, maintaining good agreement with the reference. While minor discrepancies are observed at $Re_\tau=1500$, {\newname} still significantly outperforms DFM, which exhibits significant deviations, indicating its failure to accurately reproduce temporal dynamics.
The Reynolds shear stress ($\langle u'v' \rangle$) along the wall-normal direction is shown in the third row of Fig.~\ref{fig:int_re_post} (g-i). {\newname} consistently maintains a close agreement with the reference, accurtely capturing both the magnitude and the distribution of the shear stress, particularly at higher Reynolds numbers. Conversely, DFM (\blueline) significantly under-predicts the Reynolds shear stress and exhibits a much flatter profile, highlighting its inability to model turbulent momentum transfer effectively.
Finally, the last row of Fig.~\ref{fig:int_re_post} (j-l)  shows the turbulence intensity of the spanwise velocity component  ($\langle v'v' \rangle$). {\newname} aligns well with the reference data across all Reynolds numbers and streamwise locations, demonstrating its capability to reproduce spanwise turbulent fluctuations accurately. In contrast, DFM deviates substantially, with a obvious decay of the turbulence intensity, especially near the wall. 

Overall, {\newname} demonstrates superior \emph{a posteriori} performance in capturing multiple turbulence statistics across a wide range of interpolated Reynolds numbers compared to DFM. These results highlight {\newname}'s ability to generate inflow conditions that closely match the desired turbulent statistics, offering a robust and versatile solution for unseen Reynolds number scenarios in turbulence inlet generation.



\section{Discussion}
\label{sec:discussion}

\subsection{Further analysis and comparison between {\newname} and deterministic models}
\label{sec:analyse}
\begin{figure}[!ht]
    \centering
    \captionsetup[subfloat]{farskip=-4pt,captionskip=-8pt}
    \subfloat{\includegraphics[width=.33\textwidth]{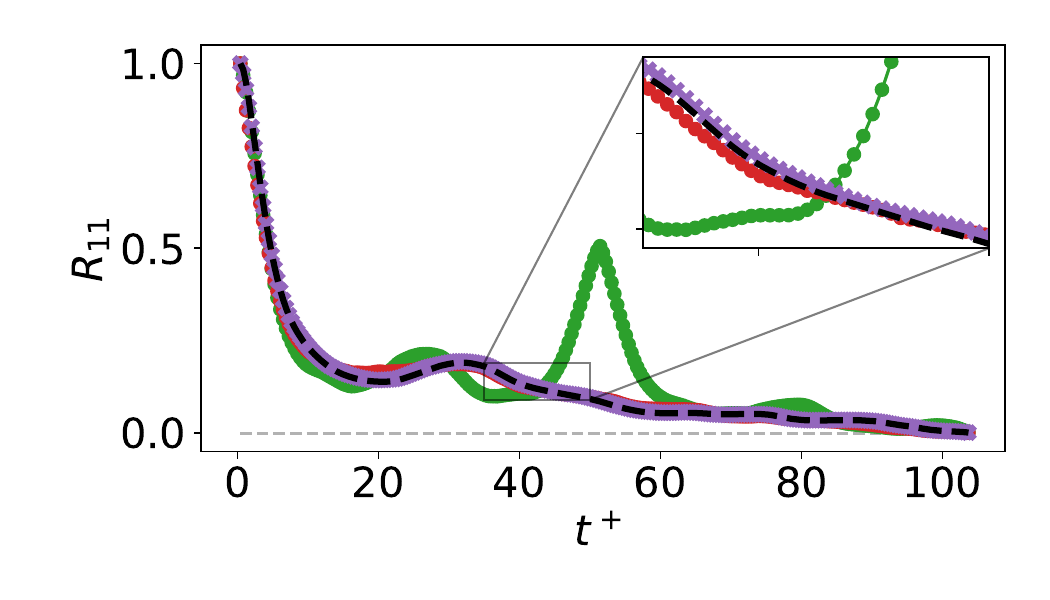}}
    \subfloat{\includegraphics[width=.33\textwidth]{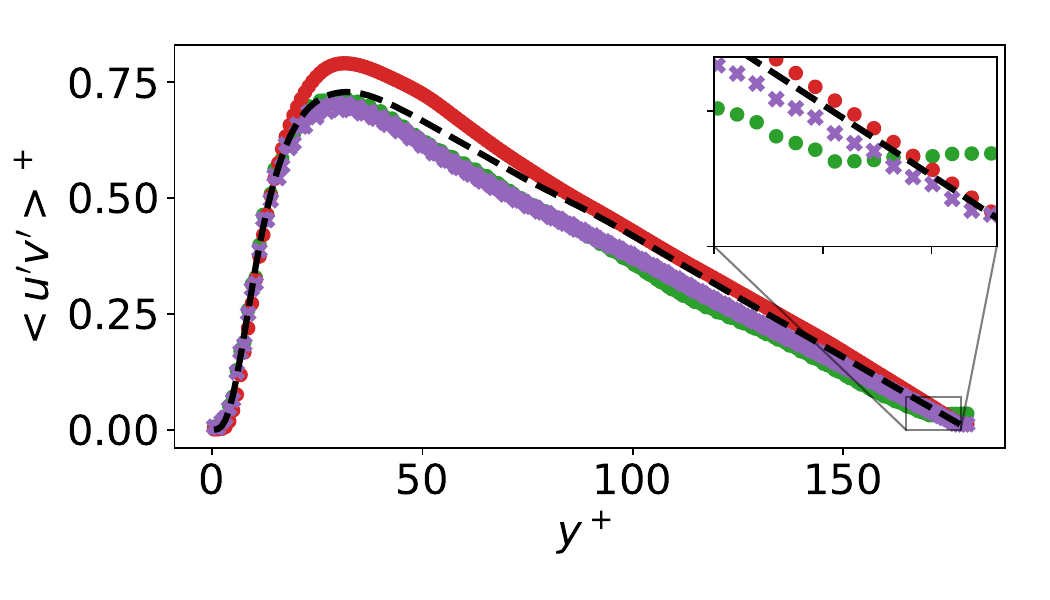}}
    \subfloat{\includegraphics[width=.33\textwidth]{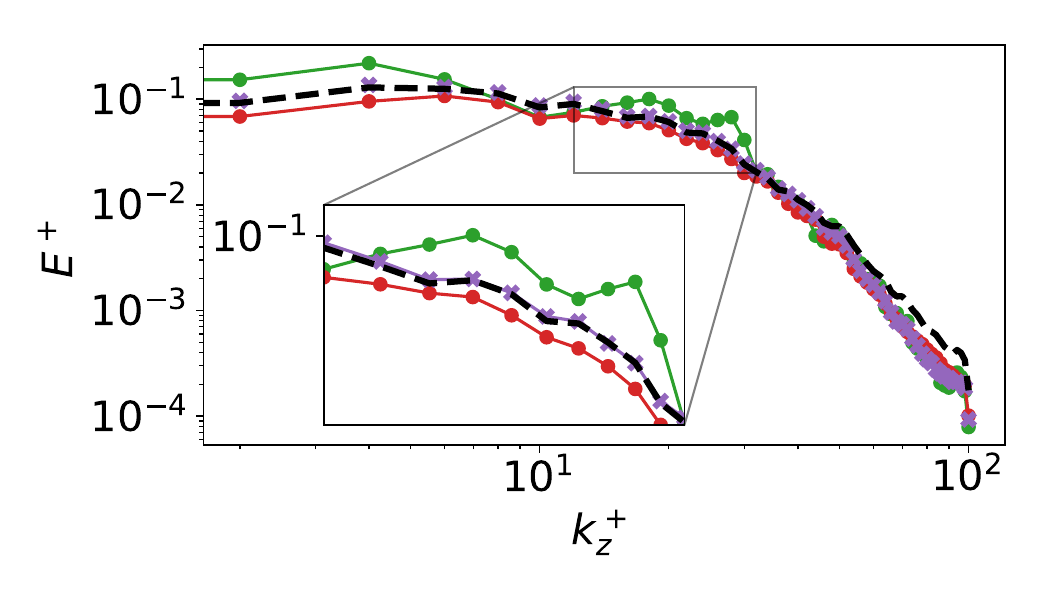}}
    \vspace{-0.8em}
    \caption{Statistics comparison in \emph{a priori} test between spatiotemporal velocity field randomly sampled from {\oldname} (\reddotline), prediction by CNN-LSTM within the training time range (\purplecross), and prediction by CNN-LSTM beyond the training time range (\greendot). Left panel: Auto-correlation ($R_{11}$) at $y^+\approx5$. Middle panel: Reynolds shear stress ($\langle u'v'\rangle$) along wall normal direction. Right panel: turbulence kinetic energy (TKE) spectra along spanwise wavenumber ($k_z^+$) at $y^+\approx5$}
    \label{fig:memorizatoin}
\end{figure}
In the Results section (Sec. \ref{sec:DNS}), we compared both the \emph{a priori} and \emph{a posteriori} performance between CNN-LSTM and {\oldname} by generating unseen data. Since our {\oldname} is a probabilistic model, we randomly sampled one spatiotemporal velocity field for comparison. While for the CNN-LSTM, we started from the first $5$ time step of the training set and rollout the model for $1200$ times steps until it reached the unseen time steps. Then we collect the following $1800$ time steps predicted by the CNN-LSTM for comparison. However, we find that CNN-LSTM has a significant performance drop when transit from training set to the unseen data. 
Figure~\ref{fig:memorizatoin} shows the statistics comparison in the \emph{a priori} test between the CNN-LSTM on training set (\greendot), CNN-LSTM on unseen data (\purplecross), random realization sampled from {\oldname} (\reddotline), and the reference data (\dashedblackline). Among all the statistics shown in Fig~\ref{fig:memorizatoin}, the CNN-LSTM on training set aligns much better with the reference data than on the testing set and comparable with our CoNFiLD. Specifically, the left panel of Fig.~\ref{fig:memorizatoin} shows the auto correlation ($R_{11}$) at $y^+\approx5$. CNN-LSTM on testing set shows a significant deviation from the reference data at $t^+\in(40,60)$, while CNN-LSTM on training set matches well with the reference data at the same temporal locations. The better captured statistics of CNN-LSTM on training set can also be observed in the Reynolds shear stress shown in the middle panel of Fig.~\ref{fig:memorizatoin}, especially near the channel center. CNN-LSTM on the testing set shows an unphysically bent curve of Reynolds shear stress ($\langle u'v'\rangle$) near channel center, while on the training set CNN-LSTM still shows a smooth striaght line which aligns with the reference data and {\oldname}. TKE (shown in right pannel of Fig.~\ref{fig:memorizatoin}) also reflects the decreased performance of CNN-LSTM on the testing set, with significant deviation at low wave number region and shows a weird bump around $k_z^+\in(12, 35)$. While on training set, CNN-LSTM even shows a slightly better agreement with reference data than {\oldname}, although both of them under-predicted the TKE in the high wavenumber region at minor level.

\begin{figure}[!ht]
    \centering
    \captionsetup[subfloat]{farskip=-4pt,captionskip=-8pt}
    \subfloat{\includegraphics[width=0.95\textwidth]{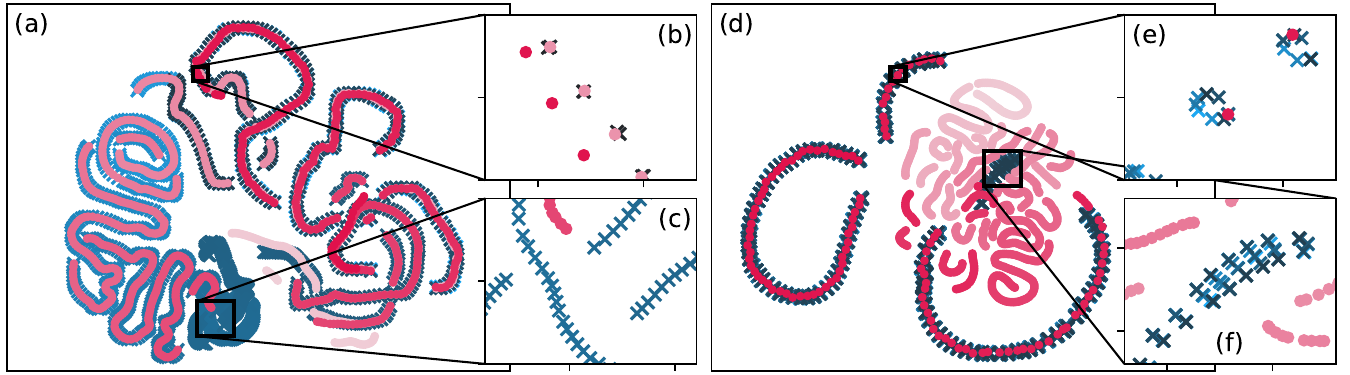}}\\
    \vspace{10pt}
    \hfill
    \subfloat{\includegraphics[width=.35\textwidth]{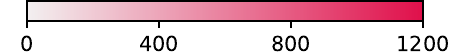}}\quad
    \subfloat{\includegraphics[width=.35\textwidth]{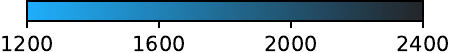}}\hfill 
    \null
    \caption{t-SNE analysis of the 2400 time snapshots of velocity field $\bm{u}$ generated by {\oldname} (a,b,c) and CNN-LSTM (d,e,f). The first 1200 steps are marked by reddish dots while the last 1200 snapshots are represented by blueish crosses. Panel b,c and panel e,f are the zoom-in views of (a) and (d), respectively. The location of the zoom-in view is indicated by the black box. }
    \label{fig:tsne_main}
\end{figure} 
To further analyse what caused the performance difference of CNN-LSTM on testing and training set, we performed an t-distributed Stochastic Neighbor Embedding (t-SNE) analysis of the generated velocity field as well as the label data. We perform $3$ separate t-SNE for each of those velocity dataset. All those three velocity fields are continuous in time. 
Specifically, since CNN-LSTM predicts from the beginning of the label data and autoregressively predicts the following $2400$ time steps, the first $1200$ steps ($\bm{\mathcal{T}}_1$) are supposed to be close to the training set while the last $1200$ steps ($\bm{\mathcal{T}}_2$) predicted by CNN-LSTM can be viewed as an extrapolation in time. Figure~\ref{fig:tsne_main} shows the $2400$ snaphosts of {\oldname} data and $2400$ time steps of CNN-LSTM predicted data projected onto the t-SNE embeddings $b_1$and $b_2$. (the t-SNE visualization of the label data to the can be found in Fig.~\ref{fig:tsne_label}). The flow field at each time step in $\bm{\mathcal{T}}_1$ are marked by reddish dots while the $\bm{\mathcal{T}}_2$ snapshots are represented by blueish crosses. Although we marked the first half and second half of the data with two different series of markers, the representation difference is only meaningful for the CNN-LSTM data. While for {\oldname}, all the data should be considered as ``newly generated data'' or unseen data since they are randomly sampled. 
The left 3 panels (a,b, and c) of figure \ref{fig:tsne_main} represent the t-SNE visualization of {\oldname} generated velocity fields, while the right 3 subfigures (d, e, and f) of \ref{fig:tsne_main} show the CNN-LSTM data in the t-SNE space. 
In general, {\oldname} shows similar behaviour in both $\bm{\mathcal{T}}_1$ (red dots) and $\bm{\mathcal{T}}_2$ (blue crosses), while the t-SNE visualization for CNN-LSTM reveals a distinct difference between the interpolation time range and extrapolation range. When we compare panel (a) and (d) of Fig.~\ref{fig:tsne_main}, CoNFiLD covers a significantly larger area than the CNN-LSTM, especially for the last half of the time steps ($\bm{\mathcal{T}}_2$).
Although both {\oldname} and CNN-LSTM have repeated samples from $\bm{\mathcal{T}}_2$ which overlap with samples from $\bm{\mathcal{T}}_1$, {\oldname} shows a significantly lower chance of generating those repeated snapshots. Comparing the zoom-in views of those repeated samples in Fig.~\ref{fig:tsne_main}b and Fig.~\ref{fig:tsne_main}e, we can observe that there are only two samples from $\bm{\mathcal{T}}_2$ around each sample from $\bm{\mathcal{T}}_1$ for the {\oldname} generated field, while for the CNN-LSTM, there are 8 samples from $\bm{\mathcal{T}}_2$ around each sample from $\bm{\mathcal{T}}_1$. Besides, the repeated samples generated by CNN-LSTM are temporally separated (marked by the brightness of blue), which indicate the CNN-LSTM keeps repeating similar velocity fields at multiple different time steps. 
Similar behaviour can also be observed in the comparison between panel (c) and panel (f) of Fig. \ref{fig:tsne_main}. These two panels show a zoom-in view of two regions where the samples from $\bm{\mathcal{T}}_2$ do not overlap with samples from $\bm{\mathcal{T}}_1$ for data generated by {\oldname} and CNN-LSTM, respectively. For those non-repeated samples, the difference in the covered regions between {\oldname} and CNN-LSTM are pronounced. The significantly wider distribution range of {\oldname} indicate much more unique samples are generated by the {\oldname} than CNN-LSTM. Besides, among those unique samples, we hardly observe {\oldname} generated ones overlap with each other, while the CNN-LSTM generated samples still highly concentrated on a single curve.
It seems that the $\bm{\mathcal{T}}_2$ samples given by CNN-LSTM exhibit an attracting limit cycle in the t-SNE space, which leads to very limited and repeated flow patterns in the physical space once the model goes beyond the training set. The limit cycle also explains the extremely over-predicted auto-correlation around $t^+\in (45,60)$ shown in figure \ref{fig:memorizatoin}(a). While {\oldname} significantly reduce the frequency of repeated samples and the samples cover a much wider range in both the t-SNE space and physical space, which is reflected by the well aligned statistics with the reference data.  

\subsection{Mean velocity profile conditioning for interpolated Re}
\begin{figure}[!ht]
    \centering
    \captionsetup[subfloat]{farskip=-2pt,captionskip=-6pt}
    \subfloat[]{\includegraphics[width=0.33\textwidth]{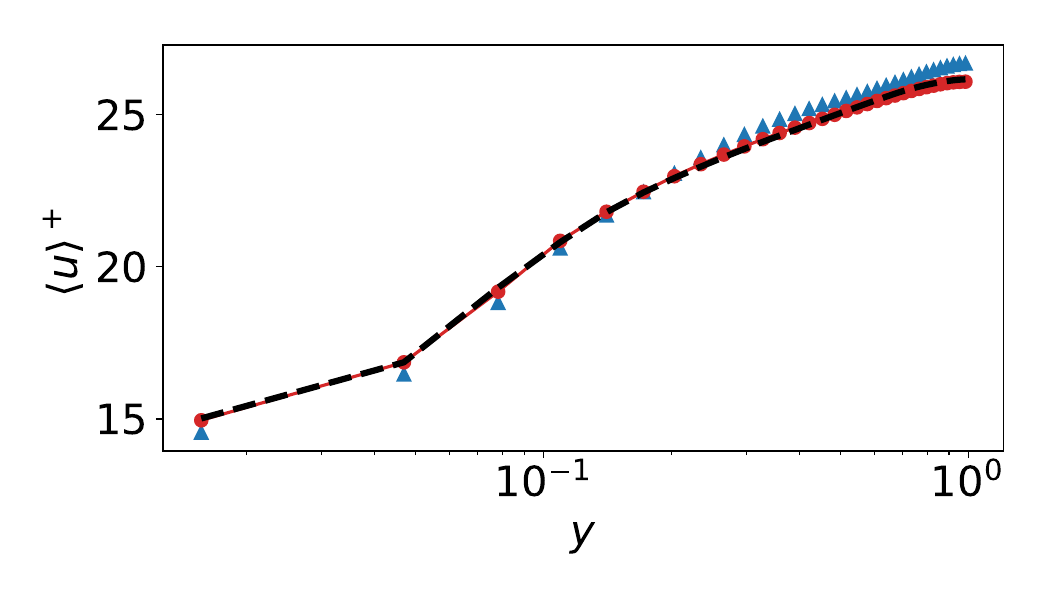}}
    \subfloat[]{\includegraphics[width=0.33\textwidth]{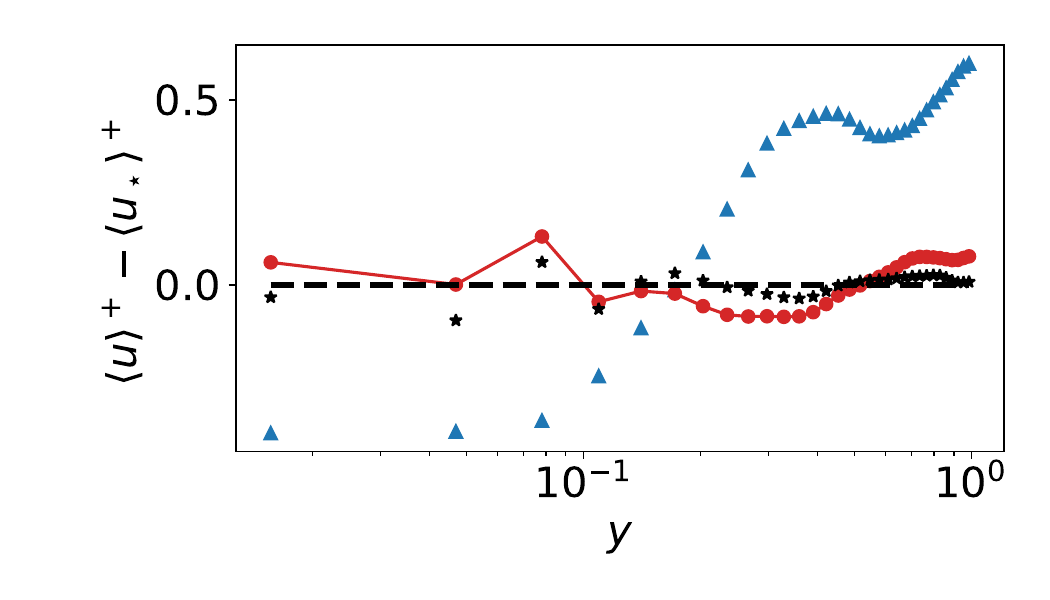}}
    \subfloat[]{\includegraphics[width=0.33\textwidth]{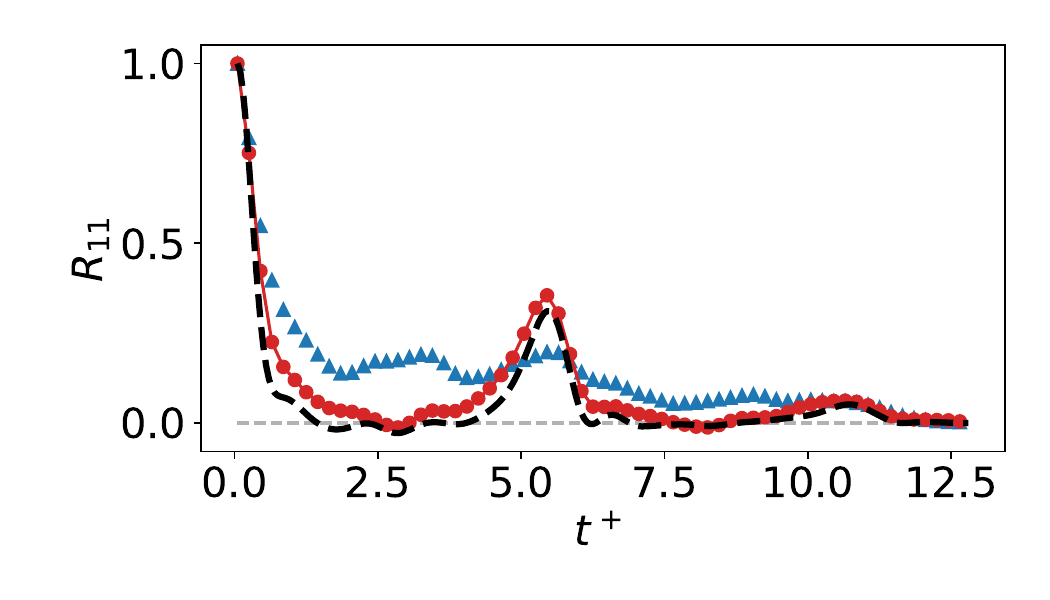}}\\
    \subfloat[]{\includegraphics[width=0.33\textwidth]{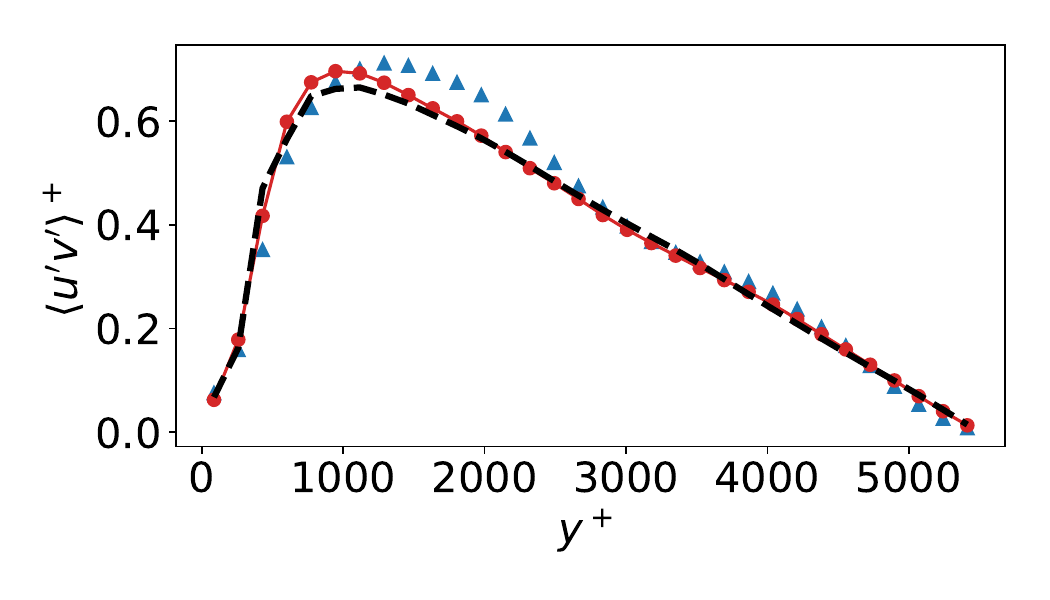}}
    \subfloat[]{\includegraphics[width=0.33\textwidth]{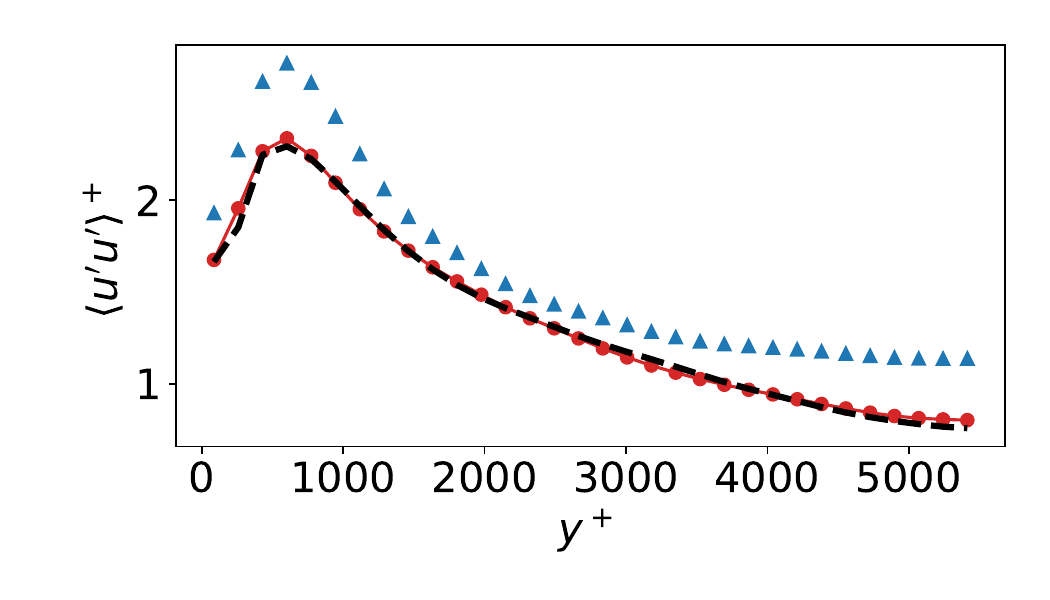}}
    \subfloat[]{\includegraphics[width=0.33\textwidth]{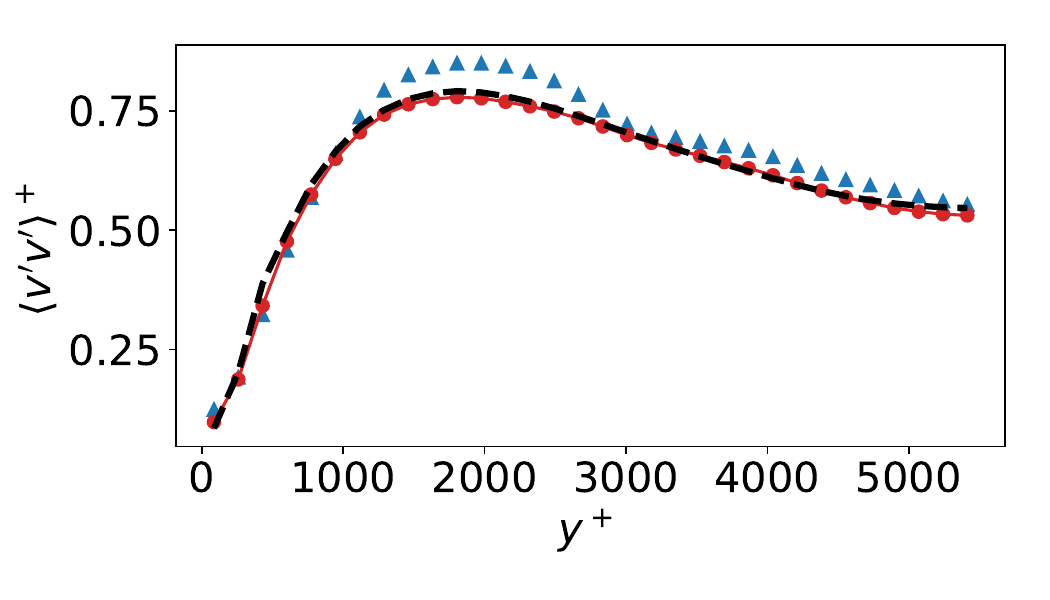}}
    \caption{Statistics comparison in \emph{a priori} test of the velocity field generated by {\newname} (\reddotline), {\newname} without conditioning on mean velocity (\bluetri), the mean velocity {\newname} conditioned on (\blackstar), and the reference data (\dashedblackline) at $Re_\tau = 5500$. (a) Mean velocity profile ($\langle u^+\rangle$) along wall normal direction. (b) Discrepancy between the mean velocity of generated flow field the mean velocity of reference data ($u_\star^+$). (c) Auto correlation ($R_{11}$) at $y^+\approx5414$ (near channel center). (d) Reynolds shear stress ($\langle u'v'\rangle$) along wall normal direction. And Turbulence intensity for velocity component $u$ (e) and $v$ (f)}
    \label{fig:inter_re_stats}
\end{figure}
When applying {\newname} to generating velocity field at unseen $Re_\tau$, it is critical to leverage mean velocity profile as posterior sampling condition for the diffusion model. In this section, we discuss about the necessity about using the extra condition information of mean velocity using the example of $Re_\tau = 5500$. It is worth noting that, it is only for the ease of implementation that we choose to use the algebraic average of the mean velocity of two adjacent Reynolds number from the training set as the conditioning information. It is also possible to use other prior knowledge to get the conditioning information (e.g. log law). Figure~\ref{fig:inter_re_stats} shows the statistics comparison between the {\newname} generated velocity (\reddotline), {\newname} without mean velocity conditioning (\bluetri), and the reference data (\dashedblackline) in \emph{a priori} test. Fig.\ref{fig:inter_re_stats}(a) shows the mean velocity comparison between them. When without the mean velocity comparison, the generated mean velocity deviates from the reference data ($u_\star$), which is shown more clearly in the panel (b). Without mean velocity condition, the model significantly under-predicts the mean velocity at near wall region while over-predicts near channel center. The discrepancy of the condition information is represented by (\blackstar) in Fig. \ref{fig:inter_re_stats}(b). Since this condition information is from the average of two adjacent Reynolds numbers ($Re_\tau=5000$ and 6000 in this case), a certain level of discrepancy between the reference data is visible. Meanwhile, when conditioned on this mean velocity, the {\newname} (\reddotline) shows a significant improvement over the case without mean velocity conditioning. Although the biased mean velocity is the only information we condition {\newname} on, all the other statistics get improved more or less over the one without conditioning. For the auto correlation ($R_{11}$), after conditioning on the mean velocity, {\newname} aligns significantly better with the reference data, which is the same case for Reynolds shear stress ($\langle u' v' \rangle$, shown in panel (d)), and turbulence intensity ($\langle u'u' \rangle$ and $\langle v'v' \rangle$ shown in panel(e) and (f), respectively). The result indicates the {\newname} is able to effectively cover a high-dimensional distribution across various flow conditions, but due to the sparsity of the training data (sparse in terms of Reynolds number), {\newname} cannot accurately retrieve the target distribution under given Reynolds number. However, as long as a weak guidance is provided, it significantly improve the accuracy to pinpoint the target distribution.


\section{Conclusion}
\label{sec:conclusion}
In this work, we demonstrate the effectiveness and robustness of CoNFiLD-inlet as the synthetic turbulence inlet generator under different flow conditions for both DNS and WMLES channel flows. When compared to traditional methods like digital filtering, CoNFiLD-inlet shows much better agreement with the precursor simulation results, and more importantly, does not require accurate prior knowledge for parameter fine-tuning during inference time. While compared with other deep learning based turbulence inlet generators, CoNFiLD-inlet shows significantly better performance especially for generating unseen scenarios. Besides, the flexibility and robustness of CoNFiLD-inlet is pronounced when we apply CoNFiLD-inlet for a wide range of Reynolds numbers in a posteriori tests without retraining the model or hyper-parameter tuning which requires prior knowledge. Moreover, further performance analysis are performed, providing some insights of why and how CoNFiLD(-inlet) works well. 

\section*{Acknowledgment}
The authors would like to acknowledge the funds from Office of Naval Research under award numbers N00014-23-1-2071 and National Science Foundation under award numbers OAC-2047127.

\section*{Compliance with Ethical Standards}
Conflict of Interest: The authors declare that they have no conflict of interest.




\clearpage
{\setstretch{1.0}\appendix

\section{Training Data}
\setcounter{figure}{0}  
\subsection{DNS}
\label{appen:dnsData}
\begin{figure}[!ht]
    \centering
    \includegraphics[width=0.95\linewidth]{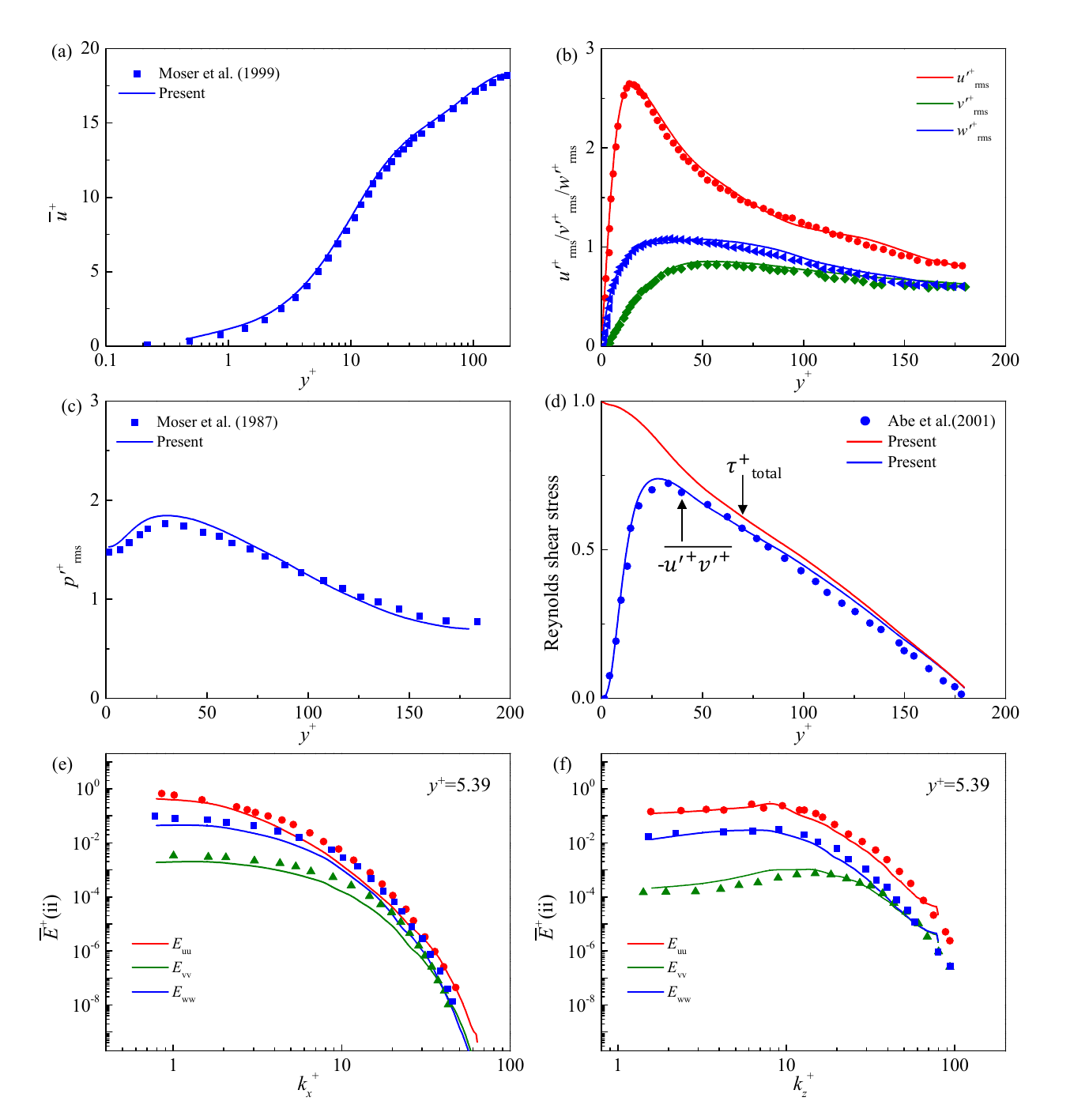}
    \caption{Statistical validation of the DNS training dataset. (a) Mean velocity profile, validated against the data from~\cite{moser1999direct}. (b) Turbulence intensity,compared with~\cite{moser1999direct}. (c) Root-mean-square values of pressure fluctuations, compared with~\cite{kim1987turbulence}. (d) Reynolds shear stress, compared with ~\cite{abe2001direct}. (e) Energy spectra along streamwise direction and (f) spanwise direction, compared with~\cite{rai1991direct}. Solid lines represent the present simulation, while scatter points denote reference data.}
    \label{fig:DNS-verify}
\end{figure}
The precursor simulation data is generated by the GPU solver Diff-FlowFsi~\cite{xiantao2024diffsi}. As shown in Fig.~\ref{fig:DNS-verify}, the simulation results are validated against the existing studies.

\subsection{WMLES} 
\label{appen:wmles_data}
Figure \ref{fig:wmles_data} shows various statistics of the training data we used in Section\ref{sec:wmles}. We also verify the mean velocity profile of our training label data against the Spalding Law of wall (Fig.~\ref{fig:wmles_data}(d)).   
\begin{figure}[!ht]
    \centering
    \captionsetup[subfloat]{farskip=-2pt,captionskip=-8pt}
    \subfloat[]{\includegraphics[width=0.48\textwidth]{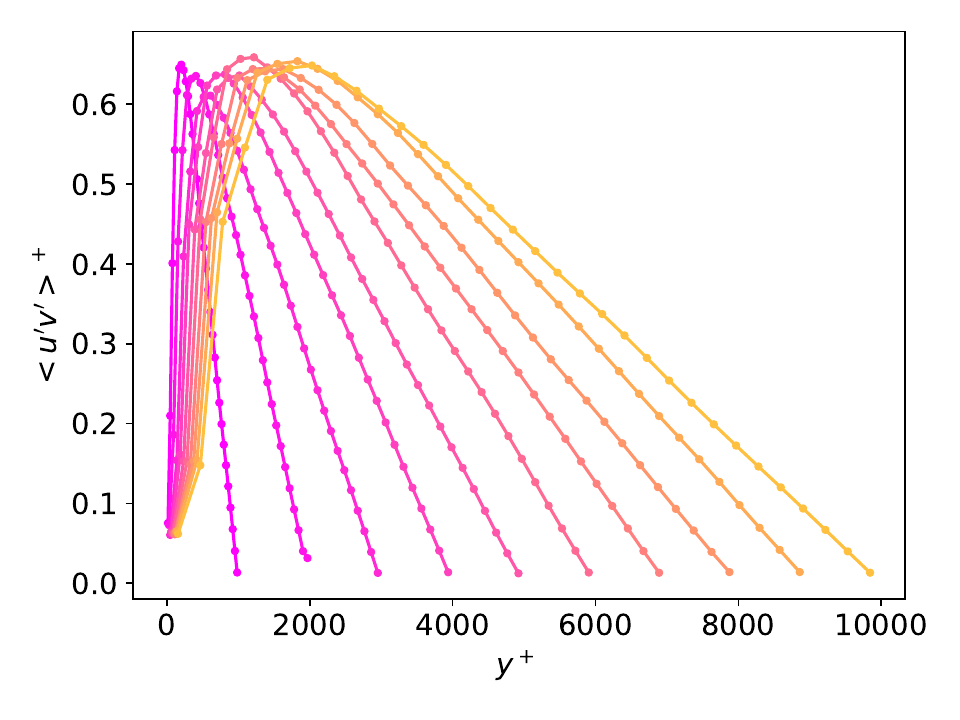}}
    \subfloat[]{\includegraphics[width=0.48\textwidth]{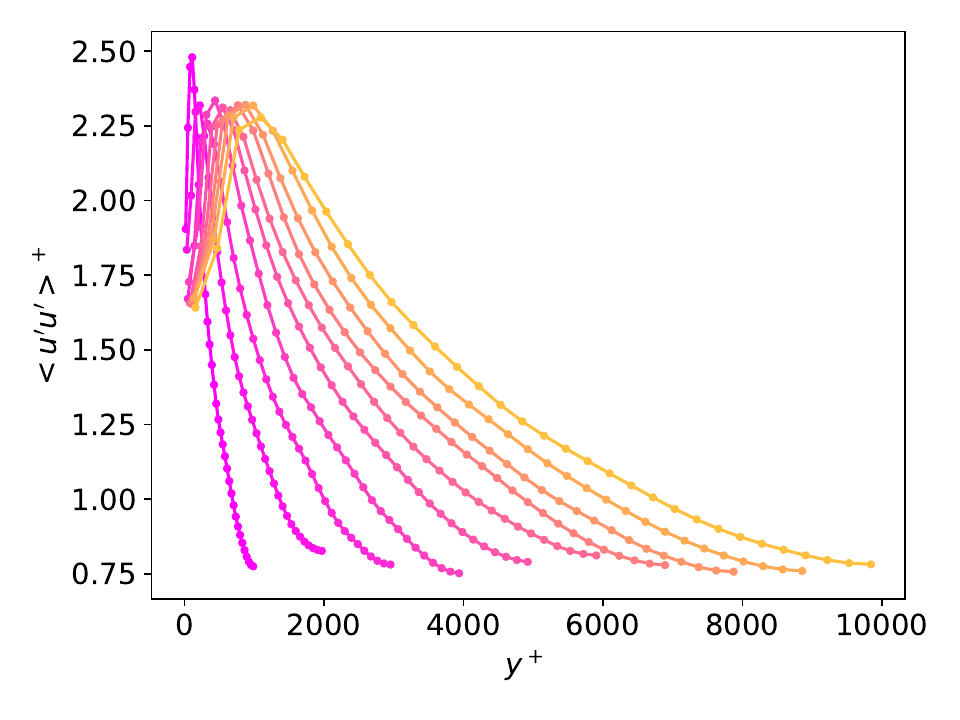}}\\
    \subfloat[]{\includegraphics[width=0.48\textwidth]{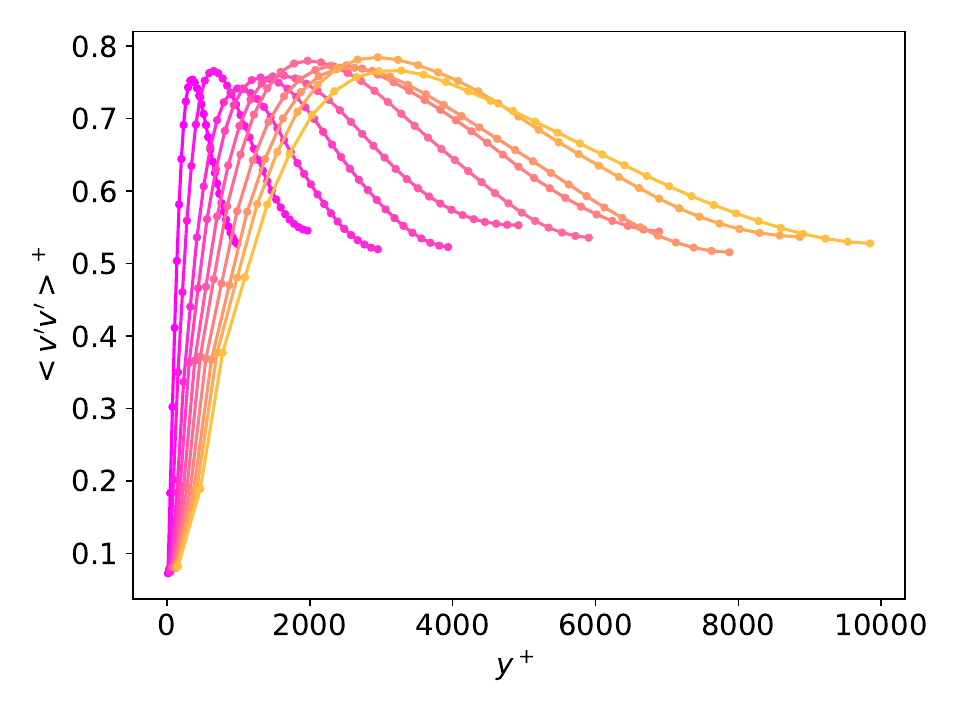}}
    \subfloat[]{\includegraphics[width=0.48\textwidth]{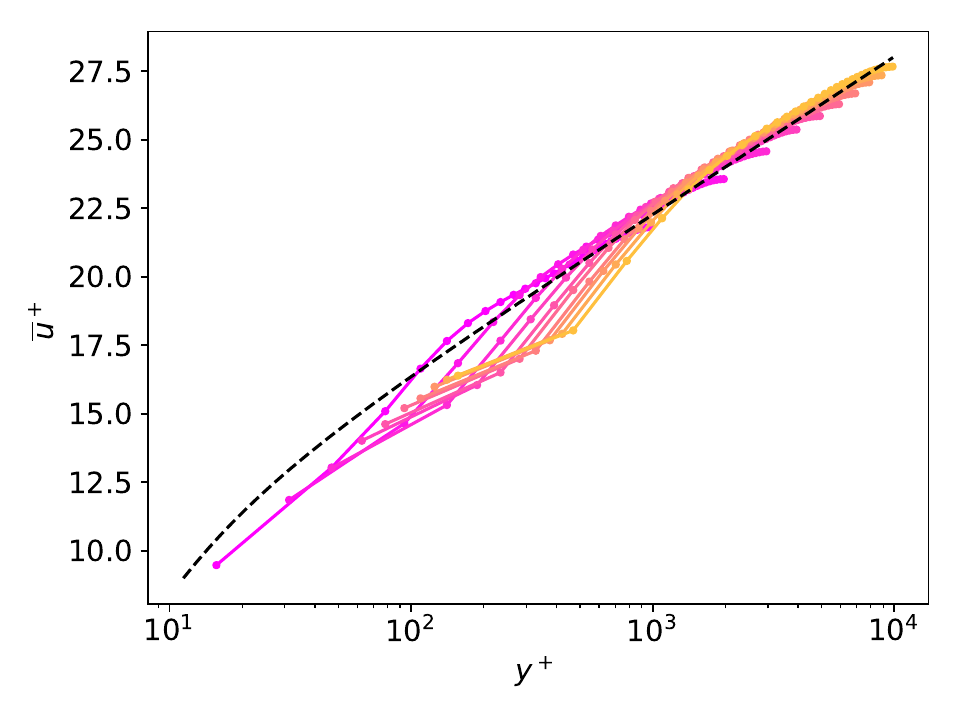}}\\
    \subfloat{\includegraphics[width=0.6\textwidth]{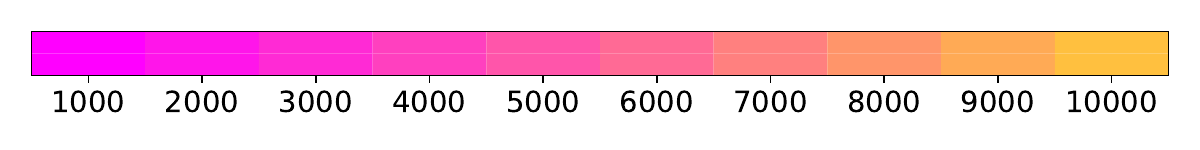}}
    \caption{Different Statistics (a) $\langle u'v'\rangle^+$, (b) $\langle u'u'\rangle^+$, (c) $\langle w'w'\rangle^+$ and (d) $\langle u\rangle^+$ of the wall-modeled LES data in the training set, with $Re_\tau = \{1000, 2000, \cdots, 10000\}$. The black dashed line in figure (d) indicates the Spalding law.}
    \label{fig:wmles_data}
\end{figure}

\section{Details about Posterior test setting}
\setcounter{figure}{0}   
\label{appen:posterior}
We use the OpenFOAM v2312 as the posterior tests solver. For DNS, the inlet BC is specified every $10$ numerical time steps, while for WMLES, we specify inlet BC every $5$ numerical time steps. The BC is then linearly interpolated to each numerical step between two adjacent time steps with specified BC, automatically handled by OpenFOAM. This temporal interpolation takes place for all the inlet BC, including \oldname/\newname, CNN-LSTM, ConvLSTM, DFM, as well as the recycling simulation data.

\section{Random ensemble sampling}
\begin{figure}[!ht]
    \centering
    \captionsetup[subfloat]{farskip=0pt,captionskip=-5pt}
    \subfloat[]{\includegraphics[width=0.33\linewidth]{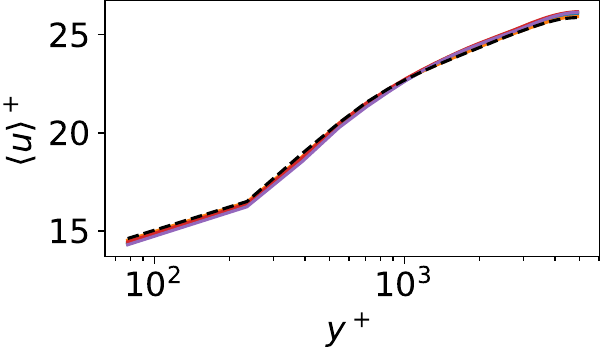}}
    \subfloat[]{\includegraphics[width=0.33\linewidth]{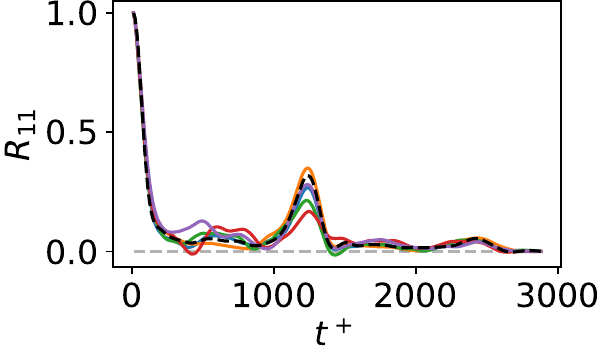}}
    \subfloat[]{\includegraphics[width=0.33\linewidth]{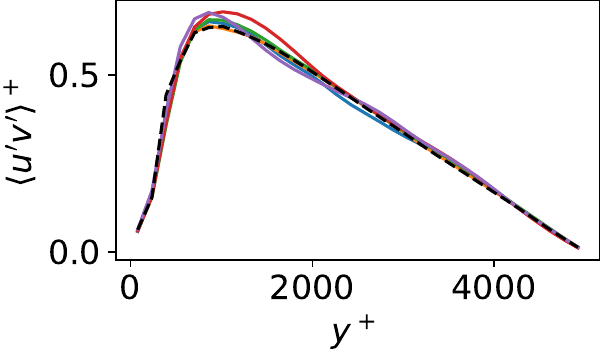}}\\
    \subfloat[]{\includegraphics[width=0.33\linewidth]{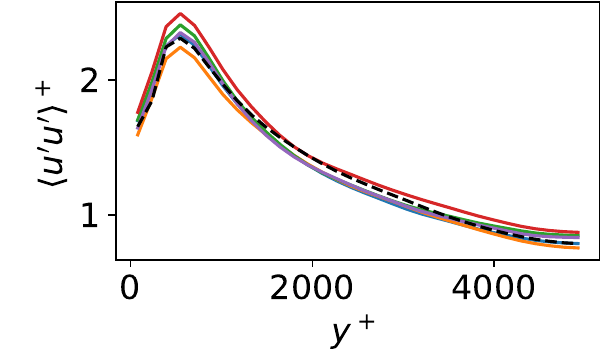}}
    \subfloat[]{\includegraphics[width=0.33\linewidth]{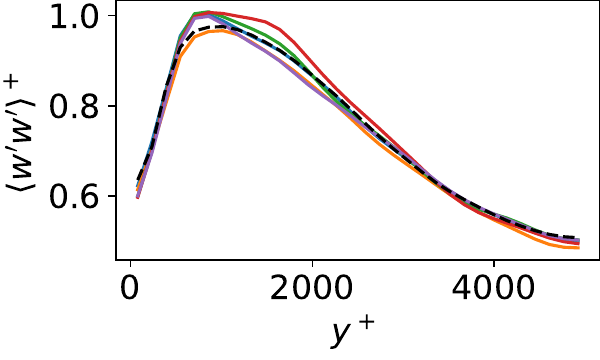}}
    \subfloat[]{\includegraphics[width=0.33\linewidth]{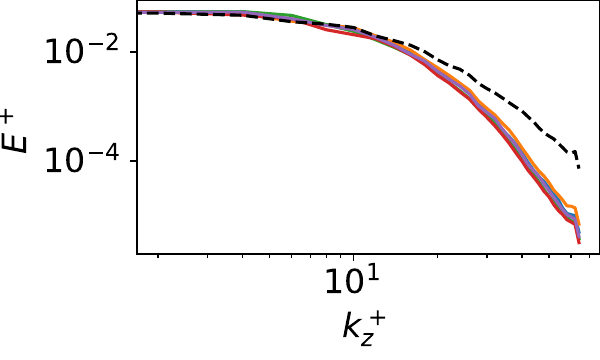}}
    \caption{Statistics of \emph{a priori} test of the $5$ random samples generate by \newname at $Re_\tau=5000$. Each sample is represented by one color while the reference data is marked by the black dashed line. Among them, we pick the visually best sample to report in the main test which is colored by orange. (a) Mean velocity profile. (b) Auto correlation $R_{11}$ at channel center. (c) Reynolds shear stress $\langle u'v'\rangle$. Turbulence intensity of $u$ (d) and $w$ (e). (f) Turbulence Kinetic Energy (TKE) along span wise wavenumber at channel center.}
    \label{fig:cherry_pick}
\end{figure}
Since \newname is a probabilistic model, all the \emph{a priori} tests and \emph{a posteriori} tests are all randomly sampled from the learned distribution. However, it is not guaranteed that all the random samples share the exactly the same statistics and filtering out the relatively low-quality samples will marginally improve the performance. Thus, in practice, we always generate $5$ random samples from the model and manually pick the sample with the best \emph{a priori} performance for \emph{a posteriori} test for all the cases we showed in this paper. Here we use the case with $Re_\tau=5000$ from the WMLES to showcase the difference between these random samples. Figure~\ref{fig:cherry_pick} shows the \emph{a priori} test results of the $5$ random samples generated by the \newname. Each sample is marked by the same color in all these subplots and the reference data is represented by the black dashed line. Among all these samples, we manually pick the best sample -- colored by orange. It is worth noting that, although we picked the visually best sample in the main text, the performance difference is not significant. Besides, since all the inference is performed on GPU, a batched generation of $5$ samples impose negligible influence in terms of generation cost. 

\section{Mesh independency}
\setcounter{figure}{0}
\begin{figure}[!ht]
    \centering
    \hfill\subfloat[]{\includegraphics[width=0.4\linewidth]{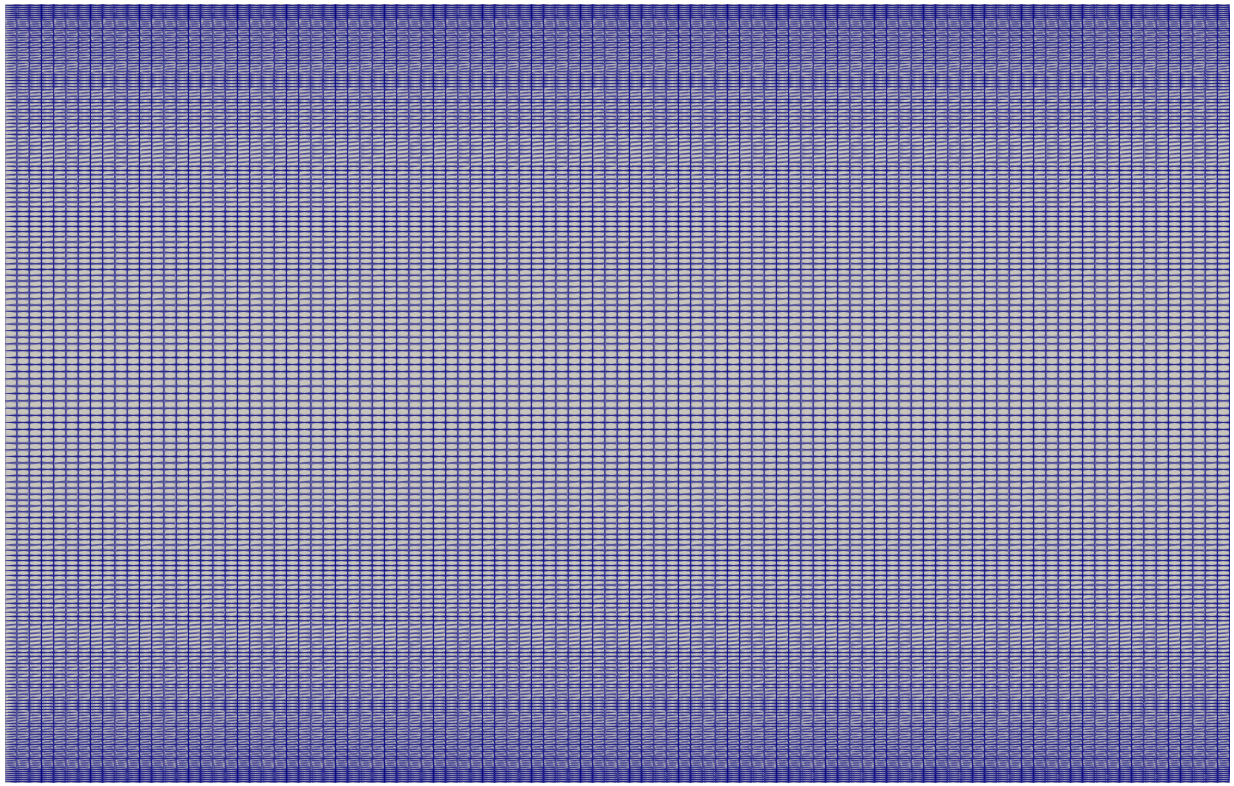}}\hfill
    \subfloat[]{\includegraphics[width=0.4\linewidth]{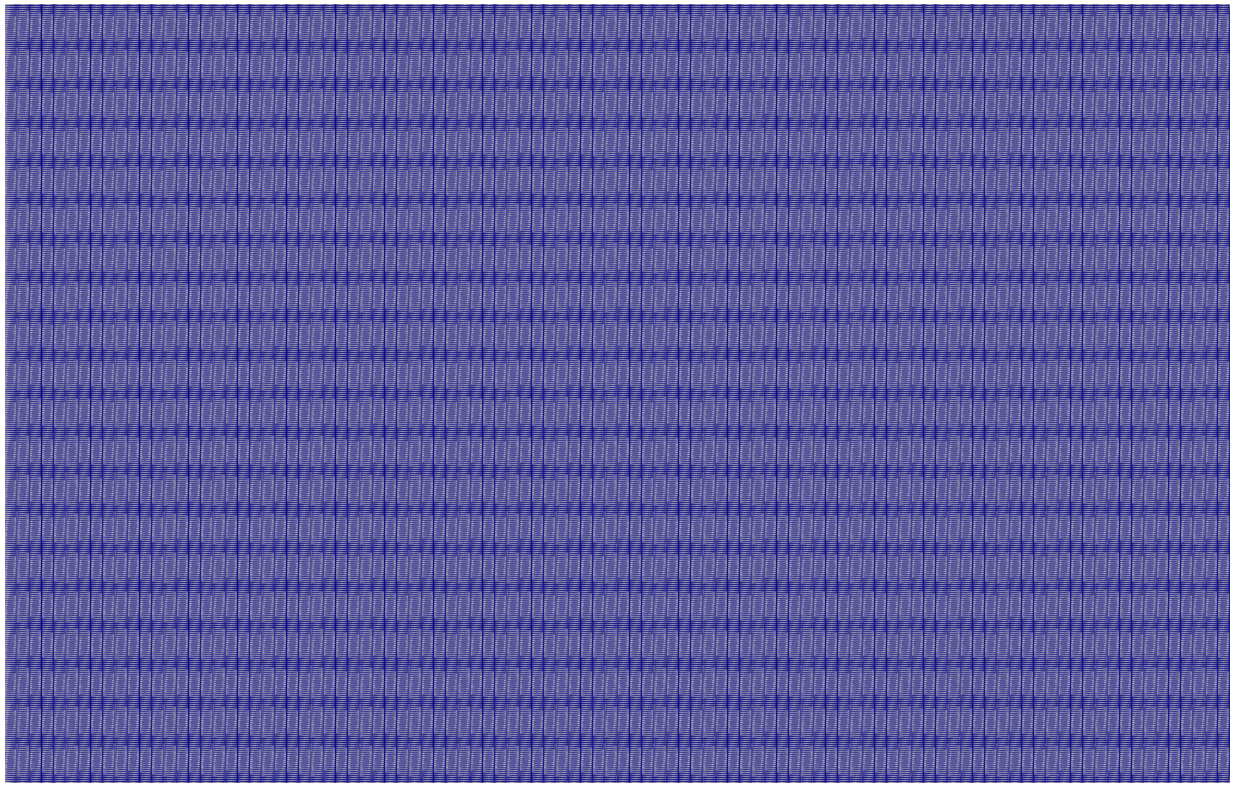}}
    \hfill
    \vspace{-1em}
    \caption{Stretching mesh (a) versus the uniform mesh (b) for the DNS $\mathbb{I}$}
    \label{fig:mesh}
\end{figure}
\begin{figure}[!ht]
    \captionsetup[subfloat]{farskip=-4pt,captionskip=-8pt}
    \centering
    \subfloat[]{\includegraphics[width=0.45\textwidth]{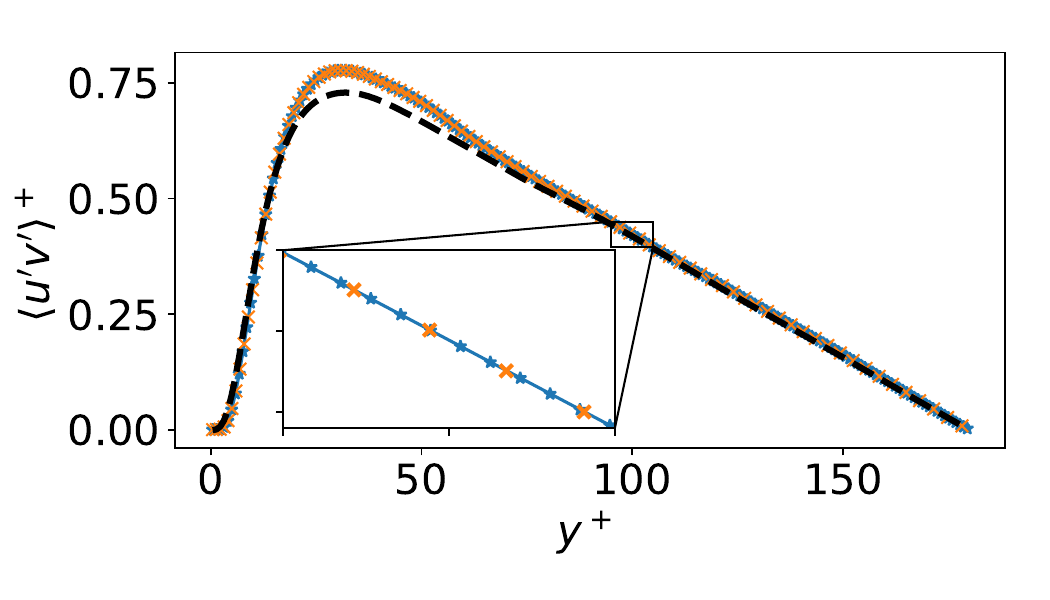}}
    \subfloat[]{\includegraphics[width=0.45\textwidth]{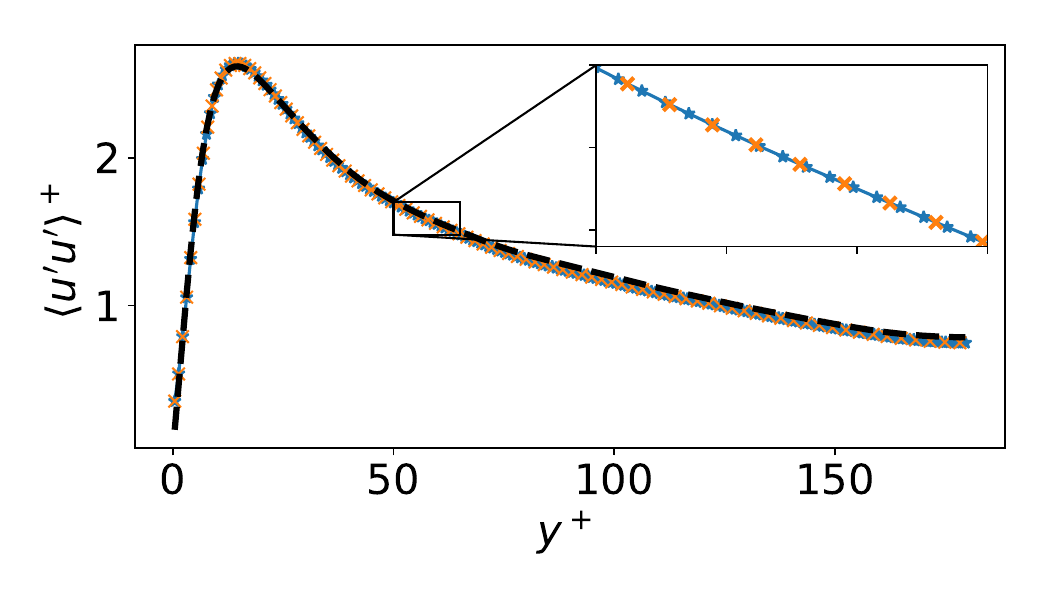}}\\
    \subfloat[]{\includegraphics[width=0.45\textwidth]{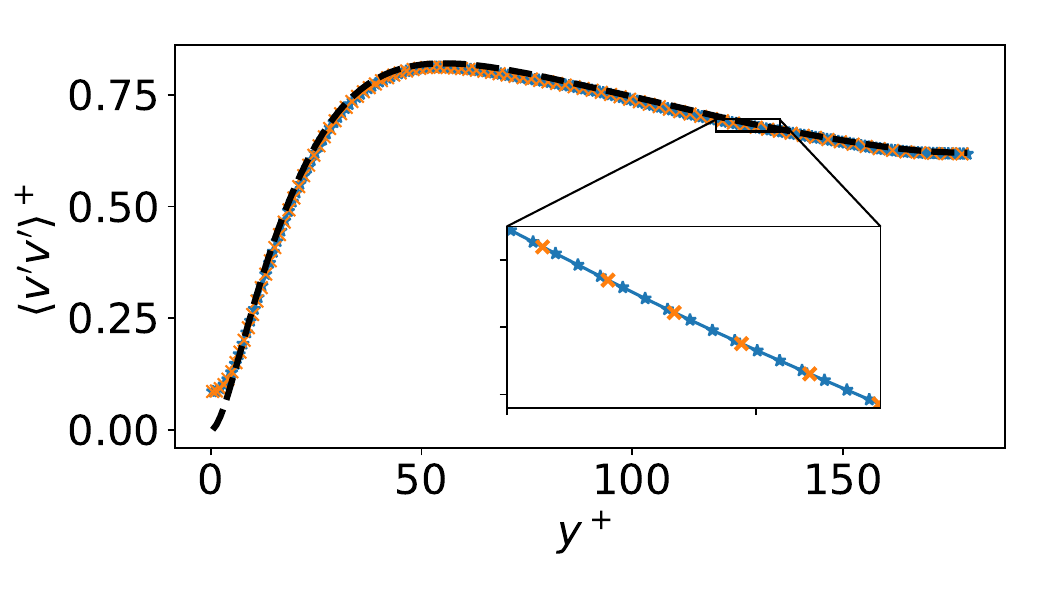}}
    \subfloat[]{\includegraphics[width=0.45\textwidth]{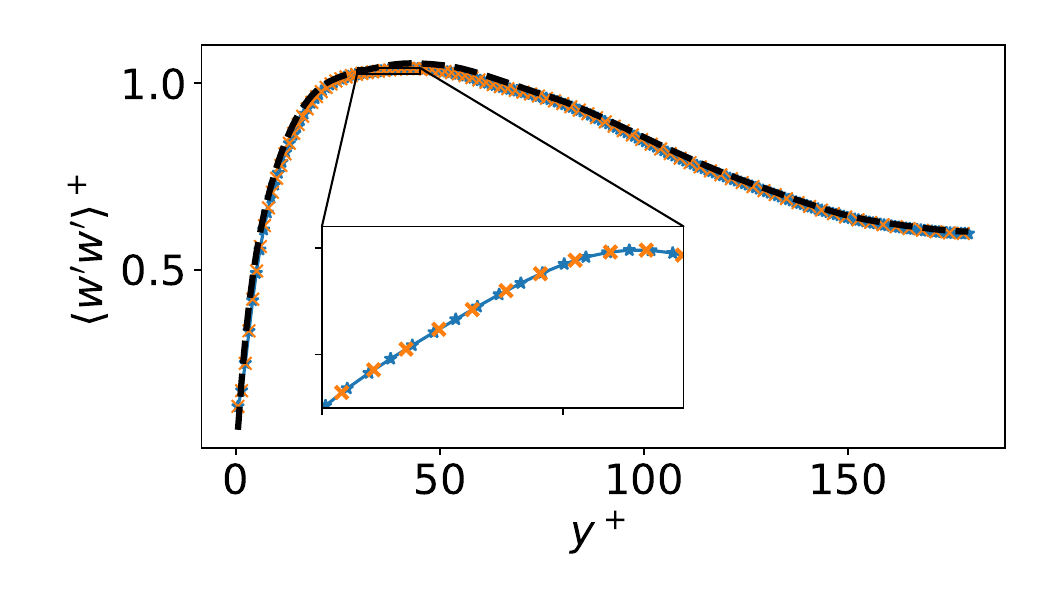}}
    \caption{\emph{A priori} test of CoNFiLD evaluated on uniform mesh (blue line with star), and stretching mesh (orange dots), compared with the reference data (black dashed line). Reynolds shear stress (a) and Turbulence intensity of components $u$,$v$,$w$ are shown in (b),(c), (d), respectively. }
    \label{fig:stretching_stat}
\end{figure}
By leveraging the CNF, \newname is essentially sampling from a mesh-irrelevant latent space when generating new realizations of the turbulence inlet. Thus \newname is able to generate inlet condition on arbitrary meshes without retraining or interpolation. To demonstrate the flexibility and effectiveness of the discretization-irrelevant property, we use the same model shown in Sec.\ref{sec:DNS} to generate velocity field on a stretching grid (shown in Fig. \ref{fig:mesh}(a), mesh size $N_y\times N_z = 192\times100$). While the model itself is only trained on velocity data collected on a uniform mesh shown in Fig. \ref{fig:mesh}(b) with mesh resolution $N_y\times N_z = 400\times100$. 

We use \oldname to generate a new sample in the latent space encoded by CNF and decode this latent on the two mesh grids shown in Fig. \ref{fig:mesh}. The \emph{a priori} test results is shown in Figure \ref{fig:stretching_stat}, with fours statistics: Reynolds shear stress $\langle u'v'\rangle^+$(a), Turbulence intensity of component $u$(b), $v$(c), and $w$(d). The blue line with stars represent the \emph{a priori} test of \oldname on the original uniform mesh, while the orange dots denote the statistics on the stretching mesh. Compared with the reference data (black dashed line), the \emph{a priori} tests on both mesh resolution show a good match. More importantly, there is almost no visible discrepancy between the the velocity field at two mesh grids. The zoom-in view of each subfigure clearly demonstrate that, on a coarser stretching mesh grid, \oldname still gives a consistent statistics even if it is completely evaluated at unseen coordinates. For all the statistics shown in figure \ref{fig:stretching_stat}, the orange dots sit right on the curve connecting the blue stars, indicating the \oldname is able to accurately represent the turbulence inlet regardless of discretization.

\section{t-SNE Visualization of the DNS training data}
\setcounter{figure}{0}
\begin{figure}[ht]
    \centering
    \captionsetup[subfloat]{farskip=-4pt,captionskip=-8pt}
    \subfloat{\includegraphics[width=0.33\linewidth]{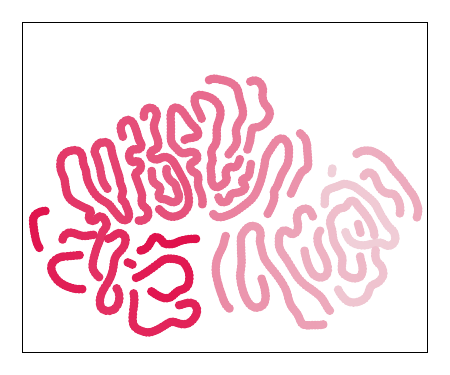}}\\
    \subfloat{\includegraphics[width=0.3\textwidth]{figs/dis/cnnlstm/tsne_label_colorbar_0.pdf}}
    \caption{Two dimensional t-SNE visualization of the training data used in the DNS inlet generator. The velocity field at each time step is projected as one dot and the color indicates the time step. }
    \label{fig:tsne_label}
\end{figure}
Figure~\ref{fig:tsne_label} shows the two-dimensional t-SNE visualization of the training dataset used in the DNS inlet generator. Each dot represents a single time step velocity field in the training set while the color indicates the temporal order of these time steps.

\section{Deep-Learning based Baseline Methods}
\label{appen:baselines}
CNN-LSTM: We use a Convolutional Neural Networks based AutoEncoder-Decoder (CNN-AE) architecture to compress the cross-sectional flow data into latent tensors. We then flatten these tensors into vectors and then train a Long-Short Term Memory (LSTM) Model to learn the temporal sequence of the latent vectors. For this, we use the architecture proposed by \cite{yousif2022physics} with certain modifications. First, the CNN-AE model within this architecture is re-implemented in Pytorch. As for the LSTM model, we use the original TensorFlow-based implementation provided by the authors. Furthermore, we only use data-driven loss for training the CNN-AE model for a fair comparision with our data-driven model.  

ConvLSTM: We propose another baseline that is solely a sequence model that operates in the high-dimensional physical space. To this end, we propose a Convolution Neural Network based LSTM model (ConvLSTM). This deep architecture is built by stacking several custom ConvLSTM cell blocks. We design these blocks such that they consists of 2 ConvLSTM cells. The first cell is customized to include a normalization layer and residual skip connection, while the second cell is standard ConvLSTM cell. The architecture was built using Haiku.

\begin{figure}[!ht]
    \captionsetup[subfloat]{farskip=-2pt,captionskip=-8pt}
    \centering
    \subfloat[]{\includegraphics[width=0.5\textwidth]{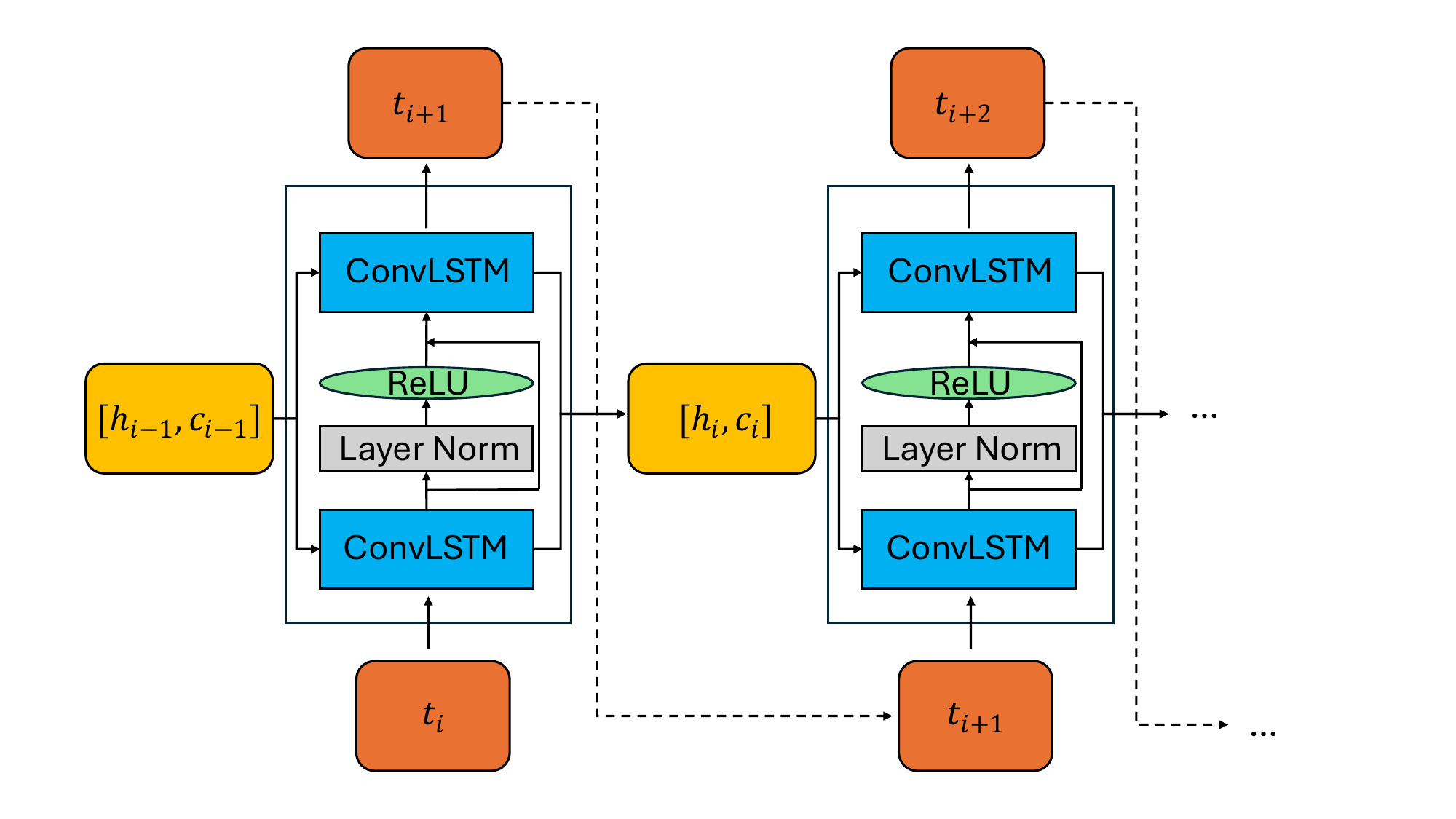}}
    \subfloat[]{\includegraphics[width=0.5\textwidth]{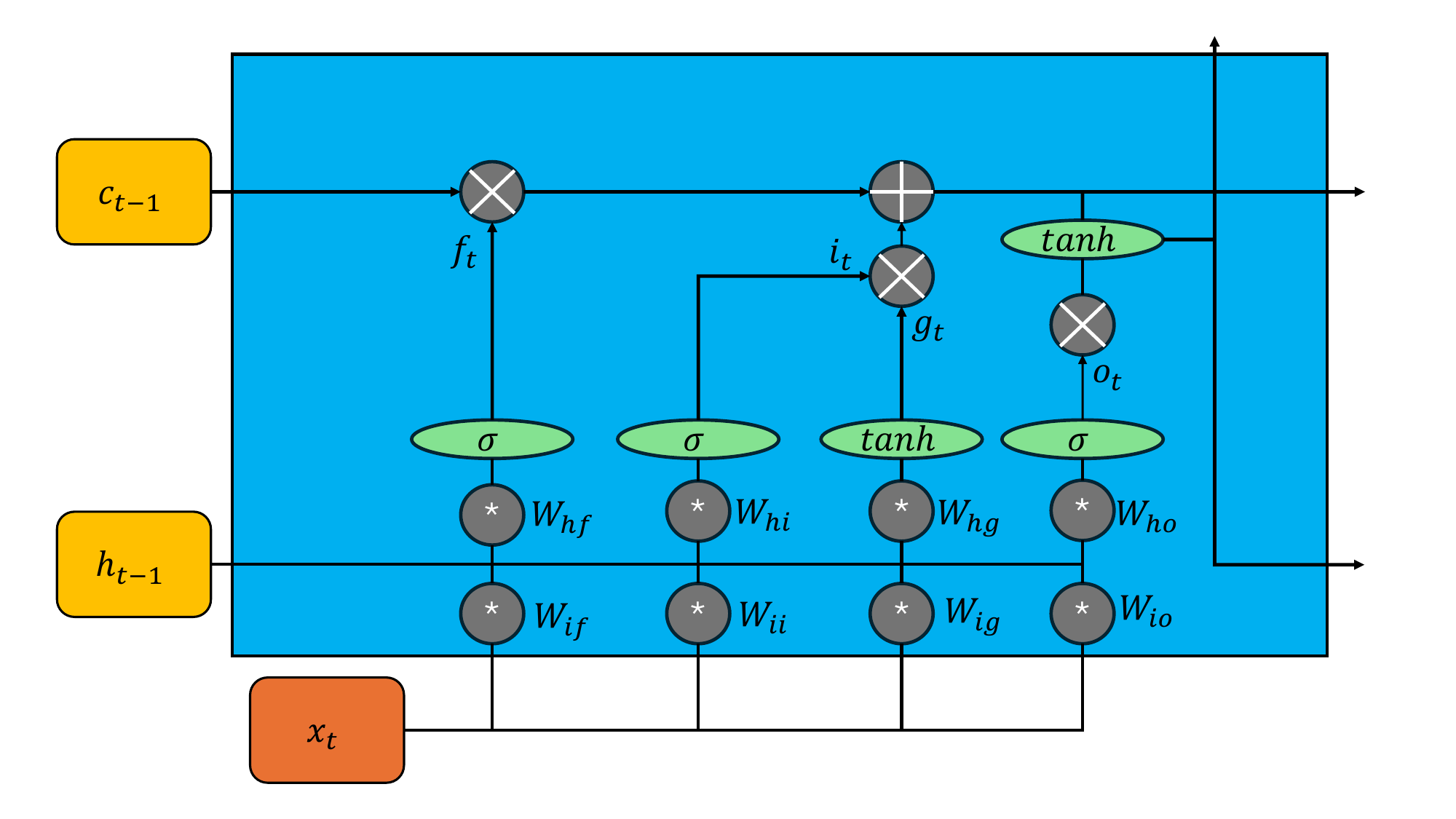}}
    \caption{Illustration of the ConvLSTM deep learning baseline. (a) shows the entire architecture, while (b) is the architecture of a single ConvLSTM block}
    \label{fig:ConvLSTM}
\end{figure}

\begin{table}[h!]
\centering
\caption{Hyper-parameters chosen for the baseline deep learning methods }
\begin{tabular}{c|c || c|c}
\toprule
\multicolumn{2}{c||}{CNN-LSTM} & \multicolumn{2}{c}{ConvLSTM}\\
\midrule
Kernel Size& (3, 5, 7) & Kernel Size & 5 \\
Base Channel count & 24 & Number of Blocks & 4\\
Channel Multiplier & (1, 2, 4, 8, 12, 6, 3, 1.5 ) & Hidden Channels List& (4, 16, 32, 16, 4)\\
Activation Function& ReLU & Activation Function& ReLU\\
Learning Rate& $1e^{-3}$ & Learning Rate& $1e^{-3}$\\
Optimizer& Adam & Optimizer& Adam\\
\bottomrule
\end{tabular}
\label{table:my_table}
\end{table}

\newpage
\section{More statistics in \emph{a posteriori} tests}
\setcounter{figure}{0}   
\subsection{More statistics of DNS inlet generator}
\label{appen:more_stat_dns}
\begin{figure}[!ht]
\captionsetup[subfloat]{farskip=-9pt}
    \centering
    \subfloat{\includegraphics[width=\textwidth]{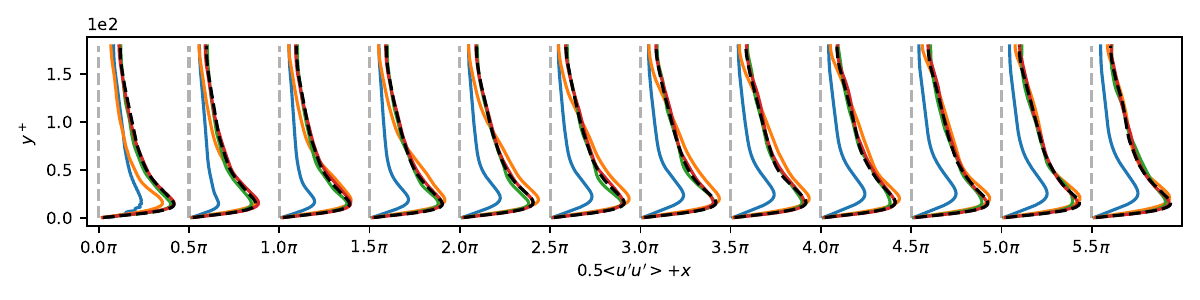}}\\
    \subfloat{\includegraphics[width=\textwidth]{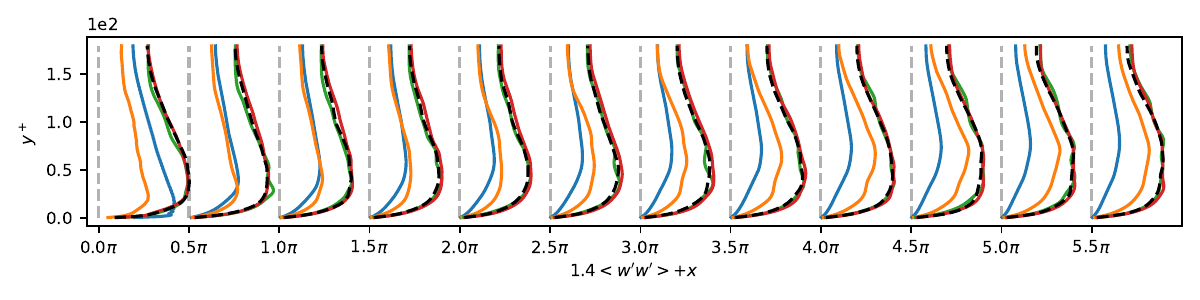}}\\
    \caption{\emph{A posteriori} test comparison in DNS of Turbulence intensity ($\langle u'u'\rangle$ top row, $\langle w'w'\rangle$ bottom row) between {\newname} (\redline), CNN-LSTM (\greenline), Digital Filtering (\blueline), ConvLSTM (\orangeline), and the reference data (\dashedblackline, i.e. precursor simulation)}
    \label{fig:extra_dns_turb_inten}
\end{figure}
\begin{figure}[!ht]
\captionsetup[subfloat]{farskip=-4pt,captionskip=-8pt}
    \centering
    \subfloat[$x=1.0\pi$]{\includegraphics[width=0.33\textwidth]{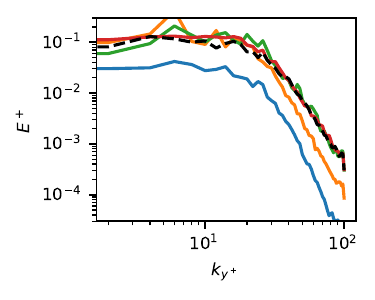}}\hfill
    \subfloat[$x=1.5\pi$]{\includegraphics[width=0.33\textwidth]{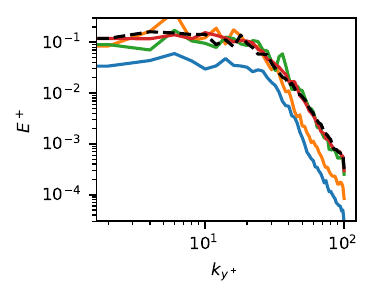}}\hfill
    \subfloat[$x=2.0\pi$]{\includegraphics[width=0.33\textwidth]{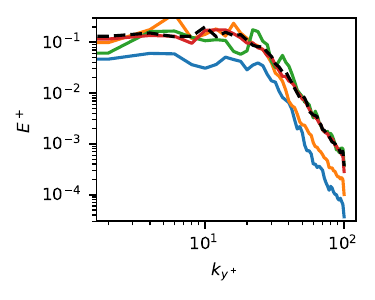}}\\
    \subfloat[$x=2.5\pi$]{\includegraphics[width=0.33\textwidth]{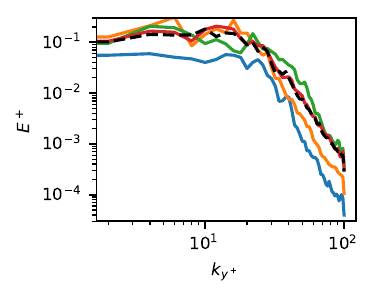}}\hfill
    \subfloat[$x=3.0\pi$]{\includegraphics[width=0.33\textwidth]{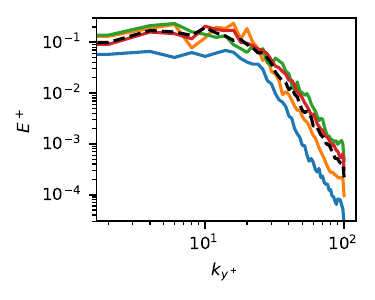}}\hfill
    \subfloat[$x=3.5\pi$]{\includegraphics[width=0.33\textwidth]{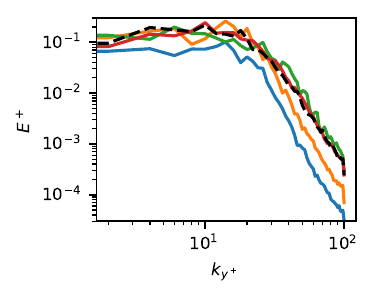}}\\
    \subfloat[$x=4.0\pi$]{\includegraphics[width=0.33\textwidth]{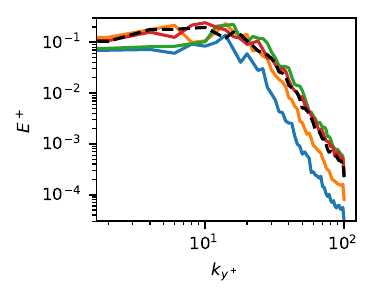}}\hfill
    \subfloat[$x=4.5\pi$]{\includegraphics[width=0.33\textwidth]{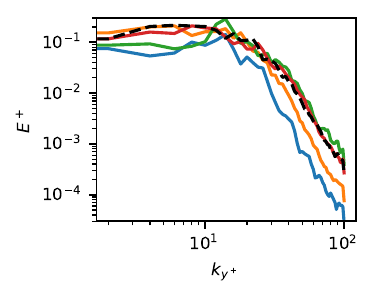}}\hfill
    \subfloat[$x=5.0\pi$]{\includegraphics[width=0.33\textwidth]{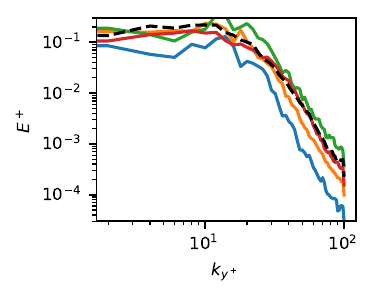}}\\
    \caption{\emph{A posteriori} test comparison in DNS of turbulence kinetic energy (TKE) between CoNFiLD (\redline), CNN-LSTM (\greenline), Digital Filtering (\blueline), ConvLSTM (\orangeline), and the reference data (\dashedblackline, i.e. precursor simulation) at different stream wise locations}
    \label{fig:extra_dns_turb_tke}
\end{figure}\clearpage

\subsection{More statistics of WMLES inlet generator at training Reynolds numbers}
\label{appen:more_stat_les}
\begin{figure}[!ht]
    \centering
    \captionsetup[subfloat]{farskip=-4pt,captionskip=-8pt}
    \subfloat{\includegraphics[width=\textwidth]{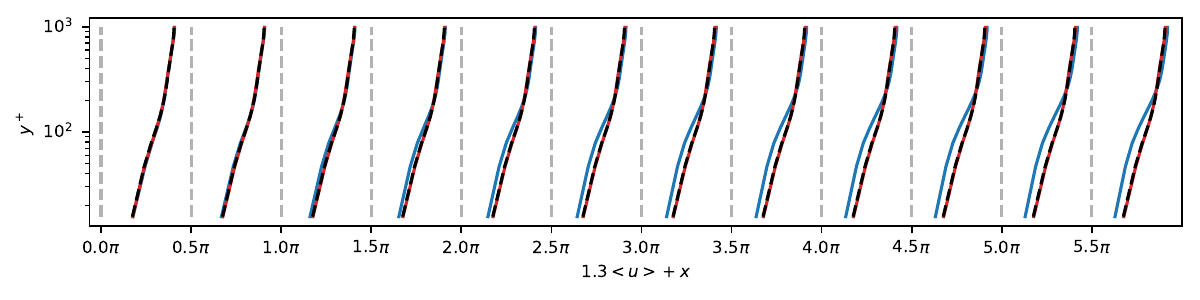}}\\
    \subfloat{\includegraphics[width=\textwidth]{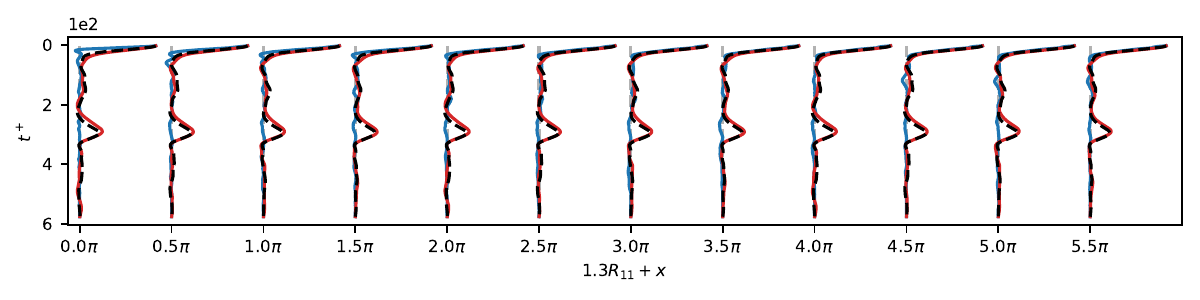}}\\
    \subfloat[$x=0$]{\includegraphics[width=.33\textwidth]{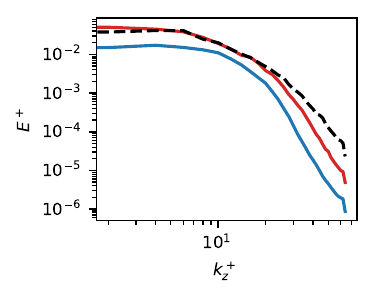}}
    \subfloat[$x=\pi$]{\includegraphics[width=.33\textwidth]{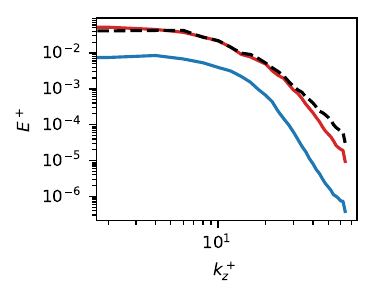}}
    \subfloat[$x=2\pi$]{\includegraphics[width=.33\textwidth]{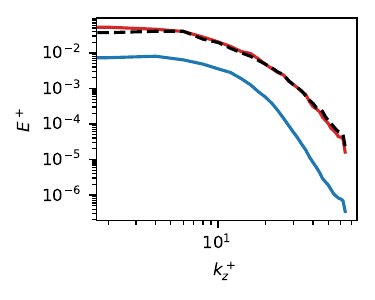}}\\
    \subfloat[$x=3\pi$]{\includegraphics[width=.33\textwidth]{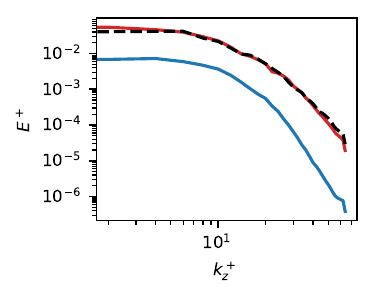}}
    \subfloat[$x=4\pi$]{\includegraphics[width=.33\textwidth]{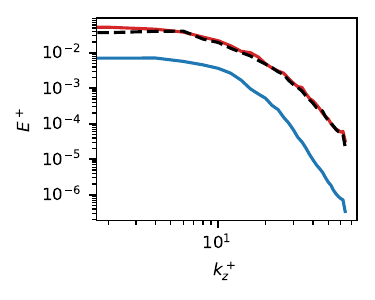}}
    \subfloat[$x=5\pi$]{\includegraphics[width=.33\textwidth]{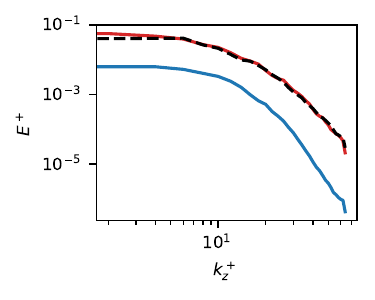}}
    \caption{\emph{A posteriori} test results of WMLES of $Re_\tau=1000$ at different stream-wise locations. Red curve represents CoNFiLD-inlet, blue curve denotes the DFM, while precursor simulation results are marked by black dashed line. First two lines: Mean velocity profile and auto correlation $R_{11}$ near channel center. Others are turbulence kinetic energy (TKE) along spanwise wave length.}
    \label{fig:wmles_extra_stat_1000_1}
\end{figure}
\begin{figure}[!ht]
    \centering
    \captionsetup[subfloat]{farskip=-9pt}
    \subfloat{\includegraphics[width=\textwidth]{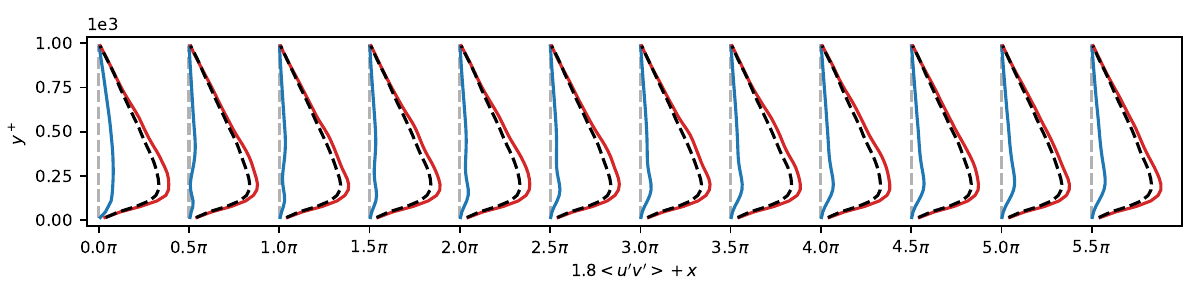}}\\
    \subfloat{\includegraphics[width=\textwidth]{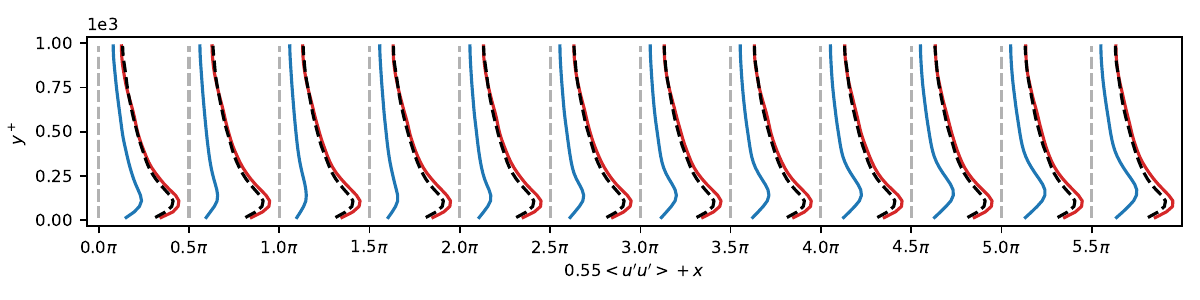}}\\
    \subfloat{\includegraphics[width=\textwidth]{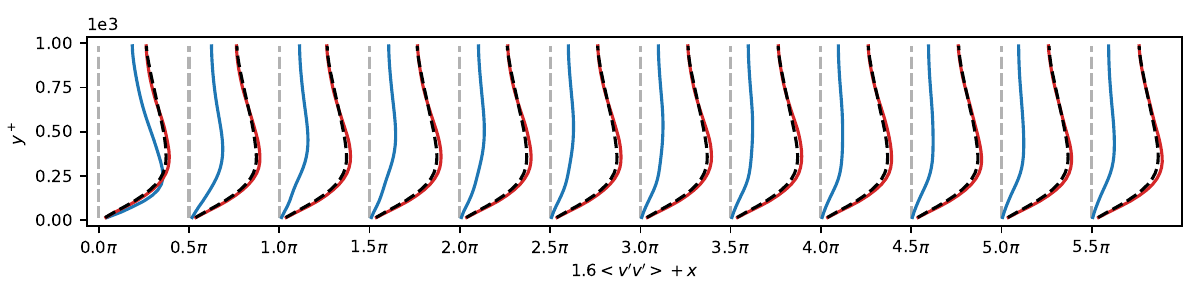}}\\
    \subfloat{\includegraphics[width=\textwidth]{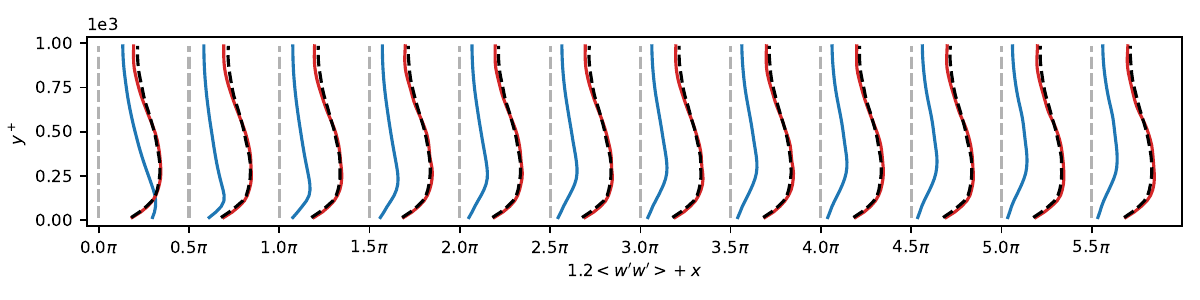}}\\
    \caption{\emph{A posteriori} test results at $12$ different stream-wise locations of WMLES of $Re_\tau=1000$. Red curve represents CoNFiLD-inlet, blue curve denotes the DFM, while precursor simulation results are marked by black dashed line. From top to bottom: Reynolds shear stress, Turbulence intensity of $u$, $v$, and $w$.}
    \label{fig:wmles_extra_stat_1000_2}
\end{figure}
\begin{figure}[!ht]
    \centering
    \captionsetup[subfloat]{farskip=-4pt,captionskip=-8pt}
    \subfloat{\includegraphics[width=\textwidth]{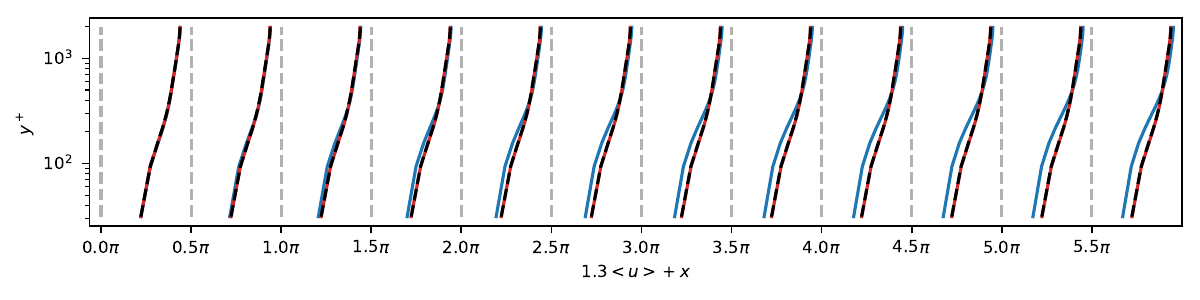}}\\
    \subfloat{\includegraphics[width=\textwidth]{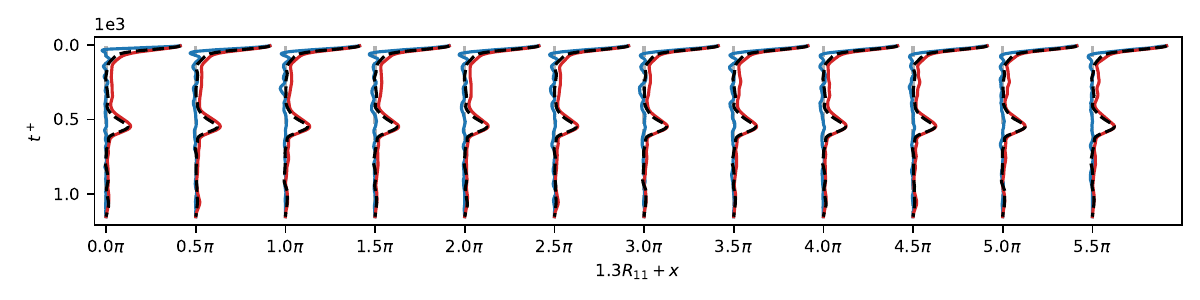}}\\
    \subfloat[$x=0$]{\includegraphics[width=.33\textwidth]{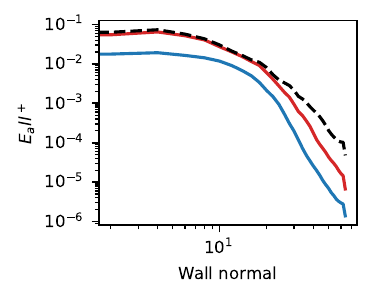}}
    \subfloat[$x=\pi$]{\includegraphics[width=.33\textwidth]{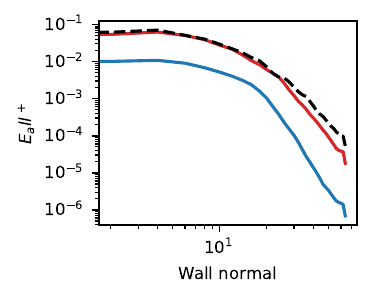}}
    \subfloat[$x=2\pi$]{\includegraphics[width=.33\textwidth]{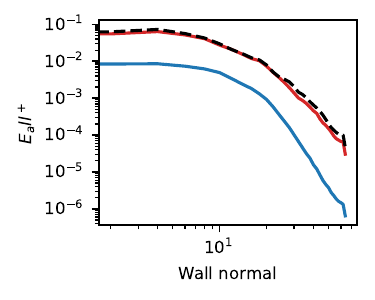}}\\
    \subfloat[$x=3\pi$]{\includegraphics[width=.33\textwidth]{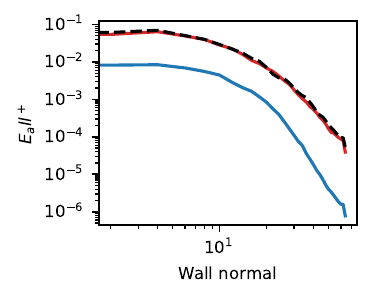}}
    \subfloat[$x=4\pi$]{\includegraphics[width=.33\textwidth]{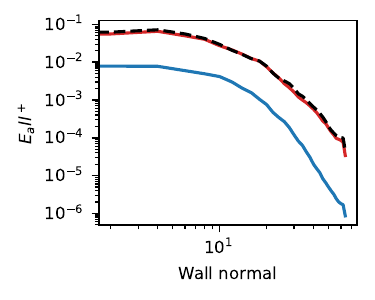}}
    \subfloat[$x=5\pi$]{\includegraphics[width=.33\textwidth]{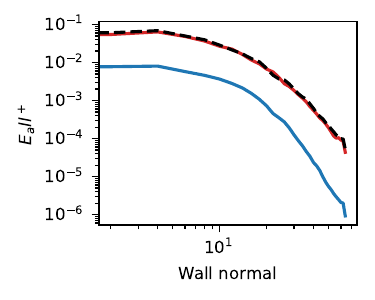}}
    \caption{\emph{A posteriori} test results of WMLES of $Re_\tau=2000$ at different stream-wise locations. Red curve represents CoNFiLD-inlet, blue curve denotes the DFM, while precursor simulation results are marked by black dashed line. First two lines: Mean velocity profile and auto correlation $R_{11}$ near channel center. Others are turbulence kinetic energy (TKE) along spanwise wave length.}
    \label{fig:wmles_extra_stat_2000_1}
\end{figure}
\begin{figure}[!ht]
    \centering
    \captionsetup[subfloat]{farskip=-9pt}
    \subfloat{\includegraphics[width=\textwidth]{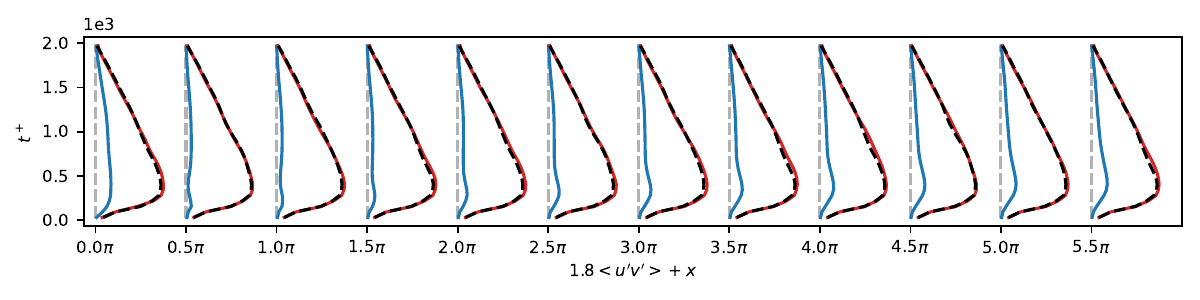}}\\
    \subfloat{\includegraphics[width=\textwidth]{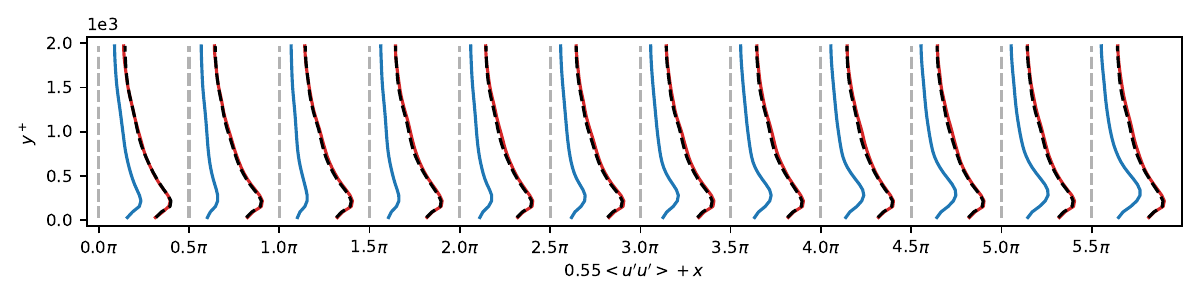}}\\
    \subfloat{\includegraphics[width=\textwidth]{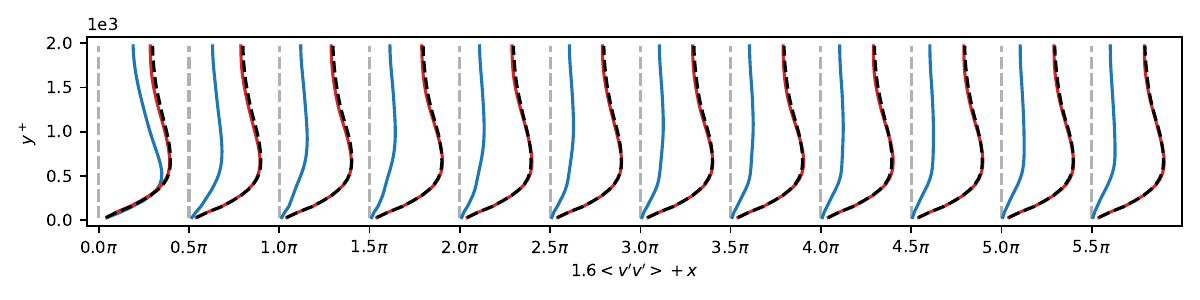}}\\
    \subfloat{\includegraphics[width=\textwidth]{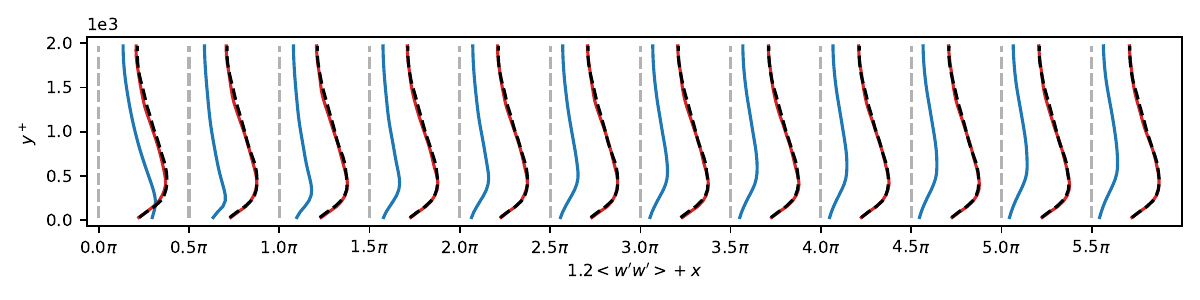}}\\
    \caption{\emph{A posteriori} test results at $12$ different stream-wise locations of WMLES of $Re_\tau=2000$. Red curve represents CoNFiLD-inlet, blue curve denotes the DFM, while precursor simulation results are marked by black dashed line. From top to bottom: Reynolds shear stress, Turbulence intensity of $u$, $v$, and $w$.}
    \label{fig:wmles_extra_stat_2000_2}
\end{figure}
\begin{figure}[!ht]
    \centering
    \captionsetup[subfloat]{farskip=-4pt,captionskip=-8pt}
    \subfloat{\includegraphics[width=\textwidth]{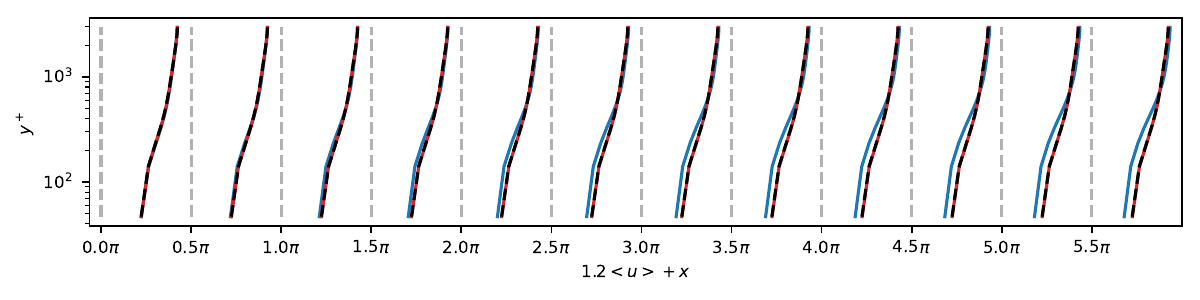}}\\
    \subfloat{\includegraphics[width=\textwidth]{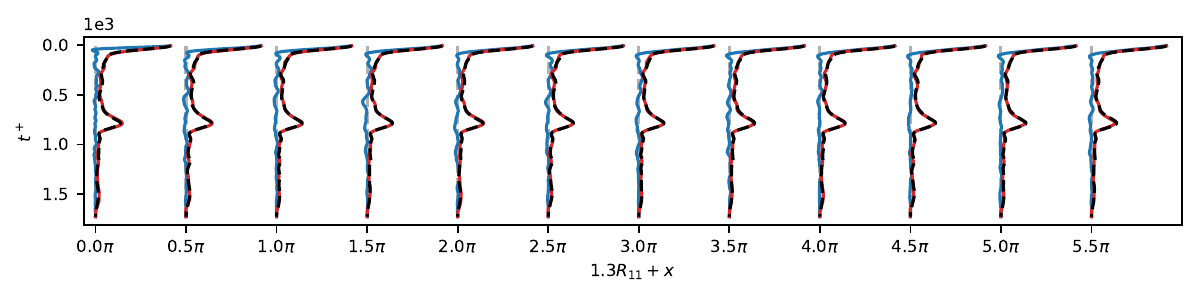}}\\
    \subfloat[$x=0$]{\includegraphics[width=.33\textwidth]{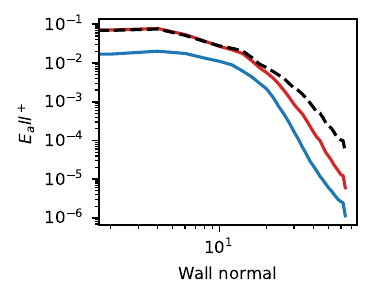}}
    \subfloat[$x=\pi$]{\includegraphics[width=.33\textwidth]{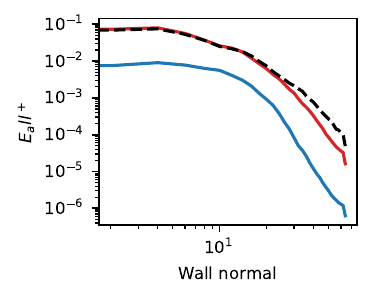}}
    \subfloat[$x=2\pi$]{\includegraphics[width=.33\textwidth]{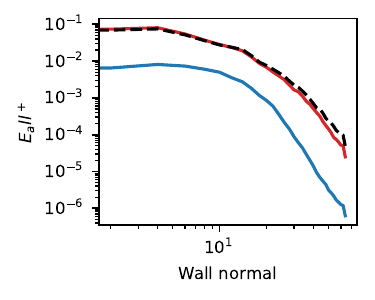}}\\
    \subfloat[$x=3\pi$]{\includegraphics[width=.33\textwidth]{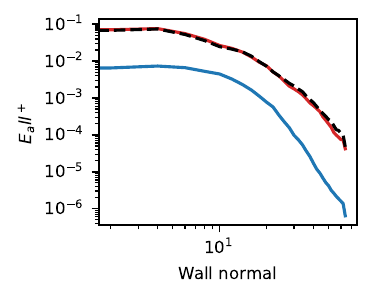}}
    \subfloat[$x=4\pi$]{\includegraphics[width=.33\textwidth]{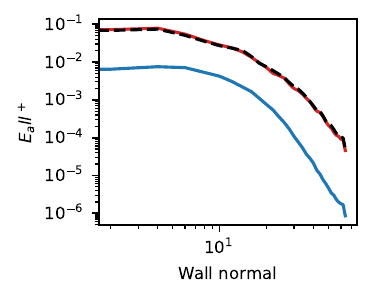}}
    \subfloat[$x=5\pi$]{\includegraphics[width=.33\textwidth]{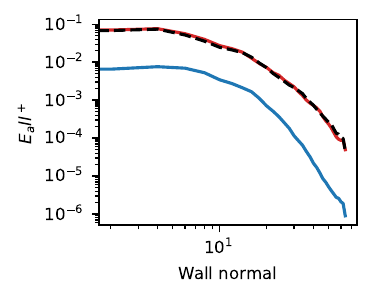}}
    \caption{\emph{A posteriori} test results of WMLES of $Re_\tau=3000$ at different stream-wise locations. Red curve represents CoNFiLD-inlet, blue curve denotes the DFM, while precursor simulation results are marked by black dashed line. First two lines: Mean velocity profile and auto correlation $R_{11}$ near channel center. Others are turbulence kinetic energy (TKE) along spanwise wave length.}
    \label{fig:wmles_extra_stat_3000_1}
\end{figure}
\begin{figure}[!ht]
    \centering
    \captionsetup[subfloat]{farskip=-9pt}
    \subfloat{\includegraphics[width=\textwidth]{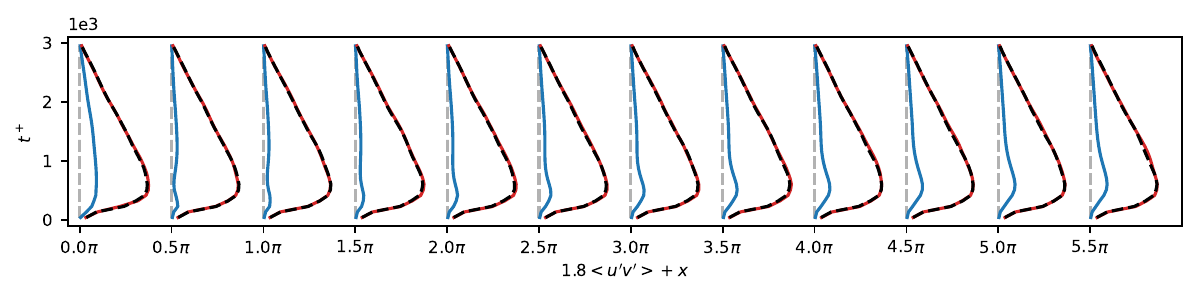}}\\
    \subfloat{\includegraphics[width=\textwidth]{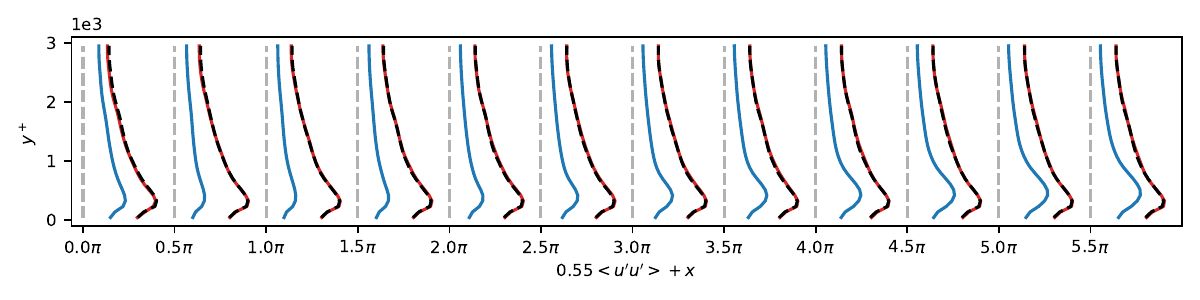}}\\
    \subfloat{\includegraphics[width=\textwidth]{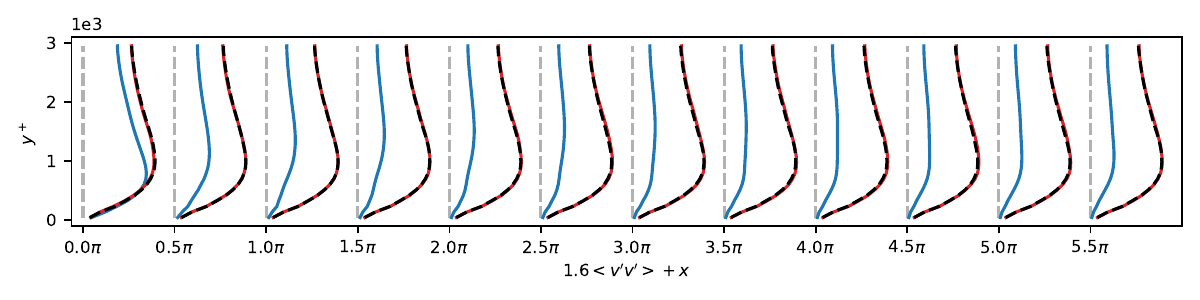}}\\
    \subfloat{\includegraphics[width=\textwidth]{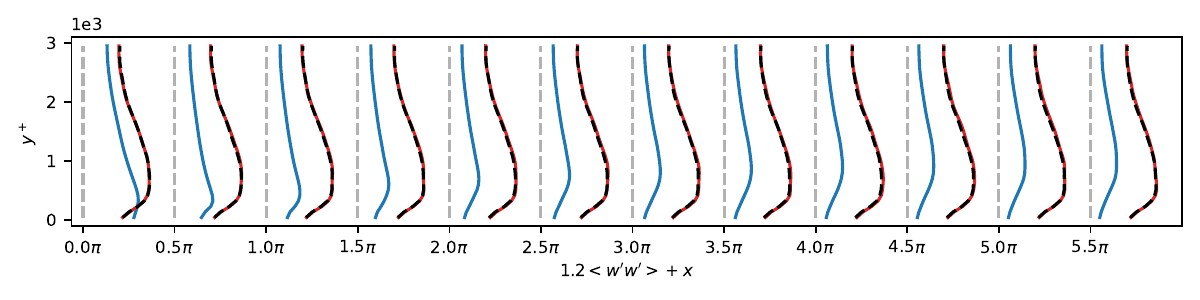}}\\
    \caption{\emph{A posteriori} test results at $12$ different stream-wise locations of WMLES of $Re_\tau=3000$. Red curve represents CoNFiLD-inlet, blue curve denotes the DFM, while precursor simulation results are marked by black dashed line. From top to bottom: Reynolds shear stress, Turbulence intensity of $u$, $v$, and $w$.}
    \label{fig:wmles_extra_stat_3000_2}
\end{figure}
\begin{figure}[!ht]
    \centering
    \captionsetup[subfloat]{farskip=-4pt,captionskip=-8pt}
    \subfloat{\includegraphics[width=\textwidth]{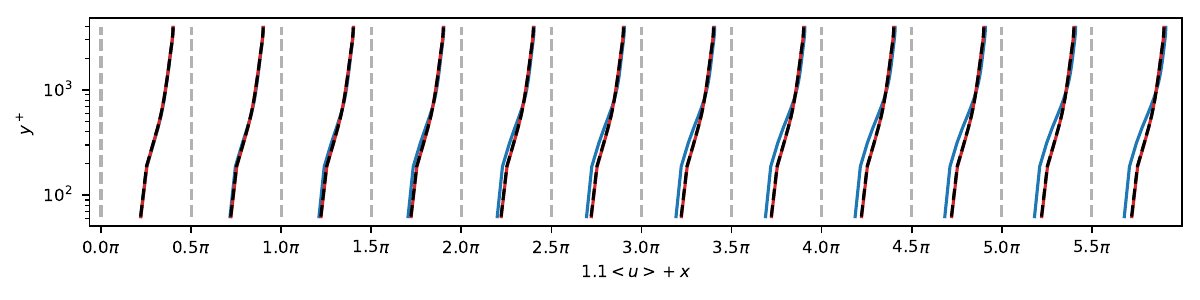}}\\
    \subfloat{\includegraphics[width=\textwidth]{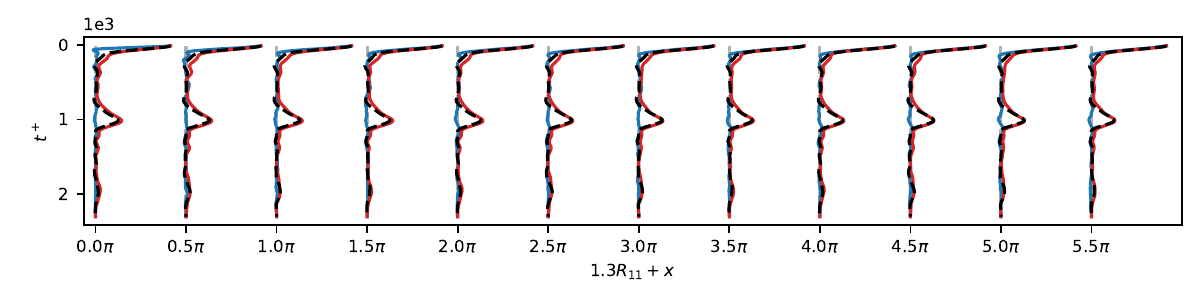}}\\
    \subfloat[$x=0$]{\includegraphics[width=.33\textwidth]{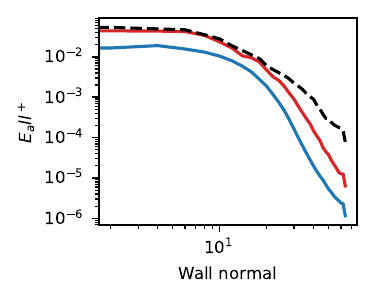}}
    \subfloat[$x=\pi$]{\includegraphics[width=.33\textwidth]{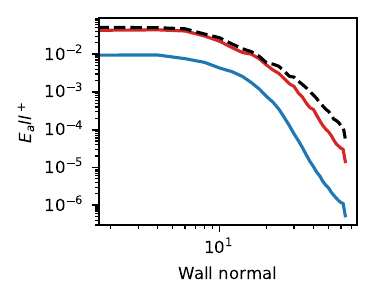}}
    \subfloat[$x=2\pi$]{\includegraphics[width=.33\textwidth]{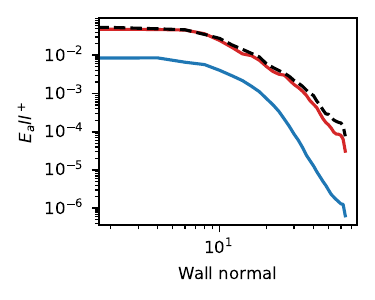}}\\
    \subfloat[$x=3\pi$]{\includegraphics[width=.33\textwidth]{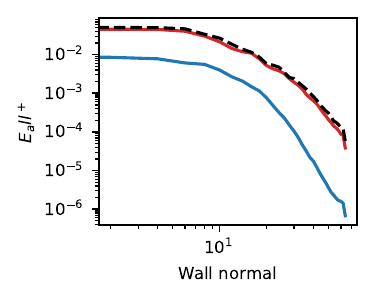}}
    \subfloat[$x=4\pi$]{\includegraphics[width=.33\textwidth]{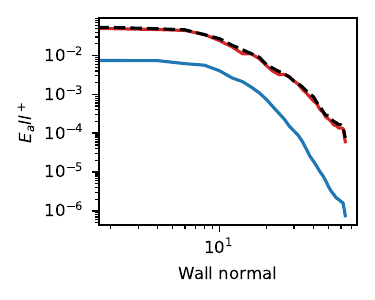}}
    \subfloat[$x=5\pi$]{\includegraphics[width=.33\textwidth]{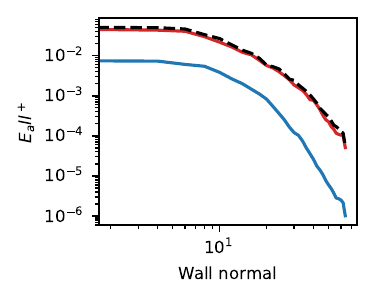}}
    \caption{\emph{A posteriori} test results of WMLES of $Re_\tau=4000$ at different stream-wise locations. Red curve represents CoNFiLD-inlet, blue curve denotes the DFM, while precursor simulation results are marked by black dashed line. First two lines: Mean velocity profile and auto correlation $R_{11}$ near channel center. Others are turbulence kinetic energy (TKE) along spanwise wave length.}
    \label{fig:wmles_extra_stat_4000_1}
\end{figure}
\begin{figure}[!ht]
    \centering
    \captionsetup[subfloat]{farskip=-9pt}
    \subfloat{\includegraphics[width=\textwidth]{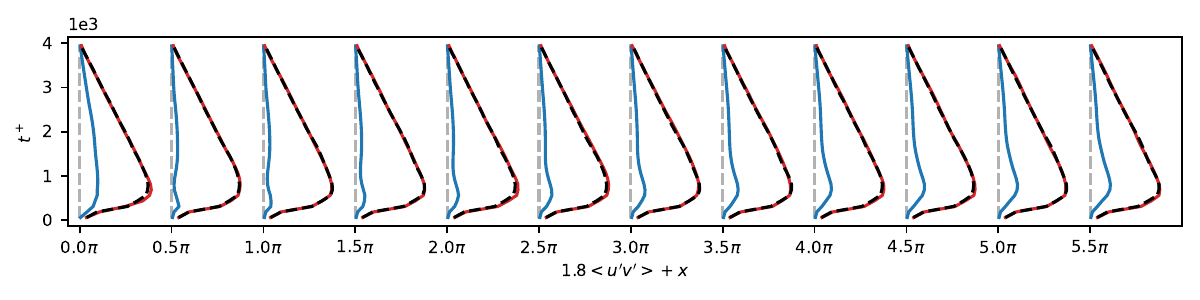}}\\
    \subfloat{\includegraphics[width=\textwidth]{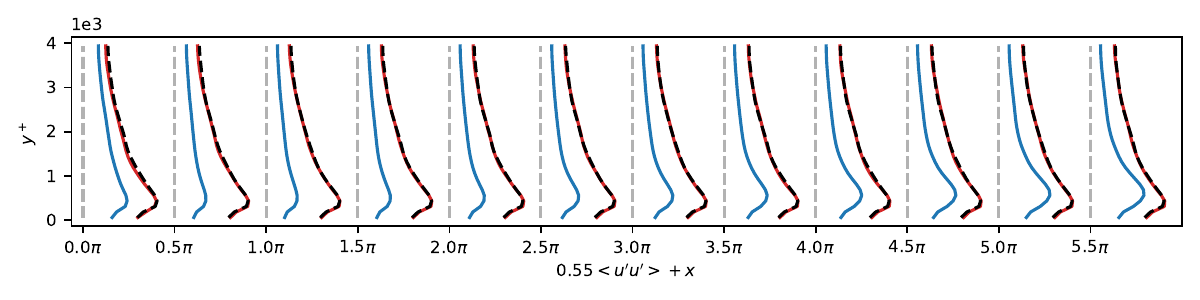}}\\
    \subfloat{\includegraphics[width=\textwidth]{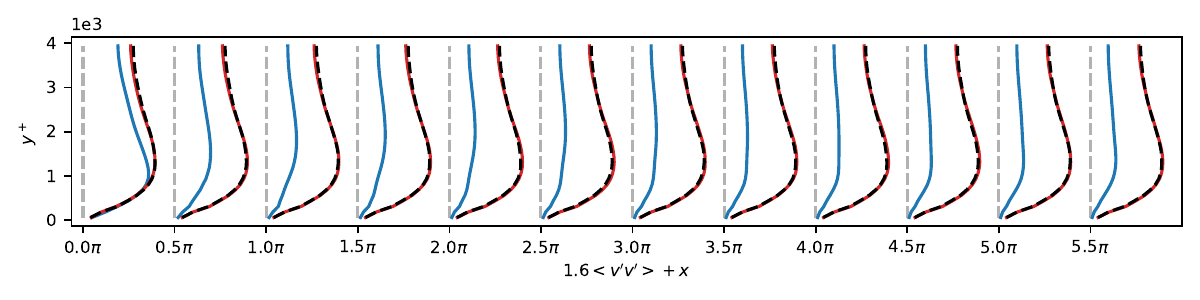}}\\
    \subfloat{\includegraphics[width=\textwidth]{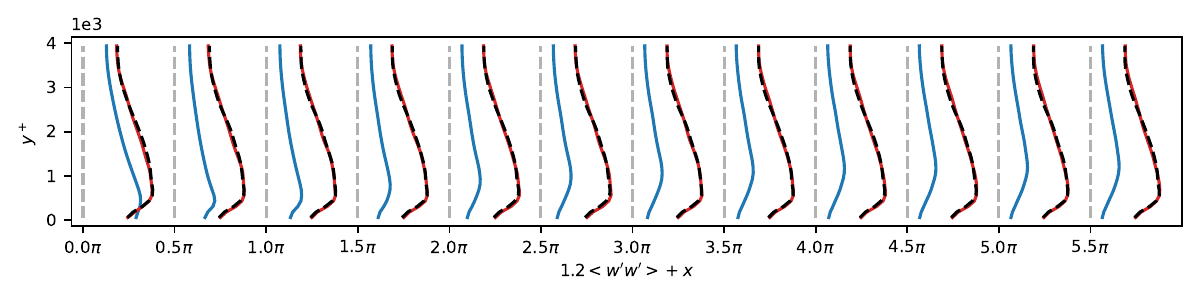}}\\
    \caption{\emph{A posteriori} test results at $12$ different stream-wise locations of WMLES of $Re_\tau=4000$. Red curve represents CoNFiLD-inlet, blue curve denotes the DFM, while precursor simulation results are marked by black dashed line. From top to bottom: Reynolds shear stress, Turbulence intensity of $u$, $v$, and $w$.}
    \label{fig:wmles_extra_stat_4000_2}
\end{figure}
\begin{figure}[!ht]
    \centering
    \captionsetup[subfloat]{farskip=-4pt,captionskip=-8pt}
    \subfloat{\includegraphics[width=\textwidth]{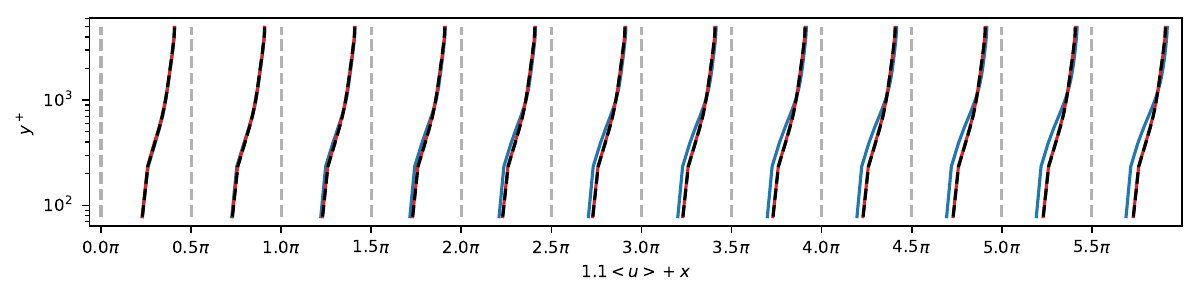}}\\
    \subfloat{\includegraphics[width=\textwidth]{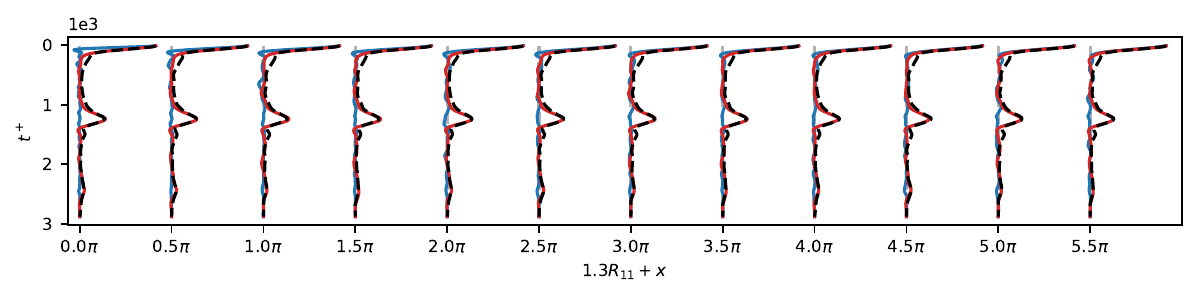}}\\
    \subfloat[$x=0$]{\includegraphics[width=.33\textwidth]{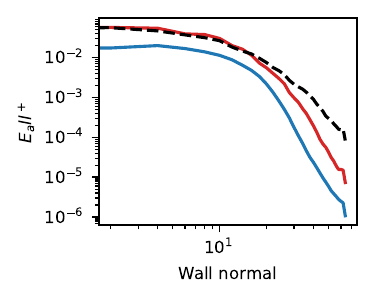}}
    \subfloat[$x=\pi$]{\includegraphics[width=.33\textwidth]{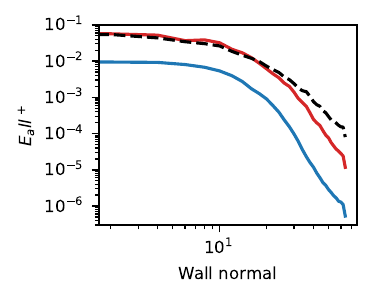}}
    \subfloat[$x=2\pi$]{\includegraphics[width=.33\textwidth]{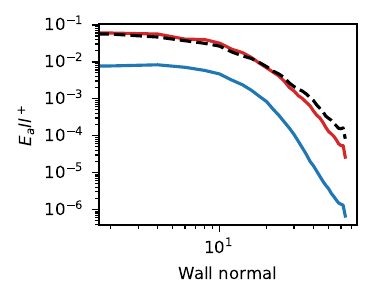}}\\
    \subfloat[$x=3\pi$]{\includegraphics[width=.33\textwidth]{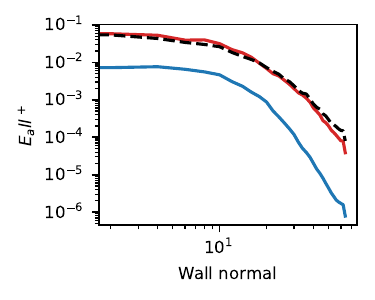}}
    \subfloat[$x=4\pi$]{\includegraphics[width=.33\textwidth]{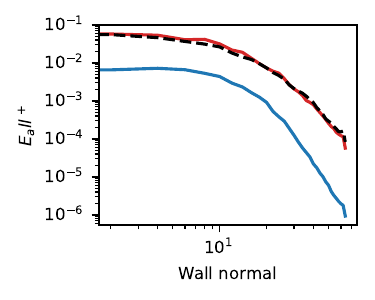}}
    \subfloat[$x=5\pi$]{\includegraphics[width=.33\textwidth]{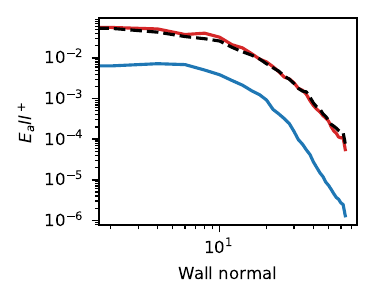}}
    \caption{\emph{A posteriori} test results of WMLES of $Re_\tau=5000$ at different stream-wise locations. Red curve represents CoNFiLD-inlet, blue curve denotes the DFM, while precursor simulation results are marked by black dashed line. First two lines: Mean velocity profile and auto correlation $R_{11}$ near channel center. Others are turbulence kinetic energy (TKE) along spanwise wave length.}
    \label{fig:wmles_extra_stat_5000_1}
\end{figure}
\begin{figure}[!ht]
    \centering
    \captionsetup[subfloat]{farskip=-9pt}
    \subfloat{\includegraphics[width=\textwidth]{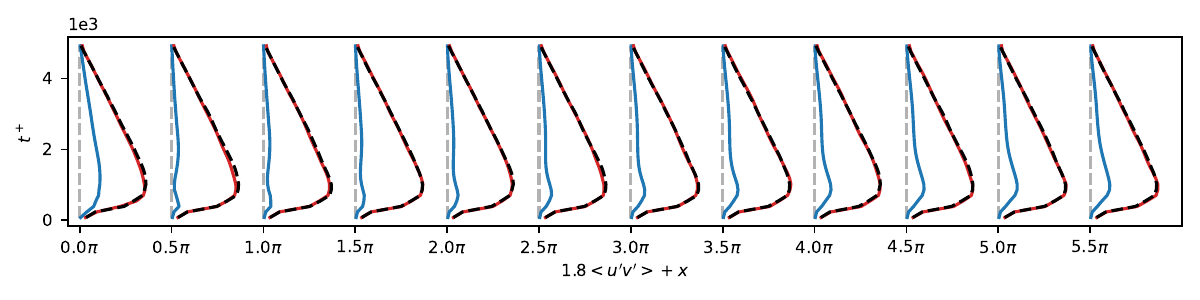}}\\
    \subfloat{\includegraphics[width=\textwidth]{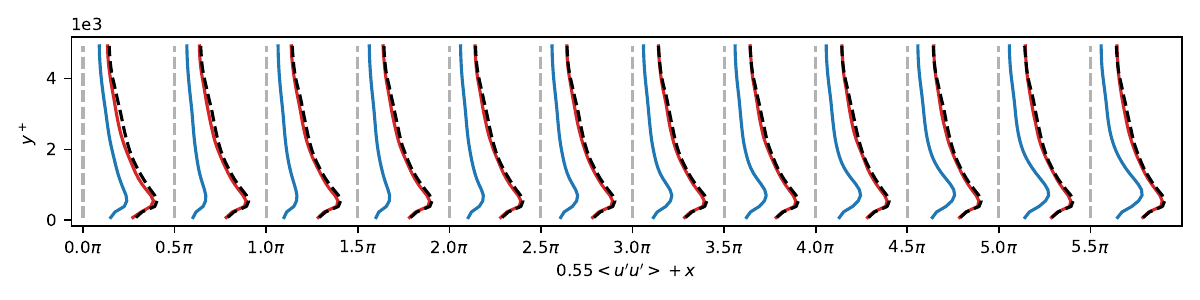}}\\
    \subfloat{\includegraphics[width=\textwidth]{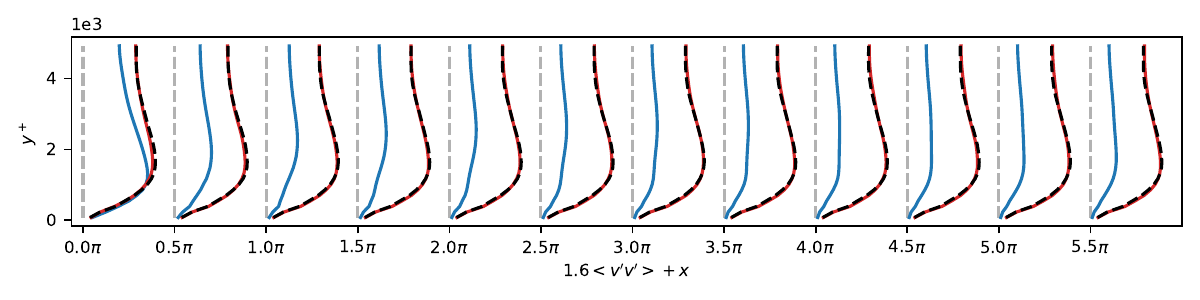}}\\
    \subfloat{\includegraphics[width=\textwidth]{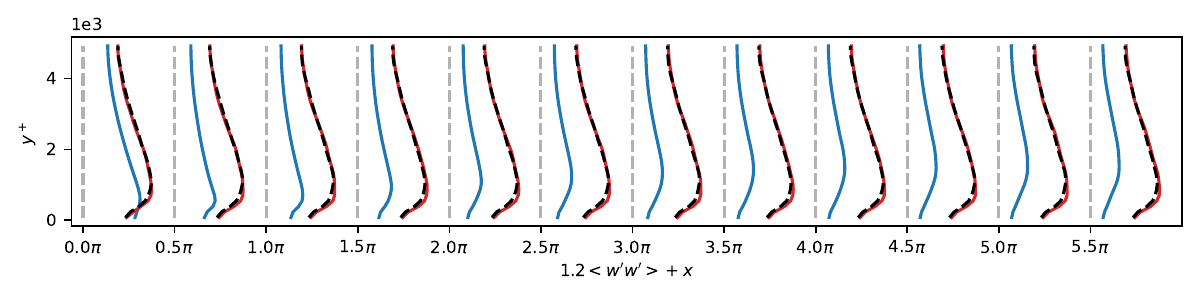}}\\
    \caption{\emph{A posteriori} test results at $12$ different stream-wise locations of WMLES of $Re_\tau=5000$. Red curve represents CoNFiLD-inlet, blue curve denotes the DFM, while precursor simulation results are marked by black dashed line. From top to bottom: Reynolds shear stress, Turbulence intensity of $u$, $v$, and $w$.}
    \label{fig:wmles_extra_stat_5000_2}
\end{figure}
\begin{figure}[!ht]
    \centering
    \captionsetup[subfloat]{farskip=-4pt,captionskip=-8pt}
    \subfloat{\includegraphics[width=\textwidth]{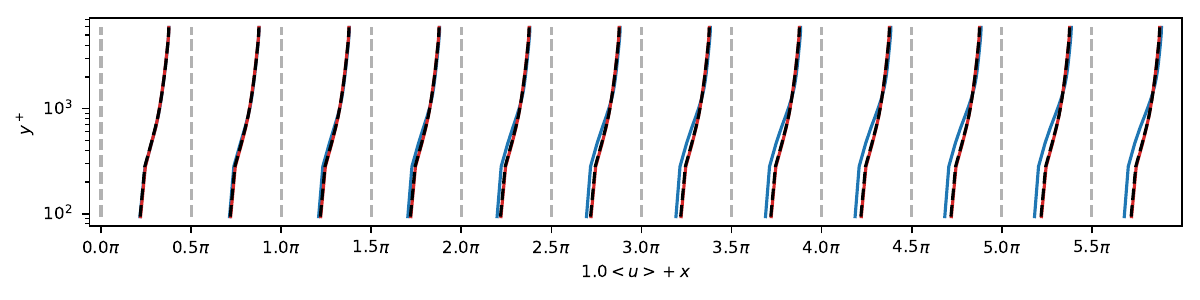}}\\
    \subfloat{\includegraphics[width=\textwidth]{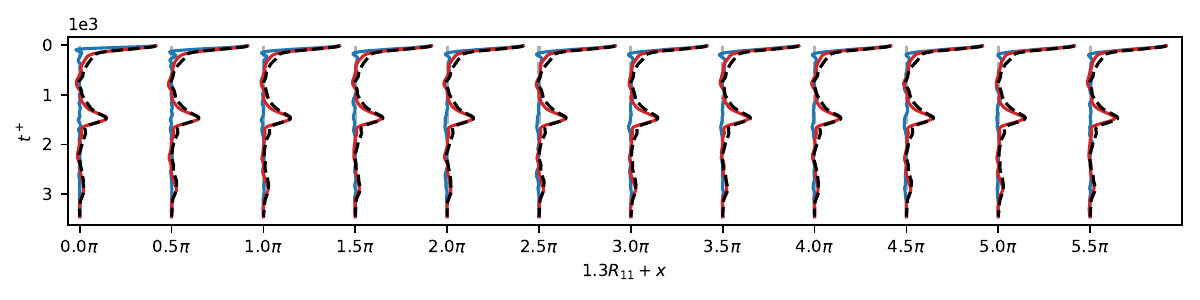}}\\
    \subfloat[$x=0$]{\includegraphics[width=.33\textwidth]{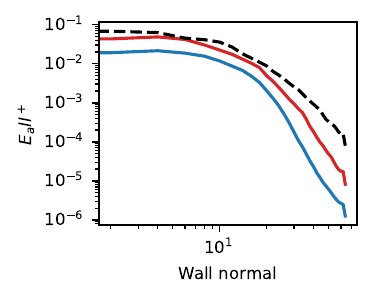}}
    \subfloat[$x=\pi$]{\includegraphics[width=.33\textwidth]{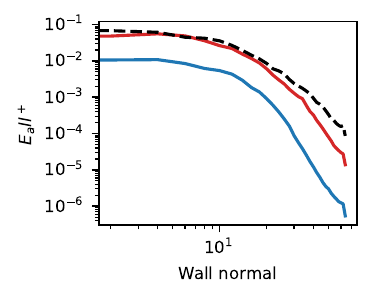}}
    \subfloat[$x=2\pi$]{\includegraphics[width=.33\textwidth]{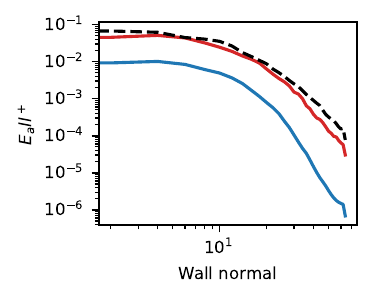}}\\
    \subfloat[$x=3\pi$]{\includegraphics[width=.33\textwidth]{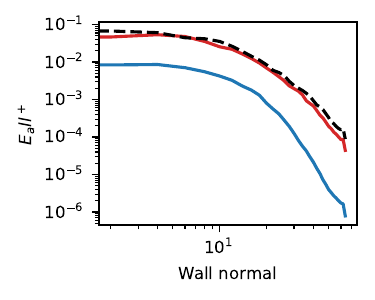}}
    \subfloat[$x=4\pi$]{\includegraphics[width=.33\textwidth]{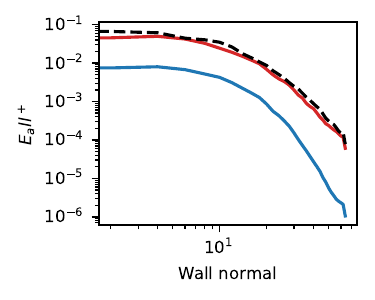}}
    \subfloat[$x=5\pi$]{\includegraphics[width=.33\textwidth]{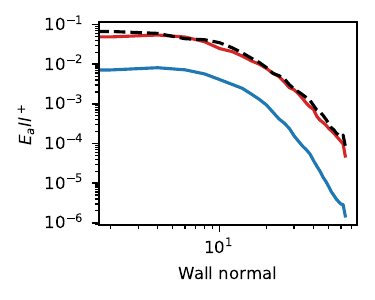}}
    \caption{\emph{A posteriori} test results of WMLES of $Re_\tau=6000$ at different stream-wise locations. Red curve represents CoNFiLD-inlet, blue curve denotes the DFM, while precursor simulation results are marked by black dashed line. First two lines: Mean velocity profile and auto correlation $R_{11}$ near channel center. Others are turbulence kinetic energy (TKE) along spanwise wave length.}
    \label{fig:wmles_extra_stat_6000_1}
\end{figure}
\begin{figure}[!ht]
    \centering
    \captionsetup[subfloat]{farskip=-9pt}
    \subfloat{\includegraphics[width=\textwidth]{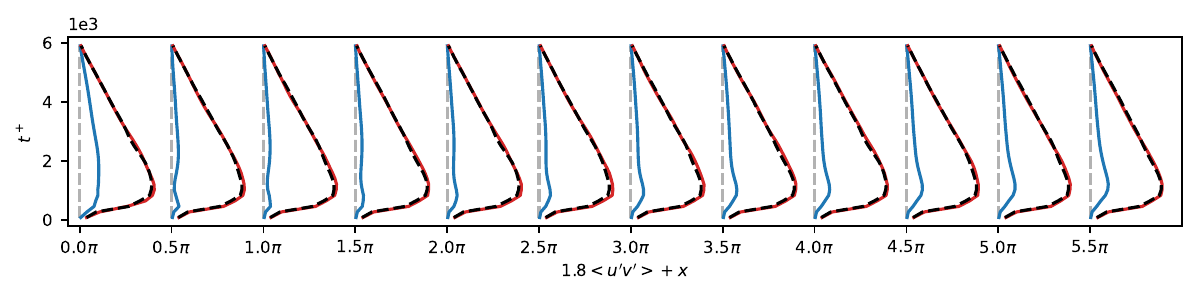}}\\
    \subfloat{\includegraphics[width=\textwidth]{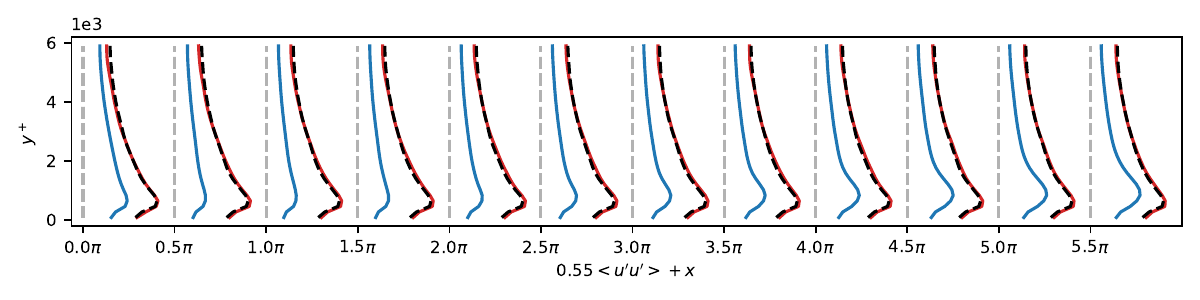}}\\
    \subfloat{\includegraphics[width=\textwidth]{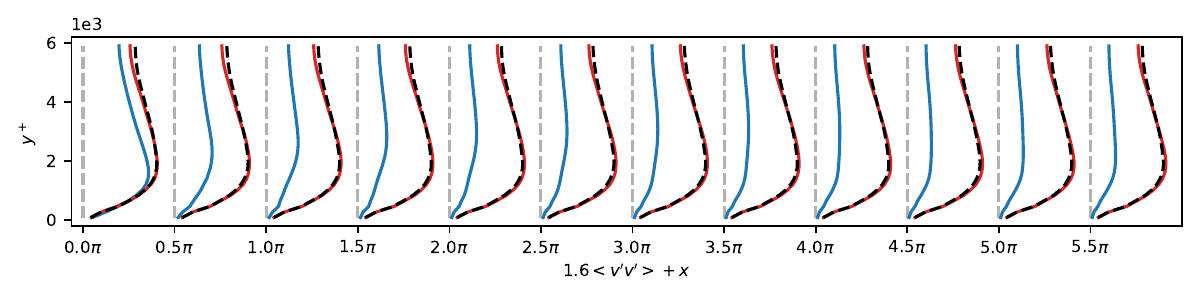}}\\
    \subfloat{\includegraphics[width=\textwidth]{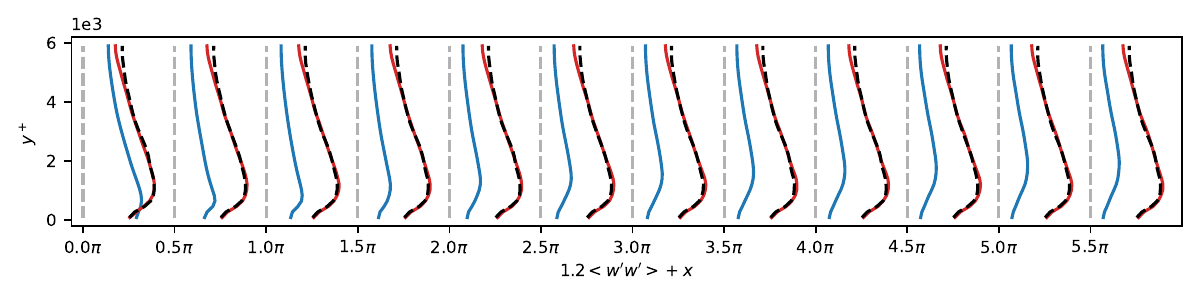}}\\
    \caption{\emph{A posteriori} test results at $12$ different stream-wise locations of WMLES of $Re_\tau=6000$. Red curve represents CoNFiLD-inlet, blue curve denotes the DFM, while precursor simulation results are marked by black dashed line. From top to bottom: Reynolds shear stress, Turbulence intensity of $u$, $v$, and $w$.}
    \label{fig:wmles_extra_stat_6000_2}
\end{figure}
\begin{figure}[!ht]
    \centering
    \captionsetup[subfloat]{farskip=-4pt,captionskip=-8pt}
    \subfloat{\includegraphics[width=\textwidth]{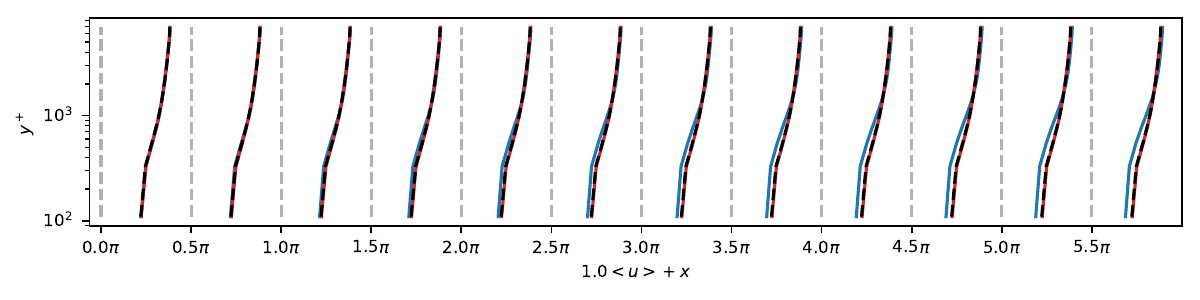}}\\
    \subfloat{\includegraphics[width=\textwidth]{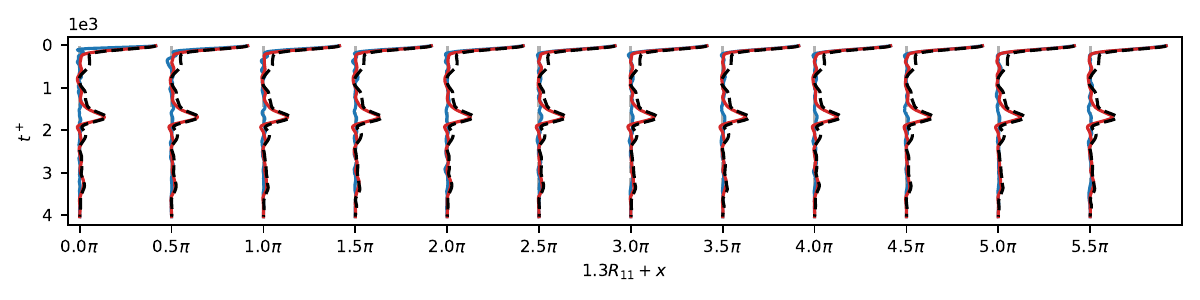}}\\
    \subfloat[$x=0$]{\includegraphics[width=.33\textwidth]{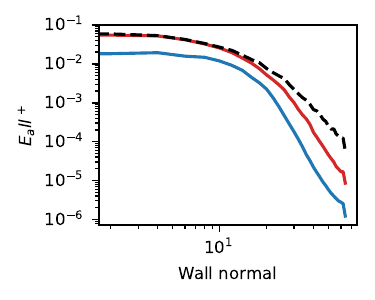}}
    \subfloat[$x=\pi$]{\includegraphics[width=.33\textwidth]{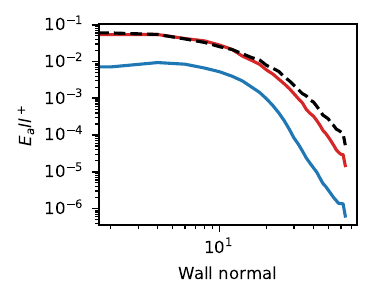}}
    \subfloat[$x=2\pi$]{\includegraphics[width=.33\textwidth]{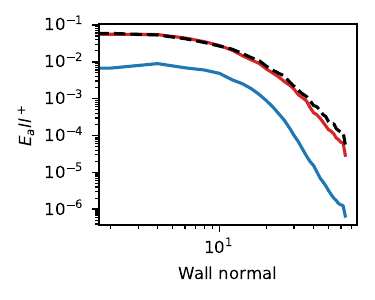}}\\
    \subfloat[$x=3\pi$]{\includegraphics[width=.33\textwidth]{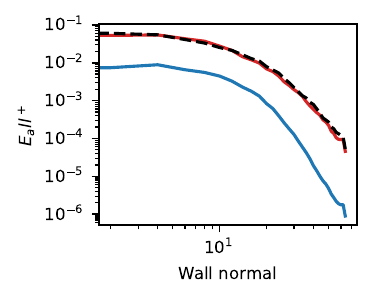}}
    \subfloat[$x=4\pi$]{\includegraphics[width=.33\textwidth]{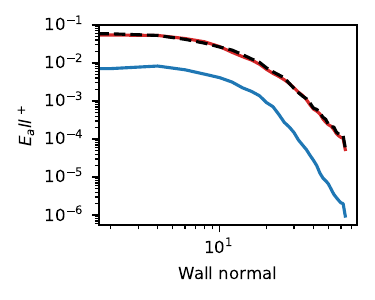}}
    \subfloat[$x=5\pi$]{\includegraphics[width=.33\textwidth]{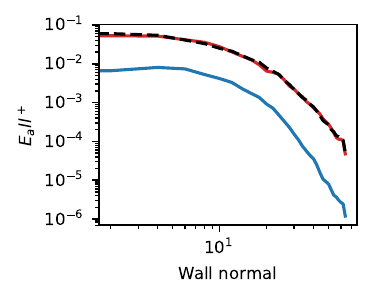}}
    \caption{\emph{A posteriori} test results of WMLES of $Re_\tau=7000$ at different stream-wise locations. Red curve represents CoNFiLD-inlet, blue curve denotes the DFM, while precursor simulation results are marked by black dashed line. First two lines: Mean velocity profile and auto correlation $R_{11}$ near channel center. Others are turbulence kinetic energy (TKE) along spanwise wave length.}
    \label{fig:wmles_extra_stat_7000_1}
\end{figure}
\begin{figure}[!ht]
    \centering
    \captionsetup[subfloat]{farskip=-9pt}
    \subfloat{\includegraphics[width=\textwidth]{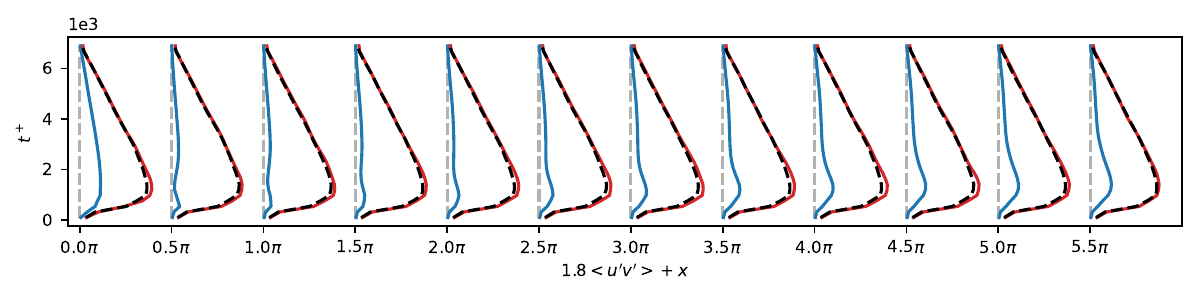}}\\
    \subfloat{\includegraphics[width=\textwidth]{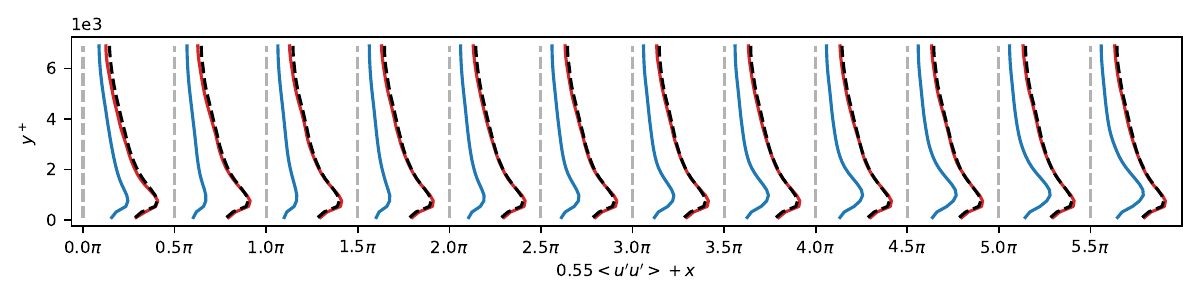}}\\
    \subfloat{\includegraphics[width=\textwidth]{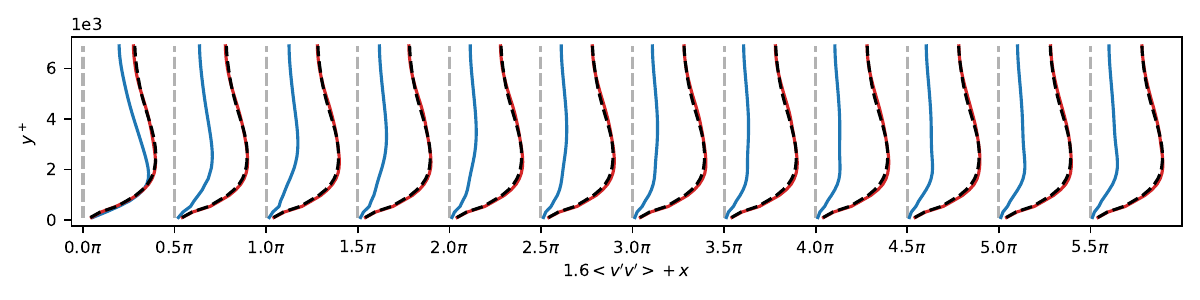}}\\
    \subfloat{\includegraphics[width=\textwidth]{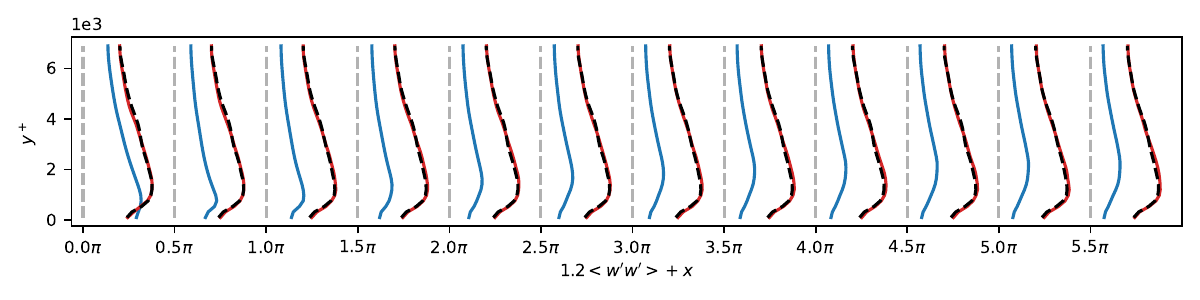}}\\
    \caption{\emph{A posteriori} test results at $12$ different stream-wise locations of WMLES of $Re_\tau=7000$. Red curve represents CoNFiLD-inlet, blue curve denotes the DFM, while precursor simulation results are marked by black dashed line. From top to bottom: Reynolds shear stress, Turbulence intensity of $u$, $v$, and $w$.}
    \label{fig:wmles_extra_stat_7000_2}
\end{figure}
\begin{figure}[!ht]
    \centering
    \captionsetup[subfloat]{farskip=-4pt,captionskip=-8pt}
    \subfloat{\includegraphics[width=\textwidth]{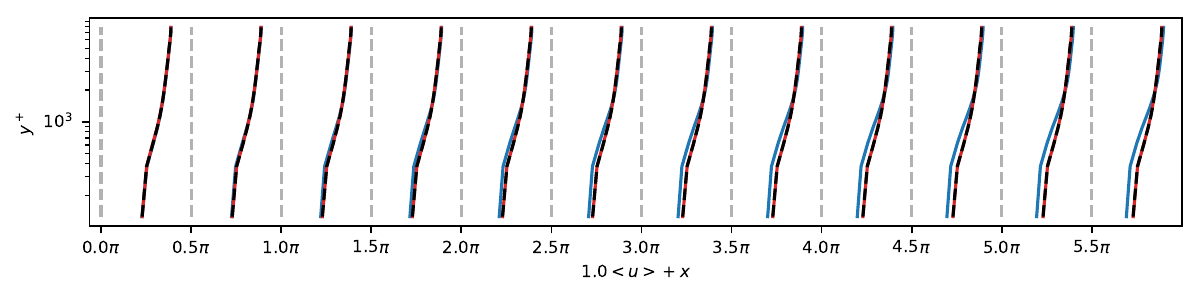}}\\
    \subfloat{\includegraphics[width=\textwidth]{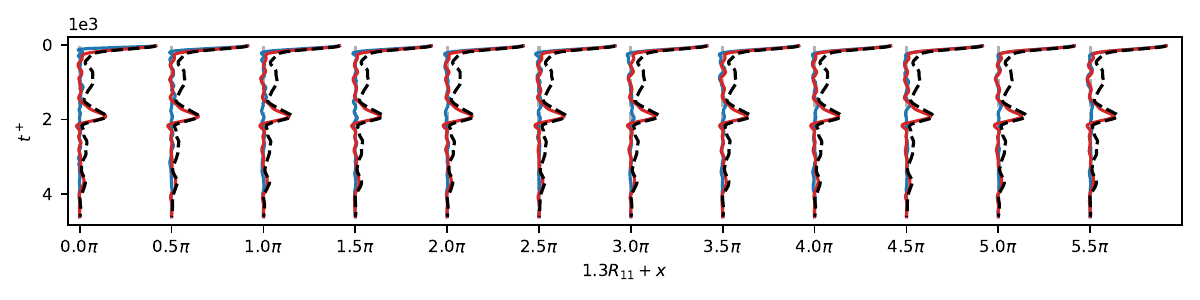}}\\
    \subfloat[$x=0$]{\includegraphics[width=.33\textwidth]{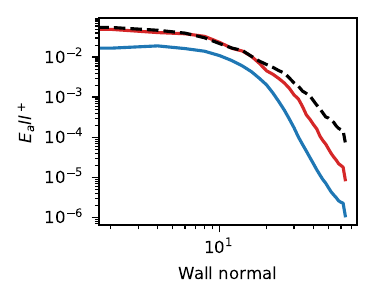}}
    \subfloat[$x=\pi$]{\includegraphics[width=.33\textwidth]{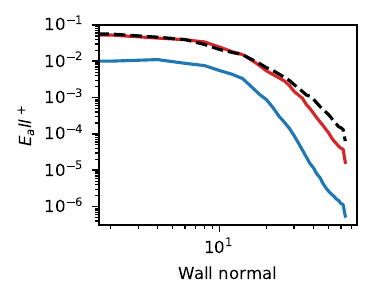}}
    \subfloat[$x=2\pi$]{\includegraphics[width=.33\textwidth]{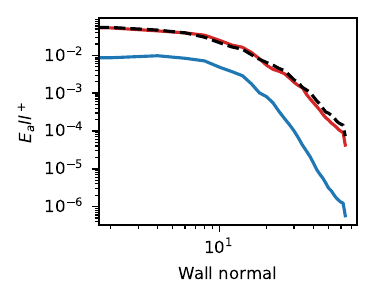}}\\
    \subfloat[$x=3\pi$]{\includegraphics[width=.33\textwidth]{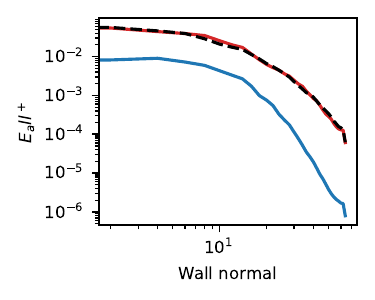}}
    \subfloat[$x=4\pi$]{\includegraphics[width=.33\textwidth]{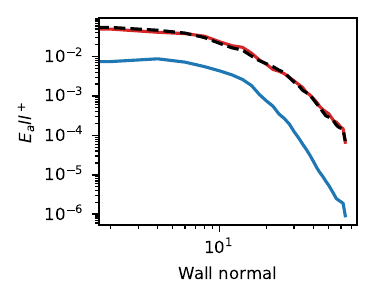}}
    \subfloat[$x=5\pi$]{\includegraphics[width=.33\textwidth]{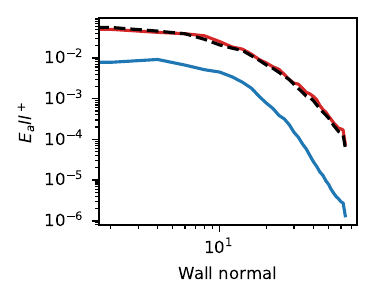}}
    \caption{\emph{A posteriori} test results of WMLES of $Re_\tau=8000$ at different stream-wise locations. Red curve represents CoNFiLD-inlet, blue curve denotes the DFM, while precursor simulation results are marked by black dashed line. First two lines: Mean velocity profile and auto correlation $R_{11}$ near channel center. Others are turbulence kinetic energy (TKE) along spanwise wave length.}
    \label{fig:wmles_extra_stat_8000_1}
\end{figure}
\begin{figure}[!ht]
    \centering
    \captionsetup[subfloat]{farskip=-9pt}
    \subfloat{\includegraphics[width=\textwidth]{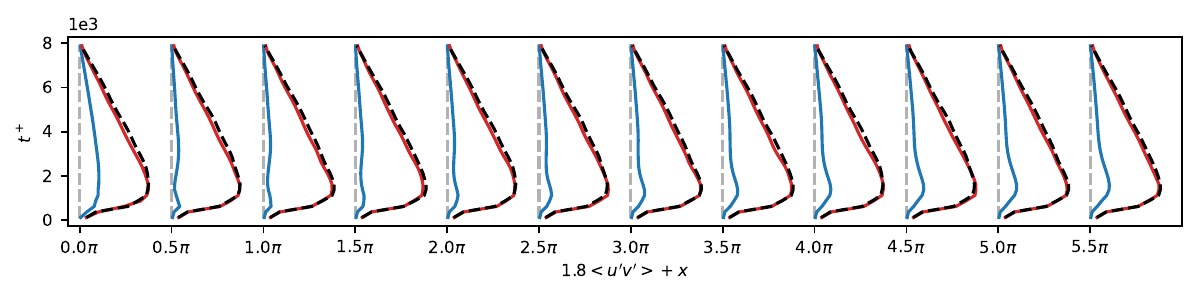}}\\
    \subfloat{\includegraphics[width=\textwidth]{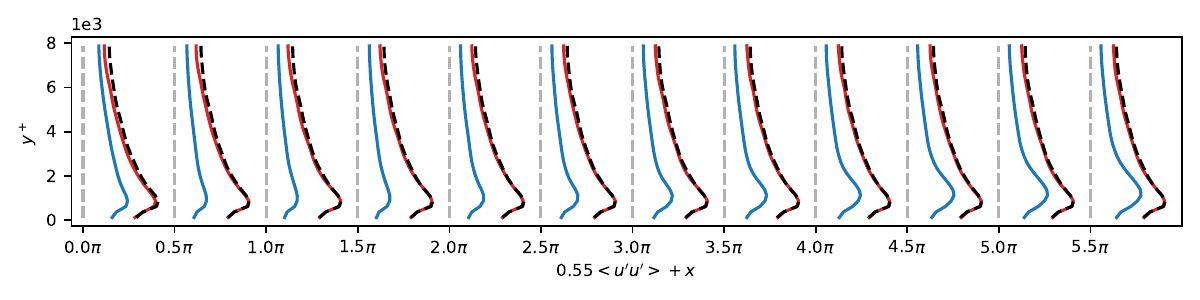}}\\
    \subfloat{\includegraphics[width=\textwidth]{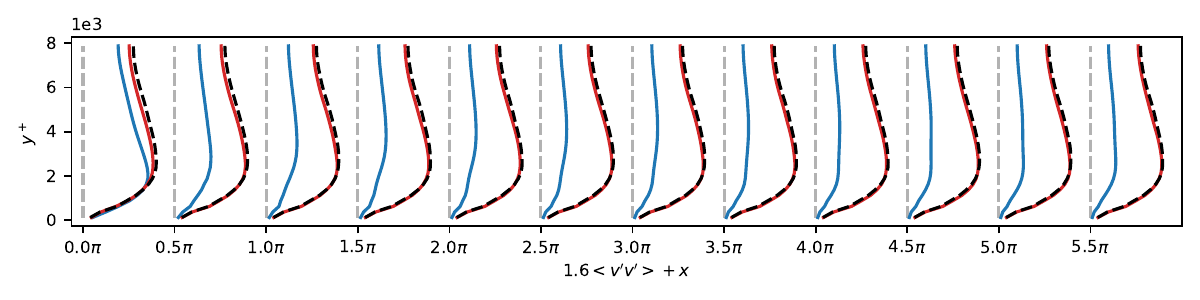}}\\
    \subfloat{\includegraphics[width=\textwidth]{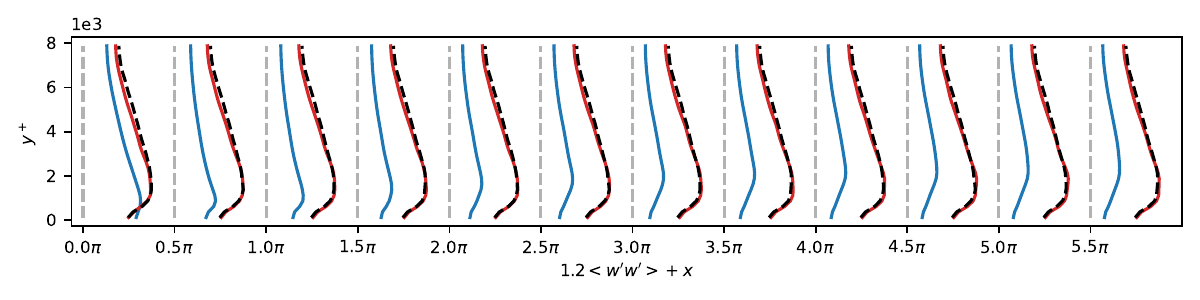}}\\
    \caption{\emph{A posteriori} test results at $12$ different stream-wise locations of WMLES of $Re_\tau=8000$. Red curve represents CoNFiLD-inlet, blue curve denotes the DFM, while precursor simulation results are marked by black dashed line. From top to bottom: Reynolds shear stress, Turbulence intensity of $u$, $v$, and $w$.}
    \label{fig:wmles_extra_stat_8000_2}
\end{figure}
\begin{figure}[!ht]
    \centering
    \captionsetup[subfloat]{farskip=-4pt,captionskip=-8pt}
    \subfloat{\includegraphics[width=\textwidth]{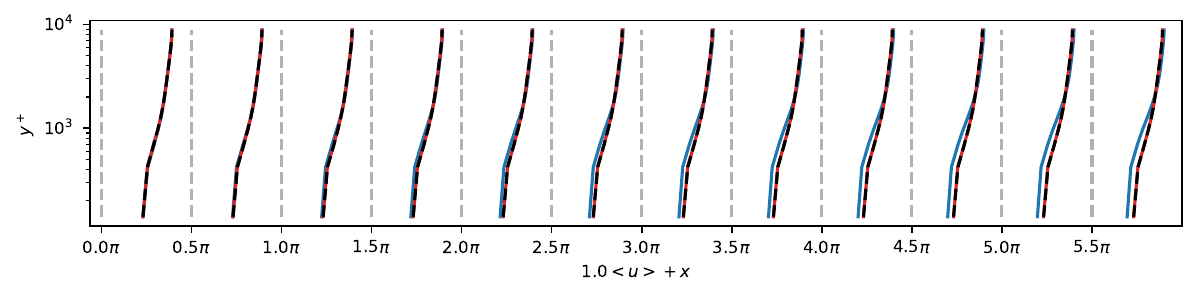}}\\
    \subfloat{\includegraphics[width=\textwidth]{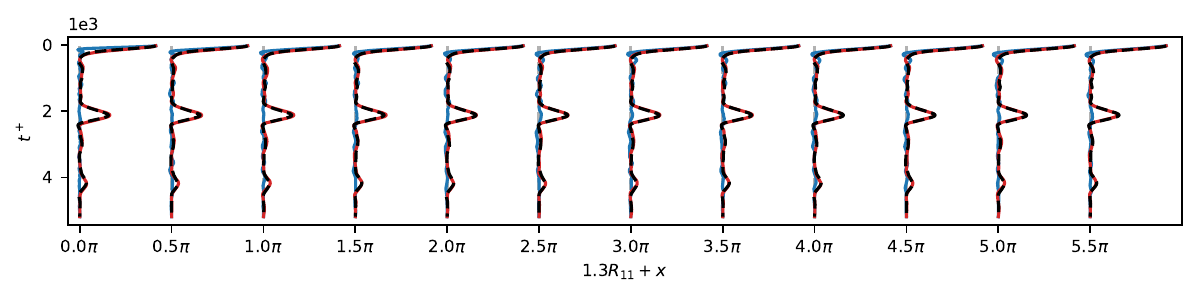}}\\
    \subfloat[$x=0$]{\includegraphics[width=.33\textwidth]{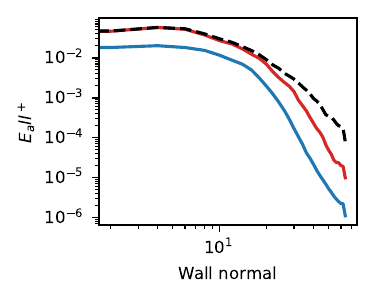}}
    \subfloat[$x=\pi$]{\includegraphics[width=.33\textwidth]{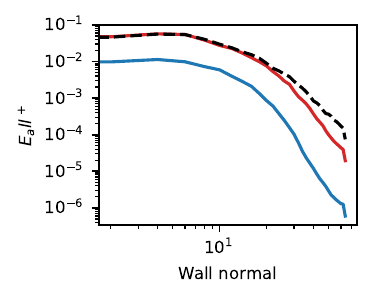}}
    \subfloat[$x=2\pi$]{\includegraphics[width=.33\textwidth]{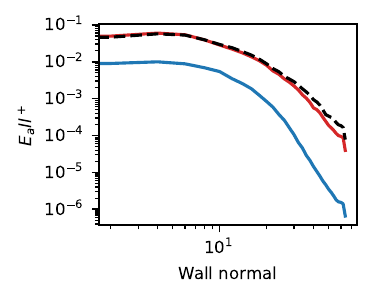}}\\
    \subfloat[$x=3\pi$]{\includegraphics[width=.33\textwidth]{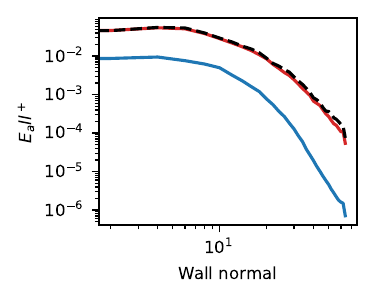}}
    \subfloat[$x=4\pi$]{\includegraphics[width=.33\textwidth]{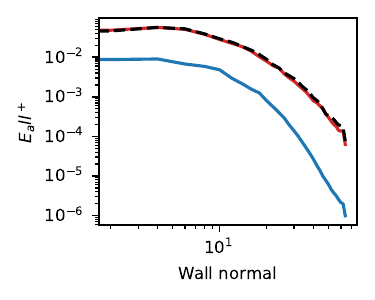}}
    \subfloat[$x=5\pi$]{\includegraphics[width=.33\textwidth]{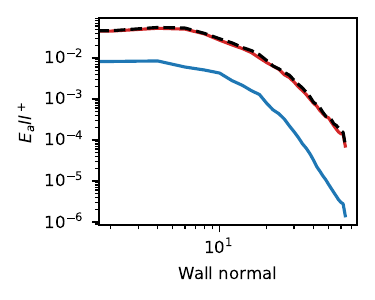}}
    \caption{\emph{A posteriori} test results of WMLES of $Re_\tau=9000$ at different stream-wise locations. Red curve represents CoNFiLD-inlet, blue curve denotes the DFM, while precursor simulation results are marked by black dashed line. First two lines: Mean velocity profile and auto correlation $R_{11}$ near channel center. Others are turbulence kinetic energy (TKE) along spanwise wave length.}
    \label{fig:wmles_extra_stat_9000_1}
\end{figure}
\begin{figure}[!ht]
    \centering
    \captionsetup[subfloat]{farskip=-9pt}
    \subfloat{\includegraphics[width=\textwidth]{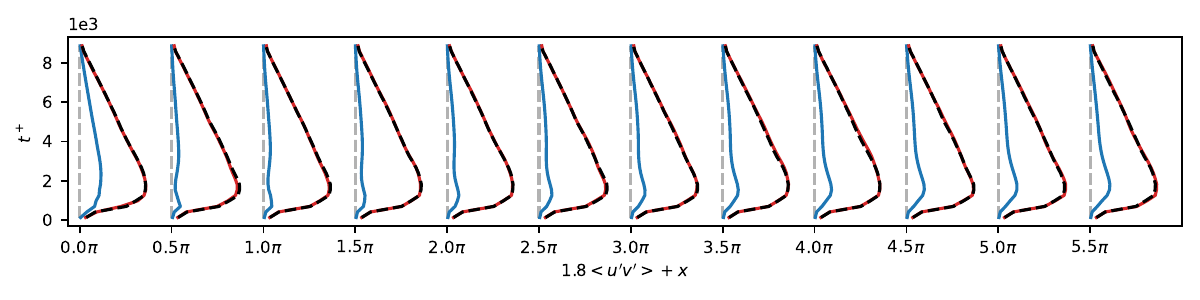}}\\
    \subfloat{\includegraphics[width=\textwidth]{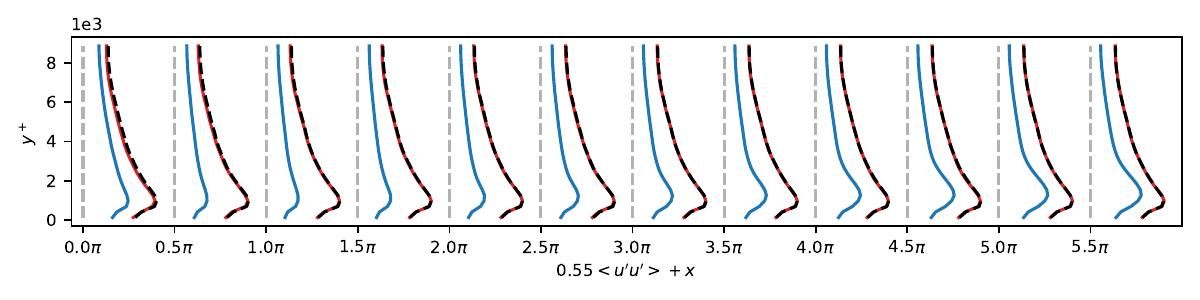}}\\
    \subfloat{\includegraphics[width=\textwidth]{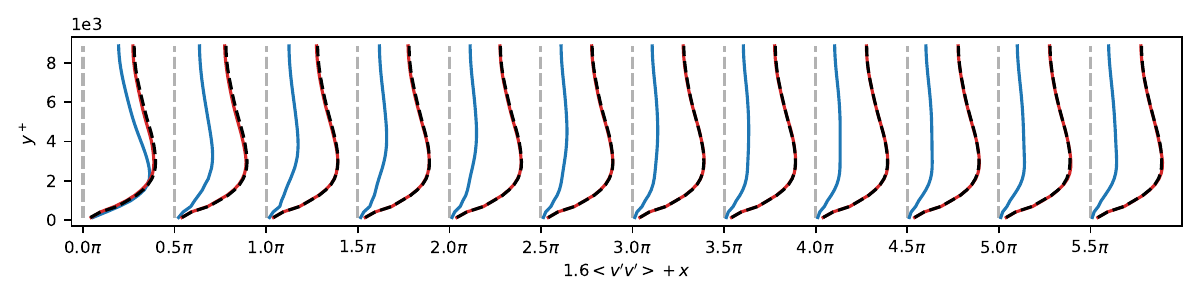}}\\
    \subfloat{\includegraphics[width=\textwidth]{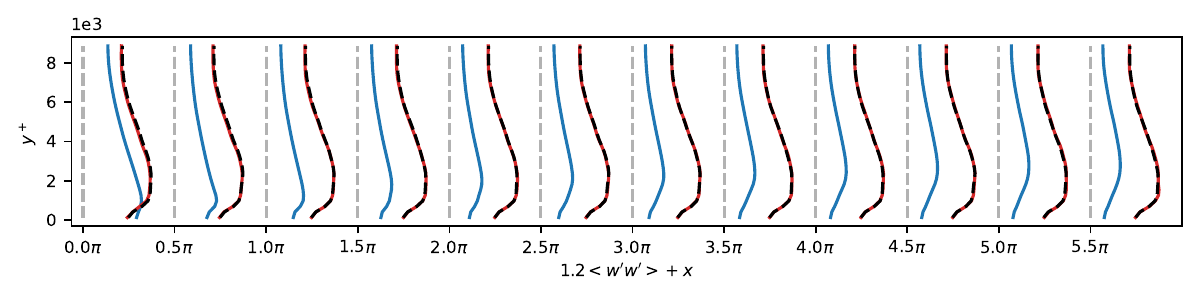}}\\
    \caption{\emph{A posteriori} test results at $12$ different stream-wise locations of WMLES of $Re_\tau=9000$. Red curve represents CoNFiLD-inlet, blue curve denotes the DFM, while precursor simulation results are marked by black dashed line. From top to bottom: Reynolds shear stress, Turbulence intensity of $u$, $v$, and $w$.}
    \label{fig:wmles_extra_stat_9000_2}
\end{figure}
\begin{figure}[!ht]
    \centering
    \captionsetup[subfloat]{farskip=-4pt,captionskip=-8pt}
    \subfloat{\includegraphics[width=\textwidth]{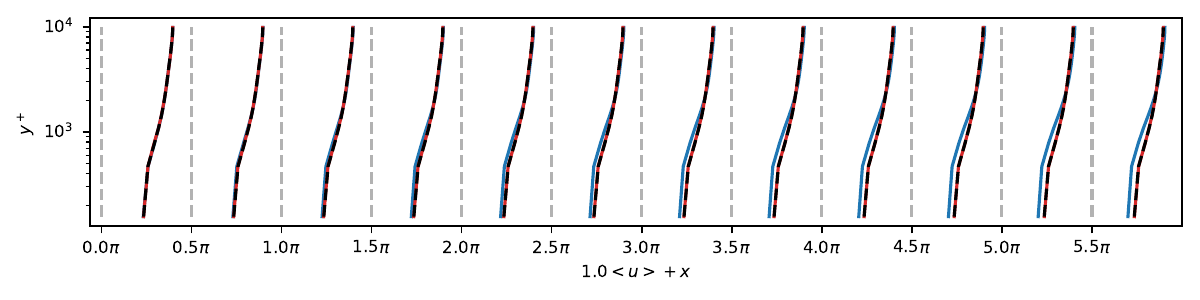}}\\
    \subfloat{\includegraphics[width=\textwidth]{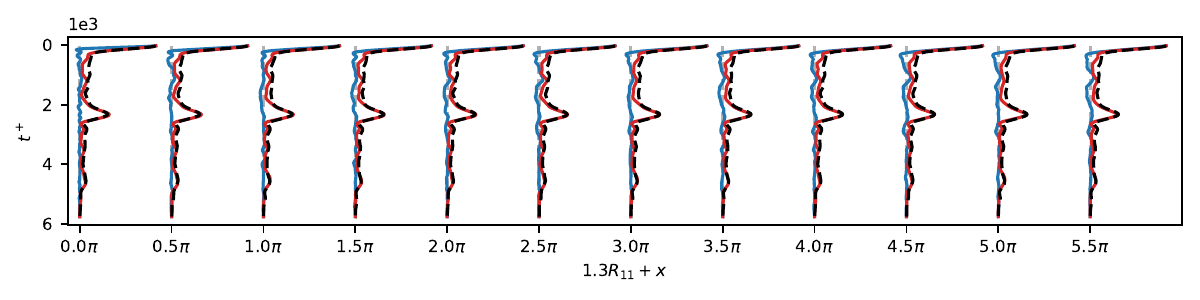}}\\
    \subfloat[$x=0$]{\includegraphics[width=.33\textwidth]{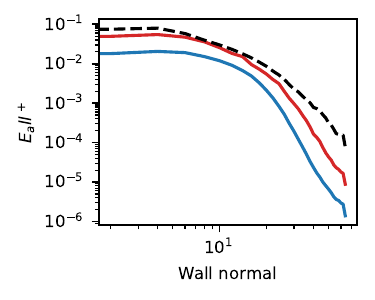}}
    \subfloat[$x=\pi$]{\includegraphics[width=.33\textwidth]{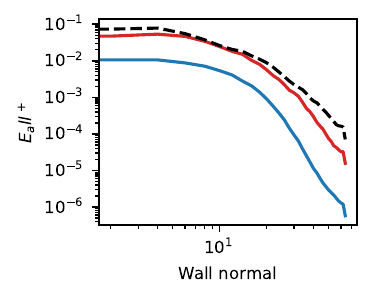}}
    \subfloat[$x=2\pi$]{\includegraphics[width=.33\textwidth]{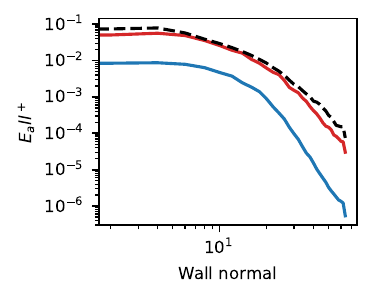}}\\
    \subfloat[$x=3\pi$]{\includegraphics[width=.33\textwidth]{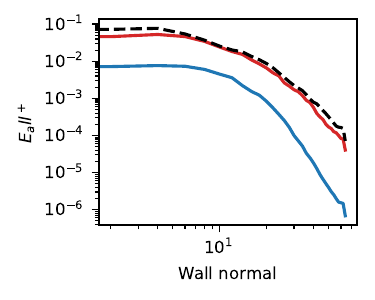}}
    \subfloat[$x=4\pi$]{\includegraphics[width=.33\textwidth]{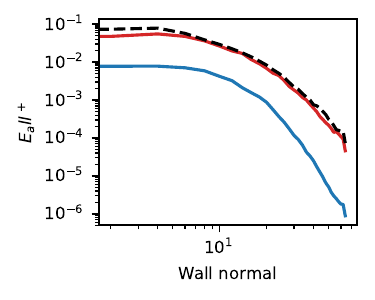}}
    \subfloat[$x=5\pi$]{\includegraphics[width=.33\textwidth]{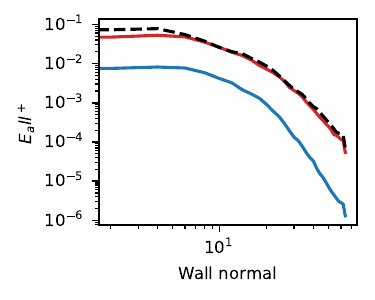}}
    \caption{\emph{A posteriori} test results of WMLES of $Re_\tau=10000$ at different stream-wise locations. Red curve represents CoNFiLD-inlet, blue curve denotes the DFM, while precursor simulation results are marked by black dashed line. First two lines: Mean velocity profile and auto correlation $R_{11}$ near channel center. Others are turbulence kinetic energy (TKE) along spanwise wave length.}
    \label{fig:wmles_extra_stat_10000_1}
\end{figure}
\begin{figure}[!ht]
    \centering
    \captionsetup[subfloat]{farskip=-9pt}
    \subfloat{\includegraphics[width=\textwidth]{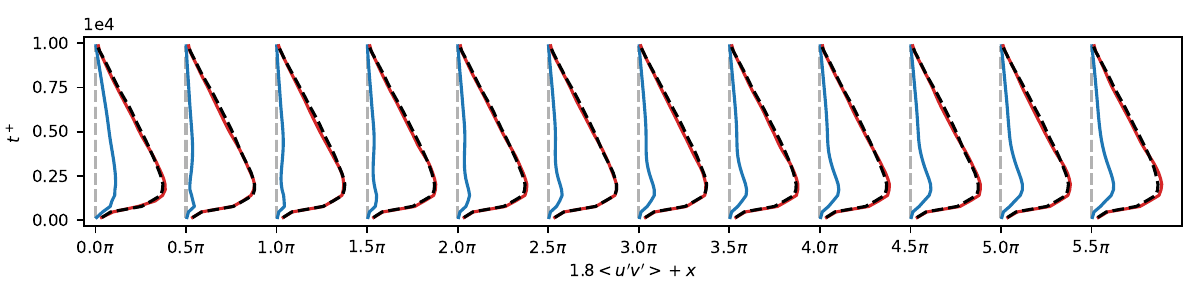}}\\
    \subfloat{\includegraphics[width=\textwidth]{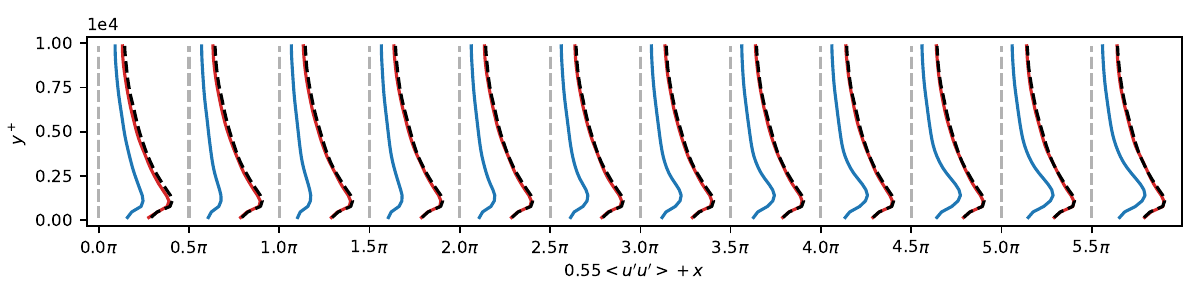}}\\
    \subfloat{\includegraphics[width=\textwidth]{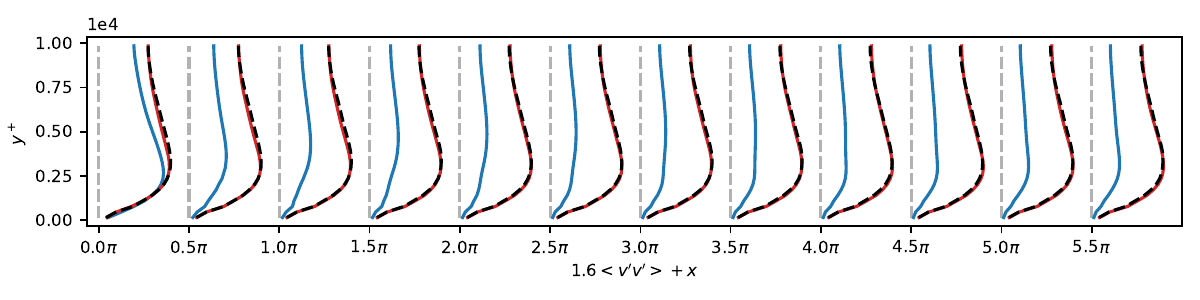}}\\
    \subfloat{\includegraphics[width=\textwidth]{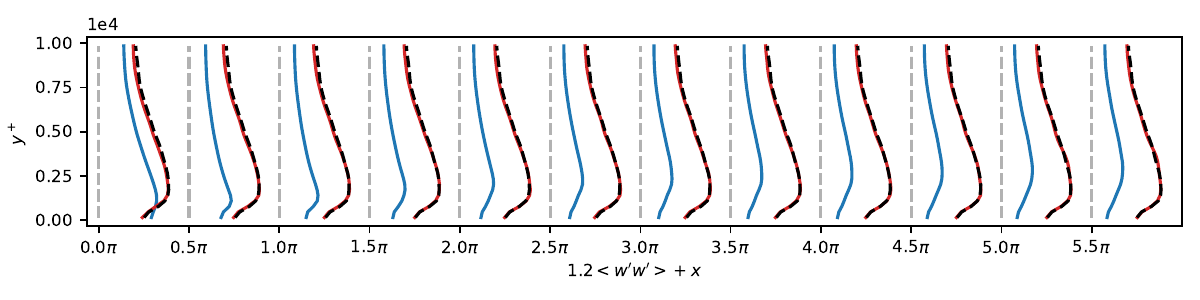}}\\
    \caption{\emph{A posteriori} test results at $12$ different stream-wise locations of WMLES of $Re_\tau=10000$. Red curve represents CoNFiLD-inlet, blue curve denotes the DFM, while precursor simulation results are marked by black dashed line. From top to bottom: Reynolds shear stress, Turbulence intensity of $u$, $v$, and $w$.}
    \label{fig:wmles_extra_stat_10000_2}
\end{figure}
\clearpage
\subsection{More statistics of WMLES inlet generator at unseen Reynolds numbers}
\begin{figure}[!ht]
    \centering
    \captionsetup[subfloat]{farskip=-4pt,captionskip=-8pt}
    \subfloat{\includegraphics[width=\textwidth]{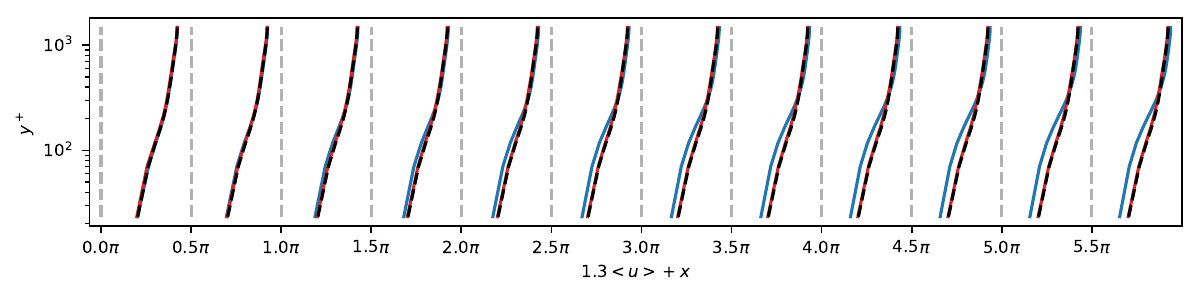}}\\
    \subfloat{\includegraphics[width=\textwidth]{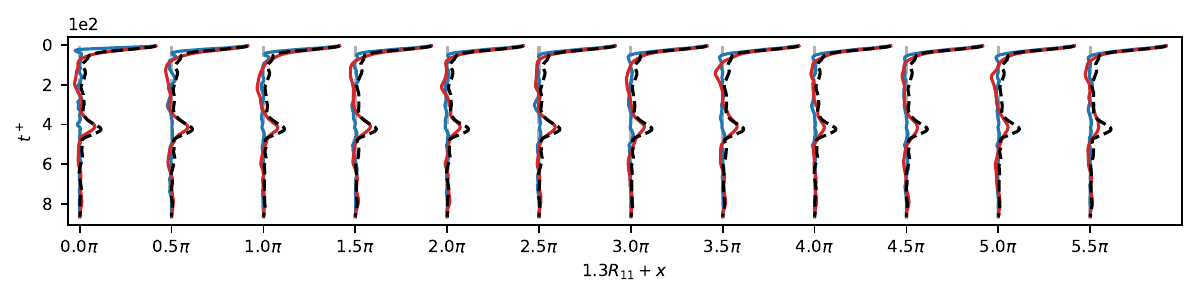}}\\
    \subfloat[$x=0$]{\includegraphics[width=.33\textwidth]{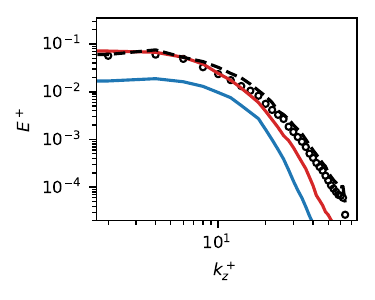}}
    \subfloat[$x=\pi$]{\includegraphics[width=.33\textwidth]{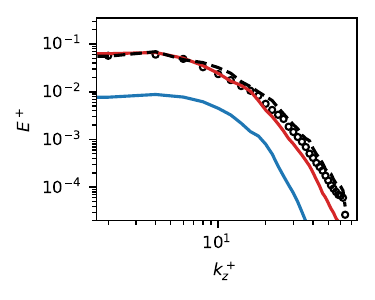}}
    \subfloat[$x=2\pi$]{\includegraphics[width=.33\textwidth]{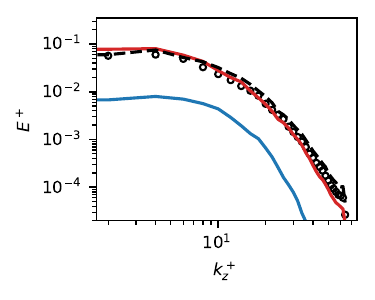}}\\
    \subfloat[$x=3\pi$]{\includegraphics[width=.33\textwidth]{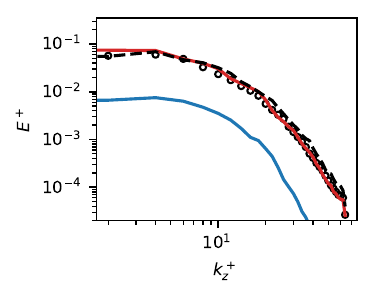}}
    \subfloat[$x=4\pi$]{\includegraphics[width=.33\textwidth]{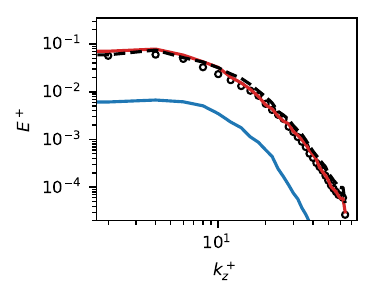}}
    \subfloat[$x=5\pi$]{\includegraphics[width=.33\textwidth]{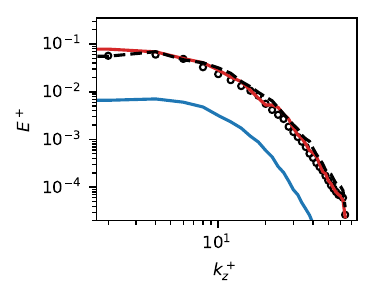}}
    \caption{\emph{A posteriori} test results of WMLES of $Re_\tau=1500$ at different stream-wise locations. Red curve represents CoNFiLD-inlet, blue curve denotes the DFM, while precursor simulation results are marked by black dashed line. First two lines: Mean velocity profile and auto correlation $R_{11}$ near channel center. Others are turbulence kinetic energy (TKE) along spanwise wave length.}
    \label{fig:wmles_extra_stat_1500_1}
\end{figure}
\begin{figure}[!ht]
    \centering
    \captionsetup[subfloat]{farskip=-9pt}
    \subfloat{\includegraphics[width=\textwidth]{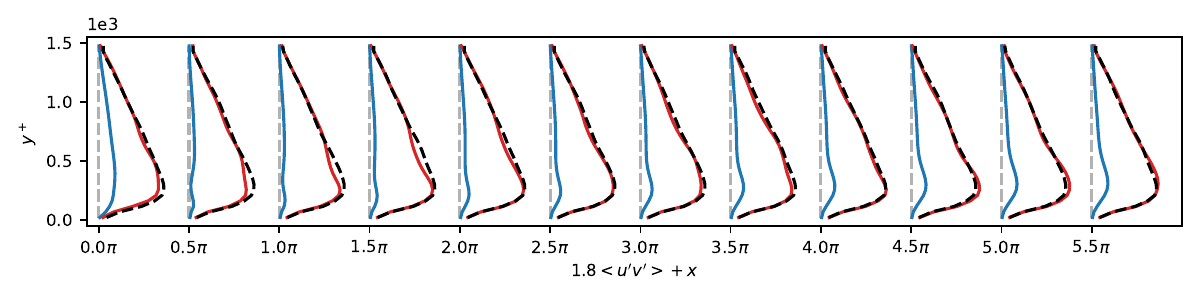}}\\
    \subfloat{\includegraphics[width=\textwidth]{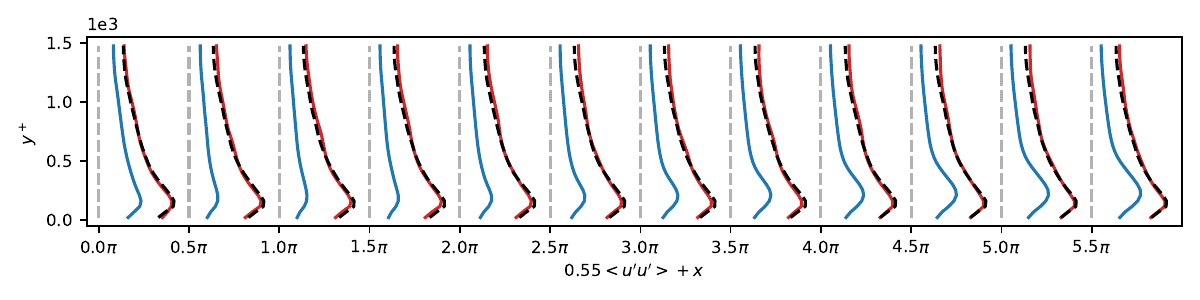}}\\
    \subfloat{\includegraphics[width=\textwidth]{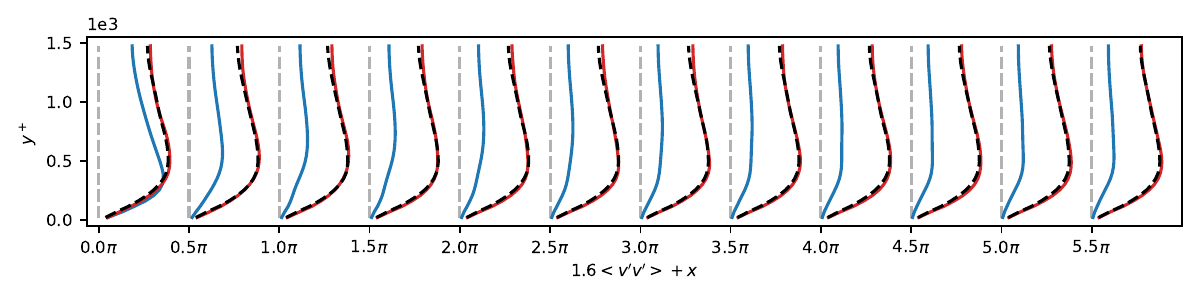}}\\
    \subfloat{\includegraphics[width=\textwidth]{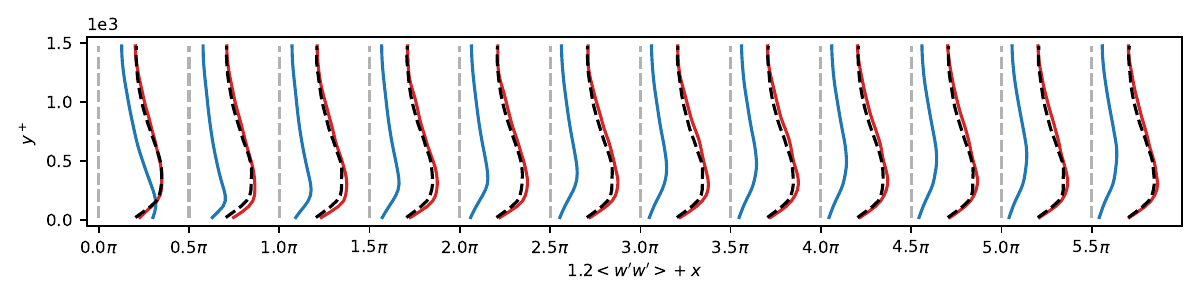}}\\
    \caption{\emph{A posteriori} test results at $12$ different stream-wise locations of WMLES of $Re_\tau=1500$. Red curve represents CoNFiLD-inlet, blue curve denotes the DFM, while precursor simulation results are marked by black dashed line. From top to bottom: Reynolds shear stress, Turbulence intensity of $u$, $v$, and $w$.}
    \label{fig:wmles_extra_stat_1500_2}
\end{figure}\begin{figure}[!ht]
    \centering
    \captionsetup[subfloat]{farskip=-4pt,captionskip=-8pt}
    \subfloat{\includegraphics[width=\textwidth]{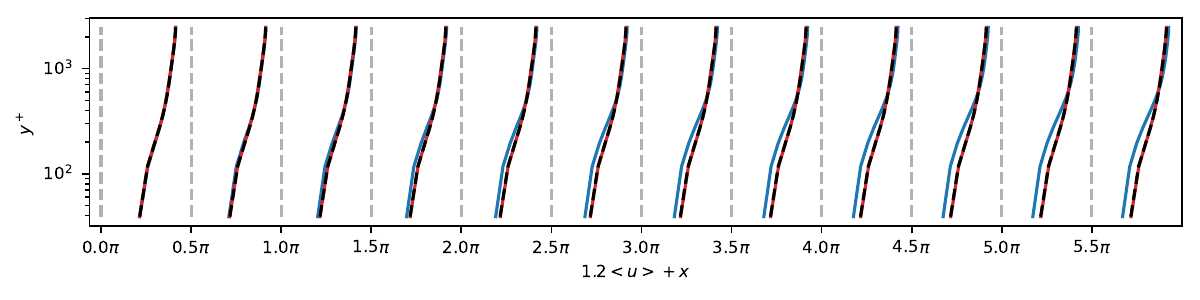}}\\
    \subfloat{\includegraphics[width=\textwidth]{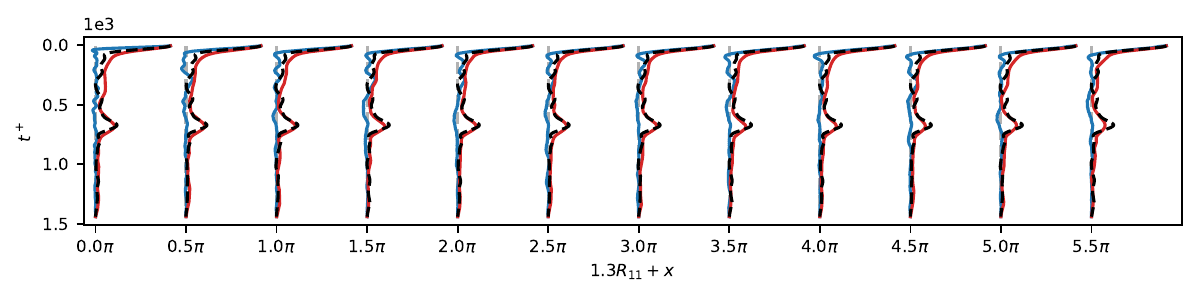}}\\
    \subfloat[$x=0$]{\includegraphics[width=.33\textwidth]{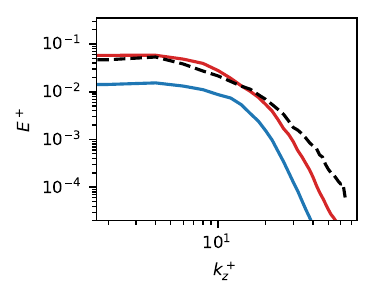}}
    \subfloat[$x=\pi$]{\includegraphics[width=.33\textwidth]{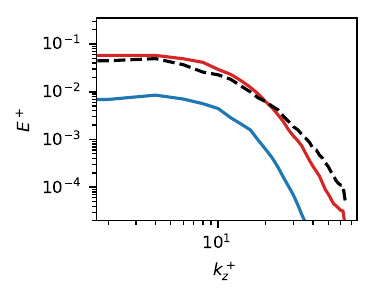}}
    \subfloat[$x=2\pi$]{\includegraphics[width=.33\textwidth]{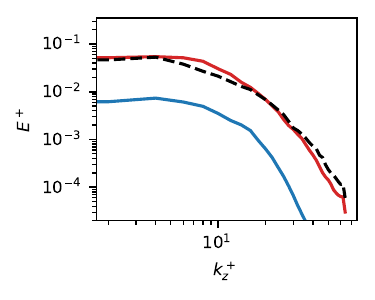}}\\
    \subfloat[$x=3\pi$]{\includegraphics[width=.33\textwidth]{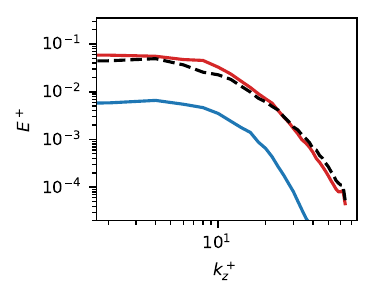}}
    \subfloat[$x=4\pi$]{\includegraphics[width=.33\textwidth]{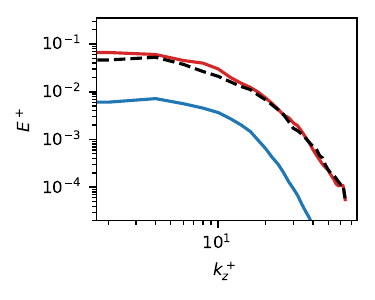}}
    \subfloat[$x=5\pi$]{\includegraphics[width=.33\textwidth]{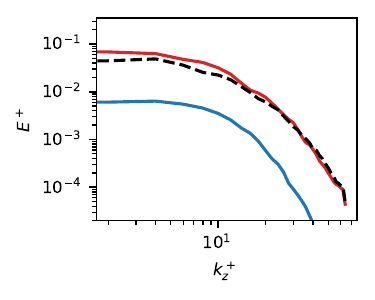}}
    \caption{\emph{A posteriori} test results of WMLES of $Re_\tau=2500$ at different stream-wise locations. Red curve represents CoNFiLD-inlet, blue curve denotes the DFM, while precursor simulation results are marked by black dashed line. First two lines: Mean velocity profile and auto correlation $R_{11}$ near channel center. Others are turbulence kinetic energy (TKE) along spanwise wave length.}
    \label{fig:wmles_extra_stat_2500_1}
\end{figure}
\begin{figure}[!ht]
    \centering
    \captionsetup[subfloat]{farskip=-9pt}
    \subfloat{\includegraphics[width=\textwidth]{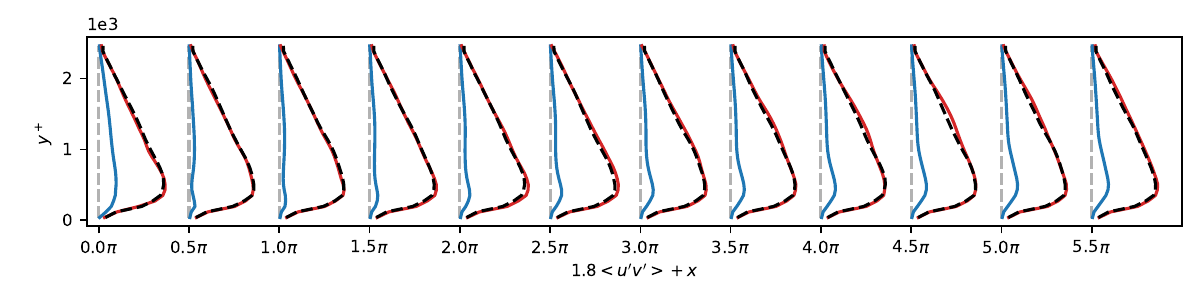}}\\
    \subfloat{\includegraphics[width=\textwidth]{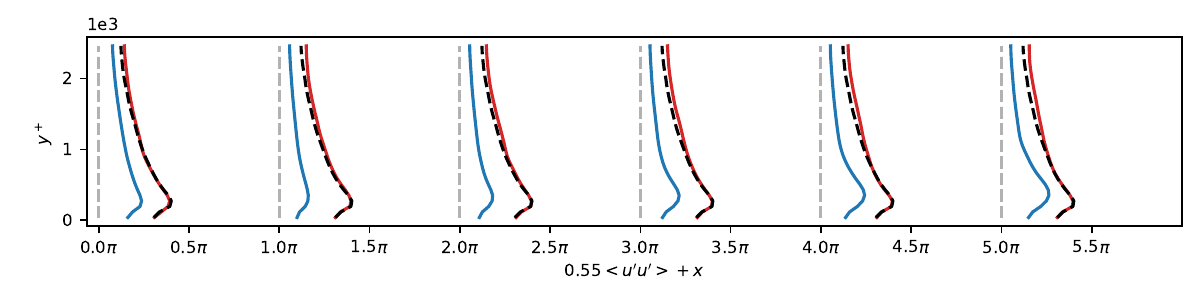}}\\
    \subfloat{\includegraphics[width=\textwidth]{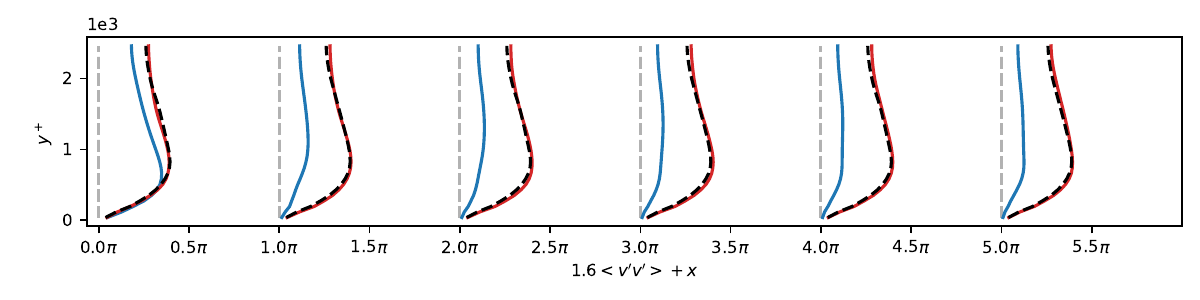}}\\
    \subfloat{\includegraphics[width=\textwidth]{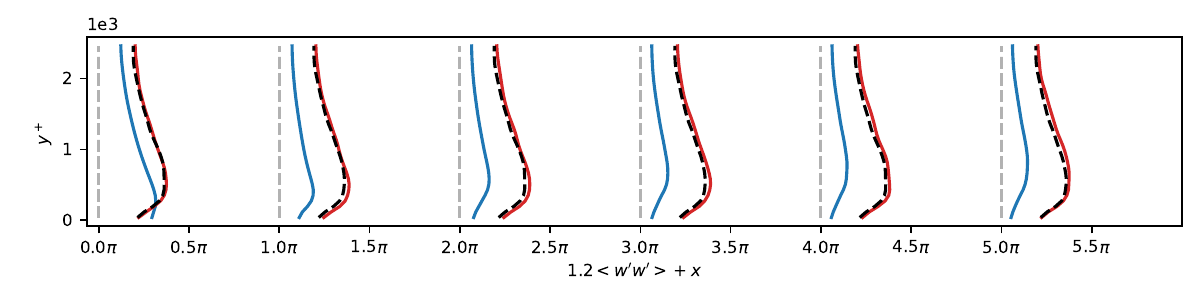}}\\
    \caption{\emph{A posteriori} test results at $12$ different stream-wise locations of WMLES of $Re_\tau=2500$. Red curve represents CoNFiLD-inlet, blue curve denotes the DFM, while precursor simulation results are marked by black dashed line. From top to bottom: Reynolds shear stress, Turbulence intensity of $u$, $v$, and $w$.}
    \label{fig:wmles_extra_stat_2500_2}
\end{figure}\begin{figure}[!ht]
    \centering
    \captionsetup[subfloat]{farskip=-4pt,captionskip=-8pt}
    \subfloat{\includegraphics[width=\textwidth]{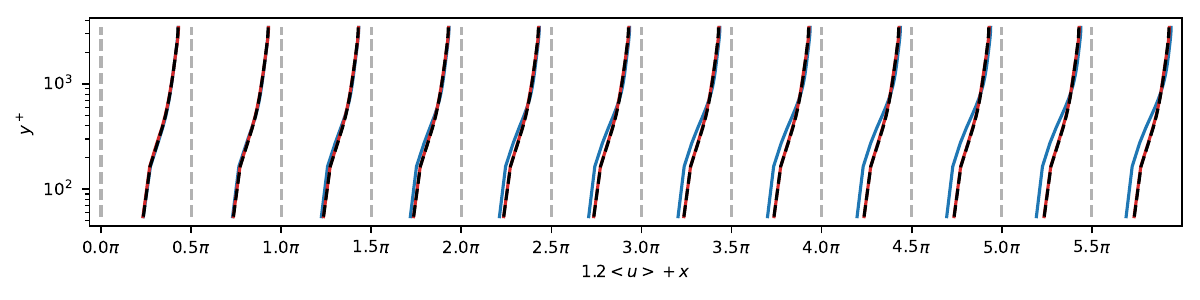}}\\
    \subfloat{\includegraphics[width=\textwidth]{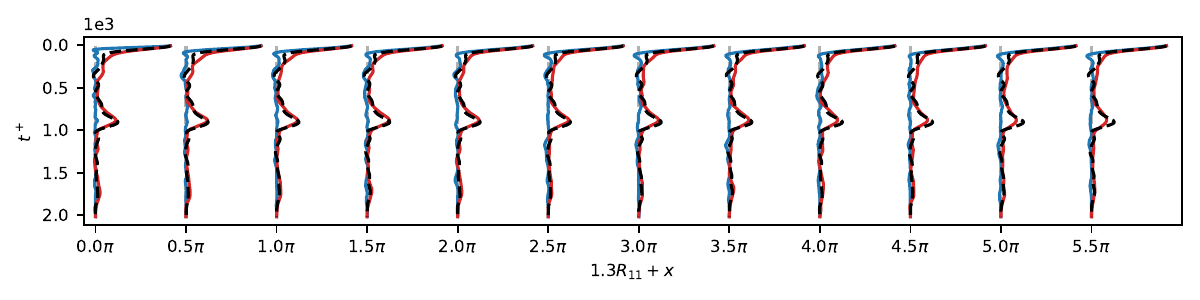}}\\
    \subfloat[$x=0$]{\includegraphics[width=.33\textwidth]{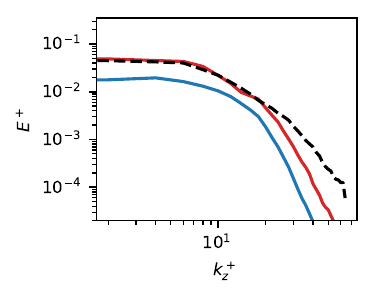}}
    \subfloat[$x=\pi$]{\includegraphics[width=.33\textwidth]{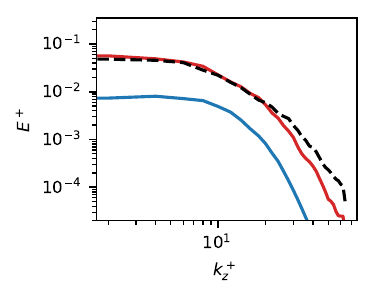}}
    \subfloat[$x=2\pi$]{\includegraphics[width=.33\textwidth]{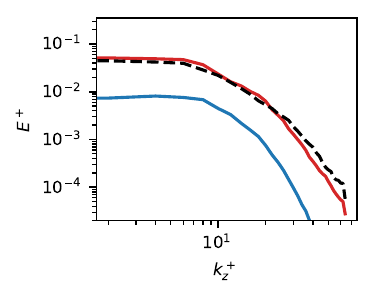}}\\
    \subfloat[$x=3\pi$]{\includegraphics[width=.33\textwidth]{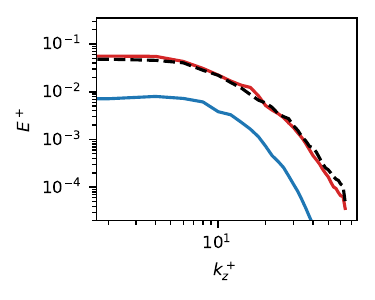}}
    \subfloat[$x=4\pi$]{\includegraphics[width=.33\textwidth]{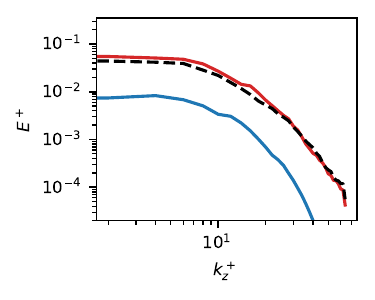}}
    \subfloat[$x=5\pi$]{\includegraphics[width=.33\textwidth]{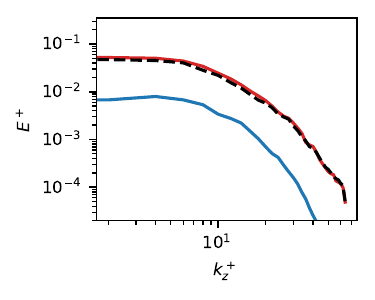}}
    \caption{\emph{A posteriori} test results of WMLES of $Re_\tau=3500$ at different stream-wise locations. Red curve represents CoNFiLD-inlet, blue curve denotes the DFM, while precursor simulation results are marked by black dashed line. First two lines: Mean velocity profile and auto correlation $R_{11}$ near channel center. Others are turbulence kinetic energy (TKE) along spanwise wave length.}
    \label{fig:wmles_extra_stat_3500_1}
\end{figure}
\begin{figure}[!ht]
    \centering
    \captionsetup[subfloat]{farskip=-9pt}
    \subfloat{\includegraphics[width=\textwidth]{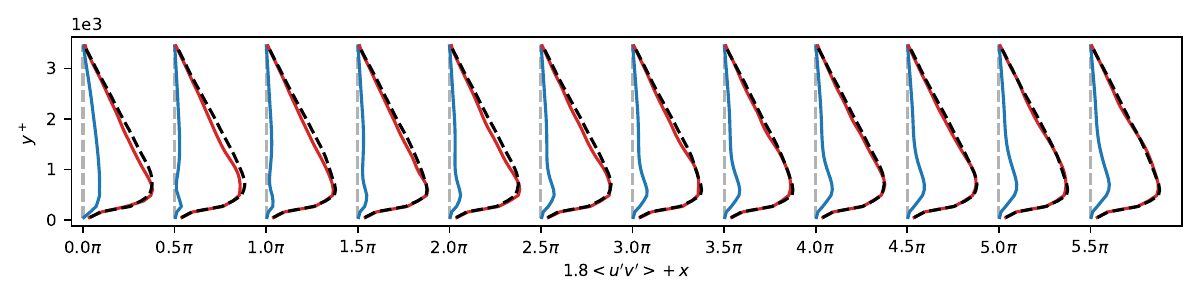}}\\
    \subfloat{\includegraphics[width=\textwidth]{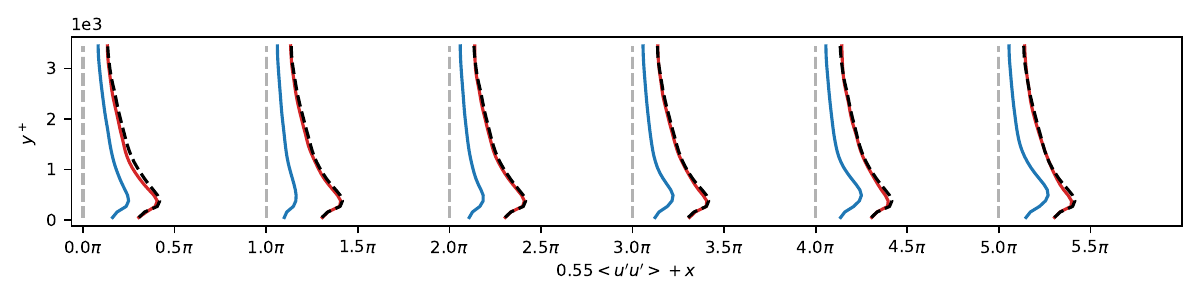}}\\
    \subfloat{\includegraphics[width=\textwidth]{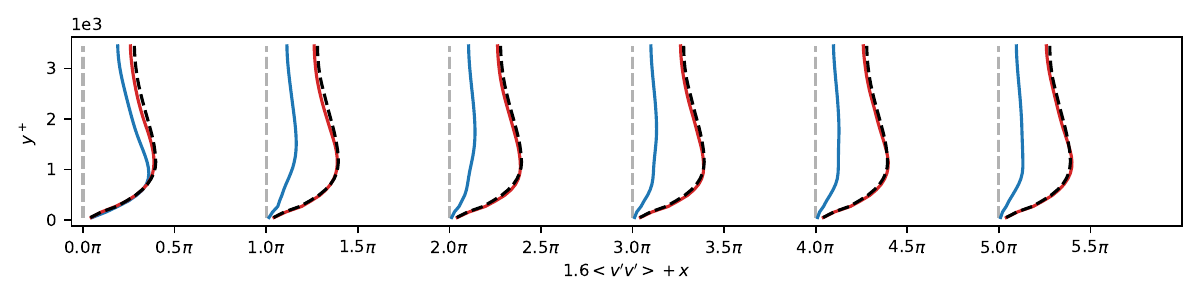}}\\
    \subfloat{\includegraphics[width=\textwidth]{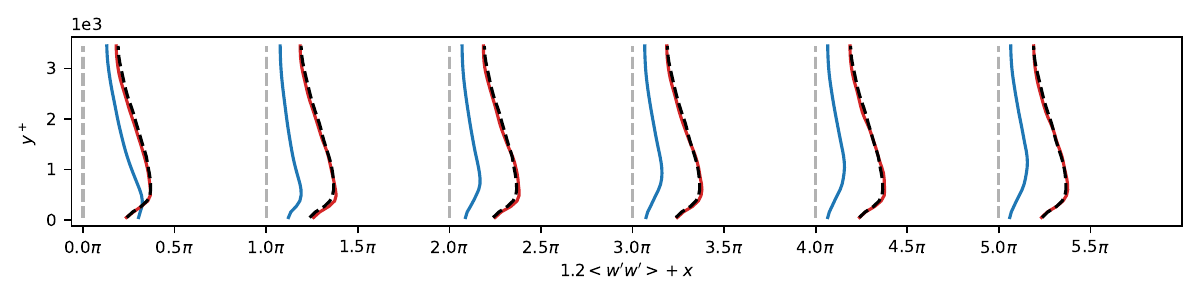}}\\
    \caption{\emph{A posteriori} test results at $12$ different stream-wise locations of WMLES of $Re_\tau=3500$. Red curve represents CoNFiLD-inlet, blue curve denotes the DFM, while precursor simulation results are marked by black dashed line. From top to bottom: Reynolds shear stress, Turbulence intensity of $u$, $v$, and $w$.}
    \label{fig:wmles_extra_stat_3500_2}
\end{figure}\begin{figure}[!ht]
    \centering
    \captionsetup[subfloat]{farskip=-4pt,captionskip=-8pt}
    \subfloat{\includegraphics[width=\textwidth]{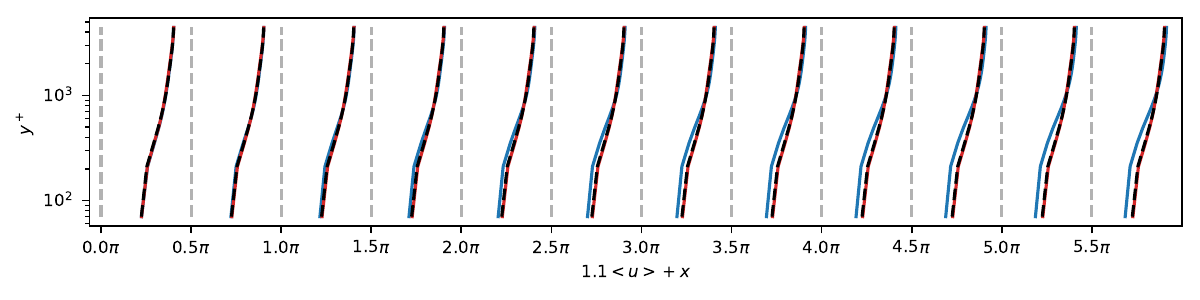}}\\
    \subfloat{\includegraphics[width=\textwidth]{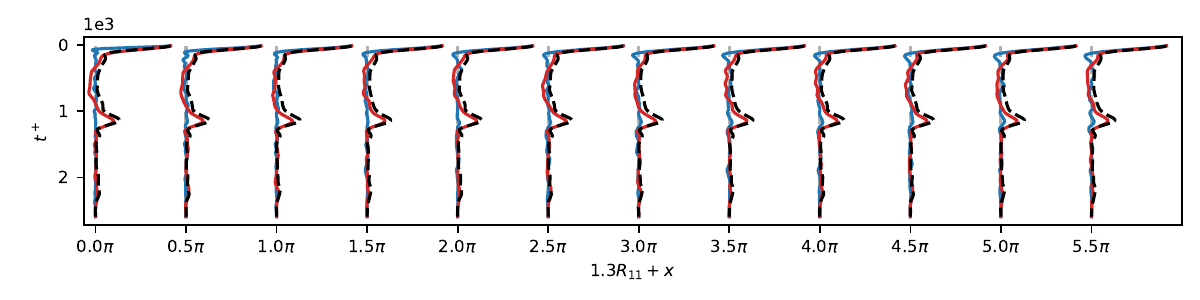}}\\
    \subfloat[$x=0$]{\includegraphics[width=.33\textwidth]{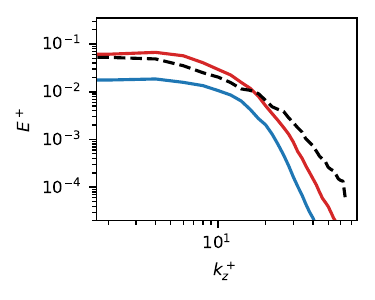}}
    \subfloat[$x=\pi$]{\includegraphics[width=.33\textwidth]{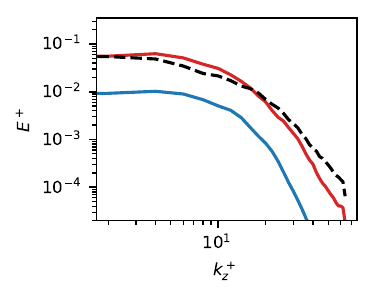}}
    \subfloat[$x=2\pi$]{\includegraphics[width=.33\textwidth]{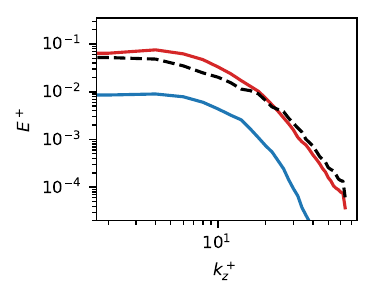}}\\
    \subfloat[$x=3\pi$]{\includegraphics[width=.33\textwidth]{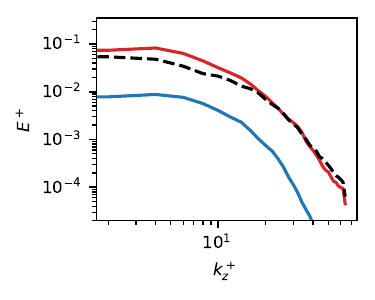}}
    \subfloat[$x=4\pi$]{\includegraphics[width=.33\textwidth]{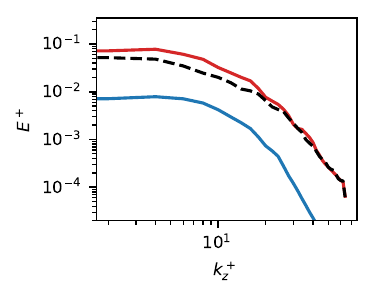}}
    \subfloat[$x=5\pi$]{\includegraphics[width=.33\textwidth]{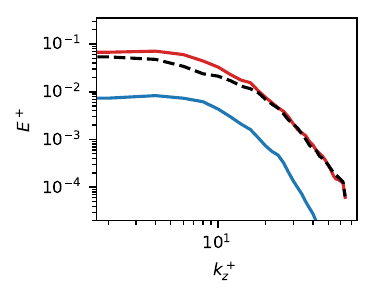}}
    \caption{\emph{A posteriori} test results of WMLES of $Re_\tau=4500$ at different stream-wise locations. Red curve represents CoNFiLD-inlet, blue curve denotes the DFM, while precursor simulation results are marked by black dashed line. First two lines: Mean velocity profile and auto correlation $R_{11}$ near channel center. Others are turbulence kinetic energy (TKE) along spanwise wave length.}
    \label{fig:wmles_extra_stat_4500_1}
\end{figure}
\begin{figure}[!ht]
    \centering
    \captionsetup[subfloat]{farskip=-9pt}
    \subfloat{\includegraphics[width=\textwidth]{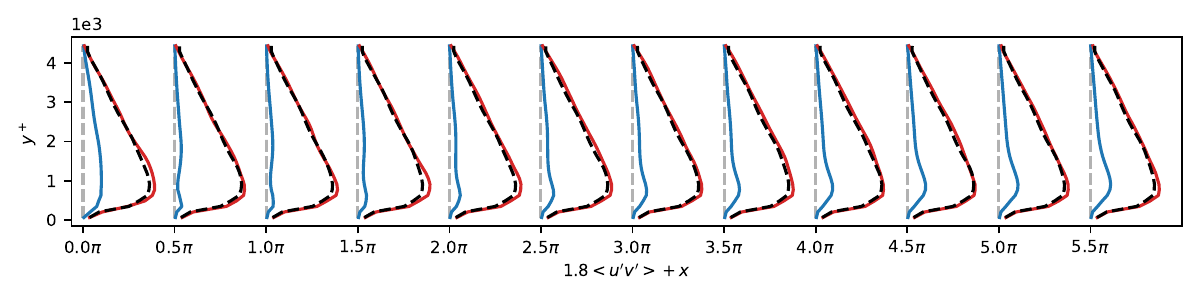}}\\
    \subfloat{\includegraphics[width=\textwidth]{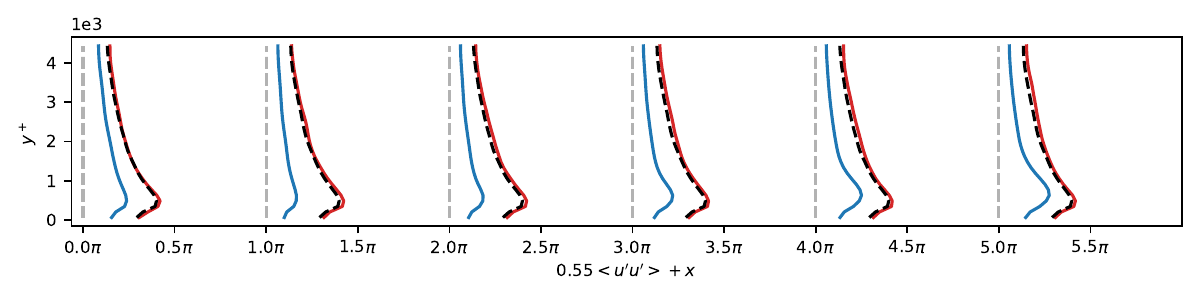}}\\
    \subfloat{\includegraphics[width=\textwidth]{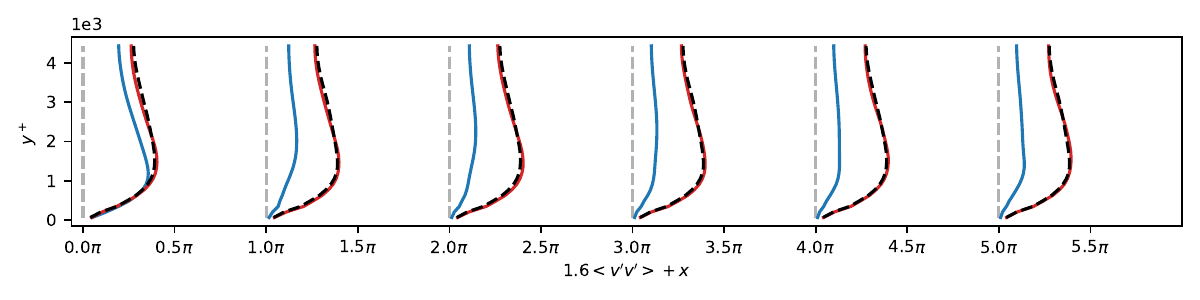}}\\
    \subfloat{\includegraphics[width=\textwidth]{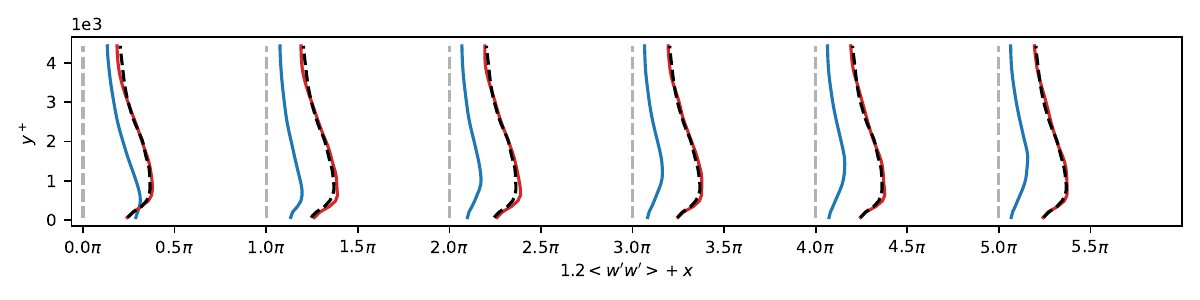}}\\
    \caption{\emph{A posteriori} test results at $12$ different stream-wise locations of WMLES of $Re_\tau=4500$. Red curve represents CoNFiLD-inlet, blue curve denotes the DFM, while precursor simulation results are marked by black dashed line. From top to bottom: Reynolds shear stress, Turbulence intensity of $u$, $v$, and $w$.}
    \label{fig:wmles_extra_stat_4500_2}
\end{figure}\begin{figure}[!ht]
    \centering
    \captionsetup[subfloat]{farskip=-4pt,captionskip=-8pt}
    \subfloat{\includegraphics[width=\textwidth]{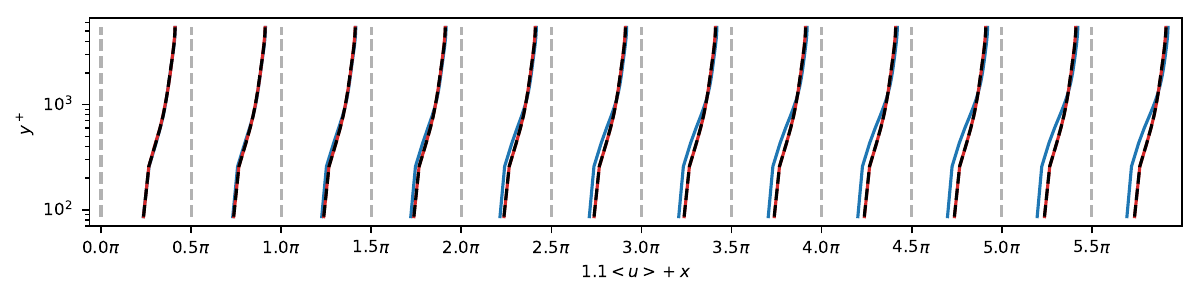}}\\
    \subfloat{\includegraphics[width=\textwidth]{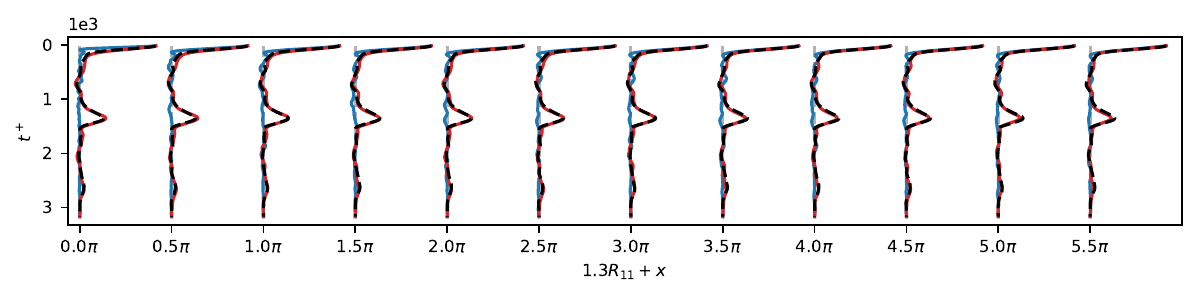}}\\
    \subfloat[$x=0$]{\includegraphics[width=.33\textwidth]{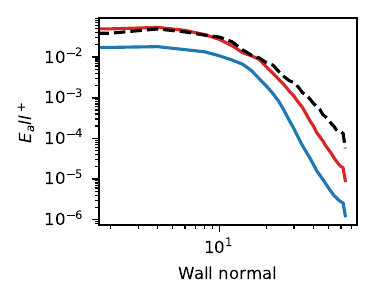}}
    \subfloat[$x=\pi$]{\includegraphics[width=.33\textwidth]{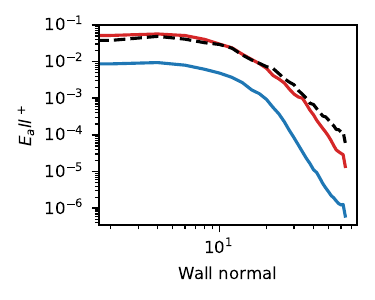}}
    \subfloat[$x=2\pi$]{\includegraphics[width=.33\textwidth]{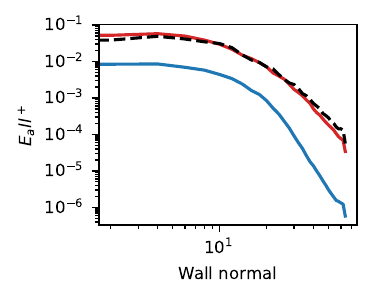}}\\
    \subfloat[$x=3\pi$]{\includegraphics[width=.33\textwidth]{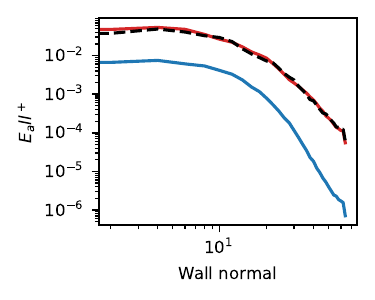}}
    \subfloat[$x=4\pi$]{\includegraphics[width=.33\textwidth]{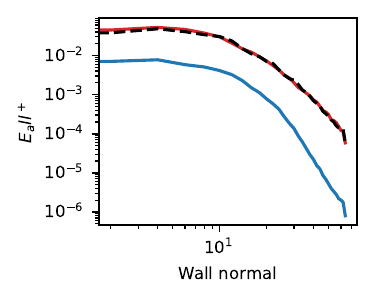}}
    \subfloat[$x=5\pi$]{\includegraphics[width=.33\textwidth]{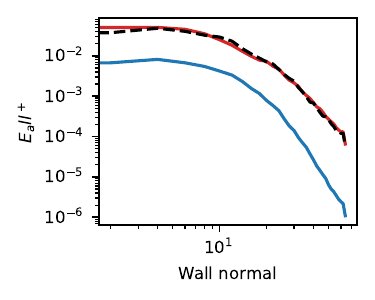}}
    \caption{\emph{A posteriori} test results of WMLES of $Re_\tau=5500$ at different stream-wise locations. Red curve represents CoNFiLD-inlet, blue curve denotes the DFM, while precursor simulation results are marked by black dashed line. First two lines: Mean velocity profile and auto correlation $R_{11}$ near channel center. Others are turbulence kinetic energy (TKE) along spanwise wave length.}
    \label{fig:wmles_extra_stat_5500_1}
\end{figure}
\begin{figure}[!ht]
    \centering
    \captionsetup[subfloat]{farskip=-9pt}
    \subfloat{\includegraphics[width=\textwidth]{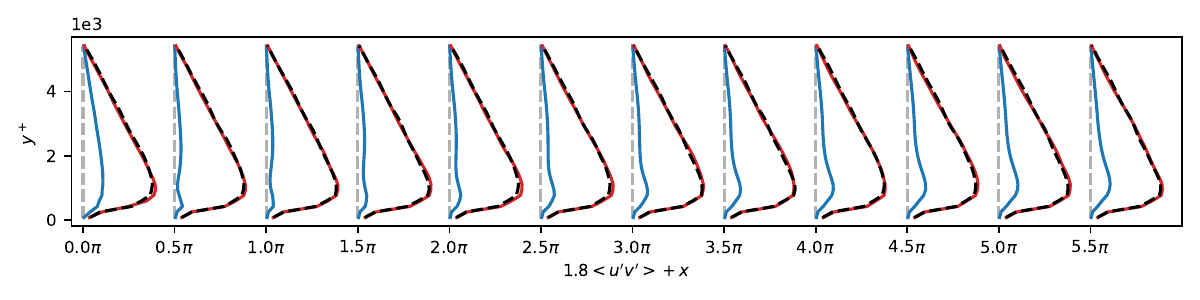}}\\
    \subfloat{\includegraphics[width=\textwidth]{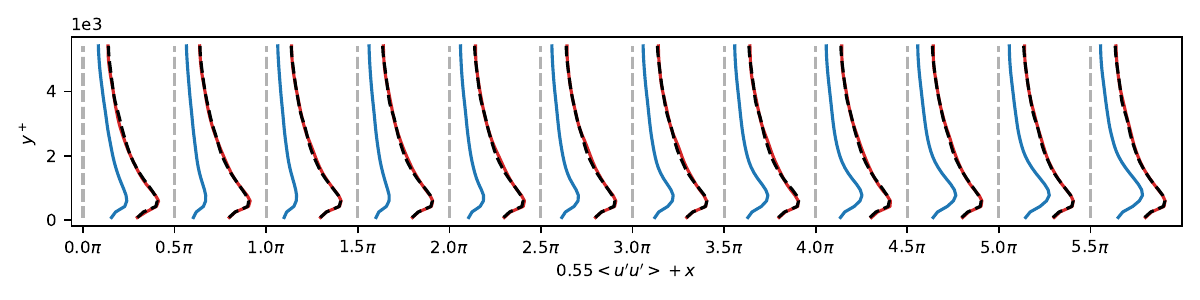}}\\
    \subfloat{\includegraphics[width=\textwidth]{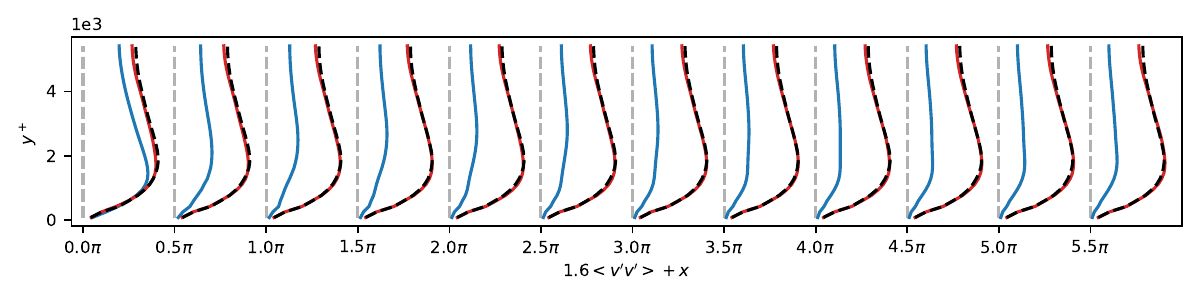}}\\
    \subfloat{\includegraphics[width=\textwidth]{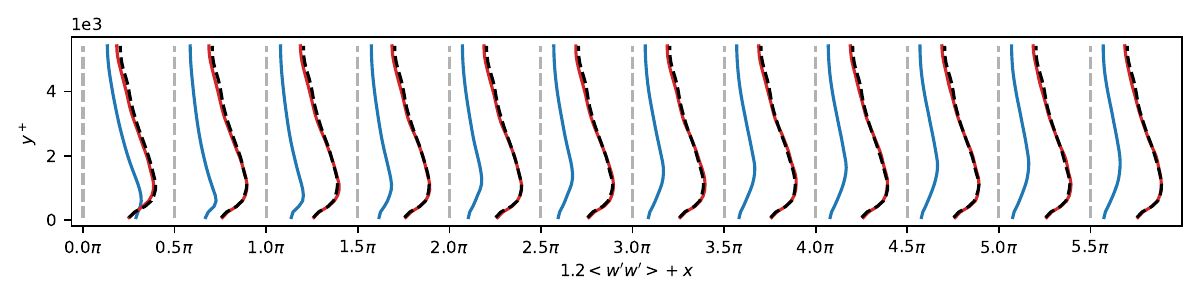}}\\
    \caption{\emph{A posteriori} test results at $12$ different stream-wise locations of WMLES of $Re_\tau=5500$. Red curve represents CoNFiLD-inlet, blue curve denotes the DFM, while precursor simulation results are marked by black dashed line. From top to bottom: Reynolds shear stress, Turbulence intensity of $u$, $v$, and $w$.}
    \label{fig:wmles_extra_stat_5500_2}
\end{figure}\begin{figure}[!ht]
    \centering
    \captionsetup[subfloat]{farskip=-4pt,captionskip=-8pt}
    \subfloat{\includegraphics[width=\textwidth]{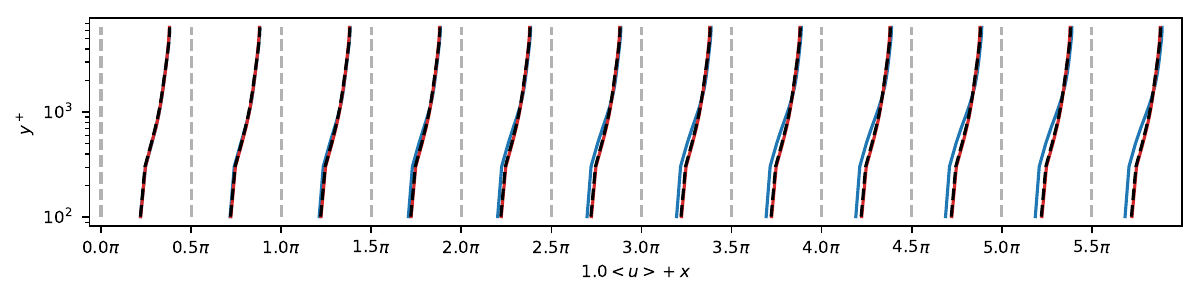}}\\
    \subfloat{\includegraphics[width=\textwidth]{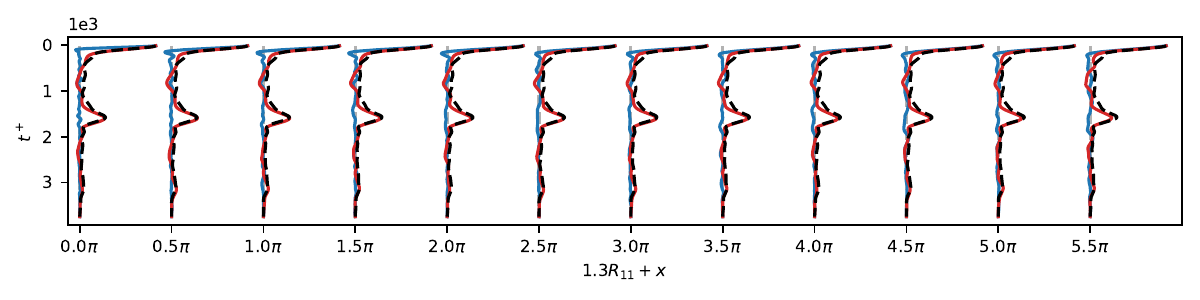}}\\
    \subfloat[$x=0$]{\includegraphics[width=.33\textwidth]{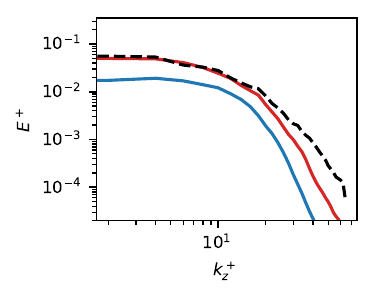}}
    \subfloat[$x=\pi$]{\includegraphics[width=.33\textwidth]{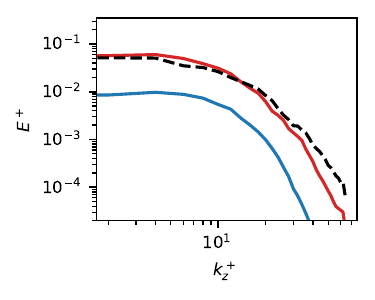}}
    \subfloat[$x=2\pi$]{\includegraphics[width=.33\textwidth]{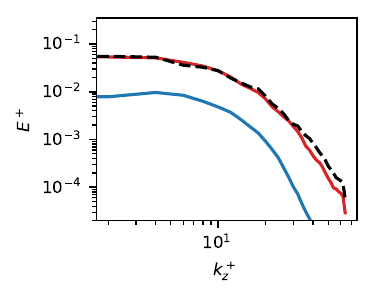}}\\
    \subfloat[$x=3\pi$]{\includegraphics[width=.33\textwidth]{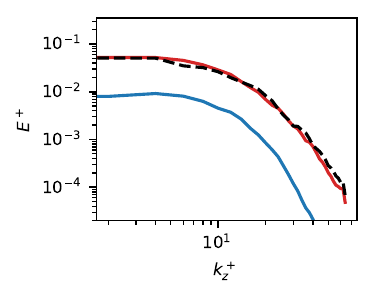}}
    \subfloat[$x=4\pi$]{\includegraphics[width=.33\textwidth]{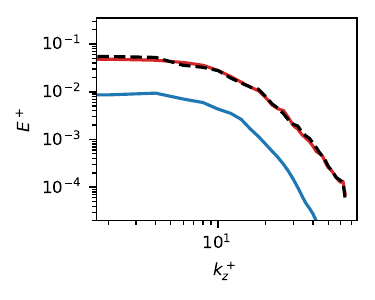}}
    \subfloat[$x=5\pi$]{\includegraphics[width=.33\textwidth]{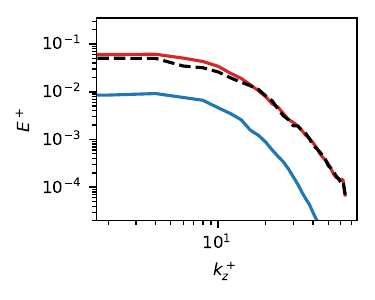}}
    \caption{\emph{A posteriori} test results of WMLES of $Re_\tau=6500$ at different stream-wise locations. Red curve represents CoNFiLD-inlet, blue curve denotes the DFM, while precursor simulation results are marked by black dashed line. First two lines: Mean velocity profile and auto correlation $R_{11}$ near channel center. Others are turbulence kinetic energy (TKE) along spanwise wave length.}
    \label{fig:wmles_extra_stat_6500_1}
\end{figure}
\begin{figure}[!ht]
    \centering
    \captionsetup[subfloat]{farskip=-9pt}
    \subfloat{\includegraphics[width=\textwidth]{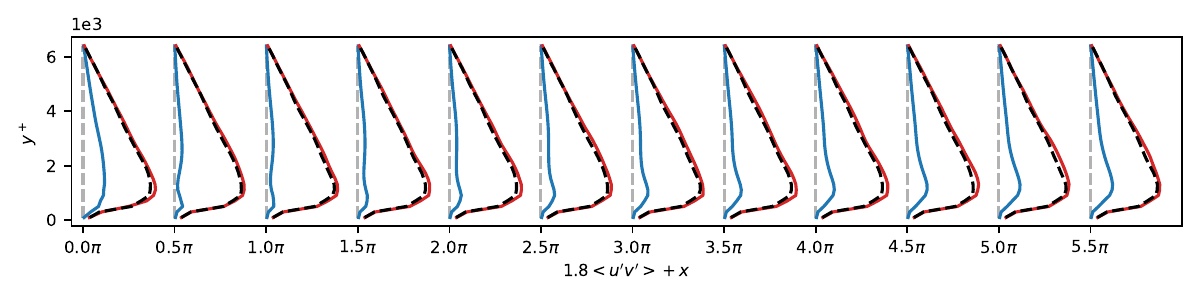}}\\
    \subfloat{\includegraphics[width=\textwidth]{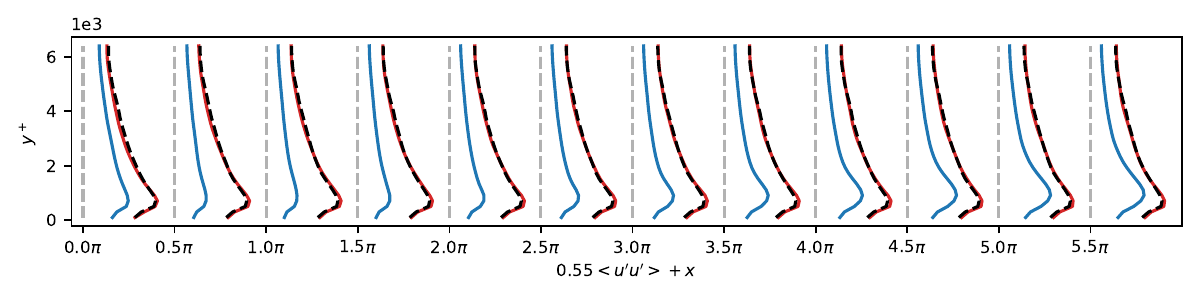}}\\
    \subfloat{\includegraphics[width=\textwidth]{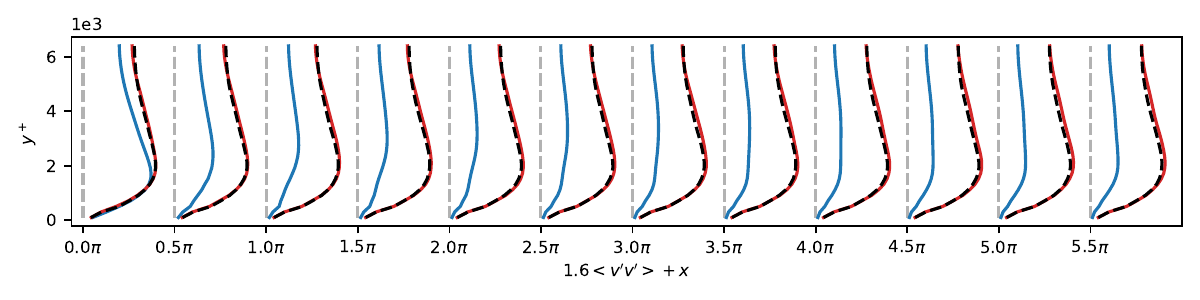}}\\
    \subfloat{\includegraphics[width=\textwidth]{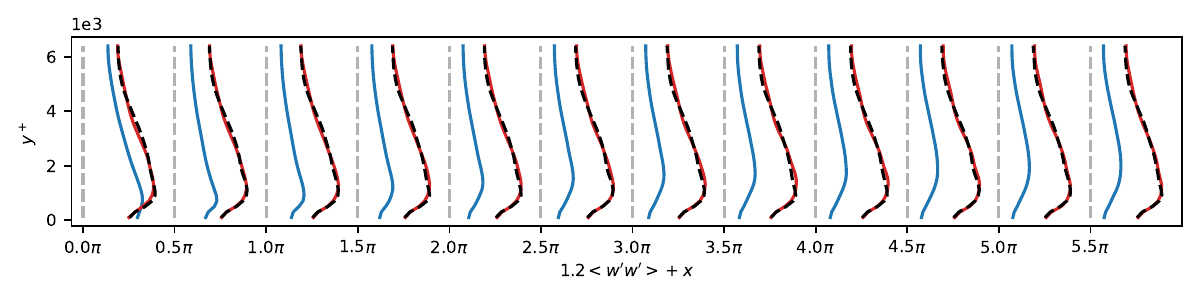}}\\
    \caption{\emph{A posteriori} test results at $12$ different stream-wise locations of WMLES of $Re_\tau=6500$. Red curve represents CoNFiLD-inlet, blue curve denotes the DFM, while precursor simulation results are marked by black dashed line. From top to bottom: Reynolds shear stress, Turbulence intensity of $u$, $v$, and $w$.}
    \label{fig:wmles_extra_stat_6500_2}
\end{figure}\begin{figure}[!ht]
    \centering
    \captionsetup[subfloat]{farskip=-4pt,captionskip=-8pt}
    \subfloat{\includegraphics[width=\textwidth]{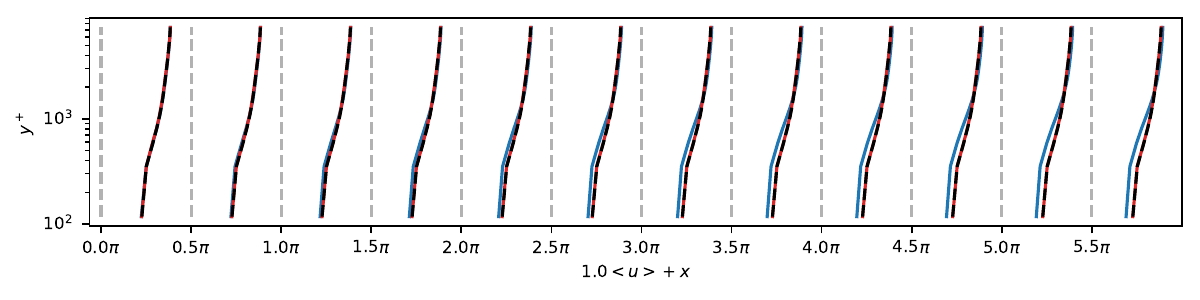}}\\
    \subfloat{\includegraphics[width=\textwidth]{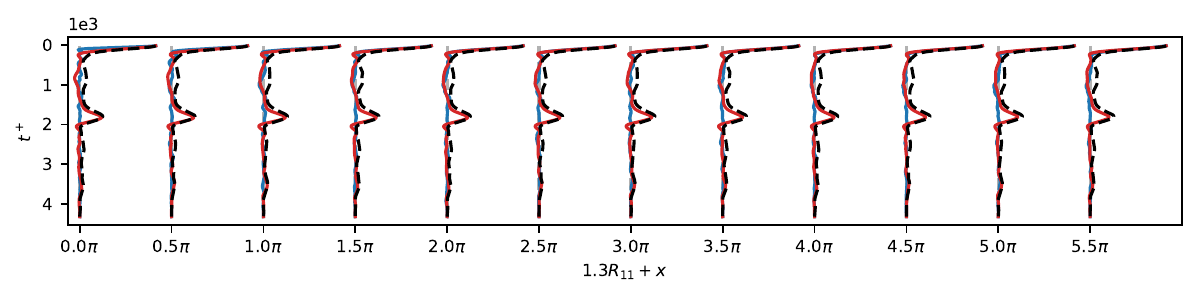}}\\
    \subfloat[$x=0$]{\includegraphics[width=.33\textwidth]{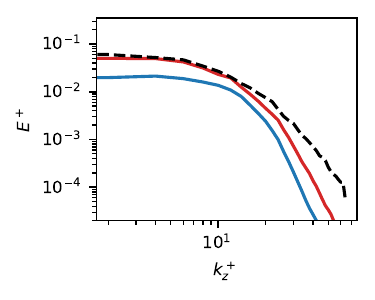}}
    \subfloat[$x=\pi$]{\includegraphics[width=.33\textwidth]{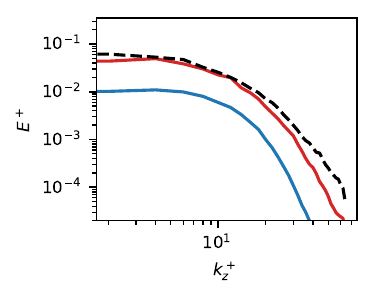}}
    \subfloat[$x=2\pi$]{\includegraphics[width=.33\textwidth]{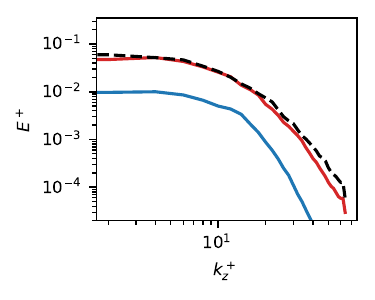}}\\
    \subfloat[$x=3\pi$]{\includegraphics[width=.33\textwidth]{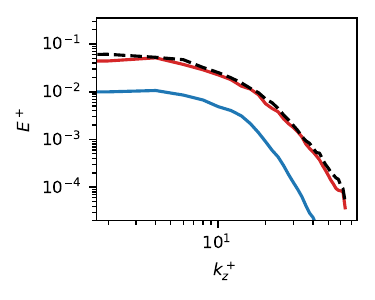}}
    \subfloat[$x=4\pi$]{\includegraphics[width=.33\textwidth]{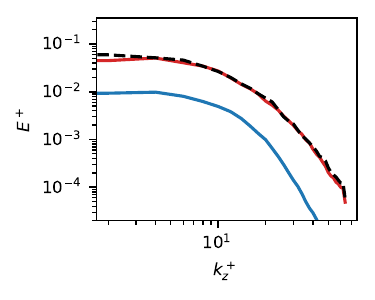}}
    \subfloat[$x=5\pi$]{\includegraphics[width=.33\textwidth]{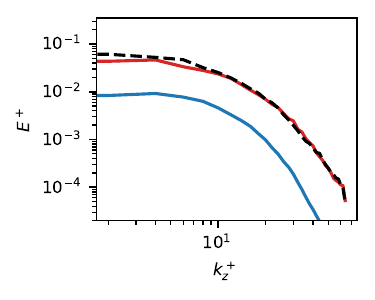}}
    \caption{\emph{A posteriori} test results of WMLES of $Re_\tau=7500$ at different stream-wise locations. Red curve represents CoNFiLD-inlet, blue curve denotes the DFM, while precursor simulation results are marked by black dashed line. First two lines: Mean velocity profile and auto correlation $R_{11}$ near channel center. Others are turbulence kinetic energy (TKE) along spanwise wave length.}
    \label{fig:wmles_extra_stat_7500_1}
\end{figure}
\begin{figure}[!ht]
    \centering
    \captionsetup[subfloat]{farskip=-9pt}
    \subfloat{\includegraphics[width=\textwidth]{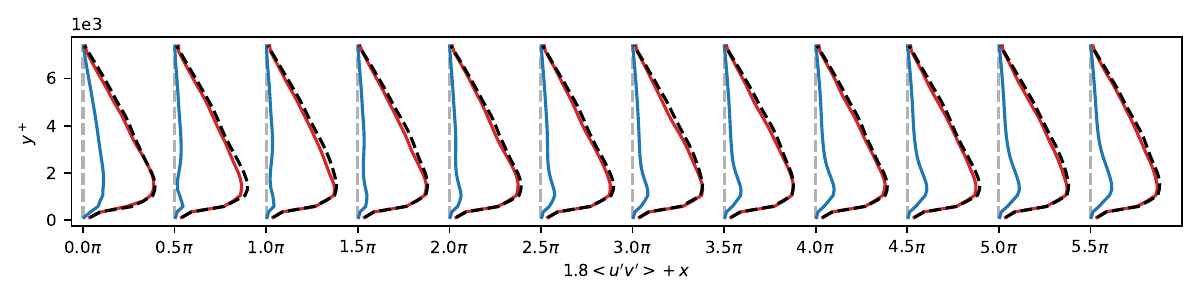}}\\
    \subfloat{\includegraphics[width=\textwidth]{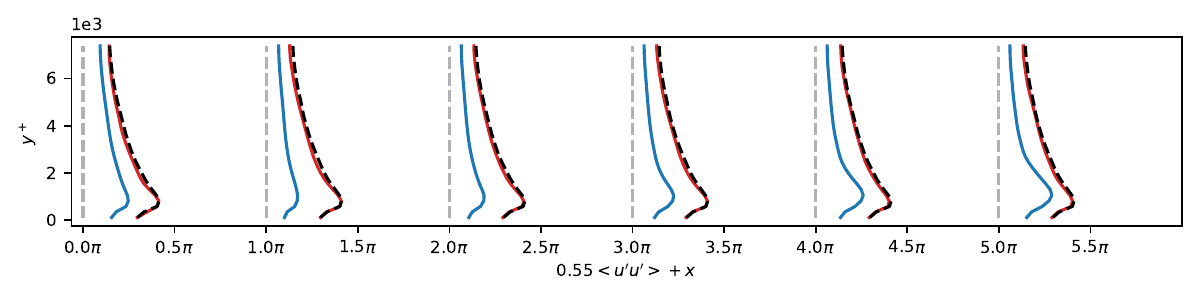}}\\
    \subfloat{\includegraphics[width=\textwidth]{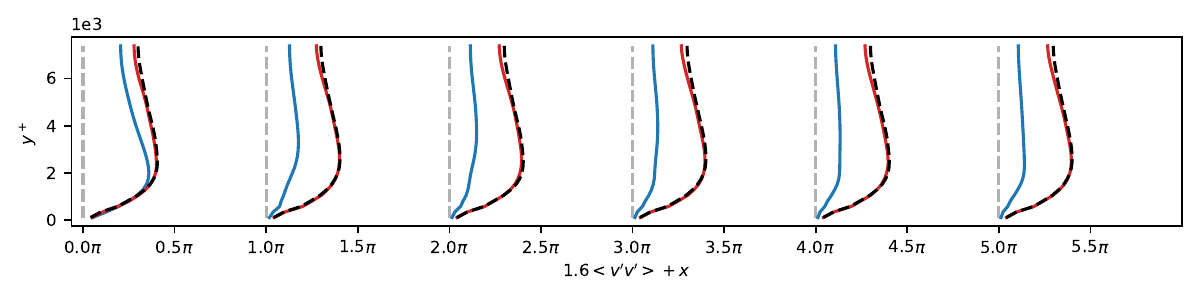}}\\
    \subfloat{\includegraphics[width=\textwidth]{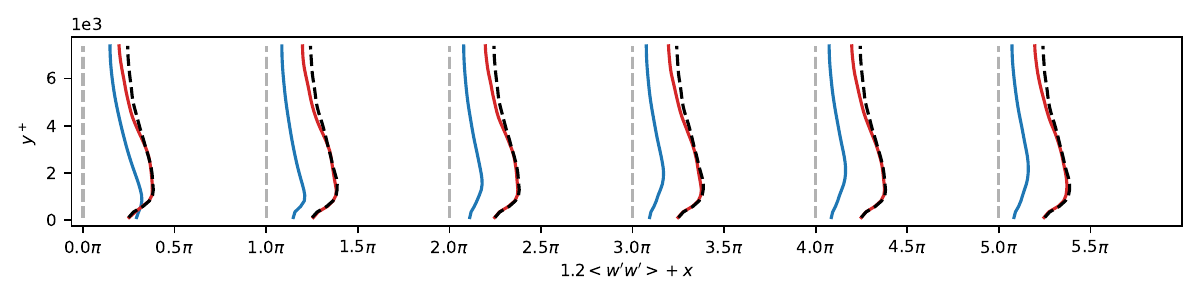}}\\
    \caption{\emph{A posteriori} test results at $12$ different stream-wise locations of WMLES of $Re_\tau=7500$. Red curve represents CoNFiLD-inlet, blue curve denotes the DFM, while precursor simulation results are marked by black dashed line. From top to bottom: Reynolds shear stress, Turbulence intensity of $u$, $v$, and $w$.}
    \label{fig:wmles_extra_stat_7500_2}
\end{figure}\begin{figure}[!ht]
    \centering
    \captionsetup[subfloat]{farskip=-4pt,captionskip=-8pt}
    \subfloat{\includegraphics[width=\textwidth]{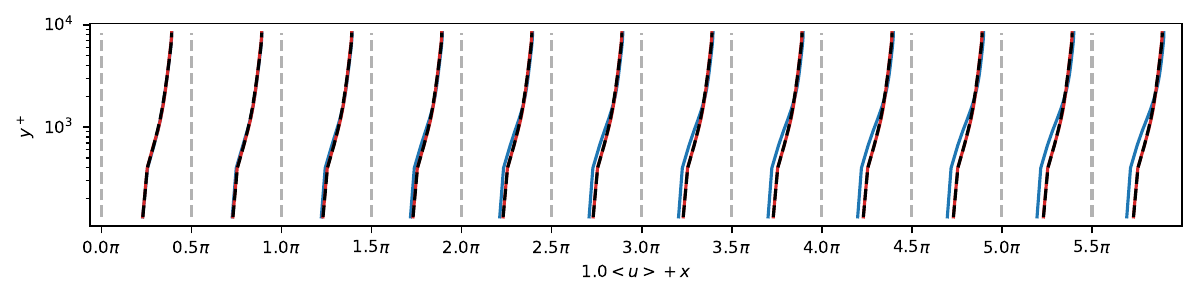}}\\
    \subfloat{\includegraphics[width=\textwidth]{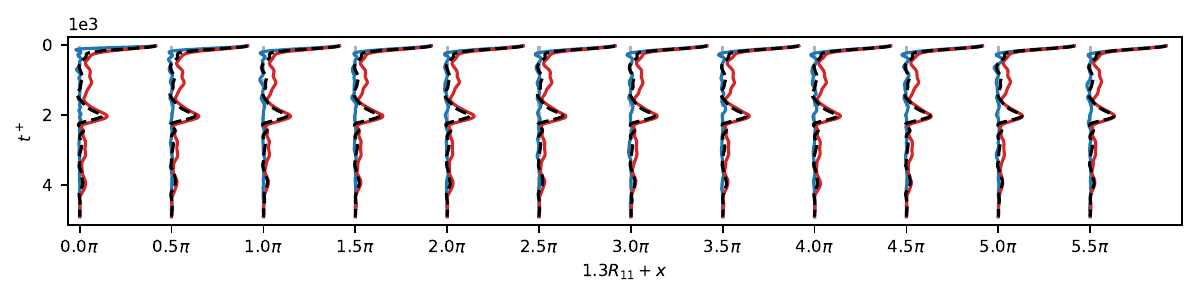}}\\
    \subfloat[$x=0$]{\includegraphics[width=.33\textwidth]{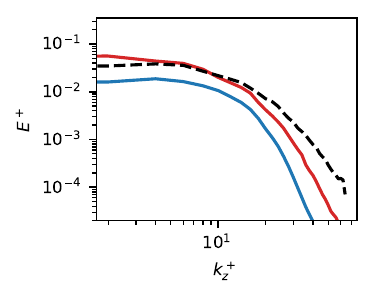}}
    \subfloat[$x=\pi$]{\includegraphics[width=.33\textwidth]{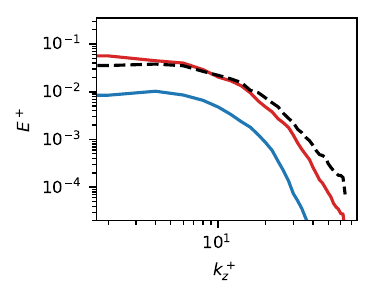}}
    \subfloat[$x=2\pi$]{\includegraphics[width=.33\textwidth]{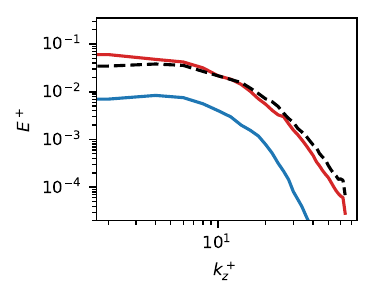}}\\
    \subfloat[$x=3\pi$]{\includegraphics[width=.33\textwidth]{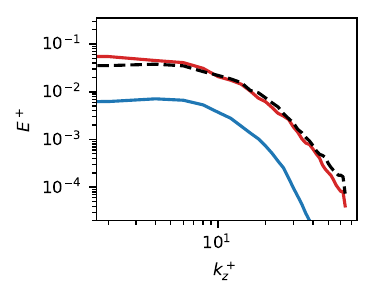}}
    \subfloat[$x=4\pi$]{\includegraphics[width=.33\textwidth]{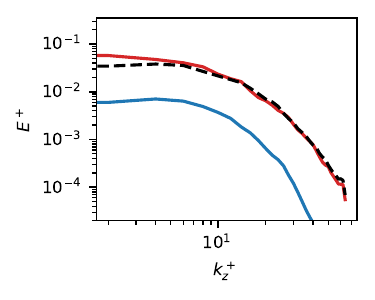}}
    \subfloat[$x=5\pi$]{\includegraphics[width=.33\textwidth]{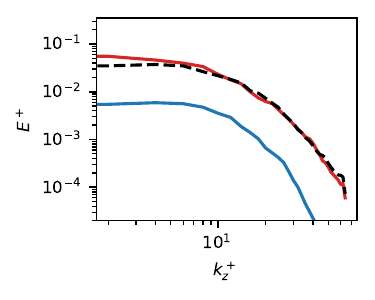}}
    \caption{\emph{A posteriori} test results of WMLES of $Re_\tau=8500$ at different stream-wise locations. Red curve represents CoNFiLD-inlet, blue curve denotes the DFM, while precursor simulation results are marked by black dashed line. First two lines: Mean velocity profile and auto correlation $R_{11}$ near channel center. Others are turbulence kinetic energy (TKE) along spanwise wave length.}
    \label{fig:wmles_extra_stat_8500_1}
\end{figure}
\begin{figure}[!ht]
    \centering
    \captionsetup[subfloat]{farskip=-9pt}
    \subfloat{\includegraphics[width=\textwidth]{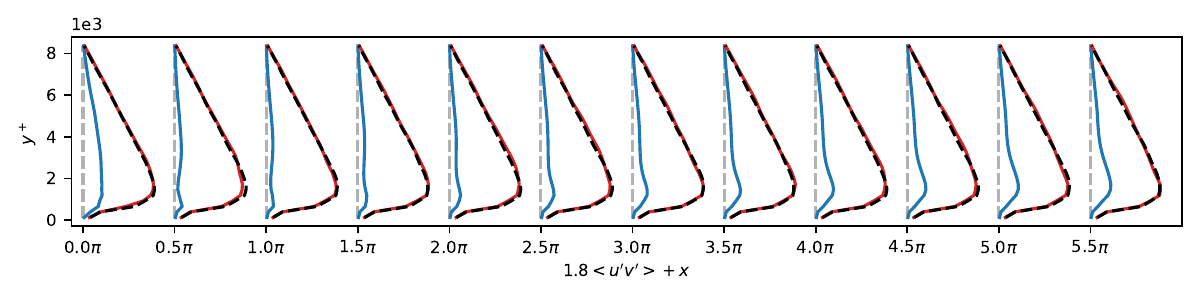}}\\
    \subfloat{\includegraphics[width=\textwidth]{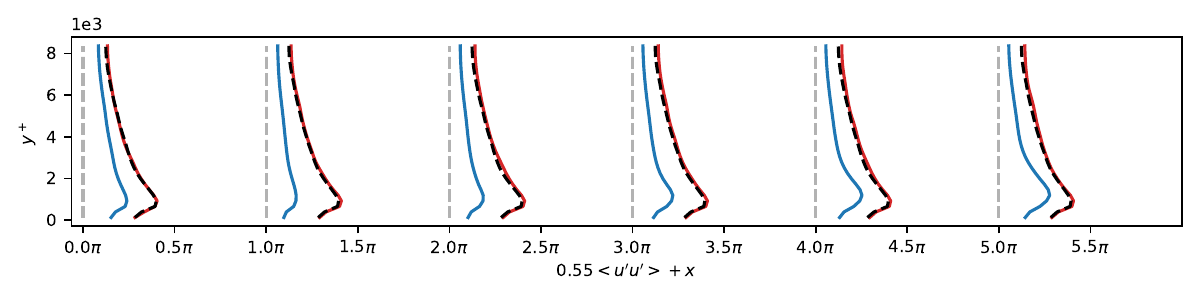}}\\
    \subfloat{\includegraphics[width=\textwidth]{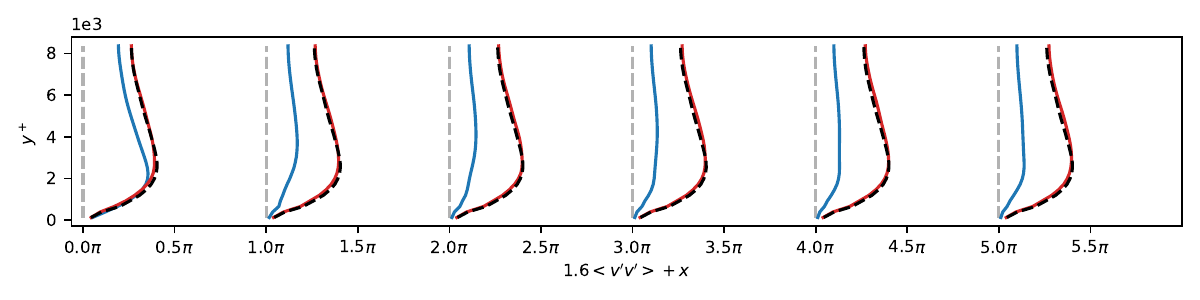}}\\
    \subfloat{\includegraphics[width=\textwidth]{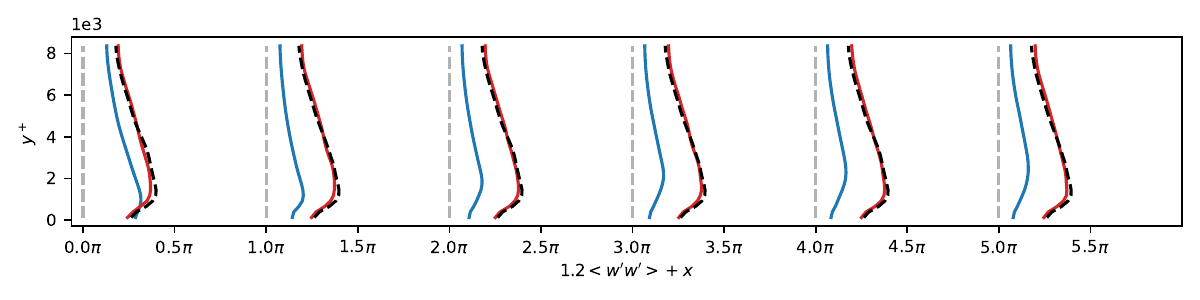}}\\
    \caption{\emph{A posteriori} test results at $12$ different stream-wise locations of WMLES of $Re_\tau=8500$. Red curve represents CoNFiLD-inlet, blue curve denotes the DFM, while precursor simulation results are marked by black dashed line. From top to bottom: Reynolds shear stress, Turbulence intensity of $u$, $v$, and $w$.}
    \label{fig:wmles_extra_stat_8500_2}
\end{figure}\begin{figure}[!ht]
    \centering
    \captionsetup[subfloat]{farskip=-4pt,captionskip=-8pt}
    \subfloat{\includegraphics[width=\textwidth]{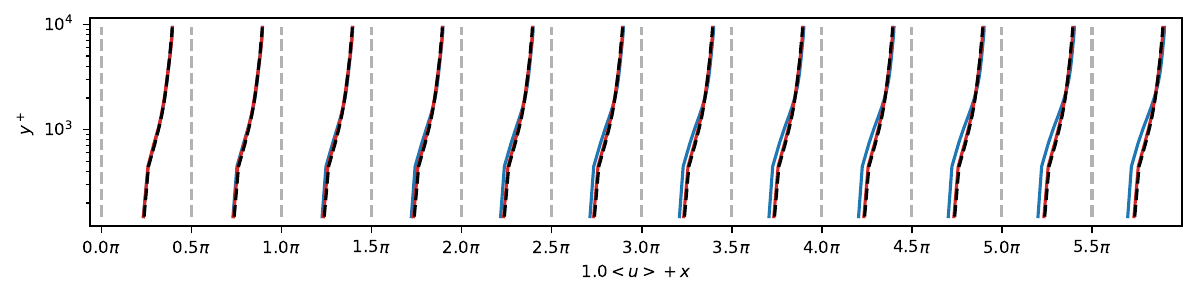}}\\
    \subfloat{\includegraphics[width=\textwidth]{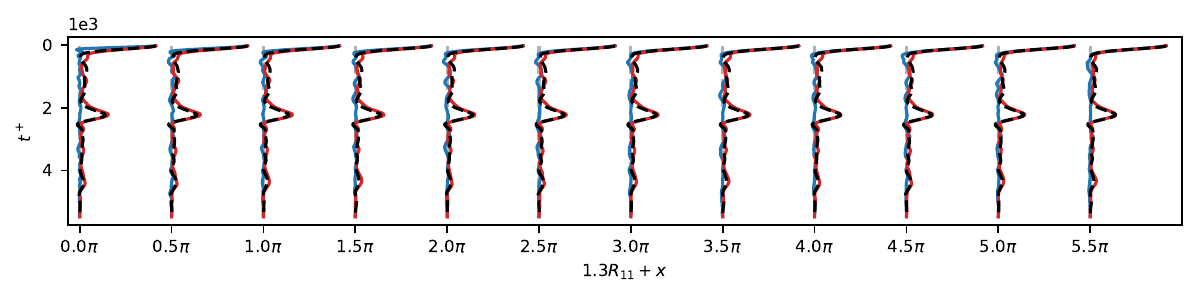}}\\
    \subfloat[$x=0$]{\includegraphics[width=.33\textwidth]{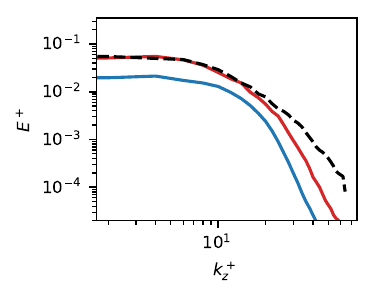}}
    \subfloat[$x=\pi$]{\includegraphics[width=.33\textwidth]{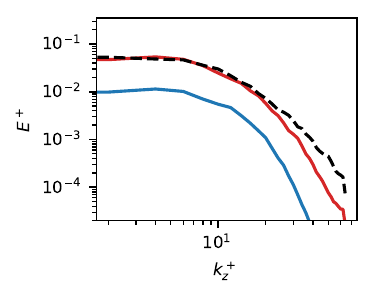}}
    \subfloat[$x=2\pi$]{\includegraphics[width=.33\textwidth]{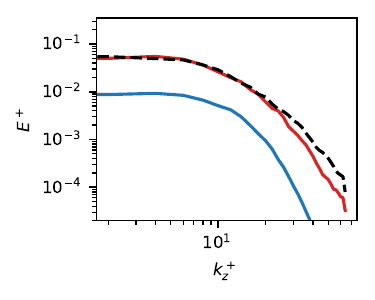}}\\
    \subfloat[$x=3\pi$]{\includegraphics[width=.33\textwidth]{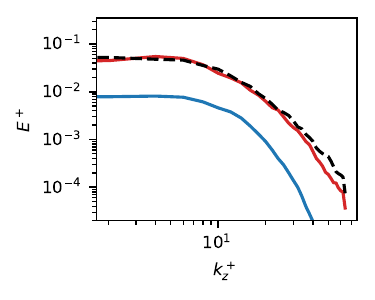}}
    \subfloat[$x=4\pi$]{\includegraphics[width=.33\textwidth]{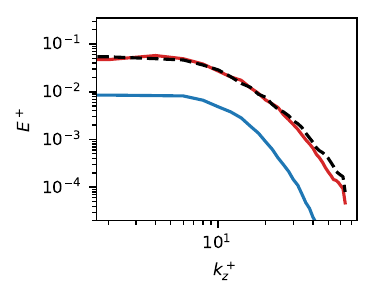}}
    \subfloat[$x=5\pi$]{\includegraphics[width=.33\textwidth]{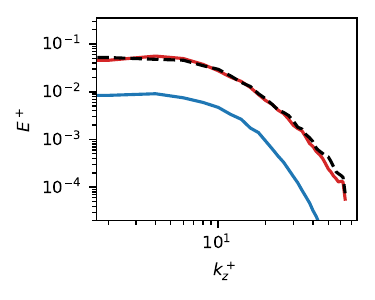}}
    \caption{\emph{A posteriori} test results of WMLES of $Re_\tau=9500$ at different stream-wise locations. Red curve represents CoNFiLD-inlet, blue curve denotes the DFM, while precursor simulation results are marked by black dashed line. First two lines: Mean velocity profile and auto correlation $R_{11}$ near channel center. Others are turbulence kinetic energy (TKE) along spanwise wave length.}
    \label{fig:wmles_extra_stat_9500_1}
\end{figure}
\begin{figure}[!ht]
    \centering
    \captionsetup[subfloat]{farskip=-9pt}
    \subfloat{\includegraphics[width=\textwidth]{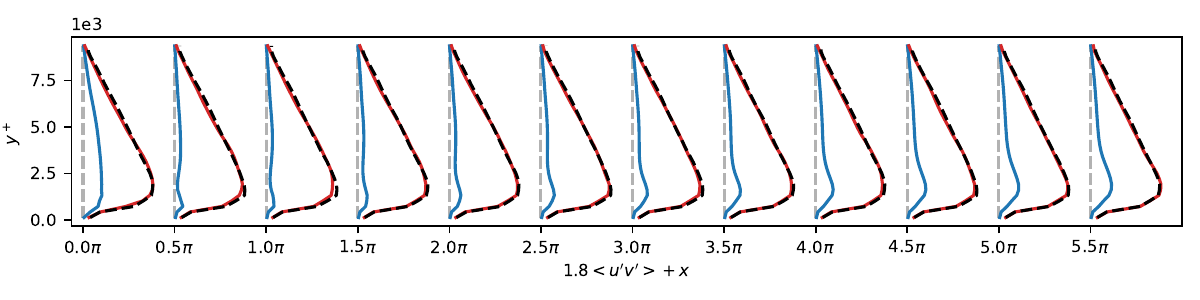}}\\
    \subfloat{\includegraphics[width=\textwidth]{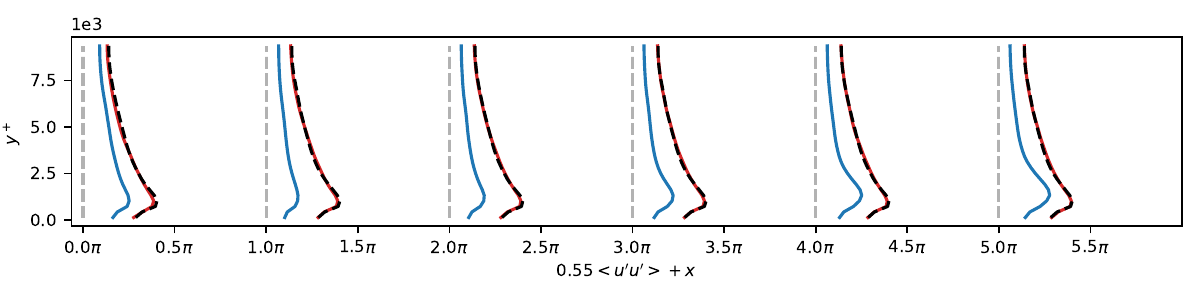}}\\
    \subfloat{\includegraphics[width=\textwidth]{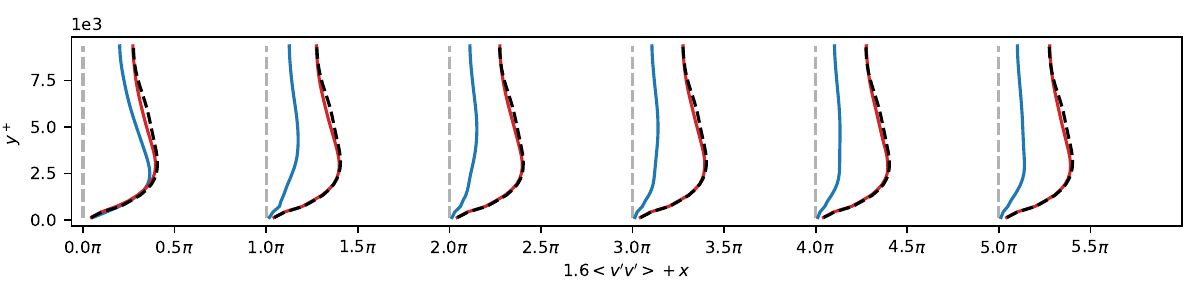}}\\
    \subfloat{\includegraphics[width=\textwidth]{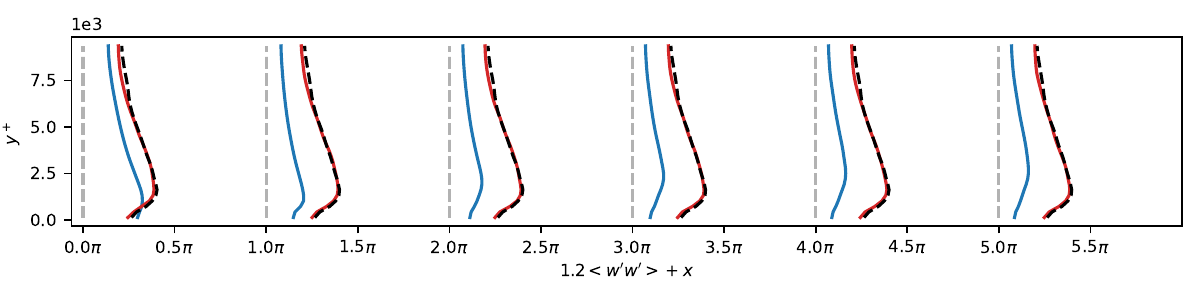}}\\
    \caption{\emph{A posteriori} test results at $12$ different stream-wise locations of WMLES of $Re_\tau=9500$. Red curve represents CoNFiLD-inlet, blue curve denotes the DFM, while precursor simulation results are marked by black dashed line. From top to bottom: Reynolds shear stress, Turbulence intensity of $u$, $v$, and $w$.}
    \label{fig:wmles_extra_stat_9500_2}
\end{figure}


}

\end{document}